\documentclass[12pt,a4paper,bibliography=totoc]{book}
\usepackage[utf8]{inputenc}
\usepackage[T1]{fontenc}
\usepackage{amsmath,amssymb}
\usepackage{amsfonts}
\usepackage{fancyhdr}
\usepackage{amssymb}
\usepackage{color} 
\definecolor{Prune}{RGB}{99,0,60}
\usepackage{mdframed}
\usepackage{multirow} 
\usepackage{multicol} 
\usepackage{scrextend} 
\usepackage{tikz}
\usepackage{graphicx}
\usepackage[absolute]{textpos} 
\usepackage{colortbl}
\usepackage{array}
\usepackage{geometry}

\usepackage{appendix}
\usepackage{blindtext,subfiles}
\usepackage{csquotes}
\usepackage{tocloft}

\usepackage{mdframed}

\usepackage{hyperref}

\def\bra{\langle}
\def\ket{\rangle}
\def\bbra{\langle\!\langle}
\def\kket{\rangle\!\rangle}

\newcommand{\trento}{T$\mathrel{\protect\raisebox{-2.1pt}{R}}$ENTo}
\newcommand{\ipglasma}{{\small IP-GLASMA}}
\newcommand{\music}{{\small MUSIC}}

\newcommand{\kompost}{K\o MP\o St}

\newcommand{\pbpb}{$^{208}$Pb+$^{208}$Pb}
\newcommand{\auau}{$^{197}$Au+$^{197}$Au}
\newcommand{\uuuu}{$^{238}$U+$^{238}$U}
\newcommand{\xexe}{$^{129}$Xe+$^{129}$Xe}
\newcommand{\ruru}{$^{96}$Ru+$^{96}$Ru}
\newcommand{\zrzr}{$^{96}$Zr+$^{96}$Zr}
\newcommand{\fm}{{\rm fm}}
\newcommand{\fmc}{{\rm fm}/c}

\newcommand{\equ}[1]{Eq.~(\ref{eq:#1})}
\newcommand{\fig}[1]{Fig.~\ref{fig:#1}}
\newcommand{\tab}[1]{Tab.~\ref{tab:#1}}

\usepackage{setspace}

\begin{document}

\begin{titlepage}

\newgeometry{left=7.5cm,bottom=2cm, top=1cm, right=1cm}

\tikz[remember picture,overlay] \node[opacity=1,inner sep=0pt] at (-28mm,-135mm){\includegraphics{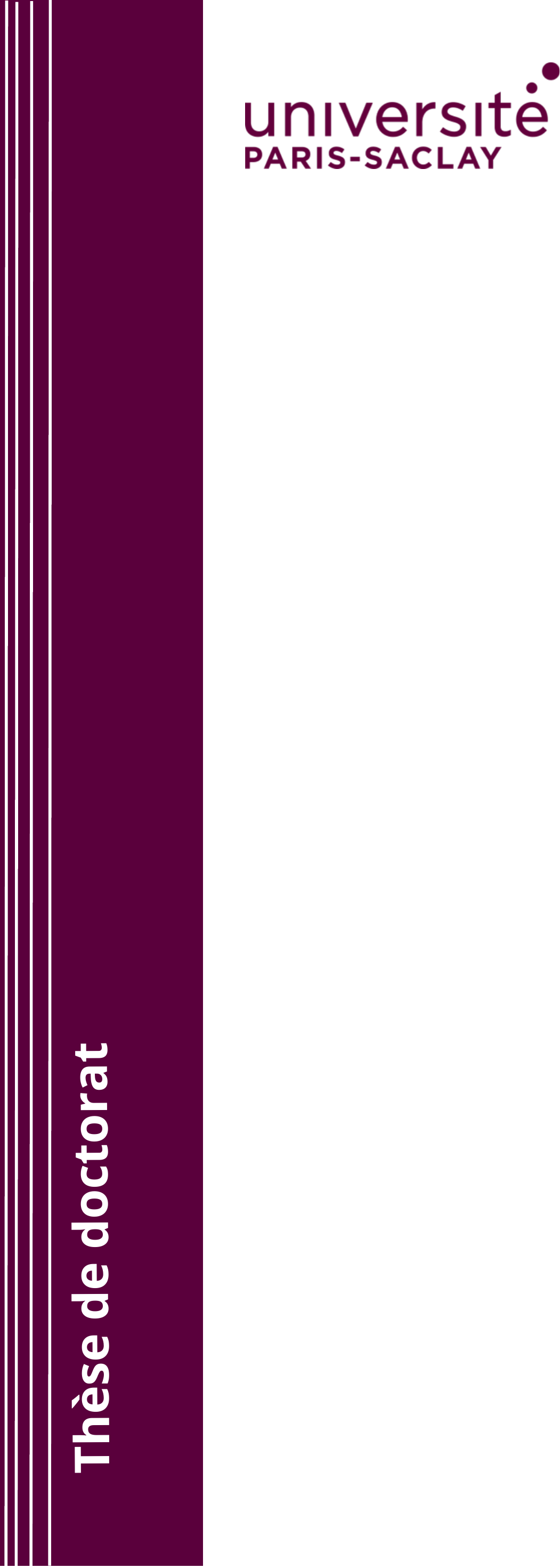}};

\fontfamily{fvs}\fontseries{m}\selectfont


\color{white}

\begin{picture}(0,0)

\put(-150,-735){\rotatebox{90}{NNT: 2020UPASP072}}
\end{picture}
 

\flushright
\vspace{15mm} 
\color{Prune}
\fontfamily{fvs}\fontseries{m}\fontsize{26}{30}\selectfont
A matter of shape:\hspace{1mm}~~~ \\
seeing the deformation \\
of atomic nuclei~~~~~~~ \\
at high-energy colliders \\


\normalsize
\vspace{1.2cm}

\color{black}
\textbf{Thèse de doctorat de l'Université Paris-Saclay}

\vspace{6mm}

École doctorale n$^{\circ}$ 564\\
Physique en \^Ile de France (PIF) \\
\vspace{5mm}
\small Spécialité de doctorat: \\ Physique \\
\vspace{5mm}
\footnotesize Unit\'e de recherche: \\ Universit\'e Paris-Saclay, CNRS, CEA \\ Institut de physique th\'eorique \\ 91191, Gif-sur-Yvette, France \\
\vspace{5mm}
\footnotesize R\'ef\'erent:\\ Facult\'e des sciences d’Orsay
\vspace{10mm}

\textbf{Th\`ese pr\'esent\'ee et soutenue en \\ visioconference totale, le 30/11/2020, par}\\
\bigskip
\Large {\color{Prune} \textbf{Giuliano GIACALONE}}

\vspace{\fill} 

\flushleft \small \textbf{Composition du jury:}
\bigskip

\scriptsize
\begin{tabular}{|p{8cm}l}
\arrayrulecolor{Prune}
\textbf{Thomas Duguet} &   Pr\'esident\\ 
Ing\'enieur de recherche, Université Paris-Saclay \\
\textbf{Hannah Elfner} &   Rapportrice \& Examinatrice\\ 
Professeur des Universit\'es, Goethe University Frankfurt \\
\textbf{Gunther Roland} &  Rapporteur \& Examinateur\\ 
Professeur des Universit\'es, \\
Massachussetts Institute of Technology \\
\textbf{Ulrich Heinz} &   Examinateur\\
Professeur des Universit\'es, Ohio State University \\
\textbf{Derek Teaney} &   Examinateur\\
Associate Professor, Stony Brook University \\
\end{tabular} 

\medskip
\begin{tabular}{|p{8cm}l}\arrayrulecolor{white}
\textbf{Jean-Yves Ollitrault} &   Directeur de th\`ese\\ 
Directeur de recherche, CNRS, Universit\'e Paris-Saclay & \\

\end{tabular} 

\end{titlepage}

\thispagestyle{empty}
\textcolor{white}{n}
\newpage

\newgeometry{top=2.cm, bottom=2.cm, left=2cm, right=2cm}

\fontfamily{cmr}\selectfont
\fontsize{14}{18}\selectfont

\renewcommand\contentsname{}

\tableofcontents


\chapter{Introduction}

\label{chap:1}

\section{High-energy nuclear physics}

\label{sec:1-1}

Over the last two decades, a novel branch of physical sciences has established itself as an active and important area of fundamental research. This is the field of \textit{high-energy nuclear physics}~\cite{wiki}. 

The name may sound a bit of an oxymoron. The adjective \textit{high-energy} implies a link with the vast field of high-energy physics, i.e., elementary particle physics. High-energy physics is devoted to studying and testing the fundamental interactions (weak, strong, and electromagnetic) that constitute the Standard Model of particle physics. This is done by means of exceptional experimental means, involving particle collider experiments, typically proton-proton collisions, performed at the most powerful accelerator facilities in the world. The term \textit{nuclear physics} denotes on the other hand the hundred-year-old effort devoted to the study of atomic nuclei, and in particular of their structure, i.e., their mass, geometry, and energy levels. These features are investigated by means of  experiments involving energy scales that are orders of magnitude lower than achieved in high-energy experiments, hence the appellative \textit{low-energy experiments}.

However, it would be reductive to state that nuclear physics is nowadays merely concerned with the study of atomic nuclei on low energy scales. I think that nuclear physics should be rather defined by its goal, which is conceptually different from that of high-energy physics. Atomic nuclei are packets of nucleons, neutrons and protons, which are in turn composed by elementary particles, quarks and gluons. The strong force of the Standard Model keeps these constituents together. While, as mentioned above, high-energy physics aims at unveiling the properties of the strong force, and of the associated quantum field theory, quantum chromodynamics (QCD), at the most fundamental level, the goal of nuclear physics is instead that of understanding the emergence of more complex forms of matter and phenomena that are shaped by this fundamental interaction.

Notorious examples of such forms of matter are atomic nuclei and neutron stars. However,  another item should nowadays be added to the list. This is the so-called \textit{quark-gluon plasma}, arguably, the weirdest of all forms of strong-interaction matter. It is a medium composed solely of quarks and gluons, and where nucleonic degrees of freedom are absent. The quark-gluon plasma is expected to emerge whenever one stuffs a (huge) lot of QCD matter inside a (very) small volume, i.e., when looking at systems that are far denser than normal nuclear matter. Since one can achieve such conditions of density only by smashing nuclei at very high energy, one has to perform high-energy experiments, i.e., collider experiments at the highest energies achievable on Earth. Instead of protons, one accelerates and smashes atomic nuclei, with the aim of producing and thus characterizing the quark-gluon plasma. Hence the name, high-energy nuclear physics.

 This branch of nuclear science emerged from the results of scattering experiments involving heavy nuclei that were conducted in the last decades of the 20th century, and it is thus a synonym of relativistic nuclear collisions, or, more common, relativistic \textit{heavy-ion collisions}. The discovery of the quark-gluon plasma was claimed in the early 2000's~\cite{Adcox:2004mh,Back:2004je,Arsene:2004fa,Adams:2005dq}, following the beginning of operation of the Relativistic Heavy Ion Collider machine at the Brookhaven National Laboratory. Since then, the field of heavy-ion collision has exploded, becoming quickly a major sub-field of nuclear research. The motivation behind the program of high-energy nuclear physics is the possibility of learning something new about QCD matter, such as the equation of state or its transport properties, under extreme conditions. This program has been highly successful. Thanks to the great amount of high-precision data coming from particle colliders, the theoretical understanding of the collision process has dramatically improved over the years. This has lead to the development of comprehensive theoretical frameworks that allow one to describe quantitatively the experimental observations, and consequently to place constraints over the physical properties of the quark-gluon medium.

In this work, I discuss a new direction of investigation which is opened by this optimal state of affairs. I argue that relativistic nuclear collisions provide us in particular with a new, powerful experimental probe of the structure of atomic nuclei, specifically, of their deformation, and that a phenomenology of nuclear structure at high energy is possible and within the reach of current experiments. Let me explain, then, how this can be done.  

\section{Macroscopic physics on nuclear scales}
\label{sec:1-2}

By creating the quark-gluon plasma in the laboratory, high-energy nuclear physics aims at characterizing strong-interaction matter in the limit of high \textit{temperature}. But what does temperature mean in this context?

\begin{figure}[t]
    \centering
    \includegraphics[width=.7\linewidth]{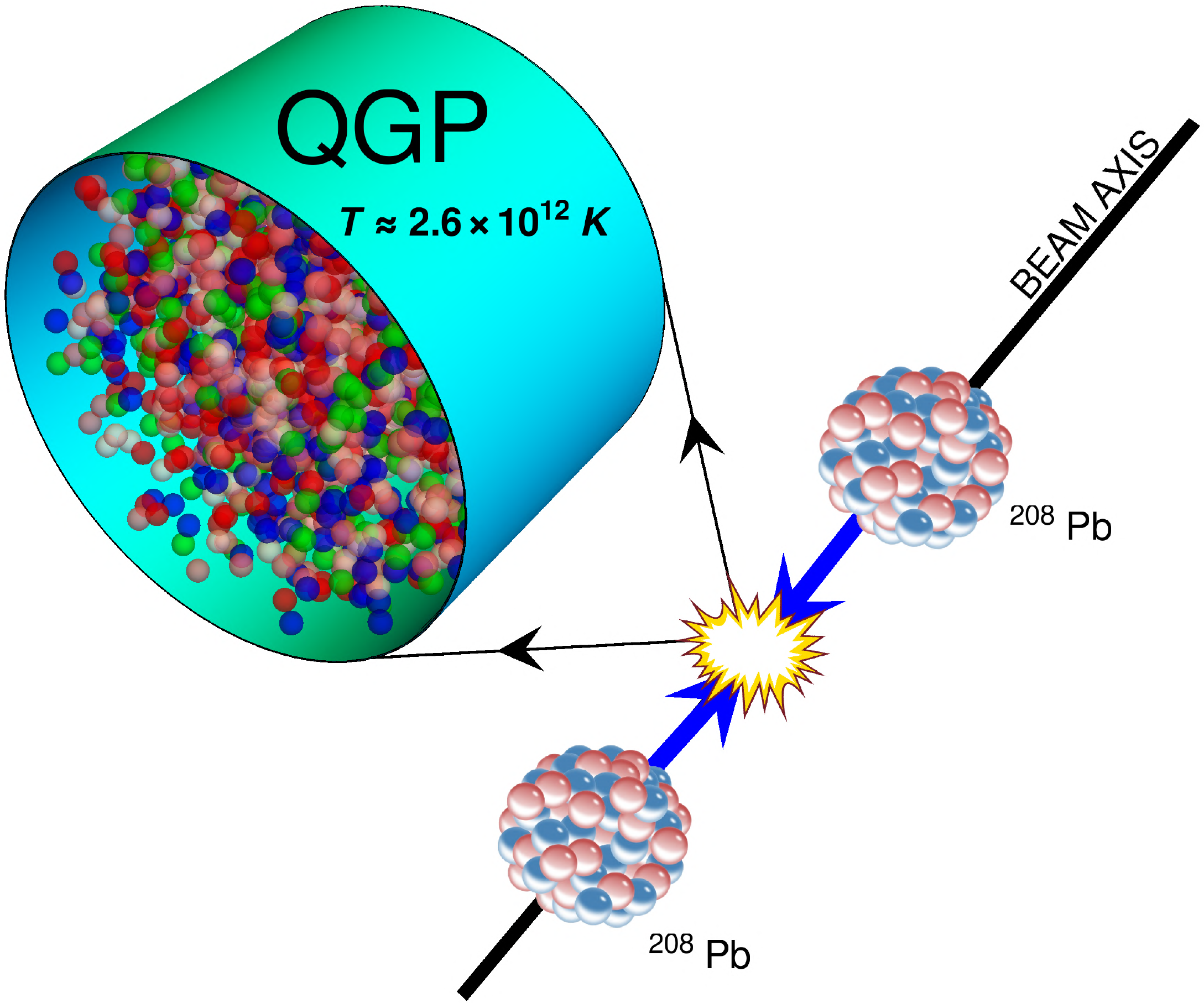}
    \caption{Collision between two $^{208}$Pb nuclei at a particle accelerator. In the interaction point, a quark-gluon plasma (QGP) is created, i.e., a region of space densely populated with quarks an gluons in thermal equilibrium at a temperature of about one trillion Kelvin. Figure from Ref.~\cite{Gardim:2019xjs}.}
    \label{fig:0_01}
\end{figure}

\paragraph{ Hydrodynamics --} In the interaction between constituent quarks and gluons in, e.g., a high-energy proton-proton collisions, there is in principle no such notion of a temperature. There in an initial state, followed by an interaction mediated by gluons, and then the emission of particles to the final state.  The situation is however quite different when one looks at a high-energy nuclear collision. Insight can be gained by looking at the illustration shown in \fig{0_01}, displaying the interaction of two nuclei accelerated along the beam pipe of a particle collider. In high-energy physics experiments, one is typically interested in elementary processes emitting a handful of particles to the final state. By contrast, in heavy-ion collisions one is interested in events where thousands of particles are detected in the final state. These events involve an incalculable amount of elementary collision processes, and it would be hopeless trying to describe them by means of perturbative QCD calculations.
 
 Nevertheless, not only such calculations are hopeless, but as a matter of fact they are also unnecessary. When a physical system is highly complex, it is usually possible to find an effective description which allows one to describe the dynamics of the bulk of particles without paying any attention to the motion of single constituents~\cite{Anderson:1972pca}. In the context of relativistic nuclear collisions and the physics of the quark-gluon plasma, this reasoning is the right path to follow. 
 
 This can be understood from simple figures. An ultrarelativistic collision between $^{208}$Pb nuclei releases typically 2000-3000 particles within a volume of order 1000~fm$^3$~\cite{Gardim:2019xjs}. This means that there are approximately $2-3$ particles per fm$^3$. Nuclear matter, e.g., the matter that makes up neutron stars and large nuclei, has about 0.16 particles per fm$^3$. The density achieved in high-energy nuclear collisions is over one order of magnitude larger than that of nuclear matter. 
 
 This has a nontrivial implication. If the particles produced over the interaction region know about each other, i.e., if they interact, then the system is in a special regime where the mean free path between two particles is negligible compared to the overall system size. This \textit{implies} that the bulk of particle motion can be described by fluid dynamical laws. Surprising as it may sound, then, the dynamics of the quark-gluon plasma, a system which is of the size of an atomic nucleus, is ruled by \textit{macroscopic} laws, involving pressure gradients, velocity fields, and temperature. I stress that this description requires the microscopic constituents to be coupled strongly enough to permit the system to reach a fair degree of local thermal equilibrium within a short time span, of typically 1 fm/$c$ (or $3\times10^{-24}~s$), following the interaction of the two nuclei. This is a nontrivial requirement. However, quarks and gluons interact via the strong force, whose associated time scale is precisely around 1~fm/$c$, and the hydrodynamic paradigm explains quantitatively all experimental observations so far made in relativistic nuclear collision experiments. The reaching of local thermal equilibrium can thus be viewed as an established experimental fact. 
 
 Today, one can in full safety claim that, as shown in \fig{0_01}, the system formed in a high-energy \pbpb{} collision is a gas of a few thousand particles in equilibrium at a temperature $T\approx 10^{12}$~K, corresponding to the temperature at which QCD predicts a gas of nucleons to melt into a plasma of de-confined quarks and gluons. This is the hottest medium ever produced in the laboratory.

\paragraph{Collectivity --}
The applicability of an effective description based on fluid dynamics in heavy-ion collisions is thus a generic consequence of the fact that such collisions produce thousands of particles. However, detectors at particle colliders can only see particles flying out of the interaction point, and do not permit one to resolve directly the quark-gluon plasma that is produced on nuclear length scales. What kind of observations do in practice confirm that the fluid paradigm is correct? The key idea is to look at the \textit{way} all the observed particles are distributed in the final states. The hydrodynamic description implies that the dynamics of the system is \textit{collective}, so that the particles detected in the final state are produced following the collective expansion of an underlying fluid that cools in vacuum. 

\begin{figure}[t]
    \centering
    \includegraphics[width=\linewidth]{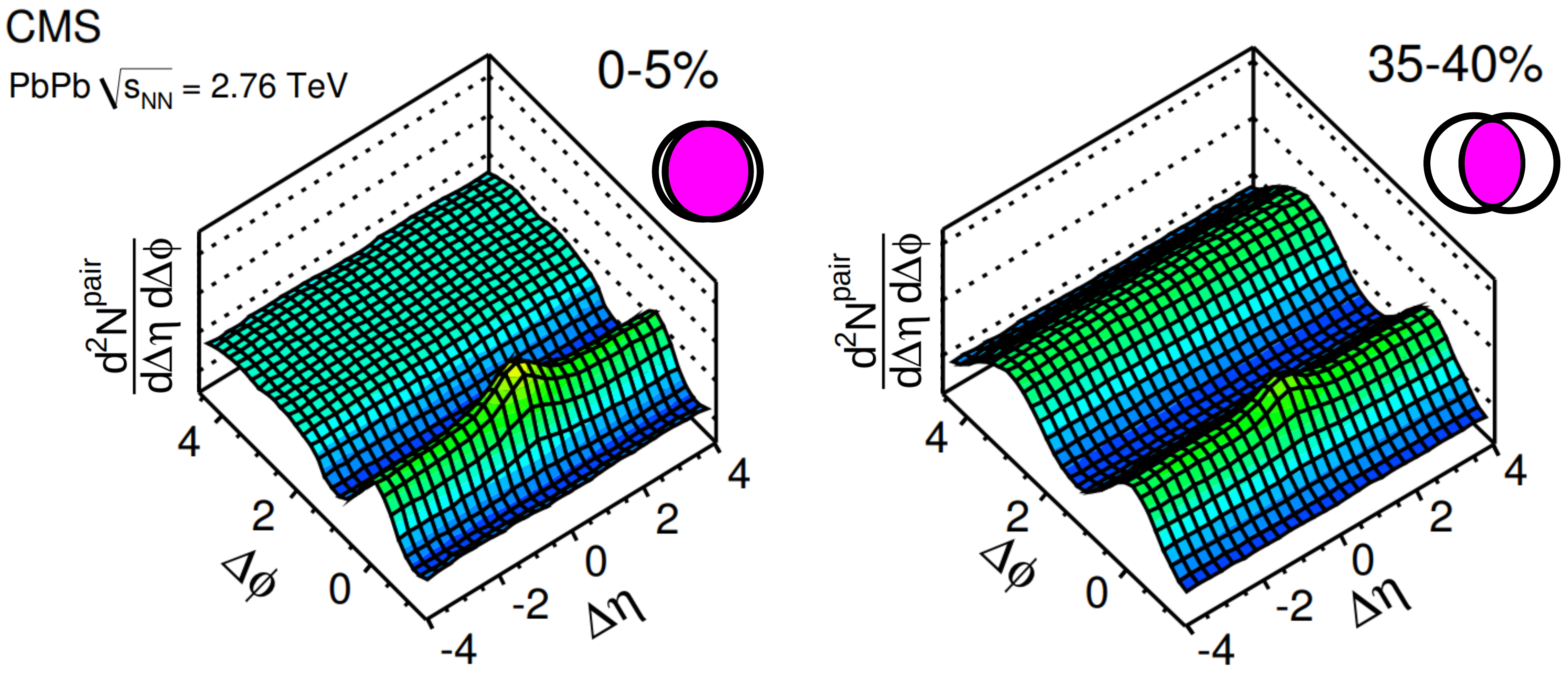}
    \caption{Number of particle pairs as a function of the relative angular distance between the considered particles in ultrarelativistic \pbpb{} collisions. Left: Central collisions. Right: Semi-peripheral collisions. Figure from Ref~\cite{Chatrchyan:2012wg}.}
    \label{fig:0_02}
\end{figure}

One can easily convince themselves that this idea makes sense by looking at basic features displayed by the angular distribution of the emitted hadrons. The relevant experimental data shown in \fig{0_02}.  The quantity which is plotted is a two-dimensional histogram, displaying the distribution of the number of pairs of particles separated by a given angular distance detected in the final states of relativistic \pbpb{} collisions. The angle $\Delta\phi$, where the \textit{azimuthal angle}, $\phi$, runs between $0$ and $2\pi$, corresponds to the angular separation between two particles in the plane orthogonal to the beam axis (as visualized in \fig{0_01}). The separation $\Delta\eta$ can be instead viewed as the angular separation between two particles along the beam axis, where $\eta=0$ corresponds to the interaction point. A separation $\Delta\eta=\pm4$ corresponds to a good approximation to pairs where the two particles are detected, respectively, at opposite ends of the detector, i.e., with a relative angle close to $\pi$ along the direction of the beam.

I focus on the left panel of \fig{0_02}. The nontrivial result is that the two-dimensional distribution is structureless along the $\Delta\eta$ direction. The only exception, i.e., the peak around $\Delta\eta\approx\Delta\phi\approx0$, has a trivial origin, and comes from the fact the probability of particle emission is in enhanced whenever two particles are \textit{collinear}, a fully generic feature of the underlying quantum field theory that governs the processes of particle emission. The same kind of peak would be observed in proton-proton, or electron-positron collisions. However, the structure stretching over the whole $\Delta\eta$ interval, which looks like a wave in the $\Delta\phi$ direction, is a feature unique of nuclear collisions, or in general of hadronic collision emitting large numbers of hadrons to the final state. The flatness of the distribution in $\Delta\eta$ implies in particular that particle pairs emitted at the two opposite ends of the detector have the same relative azimuthal angle as pairs of particles that have much smaller separations in $\Delta\eta$. The emitted particles appear thus to follow a global, collective pattern.

 Needless to say, then, that the observable presented in \fig{0_02} represents a spectacular confirmation of the effective fluid description, which naturally predicts such kind of collective phenomena in the final states of high-energy nuclear collisions. The idea that the observed correlations between particles originate solely from the underlying medium expansion amounts however to assuming that these correlations are not produced by the mechanism of particle production itself. This means that, when the quark-gluon plasma converts into hadrons, the momentum of a given hadron is chosen independently for each hadron. The combination of the hydrodynamic description with this idea of independent particle emission constitutes the so-called \textit{flow paradigm} of relativistic heavy-ion collisions.

\section{Symmetry breaking}
\label{sec:1-3}

The flow paradigm is thus motivated by simple theoretical arguments and confirmed by striking experimental observations. The question is now how one can exploit it to obtain information about the physical properties of the system created in the interaction region, i.e., of the quark-gluon plasma.

\paragraph{Elliptic flow --}
In the left panel of \fig{0_02}, one observes a second prominent feature. The distribution of pair number breaks symmetry along the $\Delta\phi$ direction, i.e., there are directions where particle emission is favored, in particular, a minimum at $\Delta\phi=\pi/2$ and a plateau around $\Delta\phi=\pi$. The origin of these kind of patterns is best deduced from the right panel of \fig{0_02}. This panel shows the same observable, but at a different collision \textit{centrality}, i.e., with the two nuclei colliding with a significant \textit{impact parameter}. The latter is defined as the spatial separation between the centers of the two nuclei in the plane orthogonal to the beam. The plot I have been analyzing in the left panel is obtained for central collisions, i.e., collisions where the impact parameter is small and the overlap of the two nuclei is almost maximal (see the illustration on top of the figure). In the right panel of \fig{0_02}, on the other hand, the impact parameter is much larger.

\begin{figure}[t]
    \centering
    \includegraphics[width=.55\linewidth]{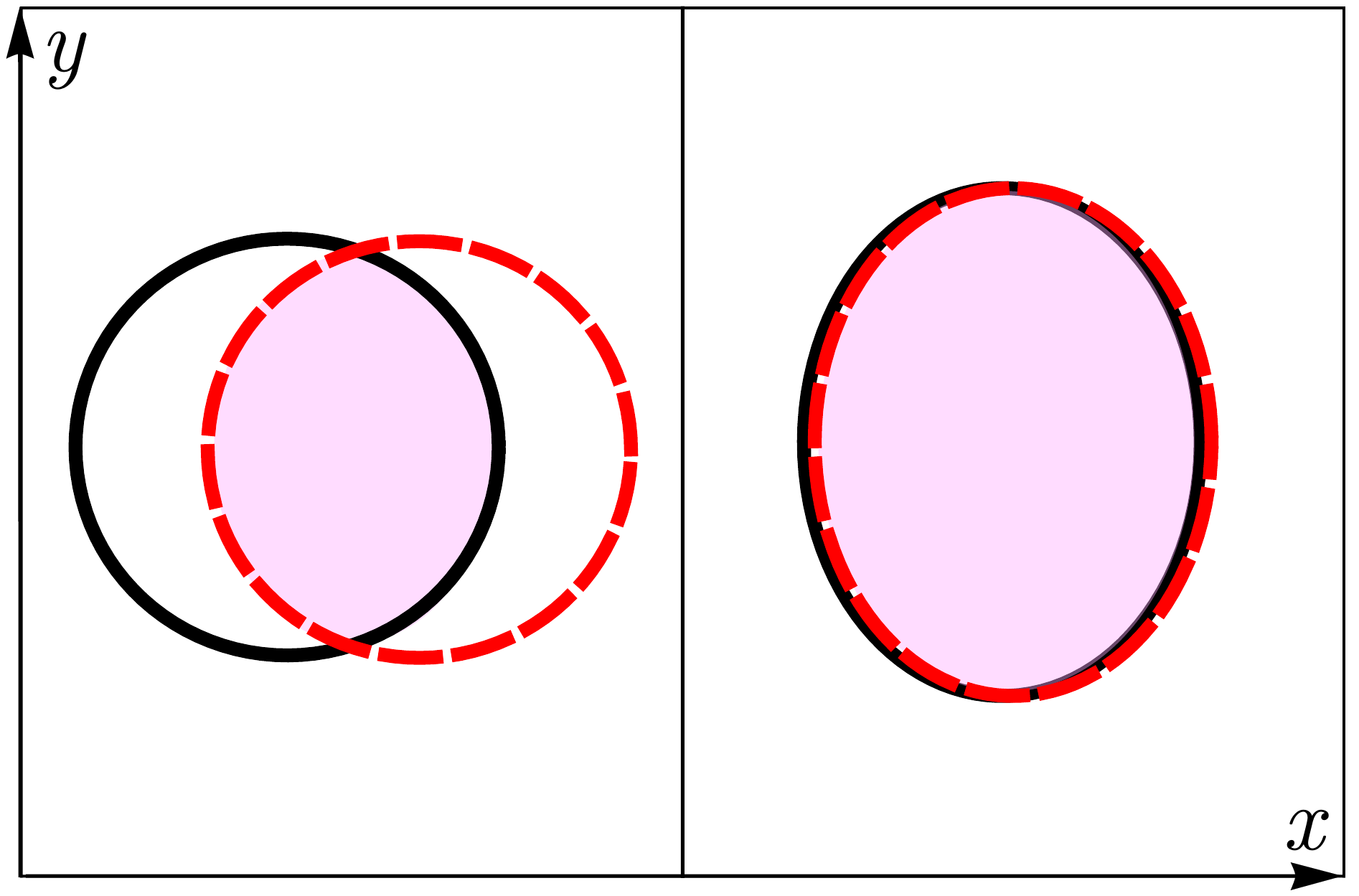}
    \caption{Collision geometry in the plane orthogonal to the beam axis. The impact parameter is along $x$. Left: Azimuthal isotropy in the region of overlap is broken by the finite impact parameter. Right: Anisotropy is generated by the deformed shape of the overlapping bodies.}
    \label{fig:0_03}
\end{figure}

Closer inspection of the distribution of particle pairs in the $\Delta\phi$ direction in \fig{0_02} reveals that not only azimuthal isotropy is broken, but the distribution acquires a pronounced $\cos(2\Delta\phi)$ modulation. This phenomenon is known as \textit{elliptic flow}.  What is its origin? Insight can be gained by looking more closely at the region of nuclear overlap. A zoom is shown in the left panel of \fig{0_03}.

The figure shows the geometry of overlap of two nuclei in the plane $(x,y)$ orthogonal to the beam axis. One immediately sees that two nuclei colliding at a finite impact parameter, as in the left panel, produce an interaction region that breaks azimuthal symmetry, and that has essentially the shape of an ellipse. This means in particular that if we call $\Phi$ the polar angle in the $(x,y)$ plane, then the interaction region has precisely a $\cos(2\Phi)$ modulation. 

Within a hydrodynamic paradigm, this feature has a striking consequence. Since the fluid is produced at rest, its dynamics is governed by pressure gradients which are determined by the geometry of the system. Breaking of symmetry in the geometry of nuclear overlap implies an imbalance in the pressure gradients that govern the hydrodynamic expansion of the system. The resulting hydrodynamic flow is asymmetric, and produces more momentum along a preferred direction (the $x$ direction, in the case of \fig{0_03}). This is ultimately carried over to the particles emitted from the fluid, and manifests as a breaking of symmetry in the azimuthal distribution of particles detected in the final state. Hence an elliptical, $\cos(2\Phi)$ modulation of the overlap region, caused by the impact parameter, yields an elliptical, $\cos(2\Delta\phi)$ modulation of the azimuthal distribution of final-state hadrons, i.e., elliptic flow. In view of this, the appearance of a visible elliptic flow as one moves from the left panel to the right panel of \fig{0_02} represents an additional spectacular experimental confirmation of the fluid description.

The conversion of initial-state anisotropy into final-state anisotropy is driven by the transport properties (speed of sound, viscosity) of the quark-gluon plasma. Consequently, if one knows the impact parameter of the detected collisions and has a good knowledge of the geometry of the quark-gluon plasma at the onset of the hydrodynamic behavior, one can use experimental data on elliptic flow to reconstruct information about the transport properties of the fluid. This is indeed a powerful method allowing theoretical calculations based on hydrodynamic simulations to achieve the goal of high-energy nuclear physics, i.e., the characterization of hot QCD matter from experimental measurements.

\paragraph{Quadrupole deformation --} The fact that the final elliptic anisotropy measured in data originates from an elliptic anisotropy in the region of nuclear overlap might ring a bell in the head of those who have some notions on the fundamental properties of atomic nuclei.  There exists in fact another well-defined origin of quadrupole asymmetry in the region of overlap, an illustration of which is given in the right panel of \fig{0_03}. Here, breaking of symmetry is not caused by the impact parameter, but solely from the fact that the two colliding nuclei have an ellipsoidal shape. This feature is neither special nor exotic. The majority of atomic nuclei are in fact nonspherical in their ground state, but present, precisely, a quadrupole deformation, one of the fundamental features of atomic nuclei investigated by theories of nuclear structure. Hence, unless the colliding species are chosen from the (actually limited) pool of stable nuclides that can be considered as spherical in their ground state (like $^{208}$Pb nuclei), one should simply expect the realization of deformed regions of nuclear overlap due to the deformed nuclear shape of the colliding bodies.

This implies that if one collided nuclei that have a quadrupole deformation in their ground state, and if one were able to select collision configurations corresponding to a vanishing impact parameter, then an excess elliptic flow should be observed due to the contamination from the collision geometries shown in the right panel of \fig{0_03}. An even more ideal situation would however be realized if from the data one were able to discern directly these geometries that maximally break azimuthal symmetry, as in these events any manifestation of elliptic anisotropy in the final state would represent a phenomenological manifestation of the quadrupole deformation of the colliding nuclei. If such observations were made, and if the transport properties of the quark-gluon plasma were known from studies of collision of spherical nuclei, one could use experimental data on elliptic flow in collisions of nonspherical nuclei as a means to obtain information about the quadrupole deformation of the colliding species. 

This defines a neat method allowing one to use high-energy nuclear experiments as probes of the deformation of atomic nuclei. This possibility has potentially far reaching implications, and is in principle of interest for a large fraction of the nuclear physics community. However, although experimental data on relativistic collisions of deformed nuclei, notably $^{238}$U nuclei, has been already collected at RHIC, little has been achieved along this direction of investigation. The reason is that, following the publication of data, it has been realized that selecting geometries of interaction corresponding to the right panel of \fig{0_03} is more difficult than originally thought, and perhaps not possible at all. As a consequence, quantitative studies of nuclear deformation at high energy have not been pursued.

\section{About this document}

\label{sec:1-4}

This document is divided into two parts.

 Chapters~$\ref{chap:2}$~and~$\ref{chap:3}$ constitute the first part. In Chapter~\ref{chap:2}, I discuss generic features of relativistic heavy-ion collisions. I outline how a collision is modeled in theoretical calculations, following the so-called Glauber Monte Carlo model. I explain, then, how the Glauber model can be coupled to a hydrodynamic description, and I give a global picture of the current understanding of the space-time evolution of a relativistic nuclear collision.  Subsequently, in Chapter~\ref{chap:3} I introduce a few observables of paramount importance in the analysis of heavy-ion collisions. These are: $i)$ the total number of particles detected in a collision event, i.e., the multiplicity; $ii)$ the elliptic flow, as discussed above; $iii)$ the average momentum of the final-state hadrons in the plane orthogonal to the beam axis, the so-called average transverse momentum. I underline the fact that these observables are amenable to a simple physical interpretation based on elementary fluid dynamic and thermodynamic laws.

The main subject of this work,  i.e., the analysis of phenomenological manifestations of nuclear structure in high-energy nuclear experiments, constitutes instead Chapters~$\ref{chap:4}$~and~$\ref{chap:5}$. In Chapter~\ref{chap:4} I perform a detailed analysis of existing experimental data on the fluctuations of elliptic flow in high-multiplicity collision at RHIC. By means of accurate theory-to-data comparisons, I show that the RHIC data provides clear evidence of the deformed, ellipsoidal shape of $^{238}$U nuclei, while suggesting, on the other hand, that $^{197}$Au nuclei are nearly spherical. This latter result turns out to be highly nontrivial. It is at variance with empirical estimates, or estimates purely based on a mean-field approximation, that can be found in the nuclear data tables. Understanding the manifestation of nearly-spherical gold nuclei in RHIC data requires thus to look at the predictions of sophisticated frameworks of nuclear structure that go beyond the mean-field picture. This opens a new direction of investigation, and establishes a deep connection between high-energy and low-energy nuclear phenomena. I then analyze LHC data on \xexe{} collisions. This data provides compelling evidence of the deformed shape of $^{129}$Xe nuclei, a result which, based on the predictions of  state-of-the-art nuclear models, suggests the first phenomenological manifestation of shape coexistence effects in high-energy experiments. In Chapter~\ref{chap:5} I overcome the problem mentioned in the previous section. I introduce a selection of collision events based on the average transverse momentum that allows one to select collision geometries corresponding to the right panel of \fig{0_03} in an experiment. The key feature is that, for collisions of deformed nuclei at high multiplicity, the configurations that one is looking for correspond to events where the temperature of the quark-gluon plasma is abnormally small. I explain that, as a consequence, the statistical correlation between elliptic flow and the average transverse momentum for events at fixed multiplicity is negative for well-deformed nuclei, a feature which is confirmed by preliminary RHIC data. I argue that this new method can serve as the basis for quantitative studies of nuclear deformation at high energy.

In Chapter~$\ref{chap:6}$ I draw my conclusions and, motivated by the results presented in Chapters~\ref{chap:4}~and~\ref{chap:5}, I make a proposal for a future experimental campaign aimed at the systematic study of nuclear structure effects in relativistic nuclear collisions. I highlight in particular the great impact that such a program would have on both high-energy and low-energy nuclear physics.

\bigskip
The results presented in this manuscript come mainly from these papers:
\begin{itemize}
    \item G.~Giacalone, J.~Noronha-Hostler, M.~Luzum and J.~Y.~Ollitrault, \textit{``Hydrodynamic predictions for 5.44 TeV Xe+Xe collisions''}, Phys. Rev. C \textbf{97}, no.3, 034904 (2018)~\href{https://arxiv.org/abs/1711.08499}{\texttt{arXiv:1711.08499}}.
    \item G.~Giacalone, \textit{``Elliptic flow fluctuations in central collisions of spherical and deformed nuclei''}, Phys. Rev. C \textbf{99}, no.2, 024910 (2019) ~\href{https://arxiv.org/abs/1811.03959}{\texttt{arXiv:1811.03959}}
    \item F.~G.~Gardim, G.~Giacalone, M.~Luzum, J-Y.~Ollitrault, \textit{``Thermodynamics of hot strong-interaction matter from ultrarelativistic nuclear collisions''}, Nature Phys. \textbf{16}, no.6, 615-619 (2020)~\href{https://arxiv.org/abs/1908.09728}{\texttt{arXiv:1908.09728}}
    \item G.~Giacalone, \textit{``Observing the deformation of nuclei with relativistic nuclear collisions''}, Phys. Rev. Lett. \textbf{124}, no.20, 202301 (2020)~\href{https://arxiv.org/abs/1910.04673}{\texttt{arXiv:1910.04673}}
    \item G.~Giacalone, F.~G.~Gardim, J.~Noronha-Hostler, J-Y.~Ollitrault, \textit{``Correlation between mean transverse momentum and anisotropic flow in heavy-ion collisions''} \href{https://arxiv.org/abs/2004.01765}{\texttt{arXiv:2004.01765}}
    \item G.~Giacalone, \textit{``Constraining the quadrupole deformation of atomic nuclei with relativistic nuclear collisions''}, Phys. Rev. C \textbf{102}, no.2, 024901  (2020) \\ \href{https://arxiv.org/abs/2004.14463}{\texttt{arXiv:2004.14463}}
\end{itemize}


\chapter{Ultrarelativistic heavy-ion collisions}

\label{chap:2}

A particle in the laboratory frame is said to be in the \textit{ultrarelativistic} regime when its energy is far greater than its mass at rest. A collision between particles is said to be ultrarelativistic when the colliding particles are ultrarelativistic. A proton has for instance a rest energy of about 1~GeV, so that a proton-proton collision happening at a center-of-mass energy of 100~GeV+100~GeV is ultrarelativistic. The same applies to nuclei. A collision between two nuclei is ultrarelativistic if the nucleons that compose the colliding nuclei are themselves ultrarelativistic. 

However, when dealing with nuclear collisions, one can define the ultrarelativistic regime by means of an equivalent, and yet insightful geometric condition. To understand this, let me quote some lines from the paper where special relativity was invented:~\cite{einstein} 
\begin{displayquote}
\textit{We consider a rigid sphere of radius $R$. [\ldots] A rigid body that has a spherical shape when measured in the state of rest thus in the state of motion -- observed from a system at rest -- has the shape of an ellipsoid of revolution with axes:
\begin{equation}
\nonumber    R\sqrt{1-\frac{v^2}{c^2}},~~R,~~R.
\end{equation}
Thus [\ldots ] at $v = c$, all moving objects -- observed from the system ``at rest'' --  shrink into plane structures.}
\end{displayquote}
In the frame of the laboratory, the colliding nuclei are Lorentz-contracted along the direction of the beampipe. A nucleus can thus be considered as a ``plane structure'' in the laboratory frame whenever the Lorentz factor, $\gamma=(1-v^2/c^2)^{-0.5}$, is large, $\gamma \gg 1$. However, how small should $R/\gamma$ be to define the ultrarelativistic regime? The dynamics of a relativistic nuclear collision involves an additional scale, corresponding to the time scale on which the system produced in the interaction region reaches thermal equilibrium. This scale is dictated by the energy scale of QCD, and is naturally of order 1 fm/$c$. In the context of nuclear collisions, I think it is thus fully appropriate to state that the ultrarelativistic regime is achieved when the thermalization process is completely decoupled from the motion of the nuclei while they cross each other. If the time taken by the nuclei to cross each other is infinitesimal compared to 1 fm/$c$, then the collision is ultrarelativistic. The bottom line is that, if the radius of a nucleus is around $R\sim10~{\rm fm}$, then having $\gamma\sim10$ is not enough to reach the ultrarelativistic limit. One needs an additional order of magnitude in the Lorentz factor, i.e., $\gamma\sim100$. This conclusion has been reached without having any knowledge of the center-of-mass energy of nucleon-nucleon interactions, which sounds like a nontrivial achievement.

In this chapter, I present an end-to-end description of the collision process, starting from the description of how nuclear collisions are performed at particle colliders, how one can model the geometry of the collision in the interaction region, and how a hydrodynamic description is coupled to such a model, eventually leading to the final observable quantity, i.e., a spectrum of hadrons.

\section{Collider experiments}

\label{sec:2-1}

To achieve the ultrarelativistic regime, one needs a lot of energy in the center of mass. This is best achieved in collider mode, with two beams of nuclei running in the accelerator ring and then crossing at an interaction point. There are only two collider facilities in the world that are able to perform nuclear collisions at ultrarelativistic energy. 

\begin{figure}[t]
    \centering
    \includegraphics[width=\linewidth]{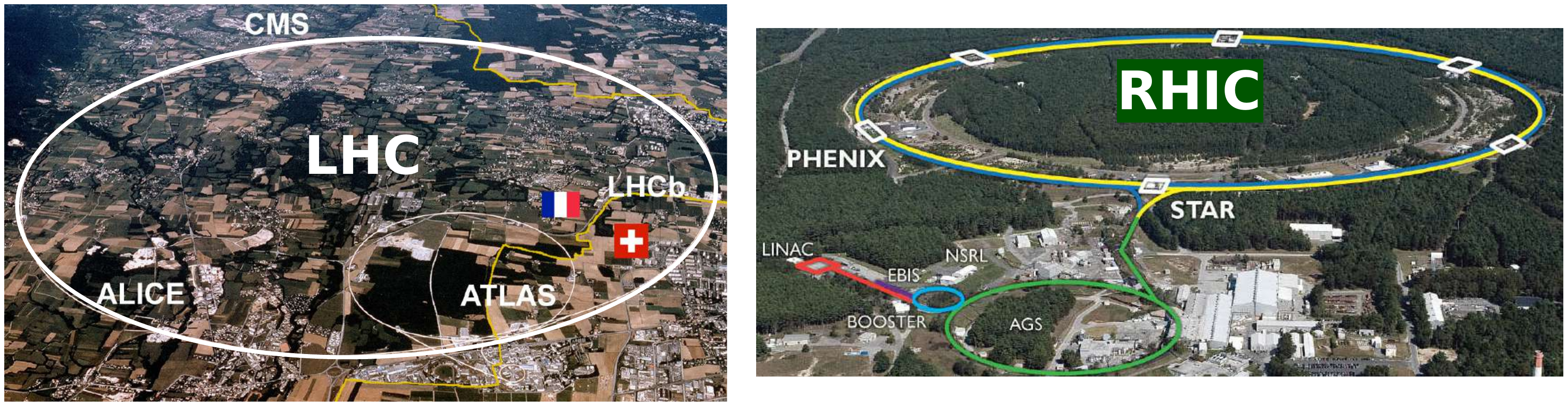}
    \caption{Aerial view of the two collider facilities that perform ultrarelativistic nuclear collisions. Left panel: the Large Hadron Collider, operated by CERN, located on the border between France and Switzerland. The length of the ring is approximately 27~km. Highlighted are the locations of the four detectors belonging respectively to the four large and infamous collaborations doing physics with this machine. Right panel: The Relativistic Heavy Ion Collider complex on Long Island, approximately 100~km east of New York City. In this work I will be mostly concerned with experimental data collected by the STAR detector, highlighted in the figure, which is currently the only detector in activity at RHIC.}
    \label{fig:2-1}
\end{figure}

The Relativistic Heavy Ion Collider (RHIC) is a synchrotron operating at the Brookhaven National Laboratory (BNL) in Upton, New York, USA (aerial view given in the right panel of ${\rm Fig.}~\ref{fig:2-1}$). It consists of an accelerator ring with a diameter of about 1.2~km. It can perform proton-proton collisions at center-of-mass energy up to 500~GeV, and nuclear collisions at a nucleon-nucleon center-of-mass energy up to 200~GeV. This machine is entirely devoted to studies of high-energy nuclear physics, and it is here that the quark-gluon plasma was discovered. RHIC is also a versatile machine, which allows to collide a lot of different species of stable nuclides (although so far only a limited number of them has been utilized in experiments). RHIC started its operation in the year 2000, and it will keep performing relativistic nuclear collision studies over the next decade.

The high-energy frontier in particle physics is currently being explored by the Large Hadron Collider (LHC), operated by European Center for Nuclear Research (CERN), in Geneva, CH (aerial view in the left panel of ${\rm Fig.}~\ref{fig:2-1}$). This synchrotron consists of a ring with a diameter of about 9~km, making it the largest particle accelerator in the world. LHC allows one to collide protons at a center-of-mass energy up to 14~TeV, and large nuclei at a nucleon-nucleon center-of-mass energy up to 5.5~TeV. Contrary to RHIC, LHC is meant to perform high-energy physics studies, and thus it mainly collides protons to perform precision tests of the Standard Model of particle physics, and to look for potential signatures of physics beyond the Standard Model. For this reason, at the LHC collisions of atomic nuclei are run only for about 1 month per year. The ALICE Collaboration, one of the four large collaborations working at the LHC, consisting of about 1000 members, is however entirely dedicated to the heavy-ion collision program. There also smaller groups (of order of 100 people) of heavy-ion physicists in the LHC's largest collaborations, ATLAS and CMS. Nuclear collisions at LHC will be performed at least until about 2030~\cite{Citron:2018lsq}. Beyond 2030, discussions are ongoing concerning the possibility of running with lighter ions for higher luminosities, as well as dedicated ambitious new detector upgrades.

\begin{figure}[t]
    \centering
    \includegraphics[width=.65\linewidth]{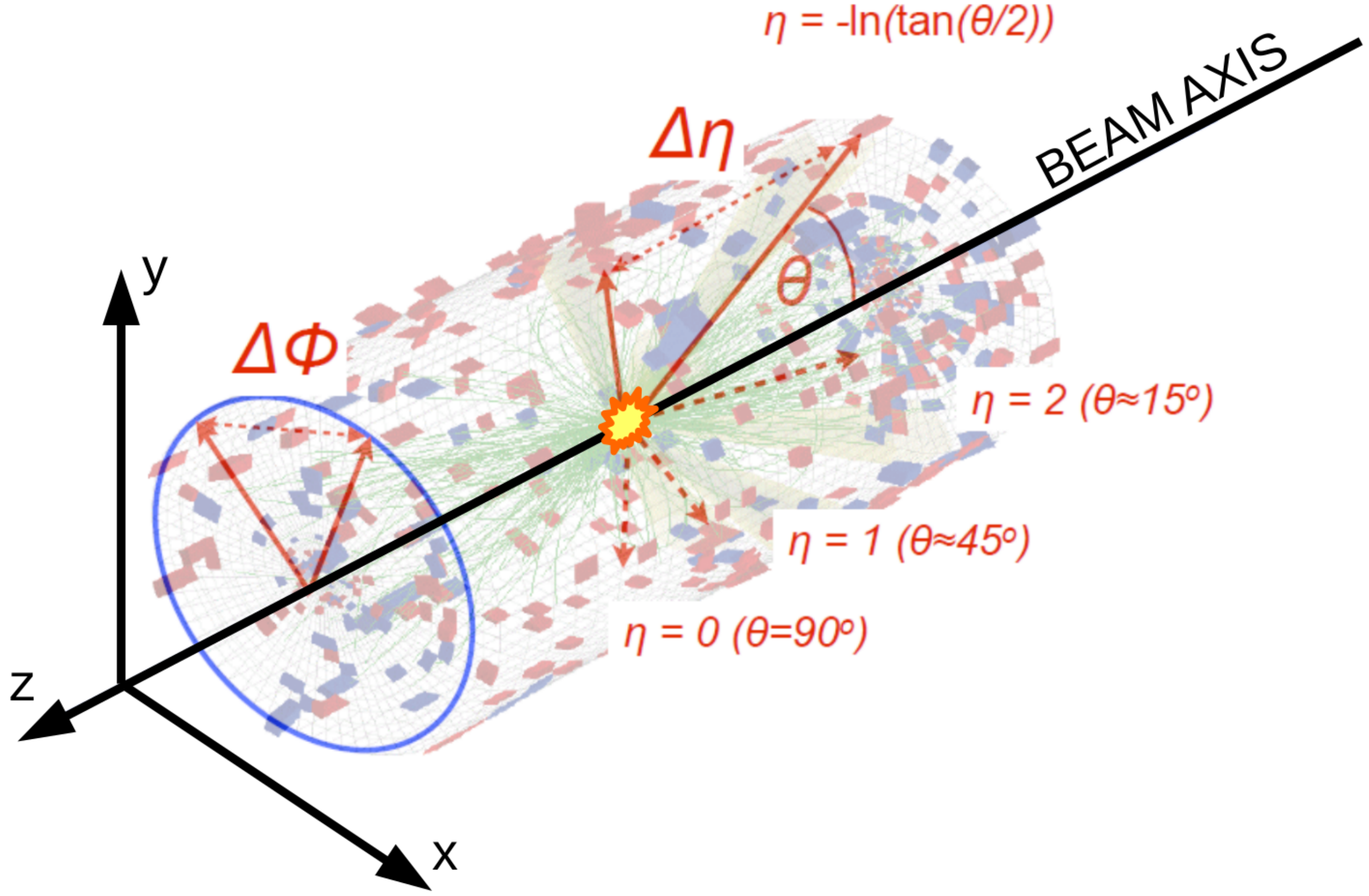}
    \caption{The geometry of a high-energy collision as observed in a particle collider detector is that of a \textit{barrel}. The detector is located around the beam axis ($z$), and covers essentially the full solid angle. The azimuthal angle, $\phi$, is defined as the angle in the plane $(x,y)$ transverse to the beam. The polar angle, $\theta$, is usually traded for a pseudorapidity, $\eta$, which is a measure of how boosted a particle is along the $z$ direction with respect to the laboratory frame. See the text for more details.}
    \label{fig:2-2}
\end{figure}

How does an ultrarelativistic nuclear collision look like? A representation of a collision occurring in a particle collider detector is depicted here in Fig.~\ref{fig:2-2}. 
The colliding objects run along the beam axis and smash in the interaction point, highlighted in the figure. The collision releases a large number of particles, depicted as green thin lines, that are collected by the detector surrounding the interaction point. Each particle is labeled by appropriate coordinates in the laboratory frame. All measurements are performed in momentum space, so that the final reconstructed object corresponding to a given particle is in fact its 4-momentum vector, $p^\mu=(p^0,{\bf p})$, where ${\bf p}=(p_x,p_y,p_z)$, and $p^0$ is equal to $\gamma m$, where $m$ is the rest mass of the particle, and I have set $c=1$. 

The common system of coordinates used in high-energy experiment analyses is Cartesian, and describes the detector as a volume in the three-dimensional $(x,y,z)$ space, as illustrated in \fig{2-2}. $z$ is the longitudinal coordinate, and runs along the direction of the beam axis; $x$ is the direction orthogonal to the beam axis pointing towards the center of the accelerator ring; $y$ is the \textit{vertical} coordinate, orthogonal to $x$ and $z$. The plane orthogonal to the beam axis, $(x,y)$, is called the \textit{transverse plane}. The momentum of a particle in this plane is called the \textit{transverse momentum}, and is defined by the following equation:
\begin{equation}
    {\bf p}_t=(p_x,p_y), \hspace{30pt} p_t \equiv |{\bf p}_t| = \sqrt { p_x^2 + p_y^2 }.
\end{equation}
Note that the total transverse momentum vector vanishes in a given collision event, i.e., $\sum {\bf p}_t = 0$, where the sum runs over all the emitted particles.

Alternatively, one can use spherical coordinates. Following \fig{2-2}, one introduces an azimuthal angle, $\phi$, i.e., the angle in the $(x,y)$ plane, and a polar angle, $\theta$, in the $(y,z)$ plane. Now, when dealing with the longitudinal component, one typically converts the polar angle into a so-called \textit{pseudorapidity}, $\eta$, which is defined by:
\begin{equation}
    \eta = - \ln \tan (\theta/2).
\end{equation}
As shown in \fig{2-2}, a particle with $\eta=0$ corresponds to $\theta=\pi/2$, while the limits of an emission collinear with the beam axis are given by $\eta=\pm\infty$.
This seemingly strange definition is in fact motivated by relativity arguments. The pseudorapidity can be written as:
\begin{equation}
\label{eq:pseudorap}
    \eta = \frac{1}{2} \ln \frac{p+p_z}{p-p_z},
\end{equation}
where $p \equiv |{\bf p}|$. This expression has to be compared to the formula for the so-called \textit{rapidity} employed by collider physicists, defined by:
\begin{equation}
\label{eq:rapidity}
    y = \frac{1}{2} \ln \frac{E+p_z}{E-p_z},
\end{equation}
where $E$ is the energy of the particle. Two comments are in order. First, the rapidity $y$ is defined in such a way that it is additive under Lorentz boosts along $z$. Hence it gives a measure of how boosted a particle is with respect to the laboratory frame ($y=0$, or \textit{midrapidity}). Second, for an ultrarelativistic particle one has $p \approx E$, which implies $\eta=y$. This clarifies the use of $\eta$ as a measure of the longitudinal particle coordinate in high-energy collisions. Note that, since the collider energy is finite, there exists a maximum value for $\eta$ (or $y$). This is the so-called the beam rapidity. It is close to 5 at top RHIC energy, and to 9 at top LHC energy.

\section{Glauber modeling of nuclear collisions}

\label{sec:2-2}

I discuss now the standard theoretical framework describing how a relativistic nuclear collision takes place in practice. This is the so-called Glauber Monte Carlo model~\cite{Miller:2007ri}. This model allows one to relate the experimental knowledge about the number of particles detected in a sample of collision events to generic properties of the \textit{geometry} of these collisions, the knowledge of which is crucial for the subsequent hydrodynamic expansion.

\subsection{Nucleon-nucleon collisions}

\label{sec:2-21}

I review the various steps that define the Glauber Monte Carlo model, and that allow one to describe the interaction between two nuclei in terms of few relevant quantities, namely, the impact parameter and the participant nucleons.

\paragraph{Nucleon positions --} 
The first step consists in shaping the colliding bodies at the time of interaction. The idea is that a nucleus is described by a density of matter $\rho(r)$, and that, on an event-by-event basis, this distribution can be used to determined the positions of the nucleons inside the nucleus. The  nucleus in the Glauber model is treated as a collection of independent nucleons, whose coordinates in space are sampled according to a single-particle density, $\rho(r)$. 

An established model of $\rho(r)$ is given by the following two-parameter Fermi distribution:
\begin{equation}
\label{eq:2pf}
    \rho(r) = \frac{\rho_0}{1+\exp \left ( \frac{r-R}{a}  \right )}.
\end{equation}
This parametrization has been employed to fit data coming from low-energy electron-nucleus scattering experiments to characterize the charge density of several nuclear species~\cite{DeJager:1987qc}.
In \equ{2pf}, $\rho_0$ is the normal nuclear matter density, which is about 0.16~fm$^{-3}$, $a$ is the skin width, or diffusiveness, that is typically of order $a\sim0.5~{\rm fm}$, while $R$ is the distance from the center of the nucleus at which the nuclear density drops by a factor 2, and is of order $R\sim 6.5~{\rm fm}$ for large nuclei. Note that \equ{2pf} represents our first encounter with nuclear structure physics in the modeling of high-energy nuclear collisions. It has a few limitations. By employing $\rho(r)$ to sample the positions of the nucleons, one is essentially assuming that the nuclear point-like matter density is the same thing as the charge density, which, strictly speaking, is wrong. Moreover, one assumes that the density of protons and the density of nucleons within the nucleus have the same shape. This is a good approximation, although it is known that the parameter $a$ for the neutron density is somewhat different from that of the proton density~\cite{Abrahamyan:2012gp,Tarbert:2013jze}, which may play a role for certain observables analyzed in heavy-ion collisions~\cite{Paukkunen:2015bwa,Helenius:2016dsk,Hammelmann:2019vwd}. 

The density in Eq.~(\ref{eq:2pf}) represents all the nuclear physics involved in the Glauber Monte Carlo model. The collision process is then entirely described in terms of constituent nucleons. As anticipated, the idea is that a colliding nucleus is given by a collection of nucleons whose coordinates are sampled according to $\rho(r)$. The sampling of nucleon positions is usually done in three dimensions, although, due to the enormous Lorentz contraction of the colliding nuclei in the lab frame, one is ultimately interested only in the transverse coordinates, $x$ and $y$. Typical model implementations~\cite{Loizides:2014vua} use as well a minimum inter-nucleon separation, say $d$, to take into account the fact that the potential energy characterizing nucleon-nucleon interactions is typically divergent on small length scales. Nuclear physics indicates that this separation should be of order $d\sim 0.5~{\rm fm}$. It is unclear to me whether or not one should include such a feature in the Glauber model, considering that the problem is already highly simplified. Note that recent hydrodynamic calculations suggest a scale $d$ larger than 1 fm~\cite{Moreland:2018gsh,Bernhard:2019bmu}, which can not represent an inter-nucleon repulsive potential. The results obtained with Glauber-kind calculations throughout this manuscript do not implement any such parameter.

\paragraph{Impact parameter --}

A crucial quantity for the determination of the geometry of the interaction region is the impact parameter, which is defined as the spatial separation between the centers of the colliding nuclei in the transverse plane. 

As mentioned in the previous sections, the impact parameter of a collision can not be controlled experimentally. This means in particular that the orientation of the impact parameter is not known, so that the quantities measured in experiments either have unknown impact parameter or are averaged over many values of it. This fact allows for a nice simplification of the implementation of the impact parameter in a theoretical calculation. In principle, the impact parameter is a 2-component vector ${\bf b}=(b_x,b_y)$. However, as its orientation is random and uniform in a sample of events, one can more simply consider that it lies along the same direction in all events. The standard choice in theoretical simulations is to take the impact parameter along the $x$ direction, ${\bf b}=(b,0)$. The direction of impact parameter is in jargon called the direction of the \textit{reaction plane}, where the reaction plane is represented by the $(x,z)$ plane. 

\begin{figure}[t]
    \centering
    \includegraphics[width=.5\linewidth]{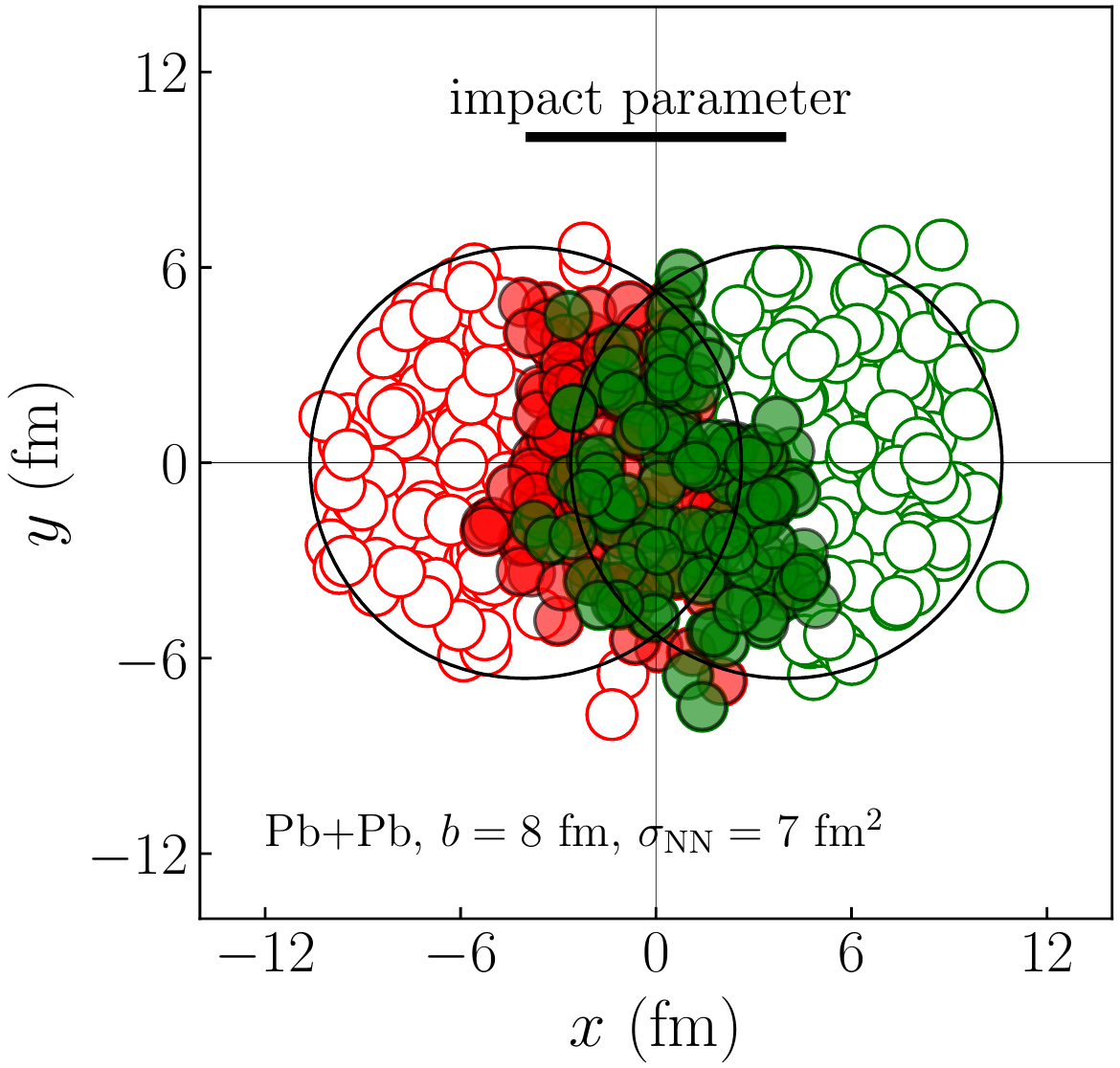}
    \caption{Transverse plane projection of a collision between $^{208}$Pb nuclei in the Glauber Monte Carlo model. The two nuclei are shown as circles of radius $R_A=6.62~{\rm fm}$, and are shifted by $\pm b/2$, with $b=8$~fm, along the $x$ direction. There is a total of $416$ nucleons, depicted as circles, and colored respectively in green or in red depending of their parent nucleus. The nuclei collide at $\sqrt{s_{\rm NN}}=5.02$~TeV, corresponding to $\sigma_{\rm NN}=7$~fm$^2$. Therefore, a green(red) nucleon is tagged as a participant nucleon whenever its distance from at least one red(green) nucleon is lower than $D=1.5$~fm, which corresponds in fact to the diameters of the small circles in the figure. Participant nucleons are highlighted as full symbols.}
    \label{fig:2-3}
\end{figure}

The probability distribution from which the impact parameter is generated on an event-by-event basis is proportional to $b$ for dimensional reasons, as I will discuss in detail below, in \equ{Pb}. Consider for the moment a collision occurring at a given value of $b$. Once the coordinates of the nucleons are sampled, one has then to shift them by $\pm b/2$ (e.g. along the $x$ axis). The result of such a procedure for $b=8$~fm is shown in ${\rm Fig.}~\ref{fig:2-3}$ in the case of a \pbpb{} collision. The two colliding nuclei are depicted as circles of radius $R_A=6.62$~fm, and are shifted along the $x$ direction by $\pm4$~fm. Each nucleus is associated with a collection of 208 nucleons, depicted respectively in red and in green. The geometry of the collision is thus determined.

\paragraph{Participant nucleons --}
What now? Now one has to select those nucleons that undergo an interaction, and that will be flagged as \textit{participant} nucleons. The idea is to look at the amount of overlap between pairs of nucleons, as follows. Pick a nucleon from a given nucleus, then check if there is at least one nucleon belonging to the other nucleus that lies within a certain distance, $D$. If yes, then the nucleon chosen at the beginning becomes a participant. 

The distance $D$ is determined by the collision energy under consideration. Since the nucleon-nucleon cross section increases with energy, the size of $D$ increases as well with the beam energy, or equivalently, with the center-of-mass energy of nucleon-nucleon interactions. This latter quantity is denoted by $\sqrt{s_{\rm NN}}$, and is equal to 200~GeV for collisions at top RHIC energy, while it is equal to 5.02~TeV for collisions at the current top LHC energy. The inelastic nucleon-nucleon cross section associated with these values of $\sqrt{s_{\rm NN}}$ is known from proton-proton collisions. One has in particular:
\begin{align}
\nonumber    \sqrt{s_{\rm NN}} = 200 ~{\rm GeV} \longrightarrow \sigma_{\rm NN} \simeq 4.2~{\rm fm}^2 , \\
\nonumber    \sqrt{s_{\rm NN}} = 2.76 ~{\rm TeV} \longrightarrow \sigma_{\rm NN} \simeq 6.4~{\rm fm}^2 , \\
\sqrt{s_{\rm NN}} = 5.02~ {\rm TeV} \longrightarrow \sigma_{\rm NN} \simeq 7.0~{\rm fm}^2.
\end{align}
Within the above-mentioned black-disk approximation for nucleon-nucleon interactions, the distance $D$ is therefore defined by:
\begin{equation}
    D = \sqrt{\sigma_{\rm NN}/\pi}.
\end{equation}

Let me go back, then, to \fig{2-3}. The nucleons represented as colored full symbols represent the participant nucleons of this specific event, i.e., those nucleons that are within a distance $\sqrt{7~{\rm fm}^2/ \pi}\approx1.5$~fm from at least one nucleon belonging to the other nucleus. Note that the size of the nucleonic balls in the figure corresponds precisely to a radius of 1.5~fm, which gives an accurate idea of the \textit{size} of a nucleon in a collision at LHC energy. The number of participant nucleons in a collision event is usually dubbed $N_{\rm part}$.

\subsection{Collision centrality}

\label{sec:2-22}

Experimentally, the impact parameter and the number of participant nucleons are not known. Collisions are sorted into classes of \textit{centrality} based on the amount of particles, or the amount of energy that they release. Clearly, a collision that produces a number of particles much higher than average corresponds to a collision in which there is a large overlap between the two nuclei, so that the number of detected particles and the impact parameter should be in a tight correlation.

\begin{figure}[t]
    \centering
    \includegraphics[width=\linewidth]{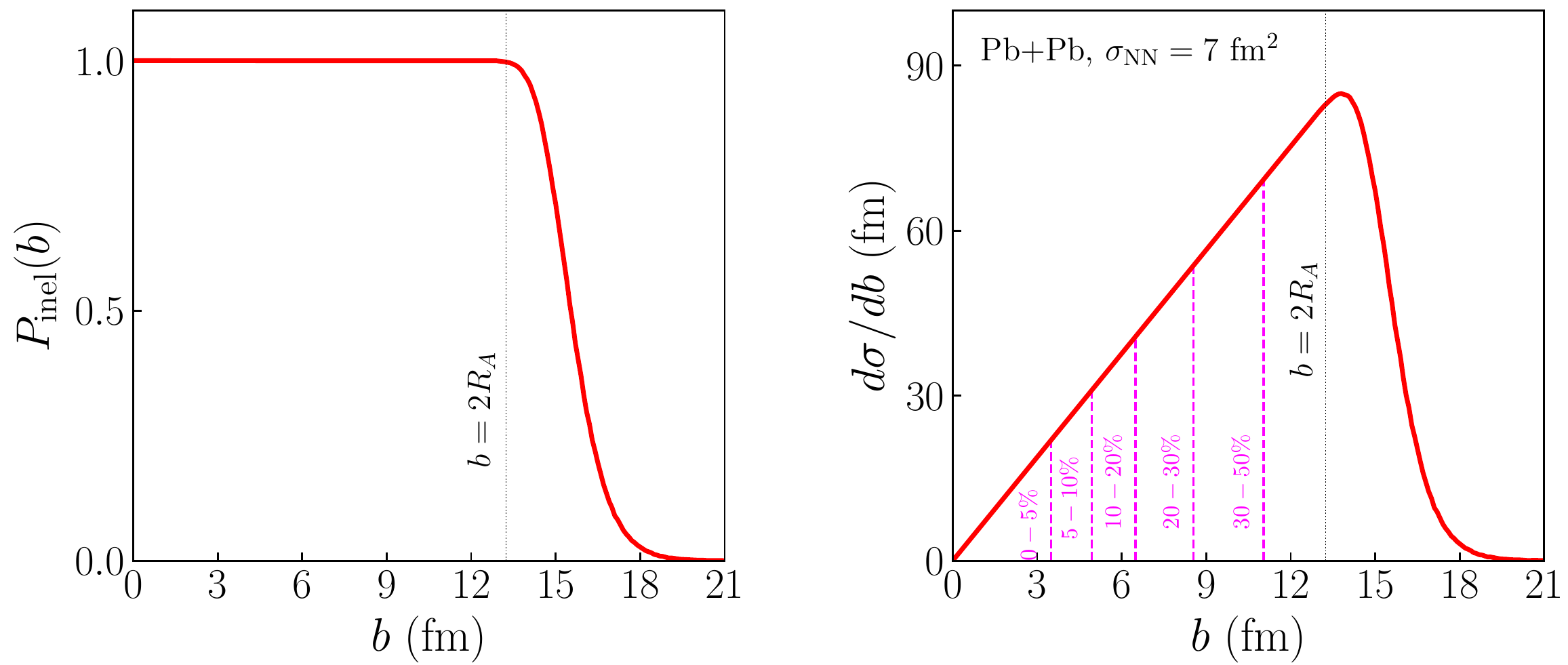}
    \caption{Left: fraction of events yielding an inelastic collision at a given impact parameter, in $^{208}{\rm Pb}+^{208}{\rm Pb}$ collisions at top LHC energy. Right: inelastic nucleus-nucleus cross section differential in $b$. This quantity can be used to define classes of true centrality, as highlighted by the vertical dashed lines. The probability distribution of $b$ is equal to $d\sigma/db$ divided by the total inelastic nucleus-nucleus cross section.}
    \label{fig:2-4}
\end{figure}

The collision impact parameter does indeed give the \textit{true} centrality of a collision. The probability density function of the impact parameter is given by the following formula:
\begin{equation}
\label{eq:Pb}
    P(b) = \frac{2 \pi b P_{\rm inel}(b)}{\sigma_{\rm inel}}.
\end{equation}
The quantity $P_{\rm inel}(b)$ is the fraction of events that yield an inelastic nucleus-nucleus collision at a given impact parameter. This quantity is plotted in \fig{2-4} for \pbpb{} collisions. Intuitively, $P_{\rm inel}$ is constant and equal to unity when $b<2R_A$, meaning that in this range of impact parameter it does never occur that two nuclei cross each other without yielding any inelastic nucleon-nucleon interactions. Beyond $b \approx 2 R_A$, $P_{\rm inel}$ decreases sharply. The fall off is not step-like because of quantum effects, i.e., the presence of participant nucleons at large impact parameter. When multiplied by $2 \pi b$, the numerator of \equ{Pb} corresponds to the probability for an inelastic nucleus-nucleus collision to occur at a given impact parameter, which I dub $d\sigma/db$. It is plotted in the right panel of \fig{2-4}. This quantity is a line of slope 2$\pi$ up to $b \approx 2 R_A$. The integral of $d\sigma/db$ over $b$ is equal to the total inelastic cross section for the nucleus-nucleus interaction, a quantity dubbed $\sigma_{\rm inel}$ in ${\rm Eq}.~(\ref{eq:Pb})$. This quantity is about $770~{\rm fm}^2$ in \pbpb{} collisions at LHC, and about $685~{\rm fm}^2$ in \auau{} collisions at RHIC.

The true centrality is then defined as the cumulative probability distribution of $b$. This means that if a collision occurs at impact parameter $b_*$, then its centrality is equal to:
\begin{equation}
\label{eq:cb}
    c_b (b_*) = \int_0^{b_*} P(b) db.
\end{equation}
Note that, as long as $b_* \leq 2R_A$, then one can consider that $P_{\rm inel}(b)$ in ${\rm Eq.}~(\ref{eq:Pb})$ is unity, and the centrality becomes:
\begin{equation}
\label{eq:cbb}
    c_b (b^*) = \frac{\pi b_*^2}{\sigma_{\rm inel}},
\end{equation}
which has a straightforward interpretation as a ratio of areas, explaining intuitively why collisions at large $b$ are more likely to occur than collisions at small $b$, and also clarifying the presence of the factor $2\pi$ in \equ{Pb}. The centrality is thus a number between $0$ and $1$. It is however more customary to express its value as a \textit{percentile}, i.e., from $0\%$ (central collisions) up to $100\%$ (peripheral collisions). Depending on their value of $c_b$, events can thus be sorted into classes of true centrality. See the right panel of \fig{2-4} for an illustration.

\begin{figure}[t]
    \centering
    \includegraphics[width=.60\linewidth]{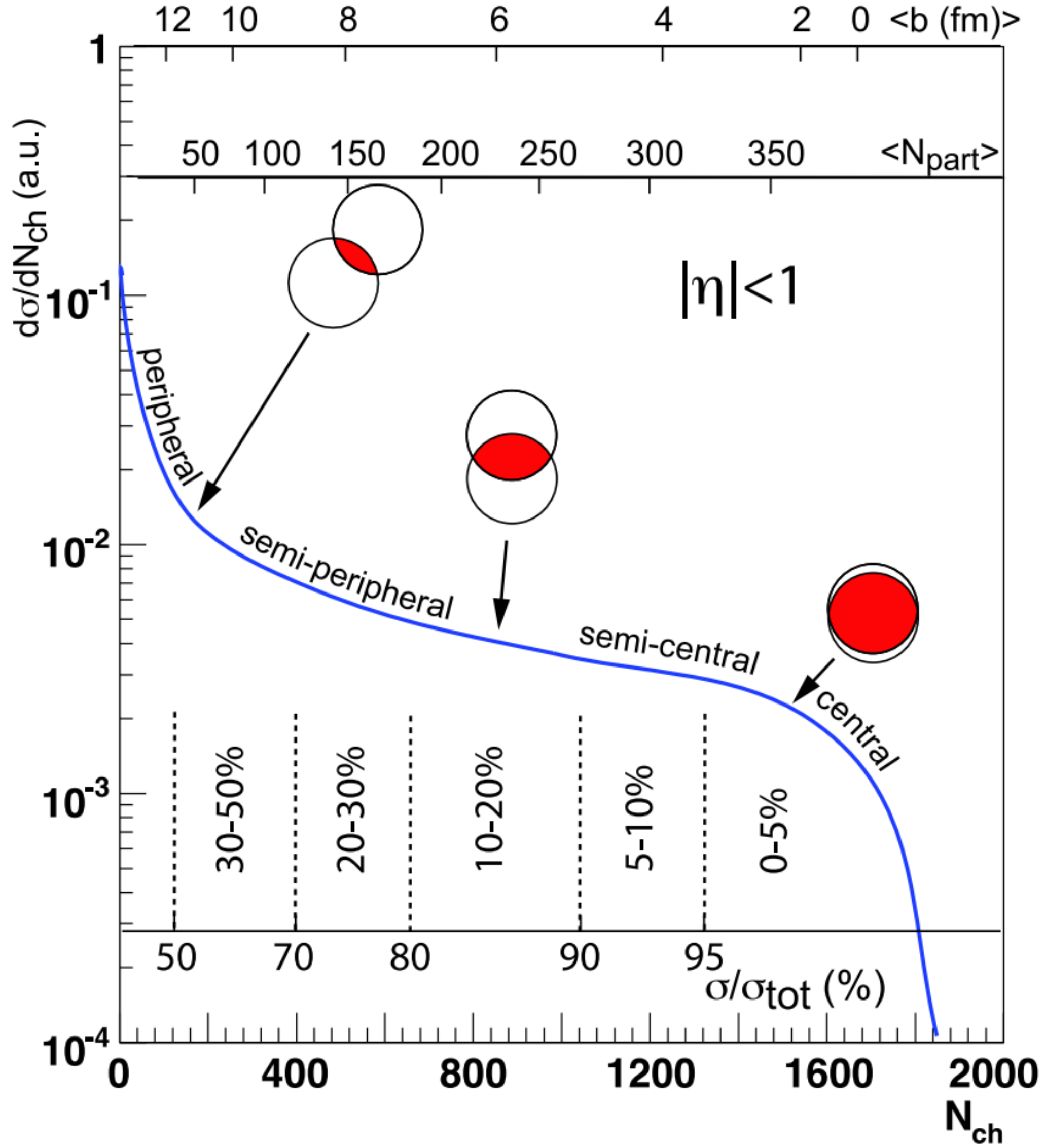}
    \caption{The solid line represents a theoretical evaluation of the histogram of charged-particle multiplicity observed in \pbpb{} collisions at LHC energy, within the pseudorapidity window $|\eta|<1$. Depending on the amount of produced particles, events are classified in centrality classes, corresponding to specific fraction of the total inelastic cross section. The Glauber model allows one to relate the measured histogram of multiplicity to the geometric properties of the interaction region, i.e., the impact parameter and the number of participant nucleons (as represented in the top of the figure). Note that a given experimentally-defined centrality corresponds to a distribution of both $b$ and $N_{\rm part}$. Figure from Ref.~\cite{Miller:2007ri}. }
    \label{fig:2-5}
\end{figure}

In an experiment, one does not know the impact parameter, so that the centrality has to defined through a different variable. I consider the case where the centrality of a collision is defined from the measured distribution of the final-state multiplicity of charged particles, dubbed $N_{\rm ch}$. An example of the histogram of $N_{\rm ch}$ resulting from \pbpb{} collision events is represented as a solid blue line in \fig{2-5}. The Glauber model does not provide a prescription to evaluate this histogram. This requires additional ingredients, as I shall discuss in the next section and in Chapter~\ref{chap:3}. Let me then assume for the time being that one has been able to reproduce the measured histogram of $N_{\rm ch}$ starting from a Glauber calculation. In \fig{2-5}, one immediately sees that collisions that yield small multiplicity have much larger probability than collisions at large multiplicity, which is expected from \fig{2-4} if $b$ and $N_{\rm ch}$ are (positively) correlated. The experimental definition of the collision centrality is given by the cumulative distribution of $N_{\rm ch}$. If one records a collision yielding multiplicity $N_*$, this corresponds to centrality:
\begin{equation}
\label{eq:cexp}
    c = \int_0^{N_*} P(N_{\rm ch}) dN_{\rm ch}.
\end{equation}
As illustrated in \fig{2-5}, one can play with the boundaries of integration in \equ{cexp} in order to define centrality classes from the histogram of $N_{\rm ch}$, much as I did for the histogram of $d\sigma/db$ in \fig{2-4}. 

Experimentally, to a given value of $N_{\rm ch}$ corresponds a distribution of both $b$ and $N_{\rm part}$, as also indicated in \fig{2-5}. Examples of plots of the distribution of $b$ at fixed $N_{\rm ch}$ can be found in Refs.~\cite{Das:2017ned,Rogly:2018ddx}. I stress that the correlation between the experimental $c$ and the true centrality, $c_b$, is in fact very tight~\cite{Broniowski:2001ei} for large collision systems, such as \pbpb{} or \auau{} collisions. The simple estimate of \equ{cbb} is in fact an excellent approximation as long as this correlation is tight. It breaks down in two cases. In very peripheral collisions, when $b$ approaches $2R_A$, which corresponds to $c \sim 70\%$ experimentally. The correlation between $c$ and $c_b$ is further lost in ultracentral collisions, when the impact parameter is essentially as large as the size of a couple of nucleons, and quantum fluctuations become the dominant effect in the determination of the system geometry. This is typically the case when $b<2~\fm$, corresponding to $c<2\%$.

\section{Hydrodynamic framework}

\label{sec:2-3}

I have thus completed the first part of my end-to-end description of a heavy-ion collision. I explained how a collision looks like experimentally, and how its geometry is described by a few simple quantities within the Glauber Monte Carlo model. Now, I want to use the output of the Glauber model as an input for hydrodynamics. The hydrodynamic expansion of the system will bring me from the participant nucleons in the region of overlap to the final-state detected hadrons.

\subsection{The initial density profile}

\label{sec:2-31}

The previous discussion left me with the picture of \fig{2-3}, i.e., with a bunch of coordinates in the $(x,y)$ plane corresponding to the location of the participant nucleons. To move on, I need to turn that information into a continuous density profile that may serve as the initial condition for the hydrodynamic expansion.

\paragraph{Thermalization --} Hydrodynamics applies if the system is in (or at least close to) thermal equilibrium. There must exist, hence, a short phase of \textit{thermalization} which brings one from the out-of-equilibrium system produced immediately after the collision to a thermal medium~\cite{Schlichting:2019abc}. The issue of thermalization in hot QCD matter is fascinating, and is made particularly timely by the observation that the hydrodynamic description of heavy-ion collisions works very well in practice. A vast literature is devoted to this problem~\cite{Berges:2020fwq}, and numerical frameworks, such as \kompost{}~\cite{Kurkela:2018vqr}, have been recently developed to include the short pre-equilibrium phase in the theoretical simulations of the collision evolution.

That being said, one should keep in mind that the inclusion of a phase of thermalization in collisions of large nuclei, while certainly needed for a complete description of the collision process, is typically of poor relevance for the phenomenological output. A central collision between large nuclei produces a system with a transverse size of order 10~fm, which is much larger than the thermalization time. As the subsequent hydrodynamic expansion is driven by pressure gradients determined by the large-scale structures of the system, i.e., structures that are significantly larger than 1~fm/$c$, the expansion has essentially no sensitivity to features produced over the short thermalization period. This is good news: if the impact of the pre-equilibrium phase were crucial in the determination of the final-state observables, then one would have a hard time performing simulations that quantitatively describe experimental data.

This situation is however different in so-called \textit{small systems}, like proton-nucleons ($pA$) and proton-proton ($pp$) collisions, or even peripheral nucleus-nucleus collisions. Recent experimental measurements show that high-multiplicity $pp$ and $pA$ collisions exhibit the same kind of collective phenomena observed in nucleus-nucleus collisions, pointing to the fact they may also reach thermal equilibrium. However, in these systems the transverse size of the medium and the thermalization time are comparable, so that corrections coming from the pre-equilibrium dynamics can in principle be significant. The problem of thermalization in small systems seems, thus, more compelling, because it may be important for understanding quantitatively the experimental observations.
 
\paragraph{The initial condition --}  For collisions of large nuclei, one can thus either model the initial condition of the hydrodynamic expansion directly, or following a short pre-equilibrium phase. In all cases, one needs a \textit{good} model.  In view of recent developments in theory-to-data comparisons, and after years of playing with models of initial conditions, a \textit{good} model of initial conditions, consistent with essentially all the phenomenology of heavy-ion collisions, can be obtained as follows.

Recall that, after the collision takes place, one is left with a bunch of coordinates for the participant nucleons. I consider now that each nucleon carries a density of participant matter. Suppose that a nucleon in the rest frame of the nucleus is described by a matter density $\rho_n (x,y,z)$. Then, in the laboratory frame, assuming an infinitely strong Lorentz boost, the nucleon becomes a transverse density, or \textit{thickness} function:
\begin{equation}
\label{eq:boosted}
    \rho_n ({\bf x}) = \int \rho_n({\bf x},z)dz.
\end{equation}
The standard prescription for the boosted nucleon density is that of a Gaussian, i.e.,
\begin{equation}
\label{eq:rhonboost}
    \int_z \rho_n({\bf x},z)dz = (2\pi w)^{-1/2} \exp \left(  -\frac{{\bf x}^2}{2w^2} \right)  ,
\end{equation}
which is normalized to return $1$ upon integration over the transverse plane, thus representing the contribution from \textit{one} participant nucleon. A participant nucleon is randomly located within the nucleus, so that its associated density is off-centered:
\begin{equation}
    \rho_i({\bf x}) = (2\pi w)^{-1/2} \exp \left(  -\frac{({\bf x}-{\bf x}_i)^2}{2w^2} \right) .
\end{equation}
The output of the Glauber model is essentially the set of coordinates ${\bf x}_i$. The Gaussian width, $w$, is not a feature of the Glauber model. It is generically chosen to be close to $0.5~\fm$. 

I label the colliding nuclei with letters $A$ and $B$. For each nucleus I construct a density of participant matter as follows:
\begin{equation}
\label{eq:tA}
    t_A = \sum_i \lambda_i \rho_{i,A},
\end{equation}
where I introduce a normalization, $\lambda_i$, that allows one to include in the model the possibility that certain participant nucleons contribute to the density more than others. Finally, the density profile of the system is a function of the kind:
\begin{equation}
\label{eq:density}
  \left (  t_A t_B \right)^\nu,
\end{equation}
where $0<\nu\leq1$. A few comments are in order:
\begin{itemize}
    \item By setting $\nu=1$ one obtains an initial density which is given by the sum of pairwise interactions between nucleons. This corresponds to the recently-developed {\small IP-JAZMA} model~\cite{Nagle:2018ybc}. Note that the amount of density released by a given nucleon-nucleon interaction depends essentially on the amount of overlap between the two participant nucleons, which sounds reasonable. 
    \item The prescription with $\nu=1$ is further consistent with the expectation of the color glass condensate (CGC) effective theory~\cite{Iancu:2003xm,Gelis:2010nm} of high-energy QCD~\cite{Kovchegov:2012mbw}. An important predictions made by this theoretical framework, first put into simple formulas by Lappi~\cite{Lappi:2006hq}, and that should nowadays be considered as textbook material (see Problem~11.8 in the recent quantum field theory textbook by Gelis~\cite{Gelis:2019yfm}), states indeed that, if $\tau=0^+$ is the time right after two infinitely-boosted nuclei cross each other, then the average energy density of the system at $\tau=0^+$ is proportional to $t_At_B$. The numerical framework which performs high-energy nuclear collisions following the prescriptions of the CGC is called \ipglasma{}~\cite{Schenke:2012wb,Schenke:2012hg}, a detailed description of which can be found in Ref.~\cite{Schenke:2020mbo}. Note that the density returned by \ipglasma{} is not strictly equivalent to that returned by the {\small IP-JAZMA} model. There are quantitative differences, because of the inclusion of additional features, in particular, sources of fluctuations in the system related to the sub-nucleonic structure of the colliding nuclei. Furthermore, the density profile associated with a nucleus is more complicated than a simple linear superimposition of nucleons, as it presents a dependence on the Bjorken-$x$ variable~\cite{Schenke:2020mbo}.
    \item If one considers that the entropy density of the system at the onset of the hydrodynamic behavior is proportional to \equ{density} with $\nu=1/2$, i.e., $\sqrt{t_At_B}$, then one obtains what I shall refer to as the \trento{} model. This model was developed by the Duke group~\cite{Moreland:2014oya}. It has been used in particular to perform comprehensive theory-to-data comparisons with the aim of inferring the most probable parameters of the model, and their mutual correlations, by means of a Bayesian analysis~\cite{Bernhard:2016tnd}. The model uses a density of the form:
    \begin{equation}
    \label{eq:generalized}
        \biggl ( \frac{t_A^p + t_B^p}{2} \biggr )^{1/p},
    \end{equation}
    which is a generalized mean. The results of the Bayesian analysis show very clearly that $p=0$ yields the best description of data. This is equivalent to the anticipated geometric mean:
    \begin{equation}
        \sqrt{t_At_B}.
    \end{equation}
    Note that this is the only combination of the kind $(t_At_B)^\nu$ that can be returned by this model. In particular, the scaling of the CGC, $\nu=1$, is not allowed by \equ{generalized}. However, let me emphasize that having $\nu=1/2$ for the entropy density at the onset of hydrodynamics is not fully at variance with $\nu=1$ for the energy density created at $\tau=0^+$. An analysis of scaling laws under the assumption of conformal symmetry shows that the process of thermalization does modify the exponent of \equ{density} as follows. If $\tau_0$ is the time at the beginning of the thermalization process, and $\tau_{\rm hydro}$ is the time at which hydrodynamics becomes applicable, then one has~\cite{Giacalone:2019ldn}:
    \begin{equation}
    \left ( \tau_{\rm hydro} s({\bf x},\tau_{\rm hydro}) \right )  \propto      \left (\tau_0\epsilon({\bf x},\tau_0) \right )^{2/3}.
    \end{equation}
    Note that, strictly speaking, $\tau_0$ is larger than $0^+$. However, assuming that $\tau_0\epsilon({\bf x},\tau_0)$ in the right-hand side remains proportional to $(t_At_B)$, which within the \ipglasma{} framework is in fact a good approximation in central collisions~\cite{Lappi:2006xc}, then the left-hand side becomes proportional to $(t_At_B)^{2/3}$. One sees that this is not too distant from $\sqrt{t_At_B}$, and shows that the great effectiveness of the \trento{} Ansatz may in fact be motivated by deeper arguments.
    \item However, in their latest Bayesian analyses~\cite{Moreland:2018gsh,Bernhard:2019bmu}, the Duke group included in the \trento{} framework the pre-hydrodynamic phase of the system, modeled as a purely free-streaming evolution. When doing so, the prescription of \trento{} turns from $\nu=1/2$ for the entropy density of the system at the beginning of hydrodynamics, to $\nu=1/2$ for the energy density of the system at $\tau=0^+$. This creates an inconsistency with the \ipglasma{} framework, that corresponds essentially to $\nu=1$, and thus a more localized profile on average. It would be useful to understand whether this inconsistency is actually required to improve the description of data in the \trento{} framework, or whether it is a mere model artifact, due to the fact that this model, starting with a generalized mean, can only return $\nu=1/2$ by construction. My suggestion for future Bayesian analyses is to constrain the shape of the density with a function of the form $(t_At_B)^\nu$. A nontrivial confirmation of the CGC picture will be achieved if the experimental data turns out to favor $\nu\approx1$.
\end{itemize}
The prescription of the CGC implies that the production of energy is a \textit{coherent} process, in the sense that the energy density is given by the sum of contributions coming from individual nucleon-nucleon interactions. This feature seems rather unattackable. Exponents $\nu<1$ in \equ{density} do not really modify this statement, but they yield a modification of the geometry of the whole collision system. At the very first instant after the interaction takes place, and considering that the interaction is ultrarelativistic, such effects seem difficult to justify, because scattering processes between quarks and gluons are typically localized semi-hard processes. The prescription with $\nu=1$ predicted by the CGC seems in general the only one that makes sense for the condition of the system immediately after the interaction occurs.

I pick now a model to exhibit a realistic example of initial condition for hydrodynamics. The prescription I shall use throughout this manuscript is the \trento{} model used in the first Bayesian analysis of the Duke group~\cite{Bernhard:2016tnd}. I consider that the entropy density at the onset of hydrodynamics ($\tau=\tau_0$) is obtained by setting $\nu=1/2$ in \equ{density}, i.e.,
\begin{equation}
\label{eq:p=0}
    s({\bf x}, \tau_0) = \frac{N_0}{\tau_0}  \sqrt{t_At_B},
\end{equation}
where $N_0$ is a global dimensionless factor that fixes the total amount of entropy in the system, which is determined by the collision energy. Further, I consider that $\lambda_i$ in \equ{tA} is randomly chosen for each participant nucleon according to the following gamma distribution:
\begin{equation}
    P(\lambda; k) =  \frac{k^k}{\Gamma(k)} \lambda^{k-1} e^{- \lambda k},
\end{equation}
which has unit mean and variance equal to $1/k$.

\begin{figure}[t]
    \centering
    \includegraphics[width=.6\linewidth]{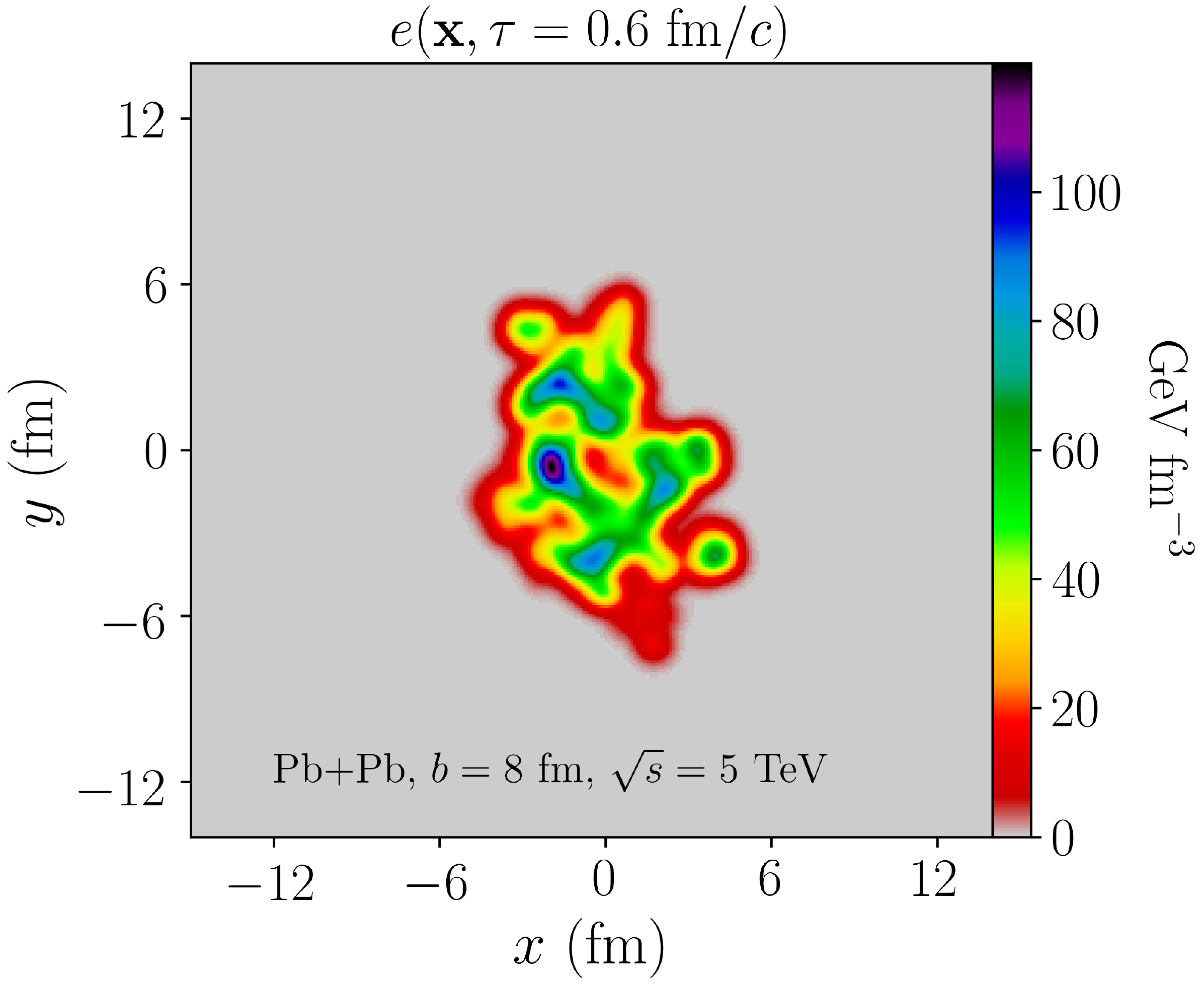}
    \caption{Profile of energy density at the onset of hydrodynamics, here occurring at $\tau=0.6~\fmc$, for a semi-peripheral ($b=8~\fm$) $5.02~{\rm TeV}$ \pbpb{} collision. The colliding nuclei and the participant nucleons are the same as in \fig{2-3}. The energy density profile is obtained with the \trento{} prescription, i.e., by applying \equ{p=0} and subsequently \equ{hoteos}.}
    \label{fig:2-6}
\end{figure}

Hydrodynamic equations are typically solved in terms of the energy density of the system, rather than the entropy density. By use of the equation of state of hot conformal QCD, I transform the entropy density given by \equ{p=0} into an energy density:
\begin{equation}
\label{eq:hoteos}
   \epsilon({\bf x}, \tau_0)= s({\bf x}, \tau_0)^{4/3} \left ( \frac{3}{4} \right )^{4/3} \left( \nu_{\rm QCD} \frac{\pi^2}{30} \right)^{-1/3},
\end{equation}
with a number of degrees of freedom $\nu_{\rm QCD}=40$ corresponding to a plasma of gluons, and two light quarks. Calculating the entropy density in \equ{p=0} from the participant nucleons shown in \fig{2-3}, \equ{hoteos} leads to the profile of energy density which is displayed in \fig{2-6}. This profile represents as an example of quark-gluon plasma created in a semi-peripheral \pbpb{} collision at top LHC energy. One should note that the density profile is by no means uniform nor smooth in the transverse plane. The energy density can vary by almost one order of magnitude within short length scales, of order e.g. 2~fm. This feature reflects the quantum nature of nuclei, and in the Glauber paradigm of nuclear collision it originates mostly from the random spatial positions of the colliding nucleons.

\paragraph{Longitudinal structure --} The expression in \equ{density} gives the initial density in the transverse plane, but what about its longitudinal structure along $z$? The important observation is that, in experimental data, the particle yields are essentially flat~\cite{ATLAS:2011ag,Chatrchyan:2011pb,Adam:2016ddh} as functions of the rapidity, $y$, given in \equ{rapidity}. This suggests that the particle production mechanism is almost independent of the longitudinal coordinate. One can motivate this observation by means of simple yet solid arguments.

The interaction between two nuclei at ultrarelativistic energy scarcely slows the interacting particles. The particles over the interaction region carry the same constant longitudinal velocity, $v_z$. A particle located at $z$ has then velocity $z/t$, where $t$ is the time in the laboratory frame and $t=0$ is the time of interaction. Hence if one performs a Lorentz boost along the $z$ direction, both $z$ and $t$ gets modified, but the value of $v_z=z/t$ is unchanged, i.e., the system is invariant under boosts along $z$. This argument was first pointed out by Bjorken~\cite{Bjorken:1982qr}. 

This picture implies that, if we define:
\begin{equation}
\label{eq:tauetas}
    \tau = \sqrt{t^2 - z^2}, \hspace{40pt} \eta_s = \frac{1}{2}\ln \frac{t+z}{t-z},
\end{equation}
which are called, respectively, \textit{proper time} and \textit{space-time rapidity}, then a boost along the $z$ direction, with coordinates $(\tau,{\bf x},\eta_s)$, leaves $\tau$ unchanged and shifts $\eta_s$ by a constant. The previous picture of boost-invariant particle production along $z$ means now that the dynamics of the system is independent of $\eta_s$. This has a nice implication. Recalling that a given particle has $v^z=p^z/E$, where $E$ is the energy of the particle, and considering that the transverse momentum of a particle emitted by the fluid contribute only a negligible amount to the rapidity of a particle, one obtains that the space-time rapidity $\eta_s$ in \equ{tauetas} coincides with the particle rapidity $y$ in \equ{rapidity}.

Hence the initial three-dimensional density profile of a heavy-ion collision is usually specified as a function of $\tau$, ${\bf x}$ and $\eta_s$, and hydrodynamic simulations are performed using these coordinates. I shall always omit the longitudinal coordinate in the following, and consider that the medium is boost invariant, i.e., the evolution of the system and the final spectrum are the same at all values of $\eta_s$. This is good enough for the practical purposes of this manuscript. I dub $\epsilon({\bf x},\tau)$ or $s({\bf x},\tau)$, respectively, the transverse energy density and transverse entropy density of the system at midrapidity, $y=\eta=\eta_s=0$. 

\subsection{Fluid expansion}

\label{sec:2-32}

The energy density at the onset of hydrodynamics in the \trento{} model for the collision considered in \fig{2-3} is thus depicted in \fig{2-6}. The subsequent expansion is ruled by conservation laws. At ultrarelativistic energy, one can safely assume the the density of baryons vanishes in the hot quark-gluon medium, due to the very large number of produced particles. The dynamics is thus ruled solely by the conservation of energy and momentum. The associated currents can be written in the form a rank-$2$ tensor, the so-called energy-momentum tensor:
\begin{equation}
    T^{\mu\nu} =
    \begin{pmatrix}
T^{00} & T^{01} & T^{02} & T^{03} \\
T^{10} & T^{11} & T^{12} & T^{13} \\
T^{20} & T^{21} & T^{22} & T^{23} \\
T^{30} & T^{31} & T^{32} & T^{33} .
\end{pmatrix}
\end{equation}
$\nu$ labels the components of the $4$-momentum, while $\mu$ labels the associated current. Thus $T^{00}$ is the density of energy; $T^{0i}$ is the density of the $i$-th component of the momentum; $T^{i0}$ is the flux of energy along direction $i$; $T^{ij}$ is the flux of $j$-th component of the momentum along direction $i$. $T^{11}$ and $T^{22}$ represent the so-called transverse pressure, $P_T$, while $T^{33}$ is the longitudinal pressure, $P_L$. 

Since one can always characterize the system by means of the energy-momentum tensor, it is useful to have an idea of what $T^{\mu\nu}$ looks like at various stages during the evolution, even before hydrodynamics is applicable. I shall use $g^{\mu\nu}={\rm diag}(1,-1,-1,-1)^{\mu\nu}$.

Immediately after the interaction of two nuclei, at $\tau=0^+$, the stress-energy tensor can be computed within the framework of the color glass condensate. According to the CGC, at $\tau=0^+$ the system is amenable to a description in terms of classical chromodelectric and chromomagnetic fields. This  system is dubbed \textit{glasma}~\cite{Lappi:2006fp}. The semi-classical methods of the CGC allow one to evaluate the full field strength tensor of the system, $F^{\mu\nu}$, which can then be used to derive the energy-momentum tensor, leading to~\cite{Lappi:2006fp}:
\begin{equation}
    T^{\mu\nu} (\tau=0^+) =
    \begin{pmatrix}
\epsilon & 0 & 0 & 0 \\
0 & \epsilon & 0 & 0 \\
0 & 0 & \epsilon & 0 \\
0 & 0 & 0 & -\epsilon 
\end{pmatrix}.
\end{equation}
The stress-energy tensor is diagonal, with all entries equal to the energy density, $\epsilon$. The distinctive feature of the glasma energy-momentum tensor is the longitudinal pressure, which comes with a negative sign, $P_L=-\epsilon$. The glasma evolves in time according to classical Yang-Mills equations, whose computation is also included in the \ipglasma{} framework. Due to the negative pressure, the total energy of the system increases during this evolution, by an amount which is proportional to $\tau$.  However, this lasts only for a short time. The longitudinal pressure does in fact vanish very quickly, on a time scale of order 0.1~fm/$c$~\cite{Gelis:2013rba}. This corresponds also to the time at which the classical fields lose coherence, and the system becomes amenable to a particle description.

The classical Yang-Mills phase is thus expected to be followed by a kinetic theory description, for the dynamics of hard quasi-particles that carry most of the energy of the system. The system is now associated with a phase space density $f(\tau, {\bf x}, {\bf p})$, and the components of the energy-momentum tensor correspond to moments of this distribution:
\begin{equation}
    T^{\mu\nu}= \int \frac{d^3 {\bf p}}{(2\pi)^3 p^0} p^\mu p^\nu f(\tau, {\bf x}, {\bf p}).
\end{equation}
The system thus follows a Boltzmann equation within the boost-invariant picture of Bjorken. Assuming conformal symmetry, which implies that $T^{\mu\nu}$ is traceless, and recalling that the longitudinal pressure is negligible at the onset of the kinetic theory description, the energy-momentum tensor must be close to:
\begin{equation}
\label{eq:tmunu2}
    T^{\mu\nu} (\tau\approx0.1~\fm) \approx
    \begin{pmatrix}
\epsilon & 0 & 0 & 0 \\
0 & \epsilon/2 & 0 & 0 \\
0 & 0 & \epsilon/2 & 0 \\
0 & 0 & 0 & 0 
\end{pmatrix}.
\end{equation}
If this energy-momentum tensor were obtained following the classical Yang-Mills phase of the \ipglasma{} framework, then \equ{tmunu2} would also contain off-diagonal terms, which are however small corrections for a large system. The explicit form of $T^{\mu\nu}$ in \equ{tmunu2} gives useful insight about the actual role of the thermalization process. Thermodynamic equilibrium is reached when the particle density is locally isotropic, i.e., when the longitudinal and the transverse pressure are equal. Thermalization is thus a process that builds up the longitudinal pressure, and which leads from $P_T=\epsilon/2$ and $P_L=0$ to $P \equiv P_L=P_T$, within a time span of order 1 fm/$c$.

Finally, at equilibrium the form of the energy-momentum tensor can be guessed from the equation of state of QCD at high temperature. High-temperature QCD has in particular $P = \frac{1}{3}\epsilon$, i.e., a speed of sound squared $c_s^2 \equiv \partial P / \partial \epsilon = 1/3$.  At equilibrium, the system is locally isotropic, and thus the form of $T^{\mu\nu}$ in the local rest frame should be close to:  
\begin{equation}
    T^{\mu\nu} (\tau \approx 1~{\rm fm} ) \approx
    \begin{pmatrix}
\epsilon & 0 & 0 & 0 \\
0 & \epsilon/3 & 0 & 0 \\
0 & 0 & \epsilon/3 & 0 \\
0 & 0 & 0 & \epsilon/3 
\end{pmatrix}.
\end{equation}
Once again, this neglects any effect coming from the physics of the first fm/$c$, which produces nonzero values for the off-diagonal terms.

I move on, then, to a brief discussion of the equations of motion that rule the evolution of $T^{\mu\nu}$ once the hydrodynamic phase sets in. 

\paragraph{Ideal hydrodynamics --} I first consider the case of an ideal fluid. Assuming that the effect of the pre-equilibrium dynamics can be neglected, the energy-momentum tensor at the beginning of hydrodynamics (in the fluid rest frame) is of the form:
\begin{equation}
    T^{\mu\nu} (\tau_0) \approx
    \begin{pmatrix}
\epsilon & 0 & 0 & 0 \\
0 & P(\epsilon) & 0 & 0 \\
0 & 0 & P(\epsilon) & 0 \\
0 & 0 & 0 & P(\epsilon) 
\end{pmatrix}
\end{equation}
where the pressure $P$ is related to the energy density via the equation of state. Now, the conservation of energy and momentum is written as:
\begin{equation}
\label{eq:conserv}
    \partial_\mu T^{\mu\nu} = 0,
\end{equation}
where there is a summation over repeated indices. Further, the energy-momentum tensor satisfies the following covariant equation:
\begin{equation}
\label{eq:tmunu}
    T^{\mu\nu} = (\epsilon+P)u^\mu u^\nu-P g^{\mu\nu},
\end{equation}
where $u^\mu$ is the $4$-velocity of the fluid, with $u^\mu u_\mu = 1$. Combined with the equation of state, \equ{conserv} and \equ{tmunu} form a closed system of equations. The dynamics of all the degrees of freedom of the fluid can thus be solved (at least numerically).

The fluid expansion decreases the temperature of the fluid elements until the point where the quark-gluon plasma description is no longer justified, i.e., parton confinement sets in and the system becomes a gas of hadrons. This occurs at the so-called \textit{critical}, or \textit{freeze-out}  temperature, $T_c$. which is of order $T_c\simeq0.15~{\rm GeV}$. In numerical codes, the freeze-out is typically implemented following a Eulerian approach. The fluid is discretized over a space-time grid. At each $\tau$, one looks at the temperature of a given fluid cell. If the temperature is below $T_c$, then one records that, at that value of $\tau$, that cell, corresponding to coordinates $(x,y)$ in space, has frozen out. Once the whole fluid has frozen out, one is left with a so-called \textit{freeze-out hypersurface}, i.e., the isothermal hypersurface corresponding to $T=T_c$.

\begin{figure}[t]
    \centering
    \includegraphics[width=\linewidth]{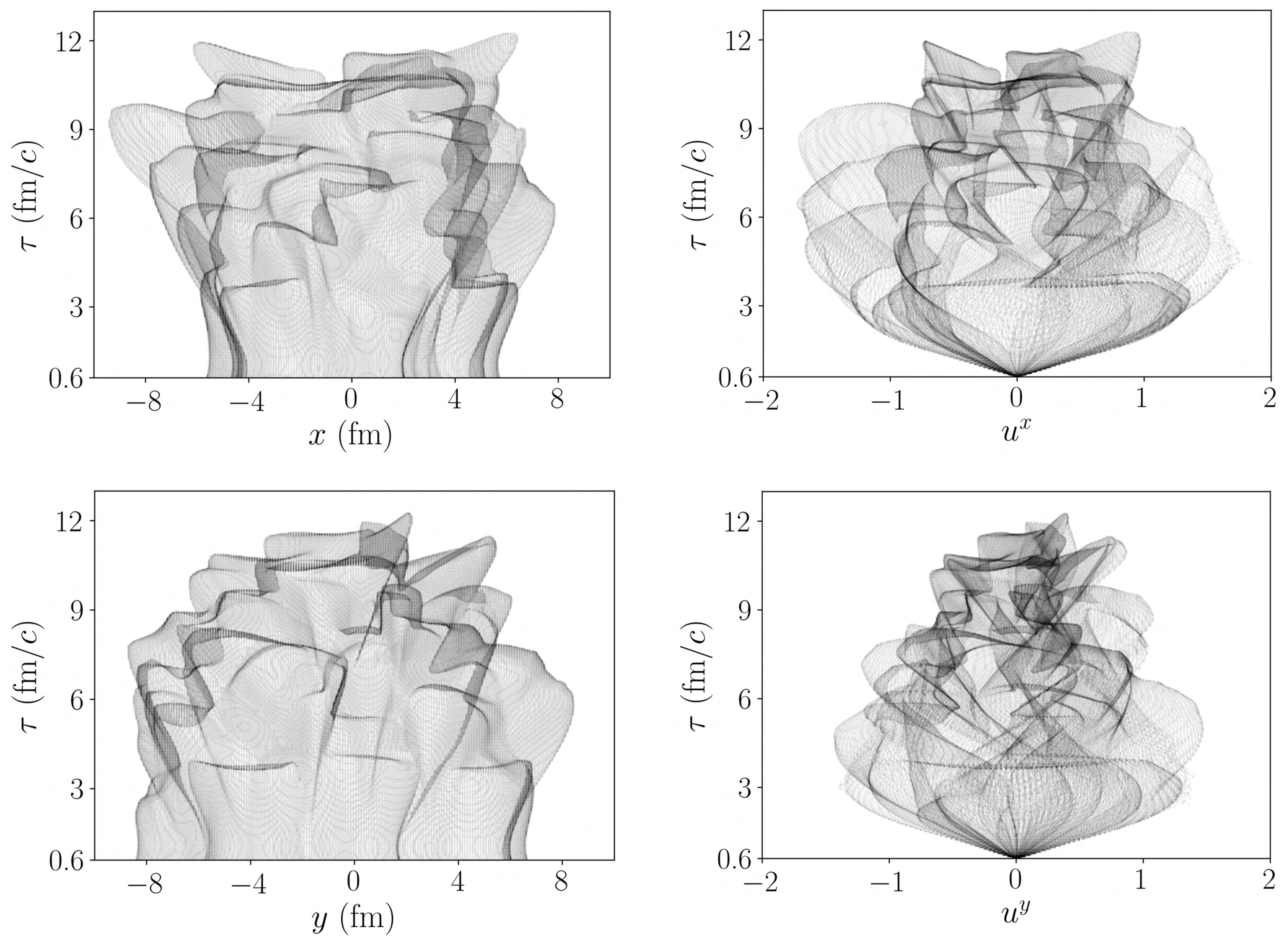}
    \caption{Freeze-out hypersurface resulting from the ideal hydrodynamic evolution of the energy density profile shown in \fig{2-6}. Left: spatial coordinates,  $x$ (top) and $y$ (bottom), of the fluid cells at freeze-out, i.e., at the time when the local temperature reaches $T_c=0.150$~GeV. Right: velocity coordinates, $u^x$ (top) and $u^y$ (bottom).}
    \label{fig:2-7}
\end{figure}

I show now an example of such a hypersurface. To do so, I take the initial condition profile given in \fig{2-6}, I assume that it corresponds to the initial condition of hydrodynamics at $\tau=0.6~\fmc$, and I evolve it with ideal fluid dynamic equations by means of the \music{} hydrodynamic code~\cite{Schenke:2010nt,Schenke:2010rr,Paquet:2015lta}. The medium has the equation of state of QCD~\cite{Borsanyi:2013bia}, and it freezes out at a temperature $T=150~{\rm MeV}$. In \fig{2-7}, I show projections of the resulting freeze-out hypersurface. Following the standard visualization of this surface that one can find in the literature, I show how $\tau$, i.e., the time at which a given cell freezes out, depends on $x$, $y$, $u^x$, and $u^y$. The simulation is boost-invariant, hence, $u^{\eta_s}=0$, and the shape of the surface is independent of $\eta_s$. Let me point out a few features. On the left of \fig{2-7}, I show $\tau$ as a function of $x$ and $y$. We can see that the initial system at $\tau=0.6~\fmc$ is more elongated in the $y$ direction (bottom panel) with respect to the $x$ direction (upper panel), reflecting the fact that I am looking at a collision occurring with a large impact parameter, $b=8~\fm$ (see \fig{2-6}). Second, one can distinctly appreciate a difference in the pattern of the flow velocities. The upper-right panel, displaying $u^x$ is significantly broader than the pattern in the lower-right panel, displaying $u^y$, where only few fluid elements have velocity larger than unity. This phenomenon corresponds precisely to the \textit{elliptic flow} discussed in Chapter~$\ref{chap:1}$, on which I shall return in greater detail in Chapter~$\ref{chap:3}$.

The freeze-out hypersurface has now to be converted into a distribution of hadrons, corresponding to the experimental observations. The idea is simply that the momentum distribution of outgoing particles leaving the fluid is the same as the distribution of particles within the fluid at the end of the hydrodynamic phase. In ideal hydrodynamics, the momentum distribution of a given species $a$ is simply a thermal distribution, $f_{\rm th}({\bf x},{\bf p})$ ($k_B=1$):
\begin{equation}
    \frac{dN_a}{d^3{\bf x} d^3{\bf p}} \propto  g_a f_{\rm th}({\bf x},{\bf p}) \propto  g_a \biggl [  1 \pm \exp \left(  - \frac{E^a}{T_c} \right ) \biggr ]^{-1},
\end{equation}
where $g_a$ is the spin degeneracy of the species $a$, $E^a = p^\mu u_\mu$ is the energy of the particle that has $4$-momentum $p^\mu$, and the $\pm$ sign in the denominator depends on whether $a$ is a fermion or a boson.  For each species, the total momentum distribution is obtained by integrating over the hypersurface:
\begin{equation}
    E_a \frac{dN_a}{d^3 {\bf p}} =  \frac{dN_a}{dy d^2{\bf p}_t} = g_a \int f_{\rm th} p^\mu d\Sigma_\mu, 
\end{equation}
where $d\Sigma_{\mu}$ is the vector normal to the freeze-out hypersurface. This integral is referred to as the Cooper-Frye formula~\cite{Cooper:1974mv}.

The outcome of the freeze-out integral is a spectrum of hadrons in momentum space, which corresponds now to the experimental observable. A final comment is however in order. The spectrum resulting from the thermal distributions of hadrons can not be compared directly to the measured one. The reason is that all kinds of hadrons are emitted from the quark-gluon plasma. Many of these hadrons undergo strong decays and only their decay products are actually observed in the detector. One needs to include the decays of resonances in the final spectrum. The effect of resonance decays is very significant, for instance, the number of stable light hadrons (pions, kaons, protons) emitted thermally at freeze-out is only half its actual value after all unstable resonances have decayed. One can also include an intermediate phase between the quark-gluon plasma and the gas of free-streaming hadrons, which, while performing resonance decays, computes as well the scattering processes that occur in the hadron gas. Codes devoted to this task are, e.g, SMASH~\cite{Weil:2016zrk} or UrQMD~\cite{Bass:1998ca,Bleicher:1999xi}. Including the rescattering of hadrons has however a minor effect on the phenomenology. It helps though hydrodynamic simulations get the right value of average transverse momentum for the heavier detected species, like protons.

\paragraph{Viscous hydrodynamics --}  
The previous discussion is valid for an ideal fluid, i.e., an inviscid medium in which there is no heat diffusion between fluid cells. However, shortly after the beginning of the heavy-ion program at RHIC, and the detection of elliptic flow,  theoretical~\cite{Teaney:2003kp,Romatschke:2007mq} studies concluded that viscous corrections, in particular the presence of a small shear viscosity of the medium, do in fact yield sizable effects on the measured elliptic flow, and thus play a role in the experimental observations.

If one aims at a quantitative understanding of data, viscous corrections to the evolution of the quark-gluon plasma have indeed to be taken into account. Constraining the viscous properties of the quark-gluon plasma from experimental data represents one of the main goals of the heavy-ion collision program.

I combine now arguments by Teaney~\cite{Teaney:2009qa} and Ollitrault~\cite{Ollitrault:2010tn} to show that the viscosity of the quark-gluon plasma is small. A viscous (nonrelativistic) fluid satisfies the Navier-Stokes equation:
\begin{equation}
\label{eq:NS}
    \rho \frac{d{\bf v}}{dt} = - \vec \nabla P + \eta \nabla^2 {\bf v},
\end{equation}
where $d/dt = \left   (   \frac{\partial}{\partial t} + {\bf v} \cdot \vec \nabla  \right )$ is the so-called material derivative, and  $\eta$ is the shear viscosity of the medium. This coefficient is zero in a perfect fluid, where the previous equations is simply equivalent to the statement that the dynamics is governed by pressure-gradient forces, i.e., $\vec F=-\vec\nabla P$. The viscous correction goes against the effect of the pressure gradients. It involves an additional gradient, and thus it scales with two powers of the inverse macroscopic length scale, say, $1/R^2$. All other terms involve only one gradient, and scale like $1/R$. The relative importance of the viscous correction over the acceleration terms is thus of order $(v \eta / \rho)/R$. As viscous hydrodynamics is defined as a small correction to the ideal-fluid scenario, one should have $(v\eta/\rho)/R \ll 1$ for a hydrodynamic description to apply.  This requirement constraints the magnitude of the viscosity. Consider now a relativistic fluid, where the mass density $\rho$ is replaced by the enthalpy density $\epsilon+P$. The condition for hydrodynamics to apply reads:
\begin{equation}
    \frac{\eta}{e+P}\frac{v}{R} \ll 1.
\end{equation}
By use of the identity $\epsilon+P=Ts$, and by trading the ratio $v/R$ for a time scale, $\tau$, the previous expression becomes:
\begin{equation}
\frac{\eta}{s} \times \frac{1}{\tau T} \ll 1.
\end{equation}
The dimensionless ratio $\eta/s$ is a convenient way to express the quality of a fluid. In ultrarelativistic heavy-ion collisions $1/\tau T$ is around $0.2$. As a consequence, $\eta/s$ should be at maximum $\eta/s\approx0.5$, otherwise the hydrodynamic description breaks down. A similar upper bound for $\eta/s$ was found recently from an estimate of the energy dissipated during the evolution the system~\cite{Giacalone:2019ldn}. These arguments show that $\eta/s$ has to be $\mathcal{O}(0.1)$ if the quark-gluon plasma can be treated as a hydrodynamic medium. The fact that experimental data are in excellent agreement with the hydrodynamic paradigm provides, thus, evidence that the created system is indeed the most perfect fluid known to mankind.

Viscous corrections modify the form of the energy-momentum tensor, and thus the space-time evolution of the medium. In the modern approach to relativistic hydrodynamics, recently reviewed by Romatschke and Romatschke~\cite{Romatschke:2017ejr}, one constructs a covariant form of $T^{\mu\nu}$ as a sum of tensor structures allowed by symmetry that are organized according to the number of gradients of $\epsilon$ and $u^\mu$ that they contain:
\begin{equation}
    T^{\mu\nu} = T^{\mu\nu}_{(0)} +  T^{\mu\nu}_{(1)} + T^{\mu\nu}_{(2)} + \ldots,
\end{equation}
where the subscript denotes the power of gradients of $\epsilon$ and $u^\mu$.
The zeroth-order truncation is the ideal energy-momentum tensor, $T_{(0)}^{\mu\nu}=(e+P)u^\mu u^\nu - g^{\mu\nu}P$, which is the only rank-2 tensor with the right symmetries that does not involve any gradient of $e$ and $u^\mu$. First-order hydrodynamics involves two additional tensor structures:
\begin{equation}
    T_{(1)}^{\mu\nu} = \eta\sigma^{\mu\nu} + \zeta (\partial_\rho u^\rho ) \Delta^{\mu\nu},
\end{equation}
where two transport coefficients appear, namely, $\eta$, the shear viscosity, which is coupled to the traceless shear-stress tensor, $\sigma^{\mu\nu}$, which is first order in gradients, and $\zeta$, the bulk viscosity, which is coupled to the fluid expansion rate $\partial_\mu u^\mu$, while $\Delta^{\mu\nu} = g^{\mu\nu}+u^\mu u^\nu$ projects onto space-like components. Modern hydrodynamic simulations include as well second-order terms, which are 11, and play a negligible role for the phenomenology of central heavy-ion collisions in which I am interested. I refer to the \music{} manual~\cite{musicmanual} for a list of all these coefficients, as well as for additional formulas related to the viscous terms. 

Viscous corrections play as well a role at freeze-out. The reason is simply that, if the fluid is viscous, then at freeze-out one can not match the momentum distribution in a fluid element to a thermal equilibrium distribution $\propto  \left ( \exp -E/T_c \right )$. The thermal distribution has itself to be modified to account for viscous corrections. The standard method to attack this issue is to transform:
\begin{equation}
    f_{\rm th}({\bf p}) \longrightarrow f_{\rm th}({\bf p})\left (1+\delta f_\eta({\bf p}) + \delta f_\zeta({\bf p})\right ),
\end{equation}
where $\delta f_\eta$ and $\delta f_\zeta$ are small correction to the equilibrium distribution. The form of the $\delta f$ corrections is essentially unknown, and relies on Ansatzes. For the shear term, $\delta f_\eta$, the most common prescription is that of Teaney~\cite{Teaney:2003kp}, which is proportional to ${\bf p}^2$. A discussion on the current status of $\delta f_\zeta$ can be found in Ref.~\cite{Byres:2019xld}.  I refer again to the \music{} manual~\cite{musicmanual} for the expressions used in theoretical simulations.  Fortunately enough, these corrections do not play a major role in the phenomenology of heavy-ion collisions, although some effects are visible~\cite{Ryu:2017qzn}.  In small collision systems, $\delta f$ corrections can on the other hand translate into sizable effects for several observables, and so this whole business may need a different kind of treatment~\cite{Almaalol:2020rnu}.

\subsection{The big picture}

\label{sec:2-33}

\begin{figure}[t]
    \centering
    \includegraphics[width=\linewidth]{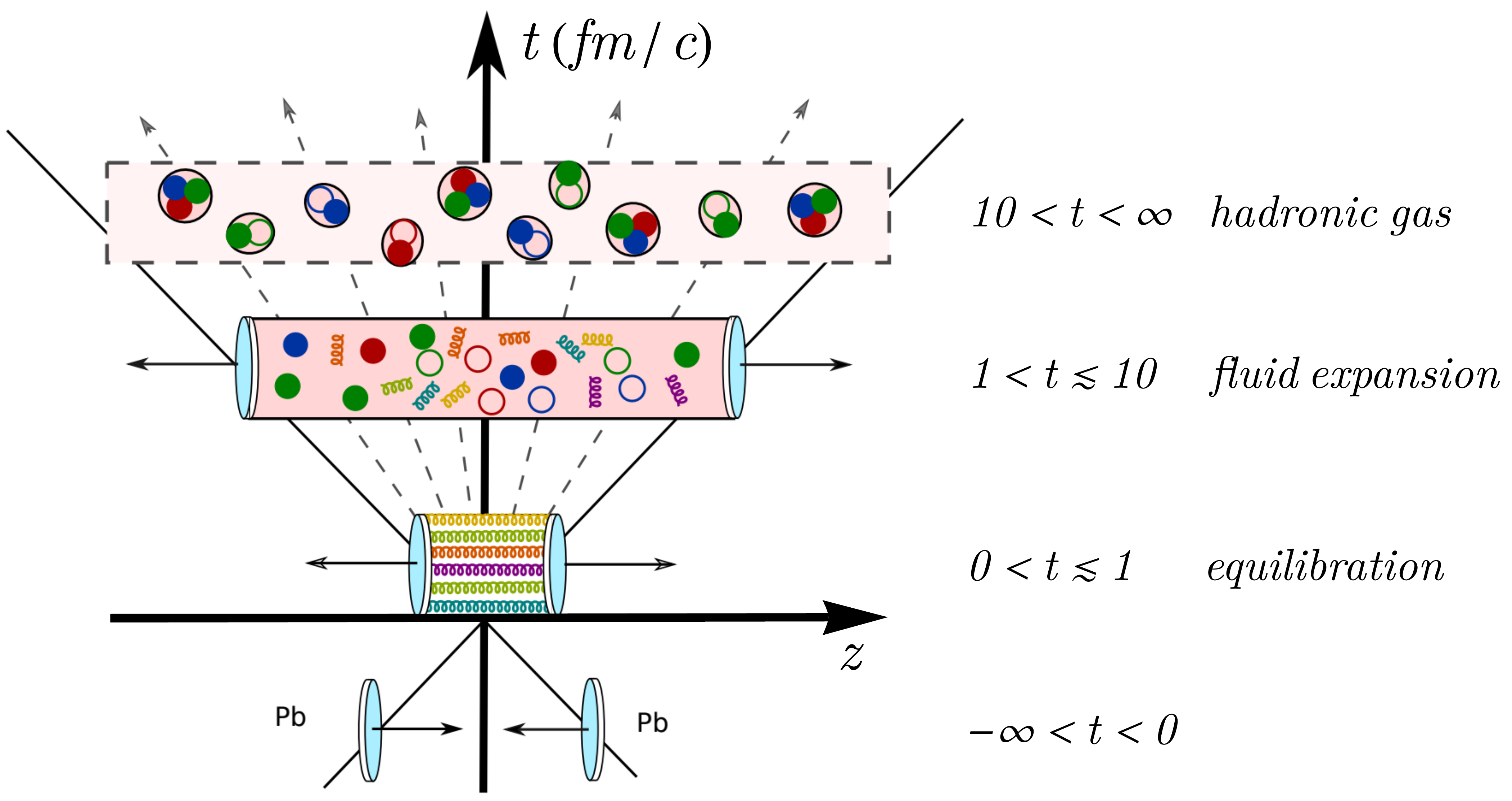}
    \caption{ Main features of the evolution of a heavy-ion collision. See the text for a detailed discussion. (courtesy Aleksas Mazeliauskas)}
    \label{fig:2-8}
\end{figure}

This concludes my end-to-end description of a heavy-ion collision. Let me give a quick summary of the salient features of the evolution of the system at $z=0$, also presented in the illustration of \fig{2-8}. The value of $t$ is in fm/$c$, and the boundaries of the proposed time intervals represent rough estimates rather than accurate figures. Also, these figures are meant to describe central collisions of large nuclei.

\begin{itemize}
    \item $-\infty < t < 0$, two nuclei, strongly Lorentz-contracted along the collision axis, approach the interaction point.
    \item $t=0$, the interaction takes place. At ultrarelativistic energy, the nuclei also cross each other at $t=0$, as the process is instantaneous.
    \item $0^+ < t < 0.1$, the system is in the glasma phase, and its evolution is dictated by classical Yang-Mills equations. The longitudinal pressure is negative: the longitudinal expansion increases the energy of the system. At the end of this phase, the longitudinal pressure is close to zero.
    \item $0.1 < t < 1$, the system has now a particle description and evolves according to a Boltzmann equation. The equilibration process builds up longitudinal pressure. Thermal equilibrium is reached when the phase space density becomes isotropic.
    \item $1 < t < 10$, the system is in the quark-gluon plasma phase, described by fluid degrees of freedom and the equation of state of hot QCD. It undergoes a collective expansion governed by the laws of relativistic viscous hydrodynamics. This is the part of the evolution process which produces the most distinct phenomenological signatures. The system is treated as a fluid until the local temperature drops below the critical temperature, of order 150~MeV. At this temperature, the fluid cells convert into hadrons. 
    \item $10< t < 100$, hadron gas cascade. The gas contains unstable hadrons that gradually decay into stable particles. Chemical equilibrium is achieved when inelastic processes no longer occur. Kinetic equilibrium is instead achieved when elastic scatterings also cease.
    \item $100 < t < \infty$, the produced stable hadrons free stream to the detector. A spectrum $d N / d^3 {\bf p}$, or $dN/(d^2{\bf p}_t dy)$ is measured. 
\end{itemize}

Before concluding this chapter, let me stress that the \ipglasma{} framework, describing the system in the temporal range $-\infty < t < 0.1$~fm/$c$, has been recently coupled to the \kompost{} framework, which further evolves the system throughout thermalization up to $1$~fm/$c$. The output has then been used as an input for the \music{} hydrodynamic code, which further evolved the equilibrated fluid all the way to the hadronic phase. This happened in 2018~\cite{Kurkela:2018wud}, i.e., 18 years following the beginning of the high-energy nuclear physics program at RHIC, and demonstrates that nowadays one is able build up end-to-end simulations whose results can eventually be compared to experimental data.


\chapter{Basics of heavy-ion phenomenology}

\label{chap:3}

The observable outcome of the space-time evolution of a heavy-ion collision, as obtained at the end of a hydrodynamic calculation, is thus a spectrum of hadrons, $dN/(d^2{\bf p}_t dy)$. An observable, $\mathcal{O}$, is a function of this quantity:
\begin{equation}
    \mathcal{O} = f \left( \frac{d N}{d^2{\bf p}_t} \right ),
\end{equation}
where I have dropped the dependence of the spectrum on $y$ since I shall always consider a boost-invariant setup. There are infinite possibilities, but some observables are more useful than others.

The goal of this chapter is very simple. I present three observables of paramount importance in the phenomenology of heavy-ion collisions, namely:
\begin{enumerate}
    \item The multiplicity, i.e., the total number of hadrons collected in the phase space available to the detector:
    \begin{equation}
    \label{eq:multiplicity}
        {\rm multiplicity} \equiv N =  \int_{{\bf p}_t} \frac{d N}{d^2{\bf p}_t}.
    \end{equation}
    \item The average transverse momentum, $\bra p_t \ket$, i.e., the first moment of the distribution of $p_t\equiv|{\bf p}_t|$:
    \begin{equation}
    \label{eq:mpt}
        \bra p_t \ket = \frac{1}{N}  \int_{{\bf p}_t} p_t \frac{d N}{d^2{\bf p}_t}.
    \end{equation}
    \item The second-order Fourier harmonic of the azimuthal part of the spectrum. In polar coordinates, one can write:
    \begin{equation}
        \frac{d^2 N}{d{\bf p}_t} = \frac{d^2 N}{p_td p_t d\phi_p},
    \end{equation}
    and extract the complex second-order Fourier coefficient:
    \begin{equation}
    \label{eq:V2}
        V_2 = \frac{1}{N} \int_{{\bf p}_t} \frac{d N}{d^2{\bf p}_t} e^{-i2\phi_p}.
    \end{equation}
    This quantity is dubbed \textit{elliptic flow}. Note that $V_{-2}=V_2^*$.
\end{enumerate}
I shall emphasize that, in spite of the apparently complicated space-time history from which they emerge, these observables have an intuitive physical origin in the hydrodynamic framework. These quantities will be at the heart of the phenomenology of nuclear deformation to be discussed later on in Chapters~$\ref{chap:4}$~and~$\ref{chap:5}$.

An important comment is in order. The thermal hadron spectrum at the end of hydrodynamics is, as a function of $p_t$, close to a thermal distribution, hence, it has typically an exponential fall-off at large $p_t$. Most of produced particles lie as a consequence at somewhat low values of $p_t$. The majority of the charged hadrons detected in the final state are pions, whose transverse momentum is on average about $0.5~{\rm GeV}$, thus giving the order of magnitude of first moment of the $p_t$ spectrum, $\bra p_t \ket$ in \equ{mpt}. The integrals that lead to the observables discussed in this chapter are thus dominated by the low-$p_t$ region of the spectrum, which is dubbed the \textit{soft} sector. In heavy-ion collisions there is also a rich phenomenology of the \textit{hard} sector. It involves the study of the rare objects that populate the high-$p_t$ tail of the spectrum. These are typically either hadrons emitted with $p_t > 3~{\rm GeV}$, or jets at much higher $p_t$, which are abundantly produced at LHC energy. This phenomenology is complementary to that of the soft sector, as it involves different energy scales and a different dimensional analysis. The goal is however the same, i.e., inferring the properties of the quark-gluon plasma, by studying in particular how the production of these energetic particles are modified by their interaction with a surrounding hot and dense medium, a problem that can be treated to a good extent by means of perturbative methods in QCD.

\section{Entropy and particle number}

\label{sec:3-1}

I discuss now the physical interpretation of the most straightforward observable of heavy-ion collisions, i..e, the multiplicity in \equ{multiplicity}. 

In the limit of high temperature, the quark-gluon plasma can be viewed as an ideal classical gas of massless particles, with zero baryon density and an equation of state close to $P=\epsilon/3$. The corresponding phase space density is given by the Maxwell-Boltzmann statistics: $\exp(-p/T({\bf x}))$, where the particle momentum, $p\equiv|{\bf p}|$, coincides with the energy per particle. The number density, $n$, and the energy density, $\epsilon$, of this medium are given by ($\hbar=1$)~\cite{Ollitrault:2007du}:
\begin{equation}
    n=\int \frac{d^3 {\bf p}}{(2\pi)^3}e^{-p/T}, \hspace{40pt}\epsilon=\int \frac{d^3 {\bf p}}{(2\pi)^3}p~e^{-p/T}.
\end{equation}
Carrying out the integrals, and summing over $\nu_{\rm QCD}$ degrees of freedom in the gas one obtains:
\begin{equation}
\label{eq:3nT}
    n=\frac{\nu_{\rm QCD}}{\pi^2} T^3, \hspace{40pt}\epsilon=3nT.
\end{equation}
Now, by use of the thermodynamic identity $\epsilon+P=Ts$, and the ideal gas law $P=nT$, one obtains for the entropy density:
\begin{equation}
\label{eq:4n}
    s=4n.
\end{equation}
The entropy of the system is thus proportional to the number of particles. 

Next, let the gas evolve according to inviscid hydrodynamics. The absence of heat diffusion between fluid cells implies that the fluid expansion is adiabatic. As a consequence, the total entropy in the fluid is conserved as a function of time. In the local rest frame (and thus in all frames), the total entropy of the system  is given by~\cite{Ollitrault:2007du}:
\begin{equation}
S = \tau  \int s u^0 d^2{\bf x}.
\end{equation}
These considerations suggest that, since the created fluid is close to perfect, then by virtue of \equ{4n} the number of particles produced at the end of hydrodynamic expansion should be proportional to the entropy, $S$, of the system. I check this explicitly in hydrodynamic simulations. I use a large batch of \pbpb{} collisions, corresponding to the calculation presented in Ref.~\cite{Giacalone:2017dud}. This calculation consists of 50000 \pbpb{} events evolved with the viscous hydrodynamic code {\small V-USPHYDRO}~\cite{Noronha-Hostler:2013gga,Noronha-Hostler:2014dqa,Noronha-Hostler:2015coa}. This calculation implements $\eta/s=0.05$, $\zeta/s=0$, and the equation of state of hot QCD~\cite{Borsanyi:2013bia}. The initial entropy profile for the hydrodynamic expansion is given by the \trento{} model of Ref.~\cite{Bernhard:2016tnd}. The profiles of entropy density returned by the \trento{} calculation correspond to the initial condition of the hydrodynamic expansion at $\tau_0=0.6~\fmc$. The effect of pre-equilibrium dynamics is neglected. The medium freezes out a temperature $T=150~{\rm MeV}$, and all resonances are thermally produced. Strong decays of resonances are also implemented~\cite{Kolb:2002ve}.

The  entropy per unit rapidity at the initial condition is defined by ($u^0=1$):
\begin{equation}
\label{eq:totS}
    S(\tau_0) =  \tau_0 \int_{\bf x}  s(\tau_0,{\bf x}),
\end{equation}
where $s(\tau_0,{\bf x})$ is the entropy density shown in \equ{p=0}, with $N_0 \approx 72$ for top LHC energy. The calculation yields a boost-invariant spectrum of hadrons, $dN/d^2{\bf p}_t$, and the charged-particle multiplicity is computed in the window $-0.5<\eta<0.5$, which corresponds to the acceptance used by the ALICE Collaboration~\cite{Adam:2015ptt}. Note that the spectrum is invariant with respect to $y$, but whenever one integrates the spectrum taken within an interval of pseudorapidity, $[\eta_-,\eta_+]$, as done in experiments,
\begin{equation}
    \frac{dN}{d\eta}\biggl|_{\eta_- < \eta < \eta_+} = \int_{{\bf p}_t} \frac{dN}{d^2{\bf p}_t d\eta}\biggl|_{\eta_- < \eta < \eta_+},
\end{equation}
then the transformation of $y$ into $\eta$ has to be properly taken into account in the definition of $dN/(d^2{\bf p}_t d\eta)$. From \equ{pseudorap} and \equ{rapidity}, this can be done by use of the equality:
\begin{equation}
    m_t \sinh y = p_t \sinh \eta,
\end{equation}
where, $m_t=\sqrt{p_t^2 + m^2}$ is the \textit{transverse mass}, $m$ being the particle rest mass.

\begin{figure}[t]
    \centering
    \includegraphics[width=.55\linewidth]{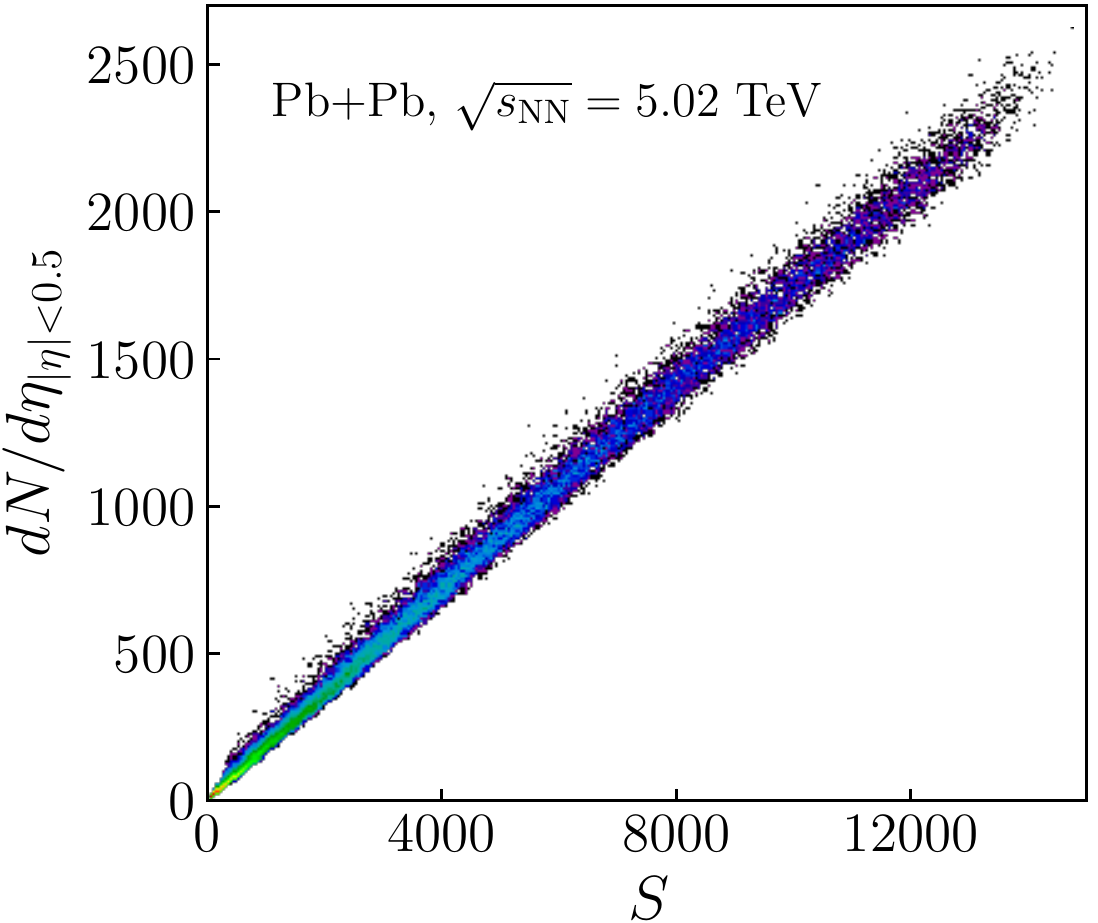}
    \caption{Joint probability distribution of total entropy per unit rapidity at the initial condition, $S$, as defined by \equ{totS}, and charged-particle multiplicity, $dN/d\eta$, in the acceptance $-0.5<\eta<0.5$, for $5.02~{\rm TeV}$ \pbpb{} collisions. Lighter colors indicate larger values of probability.}
    \label{fig:3-1}
\end{figure}

The hydrodynamic result for the correlation between the initial entropy per unit rapidity, $S$, and the final charged-particle multiplicity, $dN/d\eta$, is shown in \fig{3-1}. One observes a very strong linear correlation between these two quantities. I am showing here results for hydrodynamic simulations of \textit{minimum bias} collisions, i.e., collisions at all impact parameters. The strong linear correlation is thus observed both for low-multiplicity (peripheral) and high-multiplicity (central) events. The simple ideal gas picture where the particle number provides a measure of the entropy gives thus an accurate description of the physics at play, even in the case of a full hydrodynamic calculation involving viscous corrections, freeze-out, and resonance decays.

\begin{figure}[t]
    \centering
    \includegraphics[width=\linewidth]{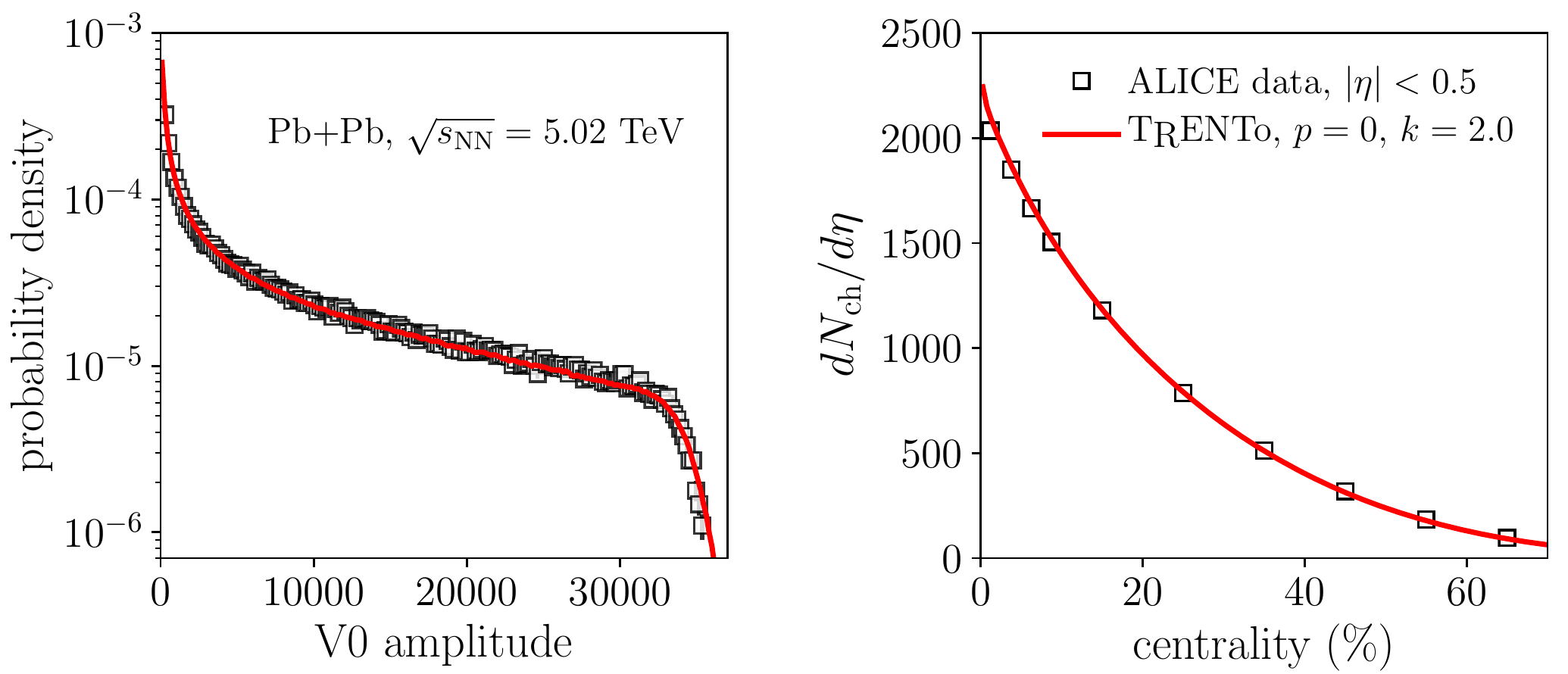}
    \caption{Left: minimum bias distribution of V0 amplitude collected by the ALICE detector in $5.02~{\rm TeV}$ \pbpb{} collisions~\cite{Adam:2015ptt}. This histogram is used to determine the centrality of the collision events. Right: centrality dependence of the average charged-particle multiplicity, $\bra dN_{\rm ch}/d\eta \ket$~\cite{Adam:2015ptt}. In both panels, red lines correspond to the results of the \trento{} model, where the multiplicity is estimated from the initial total entropy, $S$, rescaled by an appropriate factor.}
    \label{fig:3-2}
\end{figure}

A natural question is, hence, whether the \trento{} calculation alone allows one to reproduce the distributions of $dN/d\eta$ observed in experimental data. This is in fact the case, and demonstrates the remarkable goodness of the \trento{} Ansatz. I first look at the minimum bias distribution of multiplicity observed experimentally. The ALICE collaboration does not show the minimum bias distribution of the multiplicity, however, they measure a proxy of $dN/d\eta$, corresponding to the energy collected in a given calorimeter, the so-called V0 amplitude~\cite{Abelev:2013qoq}. The minimum bias distribution of this multiplicity measured by the ALICE detector is shown as squares in the left panel of \fig{3-2}, and it corresponds to the probability distribution used by the ALICE collaboration to sort events into centrality classes, as discussed in \fig{2-5}. This quantity is proportional to the actual multiplicity, $dN/d\eta$, hence, on the basis of \fig{3-1}, I should be able to describe the previous histogram by means of the entropy,  $S$, returned by the \trento{} model. The minimum bias distribution of $S$, rescaled by an appropriate factor, is shown as a red line in the figure. Agreement with data is excellent, within few percent across the full range of multiplicity. This shows that a centrality selection in the \trento{} calculation based on $S$ is fully consistent with the centrality selection performed by the ALICE collaboration.

After determining the centrality classes from the histogram shown in the left panel of \fig{3-2}, the ALICE collaboration measures the average value of $\bra dN/d\eta \ket$ as a function of collision centrality. The corresponding experimental data is shown in the right panel of \fig{3-2}. I compare data to the estimate of the average total entropy produced by the \trento{} model, $\bra S \ket$, as a function of collision centrality. The total entropy is rescaled by a factor 6, which can be viewed as the entropy per charged particle predicted by this \trento{} calculation. The result is shown as a red line in the right panel of \fig{3-2}. Agreement with data is impressive. The \trento{} result describes data with an accuracy of order 1\% across the full centrality range.  To the best of my knowledge, this is the only model of particle production which allows one to reproduce the centrality dependence of the multiplicity with such a degree of precision. One should also note that this specific result is essentially independent of the parameter $k$ and $w$ of the \trento{} model. The important feature is the square root taken in \equ{p=0}. This suggests that the agreement with data is not a coincidence, and that this model does capture to a good extent the underlying physical picture.

\paragraph{In summary --}The multiplicity of a heavy-ion collision is a measure of the entropy of the system, whose value is essentially constant across the evolution of the fluid due to quasi-ideal nature of the quark-gluon plasma. The initial entropy provides thus the natural variable to use to sort collision events into centrality classes. The initial entropy obtained in the \trento{} model provides an excellent description of the measured multiplicity distributions. 

\section{Energy and momentum}

\label{sec:3-2}

The equation for $\epsilon$ in \equ{3nT} can be written as:
\begin{equation}
    \frac{\epsilon}{n} = 3T,
\end{equation}
meaning that the energy per particle is proportional to the temperature. Now, in the ultrarelativistc regime, the momentum of a particle coincides with its energy, hence
\begin{equation}
\label{eq:3T}
    p = 3T.
\end{equation}
In the final state, a reasonable measure of the energy per particle is the first moment of the $p_t$ spectrum, i.e., $\bra p_t \ket$. It would be useful to establish a relation similar to \equ{3T}, where $p$ is replaced by $\bra p_t \ket$, and the right-hand side contains some measure of the temperature. 

If this is possible, then $\bra p_t \ket$ would give access to the thermodynamic properties of the quark-gluon plasma. As the average transverse momentum measured in \pbpb{} collisions at top LHC energy is of order 0.70~GeV~\cite{Acharya:2018eaq}, one has:
\begin{equation}
\label{eq:pt3T}
    \bra p_t \ket / 3 = 0.23~{\rm GeV}.
\end{equation}
If 0.23~GeV corresponded to the average temperature of the system at some time during the space-time history of the QGP, then $\bra p_t \ket$ could be used as a probe of the quark-gluon plasma phase, because 0.23~GeV is significantly higher than the freeze-out temperature. This correspondence has been recently established quantitatively in Ref.~\cite{Gardim:2019xjs}, whose analysis I shall reformulate here. My goal is to assign a meaning to the the right-hand side of \equ{pt3T}, and show that it corresponds to a well-defined temperature during the evolution of the system. The conclusion is that $\bra p_t \ket$ measured in heavy-ion collisions probes the thermodynamic properties of the quark-gluon medium, carrying in particular information about the equation of state of hot strong-interaction matter.

\subsection{Effective temperature: the quark-gluon plasma}

\label{sec:3-21}

I consider the quark-gluon plasma at the beginning of hydrodynamics. The fluid velocity in the midrapidity slice, $z=0$, is nonzero only along $z$, and equal to $v_z=z/t$, such that $\partial v / \partial z = 1 /t$. Conservation of energy yields:
\begin{equation}
    \frac{d\epsilon}{dt} = - \frac{\epsilon+P}{t} 
\end{equation}
where $P$ represents the longitudinal pressure, which is nonzero due to the requirement of local isotropy in the medium. The previous equation can be written as:
\begin{equation}
    d(\epsilon t) = - P dt,
\end{equation}
which implies:
\begin{equation}
\label{eq:-Pdv}
    dE = - P dV,
\end{equation}
where $E$ is the total energy per unit rapidity in the fluid, defined by:
\begin{equation}
    E(\tau) = \tau \int_{\bf x} \epsilon({\bf x},\tau),
\end{equation}
while $V$ is the volume. Equation~(\ref{eq:-Pdv}) implies that the energy of the fluid decreases during the hydrodynamic expansion due to \textit{longitudinal cooling}, i.e., the work performed by the longitudinal pressure against the longitudinal expansion of the medium. Now, this cooling is not eternal, and in fact is effective only over a short time. The reason is that, while expanding along $z$, the medium also expands freely in the transverse plane. This transverse expansion reduces the transverse pressure, but since the fluid is in equilibrium, the longitudinal pressure also decreases, because of the requirement of local isotropy. 

At a certain time, then, the pressure in the medium becomes negligible: the medium keeps expanding in the transverse direction, but with little acceleration, and longitudinal cooling is almost ineffective. As argued in Ref.~\cite{Bhalerao:2005mm}, this occurs at a time close to $\bar R/c_s$, where the transverse size $\bar R$ can be defined through:
\begin{equation}
\label{eq:Rbar}
    1/\bar{R} = \sqrt{\frac{1}{\bra x^2 \ket} + \frac{1}{\bra y^2 \ket}},
\end{equation}
where the average is weighted with the energy density of the system: $\bra \ldots \ket = \frac{1}{E} \int_{\bf x} \ldots \epsilon({\bf x}, \tau)$, and $c_s$ is the speed of perturbations in the medium, i.e., the speed of sound. As the transverse expansion merely converts internal energy into kinetic energy, it does not dissipate energy. Hence, after the conjectured time $\bar R/c_s$, the energy of the fluid, $E$, becomes approximately a constant. The entropy per rapidity in the medium, $S$, defined in \equ{totS}, is also roughly constant in time, since the expansion is nearly ideal. I conclude that $E/S$ should be roughly constant in time beyond $\tau=\bar R / c_s$, i.e., after longitudinal cooling has ended. 

\begin{figure}[t]
    \centering
    \includegraphics[width=\linewidth]{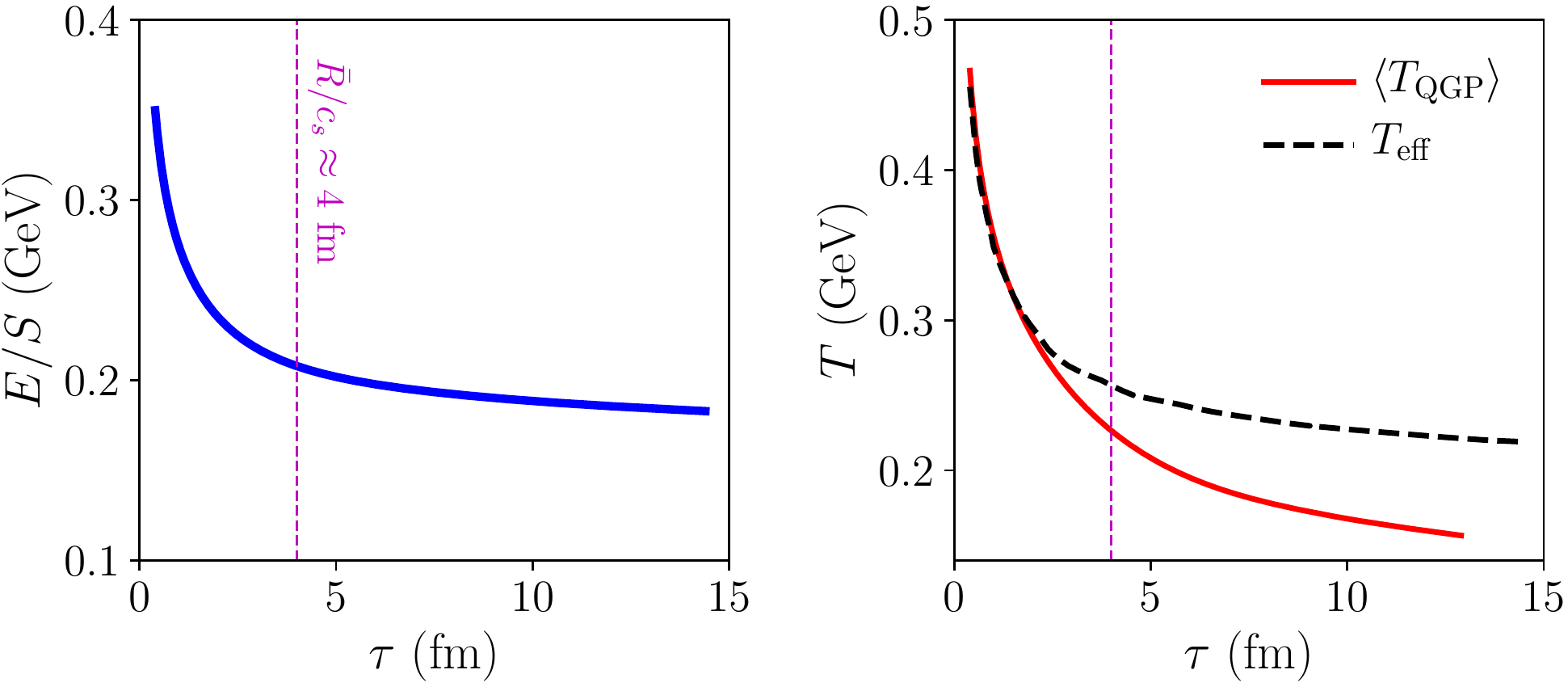}
    \caption{Left: ratio $E/S$ as a function of $\tau$ in the evolution of the average density profile of $2.76~{\rm TeV}$ \pbpb{} collisions at zero impact parameter. Right: temporal evolution of $T_{\rm eff}$ (dashed line), as defined by \equ{teff}, and $\bra T_{\rm QGP} \ket$ (solid line), as defined by \equ{Tqgp}. In both panels, the vertical line indicates $\tau\approx \bar R/c_s \approx 4$~fm.}
    \label{fig:3-3}
\end{figure}

I check this explicitly in a hydrodynamic simulation. I evolve a smooth profile of entropy density given by the so-called \textit{thickness} function of a $^{208}{\rm Pb}$ nucleus, corresponding to $\int dz \rho({\bf x}, z)$, where $\rho({\bf x},z)$ is the nuclear matter density given in \equ{2pf}. This corresponds essentially to the average entropy density profile returned by \trento{} simulations at $b=0$. The medium is evolved through the \music{} hydrodynamic code with viscous hydroydnamic equations, implementing $\eta/s=0.16$ and $\zeta/s=0$, and the QCD equation of state. The total energy per unit rapidity at the initial condition,
\begin{equation}
\label{eq:totE}
    E (\tau_0) = \tau_0 \int_{\bf x} \epsilon({\bf x},\tau_0),
\end{equation}
is normalized to 5000~GeV, corresponding to a good approximation to a central \pbpb{} collisions at $\sqrt{s_{\rm NN}}=2.76$~TeV. In the left panel of \fig{3-3}, I show the ratio $E/S$ as a function of $\tau$ in this simulation. The intuitive expectation is nicely confirmed: longitudinal cooling dissipates energy in the first few fm/$c$, while $E/S$ flattens at larger $\tau$, when the transverse expansion has washed out the longitudinal pressure. The initial profile of energy density has about $\bar R \approx 2$~fm, so that with $c_s=0.5$, which is the natural ballpark for the speed of sound of the quark-gluon plasma at LHC energy, one obtains $\bar R/c_s \approx 4$~fm. This time scale is highlighted in the figure with a vertical line. This line falls precisely around the time where the curve starts flattening.

The fact that $E/S$ is constant after $\tau\sim4~\fm$ has a nice consequence. Since $S$ is proportional to the final number of particles, then the left panel of \fig{3-3} suggests that the transverse energy per particle emitted from the medium is essentially fixed at an early time scale during the hydrodynamic phase, thus carrying information about the thermodynamic state of the fluid at $\tau\approx \bar R/c_s$. As argued at the beginning of this discussion, the transverse energy per particle should be close to the measured mean transverse momentum, $\bra p_t \ket$. With the additional consideration that resonance decays at the end of hydrodynamics do not alter the value of $E/S$, one is lead to conclude that $\bra p_t \ket$ serves as a probe of the thermodynamic state of the system deep into the quark-gluon plasma phase.

\begin{figure}[t]
    \centering
    \includegraphics[width=\linewidth]{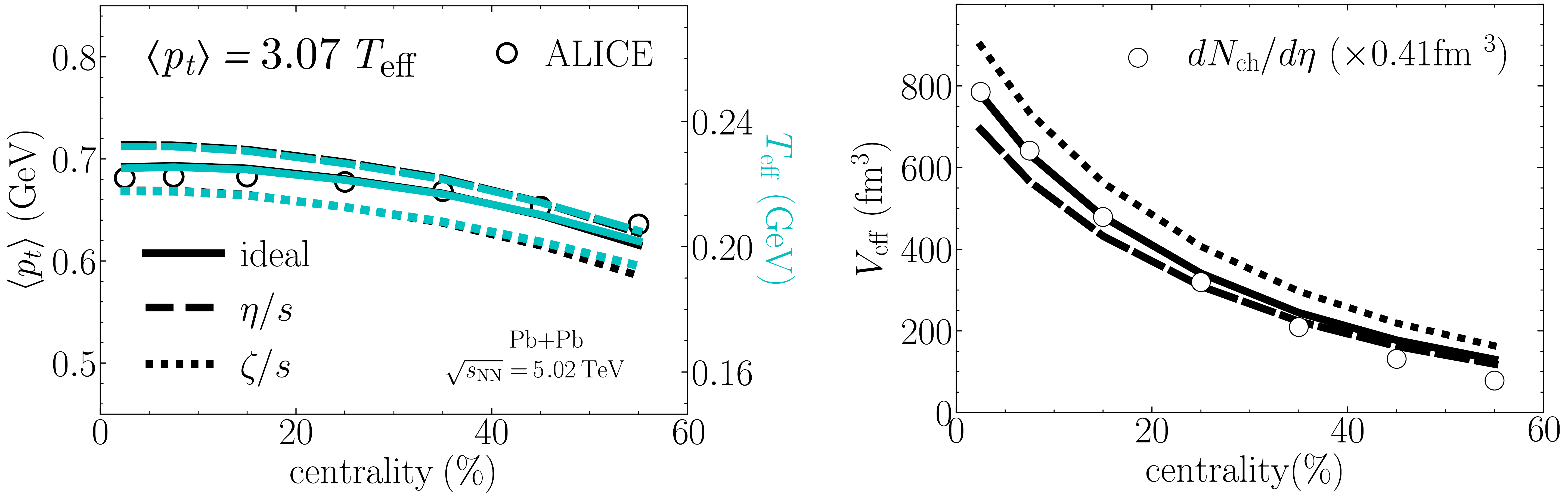}
    \caption{Left: $\bra p_t \ket$ as a function of centrality in ALICE data~\cite{Acharya:2018eaq} (symbols) and in hydrodynamic simulations (black lines) for $5.02~{\rm TeV}$ \pbpb{} collisions. The light blue line corresponds to $T_{\rm eff}$, as explained in the text. Note that the lines of $T_{\rm eff}$ and those of $\bra p_t \ket$ largely overlap, and one has in particular $\bra p_t \ket = 3.07 T_{\rm eff}$. Right: Effective volume as function of collision centrality (solid lines). Symbols are rescaled ALICE data points~\cite{Adam:2015ptt} for the charged-particle multiplicity. Different line styles indicate different hydro setups: ideal hydrodynamics with $\eta/s=0$ and $\zeta/s=0$ (solid lines), viscous hydrodynamics with $\eta/s=0.2$ and $\zeta/s=0$ (dashed lines), and viscous hydrodynamics with $\eta/s=0$ and $\zeta/s=\zeta/s(T)$ as parametrized in Ref.~\cite{Ryu:2017qzn} (dotted lines).  Figure adapted from Ref.~\cite{Gardim:2019xjs}.}
    \label{fig:3-4}
\end{figure}

\paragraph{Effective temperature --} I show that this is indeed the case. I follow the analysis of Ref.~\cite{Gardim:2019xjs}. The idea is to track the value of $E/S$ all the way until the fluid has frozen out. On the freeze-out hypersurface, the total energy and the total entropy are given, respectively, by:
\begin{align}
\label{eq:EfSf}
\nonumber     E_f = \int T^{0\mu} d\Sigma_{\mu}, \\
    S_f = \int s u^\mu d\Sigma_{\mu}.
\end{align}
An \textit{effective} description based on an equivalent uniform fluid, as introduced in Ref.~\cite{Gardim:2019xjs}, allows one to gain an intuitive understanding of the physics at play. I consider a uniform medium with volume $V_{\rm eff}$ which contains energy $E_f$ and entropy $S_f$. The medium is, hence, at a temperature $T_{\rm eff}$. One obtains a system of two equations for two variables, namely, the effective temperature and the effective volume:
\begin{align}
    S_f &= s_{\rm eff}(T_{\rm eff}) V_{\rm eff} \\
    E_f &= \epsilon_{\rm eff}(T_{\rm eff}) V_{\rm eff}.
\end{align}
Their ratio,
\begin{equation}
\label{eq:teff}
    \frac{E_f}{S_f} = \frac{\epsilon_{\rm eff}}{s_{\rm eff}}(T_{\rm eff})
\end{equation}
no longer depends on $V_{\rm eff}$, so that one can extract $T_{\rm eff}$ from the equation of state of QCD. This calculation was carried out in Ref.~\cite{Gardim:2019xjs}. By means of the \music{} code, smooth profiles of energy density, corresponding to average \trento{} profiles, tuned to reproduce the multiplicity observed in \pbpb{} collisions at LHC energy, were evolved with both ideal and viscous hydrodynamic equations. The value of $T_{\rm eff}$ obtained in hydrodynamic simulations is shown as light blue lines in \fig{3-4}, as a function of collision centrality. At top LHC energy, is of order 220~MeV, and its value depends little on the viscous corrections. In the right panel of the figure I show instead the effective volume, $V_{\rm eff}$. It is of order 800~fm$^3$ at top LHC energy in \pbpb{} collisions.

Now, from \equ{3nT} and \equ{4n}, one has $\epsilon/s \propto T$. It is thus instructive to check the dependence of $T_{\rm eff}$ on $\tau$. One can simply use the values of $E/S$ shown in \fig{3-3} and apply the equation of state at each $\tau$. The resulting curve for $T_{\rm eff}$ is shown in the right panel of \fig{3-3}, as a dashed line. We see that this curve is to a good extent the same as that of $E/S$, in the sense that it also does flatten around the same value of $\tau$. Hence, if as argued before a proxy for $E/S$, representing the energy per particle, is given by $\bra p_t \ket$, then $\bra p_t \ket$ should be essentially close to $3T_{\rm eff}$, following \equ{3T}.  In the left panel of \fig{3-3}, the black lines are the values of $\bra p_t \ket$ for different hydrodynamic setups. One sees that the picture is fully consistent, since:
\begin{equation}
\label{eq:ptteff}
    \bra p_t \ket = 3.07 T_{\rm eff},
\end{equation}
irrespective of viscous corrections and collision energy. The same calculation was also repeated in Ref.~\cite{Gardim:2019xjs} with a completely different equation of state, leading solely to a minor modification of the proportionality coefficient. In Ref.~\cite{Gardim:2020sma}, it has been further checked that \equ{ptteff} is to a good extent unaffected by the inclusion of more realistic initial conditions based on Glauber nucleons in the hydrodynamic simulations. The result is thus robust and has a clear interpretation. A given value of $\bra p_t \ket$ can be associated with a uniform quark-gluon medium at a temperature close to $\bra p_t \ket / 3$. But as the energy per particle is roughly constant following the end of longitudinal cooling, measuring $\bra p_t \ket$ in heavy-ion collisions is thus like reading a thermometer that indicates the temperature of the system into the quark-gluon plasma phase.

But I can do more. The effective equivalent-uniform-fluid description is very elegant, but, in hindsight, it is not strictly necessary. I define the average temperature within the quark-gluon plasma phase as:
\begin{equation}
\label{eq:Tqgp}
    \bra T_{\rm QGP} \ket = \frac{1}{E} \int_{\bf x} T({\bf x}) \epsilon({\bf x},\tau),
\end{equation}
where I perform the average only over fluid elements that are at a temperature larger than the freeze-out temperature. In the right panel of \fig{3-3}, I show the value of $\bra T_{\rm QGP} \ket$ as a function of $\tau$, along with the value of $T_{\rm eff}$, as returned by \equ{teff}. The striking result is that $T_{\rm eff}$ at freeze-out, and $\bra T_{\rm QGP}\ket$ at $\tau=\bar R/c_s$ are essentially the same. Beyond that time, $\bra T_{\rm QGP} \ket$ keeps decreasing and disappears around $\tau\sim 13~\fm$, where the entire system is below the freeze-out temperature, whereas the value of $T_{\rm eff}$ becomes constant, because it includes as well a contribution from the motion within the fluid cells, which is developed during the expansion. One can thus safely conclude that $T_{\rm eff}$ extracted from ALICE data corresponds approximately to the average temperature of the quark-gluon plasma at the time where longitudinal cooling ends. 

We have thus arrived to the following law: \vspace{2mm}
\begin{displayquote}
\begin{mdframed}
\textit{The value of $\bra p_t \ket$ measured in a heavy-ion collision is close to $3T$, where $T$ is the average temperature of the quark-gluon plasma at the time when longitudinal cooling becomes ineffective.}
\end{mdframed}
\end{displayquote}
 This statement is very important, as it shows that $\bra p_t \ket$ gives access to the thermodynamics of the quark-gluon plasma, in particular, to its equation of state, the extraction of which is one of the main goals of the heavy-ion collision program.

As a side remark, a similar statement holds as well for the volume of the system. I define:
\begin{equation}
    V_{\rm QGP}  = \tau \pi R^2,
\end{equation}
where $R^2$ is defined, e.g., as the mean squared radius of the entropy density profile:
\begin{equation}
\label{eq:R}
    R^2 = \frac{2}{S}  \int_{\bf x} |{\bf x}|^2 s({\bf x},\tau),
\end{equation}
where the average is taken only on fluid cells with local temperature larger than the freeze-out temperature, and the factor $2$ in the numerator ensures that the right-hand side gives precisely $R^2$ when an uniform fluid of radius $R$ is considered. With this definition, one finds indeed in simulations that $V_{\rm QGP}$ at $\tau \sim \bar R/c_s$ and $V_{\rm eff}$ coincide to a good extent.

\paragraph{Extracting the EOS --} Reference~\cite{Busza:2018rrf} states that, as of 2018, one of the main open problems in the field of heavy-ion collisions is the determination of two independent thermodynamic variables from heavy-ion data, which would give access to the EOS. In Ref.~\cite{Gardim:2019xjs} we have precisely fulfilled this task, so that I am now able to extract the EOS of the quark-gluon plasma (at $T=T_{\rm eff}$) from data. 

From \fig{3-4} I can extract the temperature. ALICE data~\cite{Acharya:2018eaq} indicates that $\bra p_t \ket = 0.685$~GeV. From \equ{ptteff}, this corresponds to a temperature:
\begin{equation}
    T \simeq \bra p_t \ket / 3.07 \simeq 222 \pm 9~{\rm MeV}.
\end{equation}
The error bar is mostly driven by the uncertainty over the freeze-out temperature. With this temperature at hand, along with the estimated volume of the system, $V_{\rm eff}$, we can evaluate all the thermodynamic quantities.

First, the number density, $n$, of the quark-gluon plasma can be evaluated as:
\begin{equation}
    n = 1.5 \frac{dN/d\eta}{V_{\rm eff}},
\end{equation}
where the factor $1.5$ is included to take into account that $dN/d\eta$ counts only the charged hadrons, which are about two thirds of the total. With $V_{\rm eff}=800~\fm^3$, and $dN/d\eta=2000$, one obtains:
\begin{equation}
    n \simeq 4~\fm^{-3}.
\end{equation}
This should be compared to the matter density of normal nuclear matter, i.e., $\rho_0=0.16$~fm$^{-3}$, showing that the quark-gluon plasma is at least a factor 20 denser than atomic nuclei, and likely about a factor 5 denser than the matter existing in the core of neutron stars, or created in neutron star mergers. Furthermore, from \equ{3nT} we can give an estimate of the number of degrees of freedom:
\begin{equation}
\label{eq:nu}
    \nu_{\rm QCD} = \frac{\pi^2 n}{T_{\rm eff}^3} \approx 30.
\end{equation}
This large value is an indication that the system is in a phase where color degrees of freedom are liberated and active, i.e., the quark-gluon plasma phase. 

\begin{figure}[t]
    \centering
    \includegraphics[width=\linewidth]{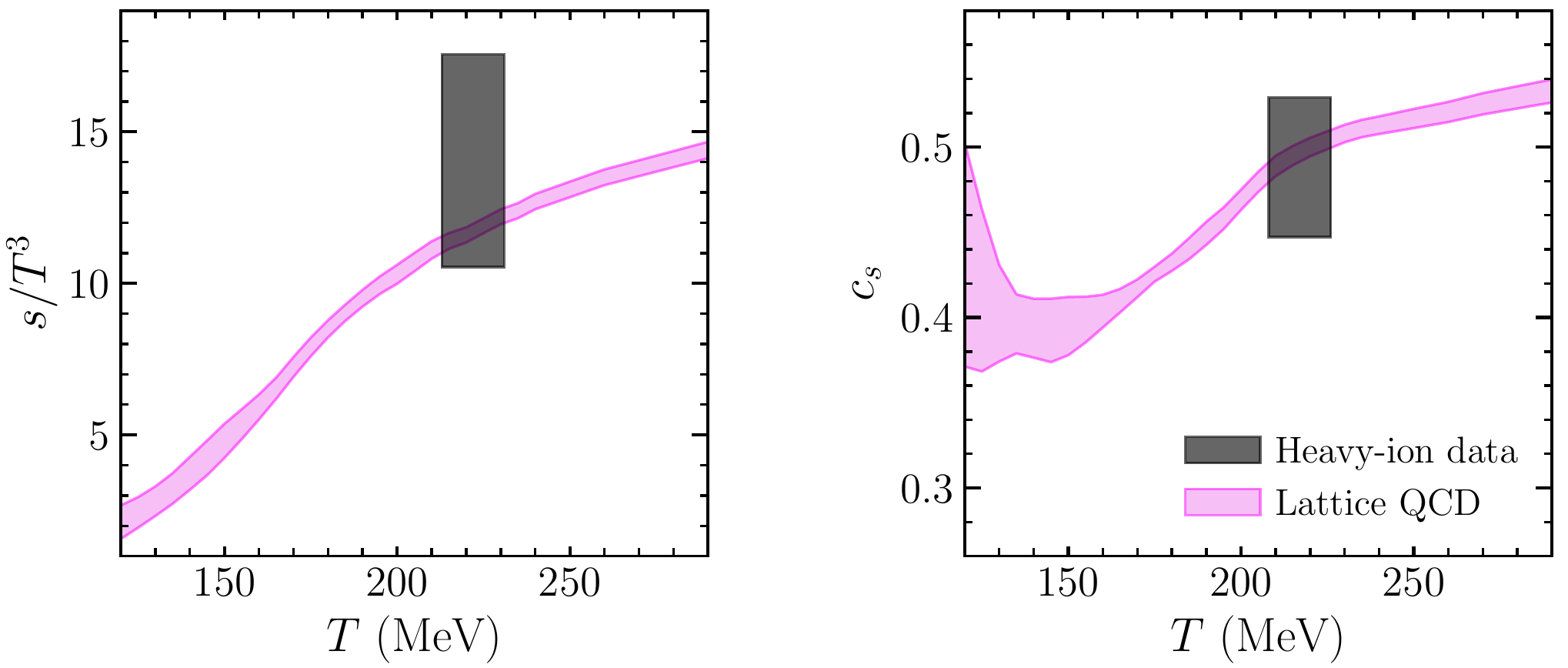}
    \caption{Equation of state of hot QCD. Left panel: entropy density scaled by $T^3$. Right panel: speed of sound. The light shaded bands in magenta represent the results of first-principles lattice QCD calculations, taken from Ref.~\cite{Borsanyi:2013bia}. The dark-shaded boxes represent, on the other hand, the extraction of these quantities from heavy-ion experimental data, as described in the text. Figure adapted from Ref.~\cite{Gardim:2019xjs}.}
    \label{fig:3-5}
\end{figure}

I extract now the entropy density, which is defined by:
\begin{equation}
    s(T_{\rm eff}) = \frac{dN/dy}{V_{\rm eff}} S/N_{\rm ch},
\end{equation}
where $dN/dy$ is the charged multiplicity per unit rapidity, and $S/N_{\rm ch}$ is the entropy per particle. The latter was recently studied and extracted from LHC data~\cite{Hanus:2019fnc}, and it is about $6.7$ in \pbpb{} collisions at top LHC energy. With $dN / dy \approx 1.15 dN / d \eta$, $dN/d\eta=2000$ and $V_{\rm eff}=800~\fm^3$, one obtains:
\begin{equation}
    s(T=220~{\rm MeV}) \simeq 20 \pm 5 \fm^{-3},
\end{equation}
where the significant uncertainty comes mostly from the fact that $V_{\rm eff}$ depends on the choice of the model of energy deposition. The entropy density as a function of temperature in hot QCD is shown in the left panel of \fig{3-5}. The light shaded band in magenta represents the result of lattice QCD calculations for the dimensionless ratio $s/T^3$~\cite{Borsanyi:2013bia}. This curve grows rather steeply around $T\approx150~{\rm MeV}$, while it is rather flat beyond $T=300~{\rm MeV}$, towards the high-temperature limit. The steep growth of the curve for $T>150~{\rm MeV}$ is a signature of parton deconfinement. Indeed, by combining \equ{4n} with \equ{nu}, one obtains:
\begin{equation}
    \frac{s}{T^3} = \frac{\nu_{\rm QCD}}{\pi^3}.
\end{equation}
Hence the growth of $s/T^3$ shown in the figure is a characteristic signature of the liberation of color degrees of freedom in the medium when $T$ becomes higher than the critical temperature. The value of $s/T^3$ extracted from ALICE data at $T=T_{\rm eff}=222\pm9$~MeV gives me the result shown as a dark shaded box in the left panel of \fig{3-5}. The estimate of the entropy density from heavy-ion collision data is thus in agreement with the results of lattice QCD, showing that the quark-gluon plasma is indeed formed in these experiments.

One can further obtain a solid estimate of the speed of sound of the fluid. I recall the following thermodynamic identity~\cite{Gardim:2019brr}:
\begin{equation}
\label{eq:cs2}
    c_s^2 \equiv \frac{dP}{d\epsilon} = \frac{sdT}{Tds} = \frac{d \ln T}{d \ln s}.
\end{equation}
Now, if $s$ is proportional to the final-state multiplicity, $dN_{\rm ch}/d\eta$, and $T_{\rm eff}$ is proportional to $\bra p_t \ket$, the relative variations of these quantities coincide, i.e.,
\begin{equation}
    \frac{dT}{T} = \frac{d\bra p_t \ket}{\bra p_t \ket }, \hspace{40pt}\frac{ds}{s} = \frac{d(dN_{\rm ch}/d\eta) }{dN_{\rm ch}/d\eta}.
\end{equation}
Hence, \equ{cs2} becomes:
\begin{equation}
    c_s^2 =   \frac{d \ln \bra p_t \ket}{ d \ln (dN_{\rm ch}/d\eta) }.
\end{equation}
As measurements of $\bra p_t \ket$ and $dN_{\rm ch}/d\eta$ are available at both $\sqrt{s_{\rm NN}}=2.76$~TeV and $\sqrt{s_{\rm NN}}=5.02~{\rm TeV}$, I can extract the value of $c_s^2$ at a value of temperature which is halfway between the values of $T_{\rm eff}$ at the two energies, which turns out to be $T_{\rm eff}\simeq217~{\rm MeV}$. By doing so, one obtains:
\begin{equation}
    c_s^2(T_{\rm eff}=217~{\rm MeV}) = 0.24 \pm 0.04. 
\end{equation}
The speed of sound of hot QCD returned by first-principles lattice calculations is shown in the right panel of \fig{3-5}. One notes, once again, the steep increase of this quantity above the critical temperature (shaded band in magenta). The dark-shaded band represents the evaluation from heavy-ion collision data, which is in nice agreement with the lattice QCD result. The speed of sound of the quark-gluon plasma extracted from heavy-ion data is thus 0.5, i.e., half the speed of light in vacuum.

\subsection{Fluctuations of $\bra p_t \ket$}

\label{sec:3-22}

I turn now my attention to an aspect of the physics of $\bra p_t \ket$ which will be crucial in the upcoming discussion of nuclear deformation effects in heavy-ion collisions. At a given collision centrality, i.e., at a fixed experimental value of the multiplicity, the value of $\bra p_t \ket$ is not constant, but it fluctuates on an event-by-event basis. Fluctuations of $\bra p_t \ket$ at a given centrality were analyzed at RHIC very shortly after the beginning of the heavy-ion program~\cite{Adams:2005ka}. A few years later, they were eventually studied by the Krakow group in the context of modern event-by-event hydrodynamic simulations to elucidate their physical origin~\cite{Broniowski:2009fm}. 
\begin{figure}[t]
    \centering
    \includegraphics[width=.6\linewidth]{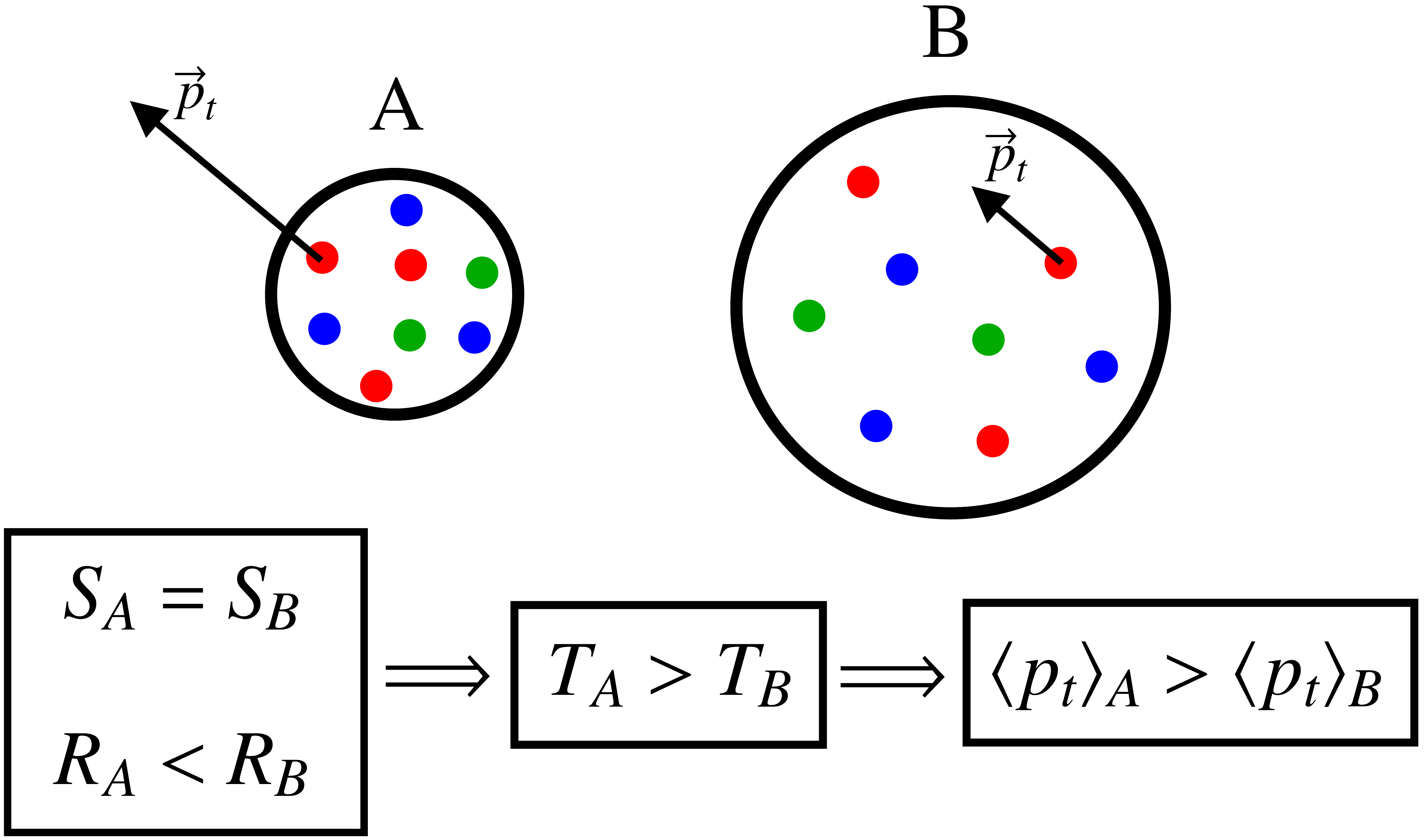}
    \caption{At fixed entropy, $S$, the origin of the fluctuations of $\bra p_t \ket$ can be ascribed to the fact that the size of the system, $R$, fluctuates. In this figure, systems $A$ and $B$ have the same entropy, but $A$ presents a smaller value of $R$. The temperature in $A$ is thus larger. For an ultrarelativistic gas, this implies in particular
that the particles in $A$ carry larger momentum, $p_t$. Figure from Ref.~\cite{Giacalone:2020awm}.}
    \label{fig:3-6}
\end{figure}

These studies show in particular that in hydrodynamics the fluctuations of $\bra p_t \ket$ are driven by the fluctuations of the system size, $R$, as defined by \equ{R}~\cite{Broniowski:2009fm,Bozek:2012fw}. This is very transparent in view of the previous result that $\bra p_t \ket$ is essentially a measure of the temperature, $T$. I refer to the illustration in \fig{3-6}. I consider two quark-gluon plasmas that share the same entropy, but that are contained within different volumes. I assume for simplicity that these systems are uniform. The medium that has a smaller volume is thus denser, and it has a larger temperature. It consequently yields to the final state particles that carry a larger value of $\bra p_t \ket$. In this setup, then, there is a one-to-one correspondence between the system size, and the other thermodynamic quantities (temperature, volume, energy).

I make an explicit check that this picture is true in event-by-event hydrodynamic simulations. These simulations are conceived in such a way to make the effect I am after more apparent. This calculation evolves, through the \music{} code, 850 profiles obtained with the same \trento{} model used for \fig{3-2} with ideal hydrodynamic equations, for \pbpb{} collisions at fixed impact parameter $b=2.5~\fm$, and fixed final-state multiplicity, i.e, fixed $S$ at the initial condition, corresponding to collisions at top LHC energy. One studies, then, the statistical correlation between $\bra p_t \ket$ and thermodynamic quantities. The left panel of the figure shows the correlation between $\bra p_t \ket$ and $E \equiv E(\tau=\tau_0)$, i.e., the total energy in the fluid at the beginning of hydrodynamics. The panel on the right shows instead the correlation between $\bra p_t \ket$ and the size, $R$.
\begin{figure}[t]
    \centering
    \includegraphics[width=\linewidth]{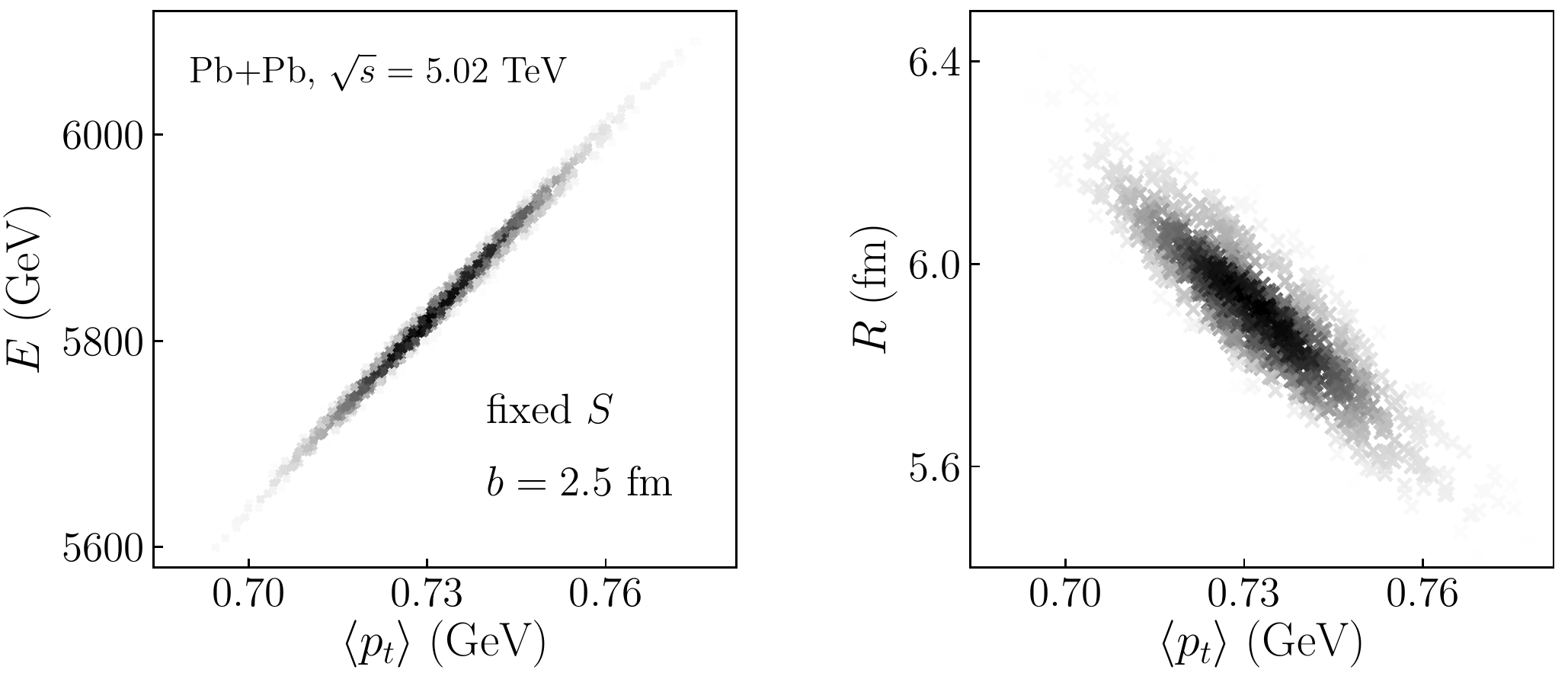}
    \caption{Results from 850 ideal hydrodynamic simulations of 5.02~TeV Pb+Pb collisions at fixed impact parameter $b = 2.5 \fm$. The events have different entropy profiles, but they all correspond to the same integrated entropy, $S$. Left: $\bra p_t \ket$ versus the initial energy per unit rapidity, $E$, defined by \equ{totE}. Right:  $\bra p_t \ket$ versus the initial size, $R$, defined by \equ{R}. Figure adapted from Ref.~\cite{Giacalone:2020dln}.}
    \label{fig:3-7}
\end{figure}

The correlation is strong and linear in both panels, confirming the proposed physical origin of the fluctuations of $\bra p_t \ket$. However, it is distinctly stronger for $E$. This is somewhat natural. The simple picture of \fig{3-6} holds for a uniform fluid, where $R$ and $T$ at fixed $S$ are in an exact one-to-one correspondence. But if one drops the assumption of an uniform fluid, and introduces spiky structures in the density profile, then $R$ and $T$ are no longer in a one-to-one correspondence~\cite{Giacalone:2020awm}, while the relation $p \sim E$ remains true, as confirmed by \fig{3-7}. One should also note that, as I shall argue later on in Chapter~\ref{chap:5}, if one relaxes the condition of fixed $S$ in \fig{3-7}, then the natural initial-state predictor of $\bra p_t \ket$ becomes precisely $E/S$. Finally, the fluctuations of $E$ and $R$ relative to their mean are of order $2\%$, while the relative fluctuation of $\bra p_t \ket$ in experimental data is around 1\%~\cite{Abelev:2014ckr}. In general it is not easy for hydrodynamic simulations to return the right magnitude of the relative fluctuation of $\bra p_t \ket$~\cite{Bozek:2017elk}, although the most recent results from the Duke group~\cite{Bernhard:2019bmu} show that a \trento{} model where \equ{p=0} is used for the energy density at $\tau=0^+$ allows to capture the fluctuations of $\bra p_t \ket$ observed in experimental data at the end of hydrodynamics.

\paragraph{In summary --} I have elucidated the origin of $\bra p_t \ket$ in a given hydrodynamic event. It corresponds approximately to $3T$, where $T$ is the average temperature of the quark-gluon plasma at the time when longitudinal cooling becomes ineffective. This allows one in particular to use $\bra p_t \ket$ as a probe of the EOS of hot QCD matter. Secondly, I have shown that the value of $\bra p_t \ket$ is not the same in all events at a given multiplicity. Fluctuations of $\bra p_t \ket$ are driven by fluctuations of thermodynamic quantities (energy, volume, etc.) at the initial condition.

\section{Momentum anisotropy from spatial anisotropy}

\label{sec:3-3}

In the field of observational cosmology, accurate measurements of the anisotropies that characterize the temperature map of the cosmic microwave background are performed. Anisotropy is quantified through a power spectrum, which is obtained with a multipole decomposition of the observed temperature map~\cite{Page:2006hz}. The study of anisotropy sheds light on the initial condition of the system, as well as on the dynamical features of the cosmological expansion. Remarkably enough, a similar kind of analysis is performed as well in the context of heavy-ion collision experiments, where the observable sky is replaced by the detector surrounding the interaction region, the map of temperature is replaced by the distribution of particle momenta, and spherical symmetry is replaced by cylindrical symmetry, due to the peculiar geometry of interaction as observed in the laboratory frame. 

Due to the latter symmetry, in heavy-ion collisions one is mostly interested in the anisotropy of particle emission in the two-dimensional plane corresponding to the midrapidity slice. One performs a Fourier expansion of the measured spectrum:
\begin{equation}
\label{eq:fourier}
    \frac{dN}{p_t dp_t d\phi_p} = \sum_{-\infty}^{+\infty} V_n e^{in\phi_p} .
\end{equation}
where the complex $V_n$ coefficients are the so-called \textit{anisotropic flow} coefficients. Flow coefficients depend in principle on the kinematic variables, $p_t$, and also on $y$ if one goes beyond the approximation of boost-invariant evolution. Here I shall only be interested in coefficients that are integrated over the entire phase space:
\begin{equation}
    V_n =  \frac{1}{N}  \int_{{\bf p}_t} \frac{dN}{d^2 {\bf p}_t} e^{-in\phi_p},
\end{equation}
where $\phi_p$ is the azimuthal angle in momentum space. The Fourier spectrum of ultrarelativistic heavy-ion collisions has been to date analyzed up to $n=9$~\cite{Acharya:2020taj}.

\subsection{Elliptic flow}

\label{sec:3-31}

Elliptic flow is the quadrupole, i.e., the Fourier coefficient corresponding to $n=2$ in \equ{fourier}:
\begin{equation}
    V_2 = \frac{1}{N} \int_{{\bf p}_t} \frac{dN}{d^2 {\bf p}_t} e^{-i2\phi_p},
\end{equation}
This quantity plays a special role in high-energy nuclear physics. The reason is that the hydrodynamic expansion of the quark-gluon plasma created in a generic heavy-ion collision is expected to yield an especially-visible elliptic flow in the azimuthal distribution of emitted particles.

\paragraph{Physical origin --}  Recall the Euler equation, corresponding to the Navier-Stokes equation for an ideal fluid ($d/dt$ is a material derivative):
\begin{equation}
\label{eq:euler}
   \rho  \frac{ d {\bf v}}{dt} = -\vec \nabla P,
\end{equation}
stating that the force per unit volume is driven by pressure gradients in the medium.  The pressure gradient scales like $1/R$, where $R$ is as usual the transverse size. This implies that, if the density profile of the system at rest, i.e., the initial condition, is not symmetric under azimuthal rotations, then its evolution is governed by a pressure gradient that is not azimuthally isotropic. The force that drives the expansion in vacuum is not the same in all directions, thus leading to an anisotropic distribution of momentum within the fluid. 

\begin{figure}[t]
    \centering
    \includegraphics[width=\linewidth]{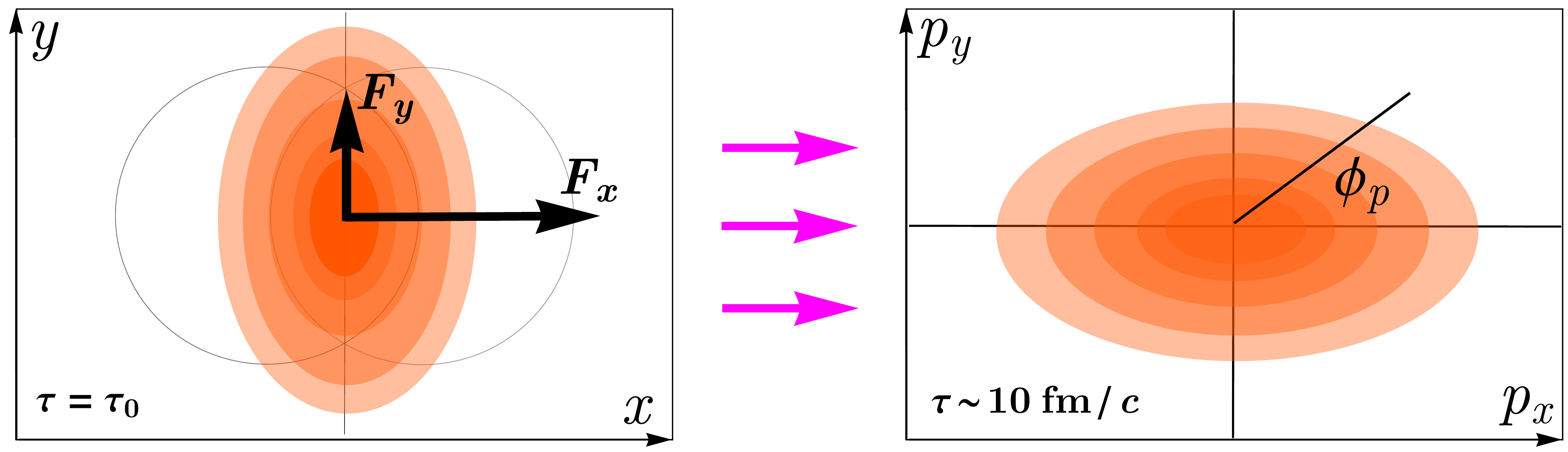}
    \caption{Build-up of flow in the medium created following the interaction of two nuclei at a finite impact parameter. Left: initial condition. The spatial density of the medium presents a distinct quadrupole anisotropy in the transverse plane. Right: Momentum density in the medium close to the end of the hydrodynamic phase. The distribution of momentum in the medium has a pronounced $\cos 2\phi_p$ asymmetry, leading to nonzero elliptic flow.}
    \label{fig:3-8}
\end{figure}
 As originally pointed out by Ollitrault~\cite{Ollitrault:1992bk}, this simple feature is crucial in heavy-ion collisions. Consider the case of an off-centered heavy-ion collision, with $b\approx10~\fm$. I follow the illustration of \fig{3-8}. The left panel is the initial condition at $\tau=\tau_0$. The colliding nuclei are depicted as circles in the $(x,y)$ plane. The medium, which is drawn as a contour plot, is created in the almond-shaped region of overlap, and acquires naturally an elliptical shape. Following \equ{euler}, in such a medium there is more acceleration along $x$ than along $y$, because of the asymmetric shape. The expansion thus builds up more flow along $x$. I look then at the right panel of the figure, showing the system in momentum space $(p_x,p_y)$ at a later time, for instance $\tau\sim10~\fm$, which is close to the end of the hydrodynamic phase. Due to the imbalance of forces in the medium at the initial condition (left panel), we see that the density of momentum in the transverse plane has also a quadrupole asymmetry. More momentum has been produced along $x$, hence, if we label $\phi_p$ the azimuthal angle in momentum space, then the distribution of momentum has precisely a $\cos 2\phi_p$ asymmetry. 
 
 Freeze-out of the medium and the subsequent decays of resonances modify only mildly the global anisotropy pattern imprinted by the hydrodynamic expansion, so that the anisotropy of the system in momentum space is carried over to the final-state particles. The previous argument thus explains why elliptic flow, as given by \equ{V2}, is important in heavy-ion collisions, and why its detection has been historically considered as a smoking-gun of the hydrodynamic behavior of the system created at RHIC. Before showing the results of experiments, I would like to follow up on the discussion of the previous sections. The idea is that elliptic flow is also an observable whose origin can be traced back to the initial state, much as the final-state multiplicity is a measure of the entropy of the system, or $\bra p_t \ket$ is a measure of the total energy at a given collision centrality.

\paragraph{Initial anisotropy --} The relevant initial-state predictor for elliptic flow is, unsurprisingly, the elliptic anisotropy of the medium at the beginning of the hydrodynamic expansion. A measure of the ellipticity of the system which takes into account the fact that the created medium is not a smooth deformed profile, but contains nontrivial structures induced by the colliding nucleons, is provided by:
\begin{equation}
\label{eq:E2}
    \mathcal{E}_2 = - \frac{\int_{{\bf x}} |{\bf x}|^2 e^{i2\phi} \epsilon(\tau_0, {\bf x}) }{\int_{{\bf x}} |{\bf x}|^2 \epsilon({\tau_0, \bf x}) },
\end{equation}
where $\phi$ is the azimuthal angle in the transverse plane, $\tau_0$ is the time at which the hydrodynamic expansion starts, $\epsilon$ is the profile of energy density, and the minus sign is a convention. In the context of heavy-ion collision, this quantity has been introduced by Teaney and Yan~\cite{Teaney:2010vd}, who rigorously derived it from a cumulant expansion of $\epsilon$, and demonstrated that $\mathcal{E}_2$ corresponds to the elliptic anisotropy of the long-wavelength modes of the system. The expression of $\mathcal{E}_2$ is derived here in Appendix~\ref{app:A}. It is interesting to note that the same expression for the ellipticity of a generic two-dimensional density profile was in fact introduced 25 years ago in the context of weak gravitational lensing, to define the elliptic anisotropy in the shape of the images of galaxies, see Eq.~(3-2) in Ref.~\cite{Kaiser:1994jb}.

\begin{figure}[t]
    \centering
    \includegraphics[width=\linewidth]{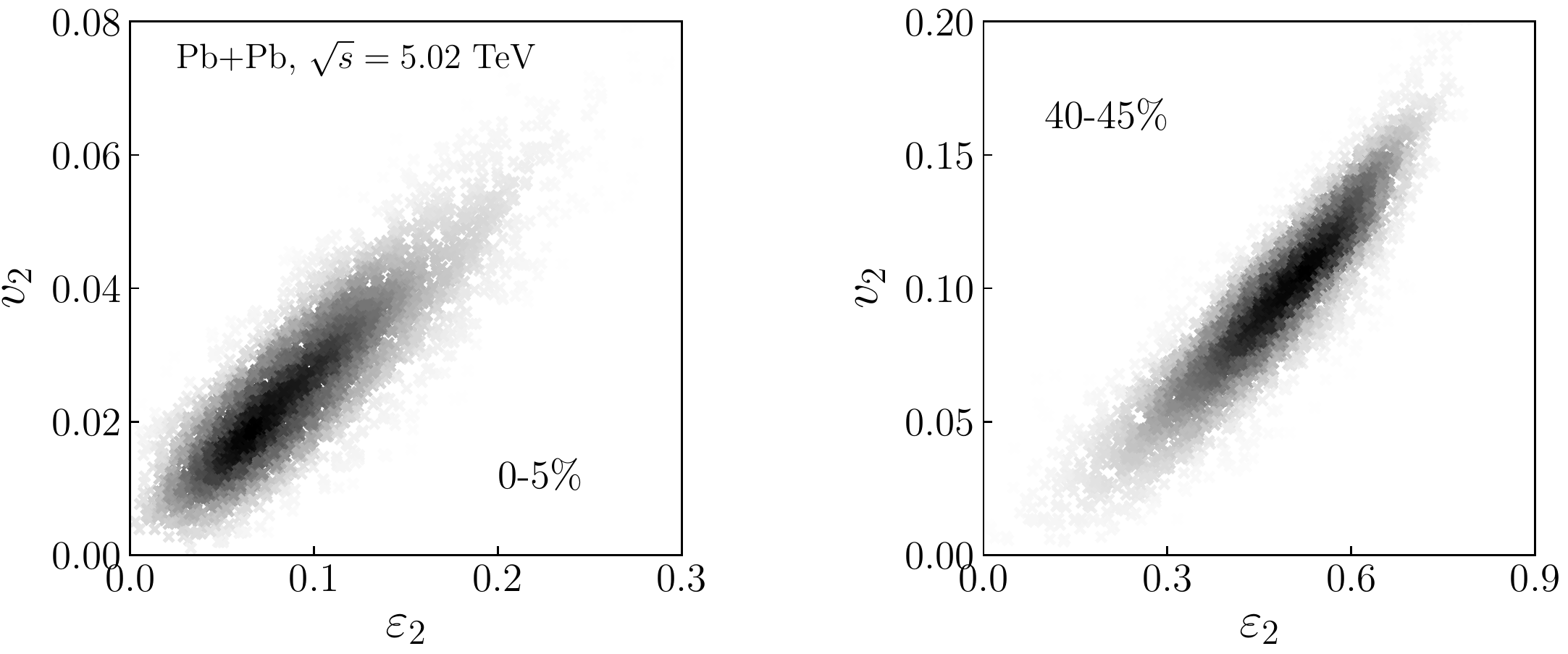}
    \caption{Correlation between $v_2\equiv|V_2|$ and $\varepsilon_2\equiv|\mathcal{E}_2|$ in narrow bins of centrality in \pbpb{} collisions at $\sqrt{s_{\rm NN}}=5.02~{\rm TeV}$. Both panels show approximately 2500 points, each point corresponding to a full hydrodynamic simulation. Left: 0-5\%, corresponding to $\bra b \ket\approx 3~\fm$. Right: 40-45\%, corresponding to  $\bra b \ket\approx 10~\fm$. }
    \label{fig:3-9}
\end{figure}

I show now in hydrodynamic simulations that the final-state elliptic flow obtained from the distribution of hadrons provides, at a given collision centrality, a measure of the initial eccentricity of the medium. I go back to the hydrodynamic simulations of \pbpb{} collisions used to draw \fig{3-1}. From the values of the initial \trento{} entropy, I sort the events into centrality classes, and in these classes I evaluate both $V_2$ and $\mathcal{E}_2$ for all events. I shall use two centrality bins for reference: 0-5\%, corresponding to central collisions at $\bra b \ket\approx 3~\fm$, and 40-45\%, corresponding to peripheral collisions, $\bra b \ket \approx 10~\fm$. I plot $v_2\equiv|V_2|$ as a function of $\varepsilon_2\equiv|\mathcal{E}_2|$, to examine their correlation. The results are displayed in \fig{3-9} 
 
 For both collision centralities, I remark a strong correlation between the final elliptic flow and the initial eccentricity. The statistical correlation between these quantities, as given e.g. by the Pearson correlation coefficient is larger than $0.9$ in both panels. The correlation is thus mostly linear. If one defines a coefficient of linear response:
 \begin{equation}
     \kappa_2 = \frac{ \bra  V_2\mathcal{E}_2^* \ket}{\bra \mathcal{E}_2\mathcal{E}_2^* \ket},
 \end{equation}
where angular brackets denote an average over events in the centrality bin, one has to a very good approximation:
\begin{equation}
\label{eq:linearresp}
    v_2 = \kappa_2 \varepsilon_2,
\end{equation}
as I shall show explicitly in a moment. The coefficient $\kappa_2$ is a property of the medium. It depends on the viscosity of the quark-gluon plasma, as well as on its equation of state.

\paragraph{Viscosity --} 
I make now a short digression to show that, due to simple dimensional arguments, elliptic flow represents a neat probe of the viscosity of the quark-gluon plasma. I recall the nonrelativistic Navier-Stokes equation, where I include now both the shear viscosity, $\eta$, and the bulk viscosity, $\zeta$~\cite{rezzolla}:
\begin{equation}
\label{eq:NSfull}
    \rho \frac{d{\bf v}}{dt} = \underbrace{-\vec \nabla P}_{1/R} + \underbrace{ \eta \vec \nabla^2 {\bf v} + \vec \nabla \biggl [ \vec \nabla \cdot {\bf v} \biggl ( \zeta + \frac{2}{3}\eta \biggr ) \biggr ]  }_{1/R^2}.
\end{equation}
 The viscous corrections go against the pressure-gradient force, i.e., against the development of anisotropic flow. As these corrections scale with two inverse powers of the system size, they are more important at large centrality than at small centrality, and in smaller colliding systems. Viscous corrections to elliptic flow can then be studied systematically by varying the size of the system, which makes $V_2$ a powerful probe of the role of $\eta$ and $\zeta$. One should however note that the viscosity does not break up the linear relation between $\varepsilon_2$ and $v_2$ discussed in \fig{3-9}. The effect of the viscous corrections is simply that of damping the value of the response, $\kappa_2$, in \equ{linearresp}.

\begin{figure}[t]
    \centering
    \includegraphics[width=.75\linewidth]{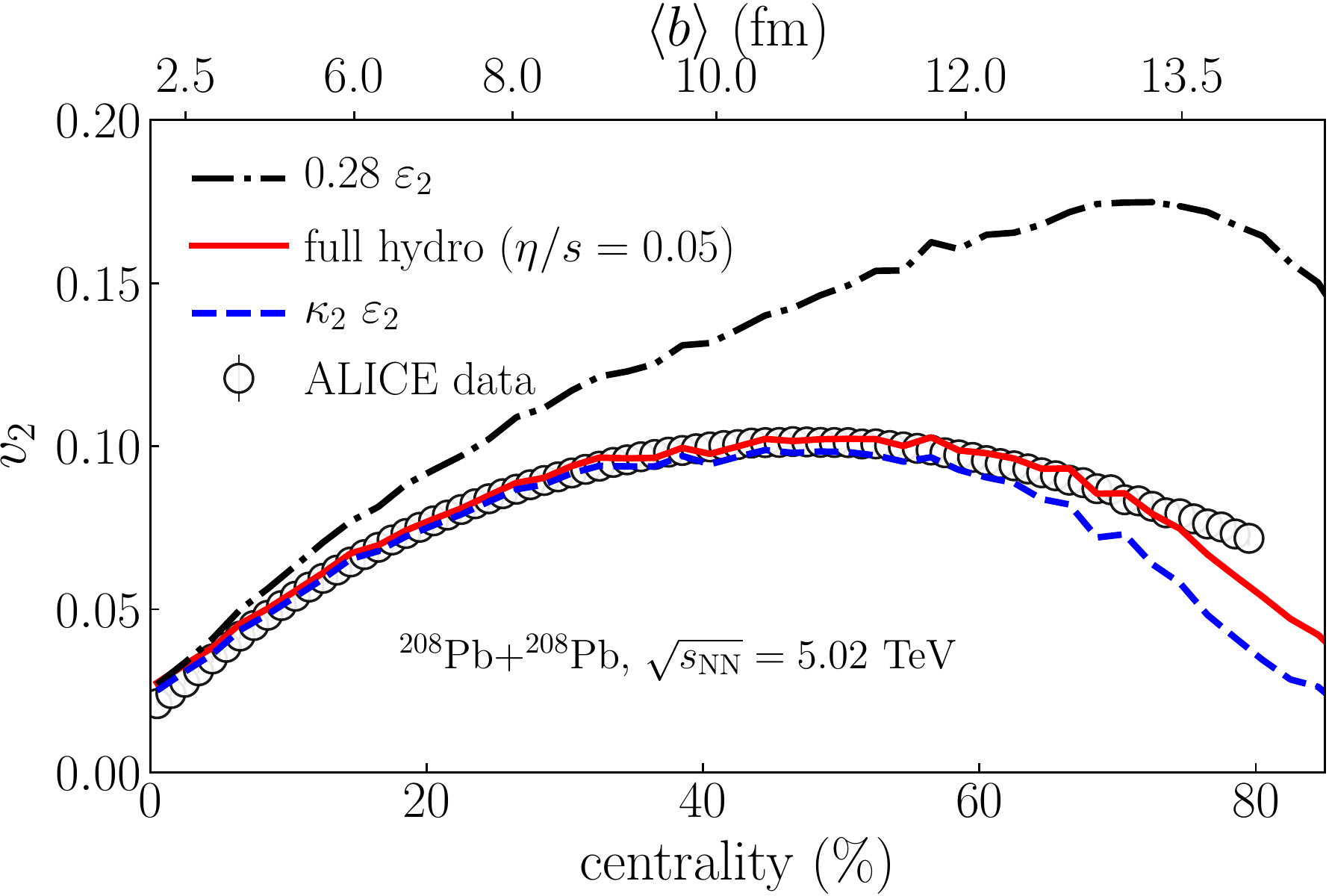}
    \caption{Centrality dependence of the rms elliptic anisotropy in \pbpb{} collisions at $\sqrt{s_{\rm NN}}=5.02~{\rm TeV}$. Dot-dashed line: rms eccentricity of the medium, $\varepsilon_2$, rescaled by a factor $0.28$. Solid line: rms elliptic flow, $v_2$, resulting from full viscous hydrodynamic simulations. Dashed line: results for $\kappa_2\varepsilon_2$, where $\varepsilon_2$ is the rms eccentricity shown as a dot-dashed line, while $\kappa_2$ is defined by \equ{linearresp}. Symbols are ALICE data~\cite{Acharya:2018lmh} on the rms elliptic flow. The axis on top of the plot displays the average value of the impact parameter, computed with \equ{cb}, corresponding to a few selected centrality percentiles.}
    \label{fig:3-10}
\end{figure}

\paragraph{Phenomenology --} 
Time to look at experimental data. In the upcoming Sec.~\ref{sec:3-4} I shall explain that, experimentally, one can not measure elliptic flow, $V_2$, or its magnitude, in each event, due to the fact that the number of particles detected in one event is not sufficient to reconstruct a Fourier series. In experiments, one can only measure the even moments of the distribution of $|V_2|$, i.e., moments of $V_2V_2^*$, which do not depend on the phase of the complex number. The standard measure of elliptic flow is the rms value:
\begin{equation}
\label{eq:rmsv2}
    v_2 \equiv \sqrt{ \bra |V_2|^2 \ket},
\end{equation}
where the angular brackets denote an average over events, typically corresponding to a given class of centrality. Note that, if \equ{linearresp} is valid on an event-by-event basis, then the rms $v_2$ defined by \equ{rmsv2} is in a linear correlation with the rms value of the initial ellipticity:
\begin{equation}
    \sqrt{ \bra |V_2|^2 \ket} = \kappa_2 \sqrt{ \bra |\mathcal{E}_2|^2 \ket}.
\end{equation}
I analyze now these quantities as a function of collision centrality in $5.02~{\rm TeV}$ \pbpb{} collisions, by means of the same hydrodynamic simulations used to draw \fig{3-1} and \fig{3-9}. The results are shown in \fig{3-10}.

The black dash-dotted line represents the rms $\varepsilon_2$, rescaled by a constant factor $0.28$. This quantity grows by essentially one order of magnitude from central to peripheral collisions. This is related to the increasing collision impact parameter, which leads to a region of nuclear overlap that possesses a more and more enhanced elliptical asymmetry. The red solid line shows instead results for the rms $v_2$ computed at the end of the full viscous hydrodynamic simulations~\cite{Giacalone:2017dud}. The elliptic flow coefficient grows again by a large factor from central to peripheral collisions, although it is not simply given by a global rescaling of the curve of $\varepsilon_2$. In particular, in peripheral collisions the rms elliptic flow tends to flatten, and presents a decreasing trend much earlier than $\varepsilon_2$. This is an effect of the above-mentioned viscous damping, which becomes sizable at large centrality, due to the smaller system size [recall \equ{NSfull}]. The blue dashed curve shows the rms $ v_2 = \kappa_2 \varepsilon_2$, where $\kappa_2$ is calculated, at each centrality, from \equ{linearresp}. I note an excellent agreement between the red solid curve, the rms $v_2$, and the blue dashed curve, i.e., the estimate of linear response theory. The response is thus linear all the way to peripheral collisions, where nonlinearities become more sizable. 

To a very good approximation, then, the role of hydrodynamics is that of providing, at each centrality, a response coefficient $\kappa_2$, whose magnitude and centrality dependence depends on the properties of the medium, i.e., viscosity and equation of state. The centrality dependence of $\kappa_2$ can be essentially read by eye off this plot by comparing the curve for the rescaled rms $\varepsilon_2$ with the curve for $\kappa_2 \varepsilon_2$. Finally, \fig{3-10} shows, as symbols, recent ALICE data on the rms $v_2$. The agreement between data and the the red solid curve, i.e., the full hydrodynamic prediction, is good. If the value of $\kappa_2$ were known beforehand, then, one could simply try to describe experimental data from the \trento{} model. This procedure would work within an accuracy of order $5\%$ in central and semi-central collisions, which is excellent. In the following, I shall indeed exploit this fact to perform both comparisons with data and new predictions by means of the sole knowledge of $\varepsilon_2$ at a given collision centrality.

\subsection{The role of fluctuations}
\label{sec:3-32}

A nontrivial result displayed by \fig{3-10} is the fact that elliptic flow does not vanish in the limit of central collisions, $b\rightarrow0$. The origin of elliptic flow in central collisions is easy to guess from the analysis performed in this manuscript. As pointed out in \fig{2-6}, the quark-gluon plasma does not have a smooth profile of density, but is rather a spiky landscape with valleys and peaks. These structures, which in heavy-ion collisions are mostly caused by event-by-event fluctuations in the positions of the colliding nucleons, generate anisotropy. Hence elliptic flow is quite significant even in collisions at zero impact parameter. In addition, the hadron distribution acquires a full spectrum of Fourier modes:
\begin{equation}
\label{eq:Vn}
    V_n = \frac{1}{N} \int_{{\bf p}_t} \frac{dN}{d^2 {\bf p}_t} e^{-in\phi_p}.
\end{equation}

\begin{figure}[t]
    \centering
    \includegraphics[width=.75\linewidth]{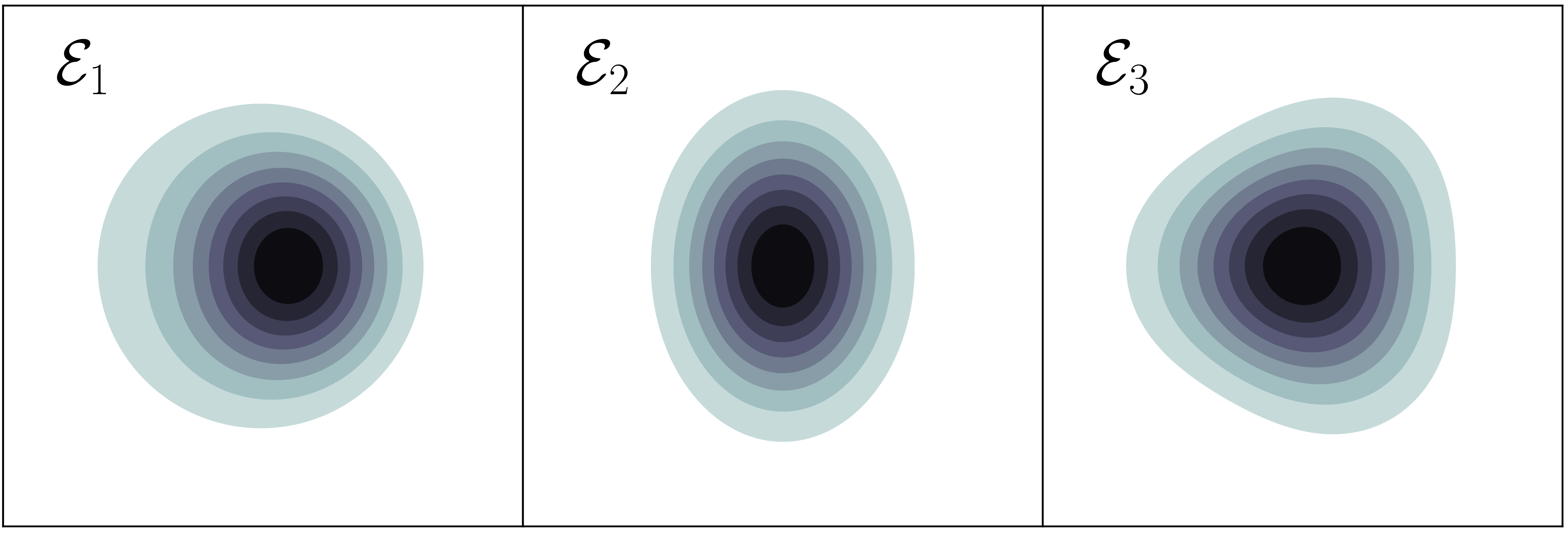}
    \caption{Anisotropies of order $n=1,2,3$. Left: dipole asymmetry. The profile has $\varepsilon_1=0.22$, $\varepsilon_2=0$, $\varepsilon_3=0$. Center: ellipticity. The profile has $\varepsilon_1=0$, $\varepsilon_2=0.28$, $\varepsilon_3=0$. Right: triangularity. The profile has $\varepsilon_1=0$, $\varepsilon_2=0$, $\varepsilon_3=0.22$.}
    \label{fig:3-11}
\end{figure}

In the hydrodynamic paradigm, these coefficient have also a simple geometric origin. One can play the same game as in the case of elliptic flow, and define:
\begin{equation}
\label{eq:E1E3}
    \mathcal{E}_1 = - \frac{\int_{{\bf x}} |{\bf x}|^3 e^{i\phi} \epsilon(\tau_0, {\bf x}) }{\int_{{\bf x}} |{\bf x}|^3 \epsilon({\tau_0, \bf x}) },
    \hspace{40pt}
    \mathcal{E}_3 = - \frac{\int_{{\bf x}} |{\bf x}|^3 e^{i3\phi} \epsilon(\tau_0, {\bf x}) }{\int_{{\bf x}} |{\bf x}|^3 \epsilon({\tau_0, \bf x}) },
\end{equation}
where $\mathcal{E}_1$ is the \textit{dipole asymmetry}, and $\mathcal{E}_3$ is the \textit{triangularity}, originally introduced by Alver and Roland~\cite{Alver:2010gr}. The geometric interpretation of $\mathcal{E}_1$ and $\mathcal{E}_3$ is displayed, for sake of clarity, in \fig{3-11}. Much as $V_2$ is a linear response to the initial $\mathcal{E}_2$, the final $V_1$ and $V_3$ are a linear response to the initial $\mathcal{E}_1$ and $\mathcal{E}_3$. The expressions of $\mathcal{E}_1$ and $\mathcal{E}_3$ were also obtained with the rigorous derivations of Teaney and Yan~\cite{Teaney:2010vd}, and a derivation of these quantities is also proposed here in Appendix~\ref{app:A}. It is once again interesting to note that these quantities were also shown about 15 years ago in the context of weak gravitational lensing, with the aim of characterizing anisotropies in the shapes of galaxies beyond the elliptical one (see Eqs.~(26) and (27) of Ref.~\cite{Okura:2006fi}). Finally, anisotropies of order $n>3$, like $V_4$ or $V_5$, have also a geometric origin, $\mathcal{E}_{n>3}$, which can be evaluated as in the Teaney-Yan paper. However, the expressions are not as straightforward as for $n<4$, due to the fact that higher-order harmonics receive important contributions from the mixing of lower-order terms allowed by symmetry \cite{Teaney:2012ke,Yan:2015jma,Giacalone:2018wpp}. For instance $V_4$ receives a contribution from $V_2^2$, which is in fact dominant in peripheral collisions, while $V_5$ receives a contribution from $V_2V_3$, and so on.

\begin{figure}[t]
    \centering
    \includegraphics[width=.8\linewidth]{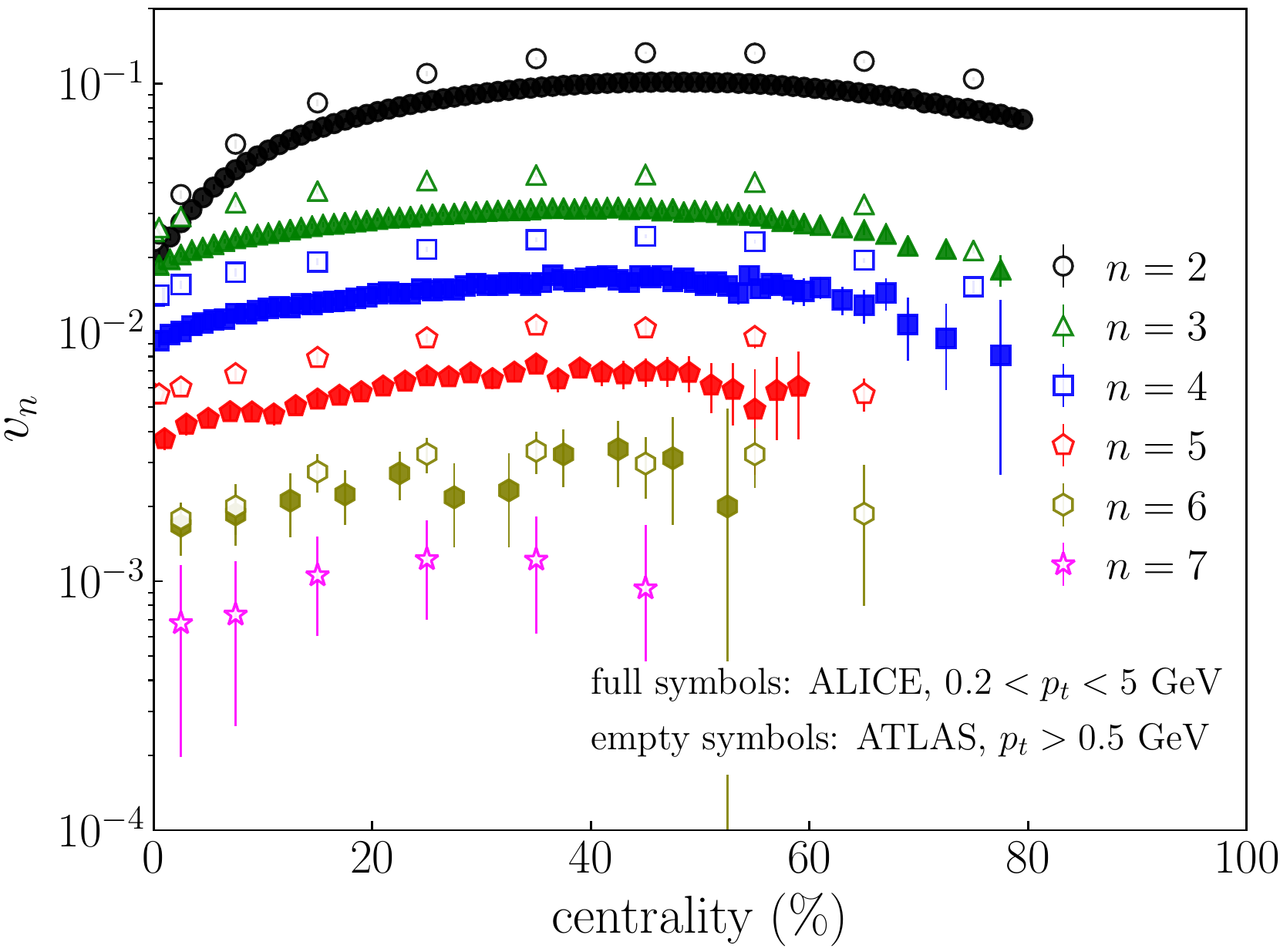}
    \caption{Centrality dependence of the rms Fourier anisotropies of the azimuthal momentum distribution of hadrons produced in $5.02~{\rm TeV}$ \pbpb{} collisions. Full symbols: ALICE data~\cite{Acharya:2018lmh} ($0.4<|\eta|<1$, $0.2<p_t<5$~GeV). Empty symbols: ATLAS data~\cite{Aaboud:2018ves} ($1.0<|\eta|<2.5$, $0.5<p_t<60$~GeV). }
    \label{fig:3-12}
\end{figure}

In \fig{3-12} I show experimental LHC data on the rms values of $v_n$, for $n=2\ldots7$. I show both ALICE and ATLAS data, to assess the role played by the $p_t$ cuts implemented in the integral of the spectrum in \equ{Vn} (indicated in the figure), which turns out to yields a vertical shift for all $n$. The feature of the experimental data that jumps to the eye is the fact that the Fourier spectrum is rather strongly-ordered (note the log scale for the vertical axis). This is nontrivial. In central collisions, for example, the value of the initial-state anisotropy $\varepsilon_n$ is of the same order of magnitude for all values of $n$. The splitting between harmonics in central collision is a consequence of hydrodynamics, in particular of viscous hydrodynamics, as the damping of higher-order harmonics is enhanced by the viscous corrections, for the same dimensional arguments pointed out previously. As expected, elliptic flow is the only coefficient that grows by almost one order of magnitude from central to peripheral collisions, due to the increasing impact parameter. The other coefficients display instead a mild increase with the centrality percentile. This is intuitive. The size of the system decreases with the impact parameter, and a smaller system size is associated with larger density fluctuations, i.e., larger anisotropy for all values of $n$.

\paragraph{In summary --} The azimuthal distribution of hadrons measured in heavy-ion collisions is characterized by a spectacular degree of anisotropy, and a full spectrum of nonzero Fourier harmonics. In hydrodynamics, final anisotropy in momentum space originates from spatial anisotropy in the density profile at the initial condition. Elliptic flow is the largest anisotropy, as the corresponding spatial anisotropy, $\varepsilon_2$, is mostly induced by the collision impact parameter. At a given collision centrality, $v_2$ and $\varepsilon_2$ are in a linear relation, $v_2=\kappa_2\varepsilon_2$.

\section{Observables as multi-particle correlations}
\label{sec:3-4}

Before closing this chapter, I briefly discuss the experimental method that is used in practice to measure the observables analyzed in the previous sections. The point is that measuring $v_2$ or $\bra p_t \ket$ is not as straightforward as it may seem. In theory, e.g., at the end of a hydrodynamic simulation, one is left with a continuous spectrum of hadrons in phase space, that can be integrated to obtain the relevant observables, as in \equ{mpt} and \equ{V2}. However, the outcome of a real collision consists of a collection of $N$ hadrons, and integrals are performed by means of discrete sums. This implies that, since $N$ is finite, any observable comes necessarily with a statistical uncertainty of order $1/\sqrt{N}$. At current particle collider experiments, this statistical fluctuation is as large as the genuine \textit{dynamical} fluctuation that one would like to understand. Meaningful observables involving $v_2$ and $\bra p_t \ket$ can however be constructed to overcome this difficulty.

\paragraph{$\boldsymbol{\bra p_t \ket}$ -- }  In an experiment, the average transverse momentum is obtained by replacing the integral in \equ{mpt} with a sum over the particles of the event under consideration:
\begin{equation}
    \frac{1}{N} \int_{{\bf p}_t} p_t \frac{dN}{d^2{\bf p}_t} \longrightarrow \frac{1}{N} \sum_{i=1}^N p_{t,i},
\end{equation}
where $N$ is the number of detected hadrons and $p_{t,i}$ is the transverse momentum of the $i$-th hadron. The relative dynamical fluctuation of $\bra p_t \ket$ in central collisions is of order 1\%~\cite{Abelev:2014ckr}. Therefore, when computed in a single event with $N\sim1000$, this quantity is almost meaningless. To remove the large influence of statistical fluctuations one has to average over a large batch of events, typically, the events belonging to a given class of centrality:
\begin{equation}
    \bbra p_t \kket \equiv \biggl \bra \frac{1}{N} \sum_{i=1}^N p_{t,i} \biggr \ket = \frac{1}{N_{\rm ev}} \sum_{j=1}^{N_{\rm ev}} \frac{1}{N_j} \sum_{i=1}^{N_j} p_{t,i},
\end{equation}
where $N_{\rm ev}$ is the number of events belonging to the considered centrality class, $N_j$ is the multiplicity of the $j$-th event, and $i$ runs over the particles collected in this event. This correctly gives the average value of $\bra p_t \ket$ at a given centrality.

In Chapter~\ref{chap:5} I shall use as well the second centered moment of the $\bra p_t \ket$ distribution, i.e., the variance, at a given centrality. This quantity is given by:
\begin{equation}
\label{eq:variance}
    \biggl \bra  \bigl ( \bra p_t \ket - \bbra p_t \kket \bigr)^2  \biggl \ket =  \left\langle \frac{\sum_{i,j} \left( p_{i}-\langle\!\langle p_t\rangle\!\rangle\right)\left( p_{j}-\langle\!\langle p_t\rangle\!\rangle\right)}{N^2}   \right\rangle,
\end{equation}
where outer brackets denote the average over events in a given class of centrality. and $i,j$ labels a pair of particles in a given event. The variance thus defined contains both the dynamical component of the fluctuation of $\bra p_t \ket$, as well as a trivial statistical component due to the finite value of $N$. This latter contribution comes from pairing particles with themselves, i.e., from the terms having $i=j$. Subtracting these terms from \equ{variance} yields the so-called \textit{dynamical} fluctuation:
\begin{equation}
\label{eq:ptvar}
\sigma^2_{\rm dynamical}(\bra p_t\ket) =   \left\langle \frac{\sum_{i,j\not =i} \left( p_{i}-\langle\!\langle p_t\rangle\!\rangle\right)\left( p_{j}-\langle\!\langle p_t\rangle\!\rangle\right)}{N\left(N-1\right)}   \right\rangle.
\end{equation}  
Here the careful reader should have noticed that, if the statistical component of the fluctuation is entirely contained in the diagonal terms, then this assumes that the momenta of two distinct particles $i$ and $j$ are independent variables. This assumption does in fact underlie the interpretation of multi-particle observables such as \equ{ptvar}, and is at the heart of the phenomenology of relativistic nuclear collisions. The idea is precisely that, from the freeze-out hypersurface, particles are emitted independently from an underlying distribution of ${\bf p}$, i.e., the spectrum. This idea goes under the name of \textit{flow paradigm}, and all experimental evidence points to the fact that this idea is correct. The flow paradigm implies in particular that the distribution of particle pairs emitted at freeze-out is equal to the product of single-particle distributions~\cite{Luzum:2011mm}:
\begin{equation}
    \frac{dN_{\rm pair}}{d^3{\bf p_1}d^3{\bf p}_2} = \frac{dN}{d^3{\bf p}_1} \frac{dN}{d^3{\bf p}_2},
\end{equation}
and this generalizes to triplets, quadruplets, etc., of particles. The hydrodynamic description is thus a single-particle description, where correlations originate solely from fluctuations of the single-particle distribution. One should note that, on the other hand, a single-particle description does not imply hydrodynamics. For instance, calculations of particle production in the dilute-dense approach of the color glass condensate also assume independent particle emission, although there is not a fluid involved~\cite{Dusling:2017aot}.

An additional comment is in order. A measurement of \equ{ptvar} in heavy-ion collisions does not return the correlation due to genuine dynamical effects. Other phenomena can contribute to the two-particle correlations, most notably, the decays of resonance, which naturally produce particles with correlated momenta, as well as the hadronization of jets. All such phenomena do however yield correlations of particles across small ranges of rapidity. In the experimental analysis, hence, one can suppress the contribution of these background effects, in jargon called \textit{non-flow} contributions, by imposing a gap in the values of $y$, or $\eta$, between the hadrons used to evaluate the correlations. Methods involving multiple rapidity gaps have also been invented~\cite{Jia:2017hbm}. Once this correction is performed, the final result can be considered as the genuine long-range correlation of collective origin.

\paragraph{$\boldsymbol{v_2}$ -- } The experimental definition of elliptic flow in a given event reads:
\begin{equation}
 V_2 =     \frac{1}{N} \sum_i e^{-i2\phi_{p,i}}, 
\end{equation}
where $\phi_{p,i}$ is the azimuthal direction of the momentum of particle $i$. Measuring such a quantity in a single event does basically amount to measuring statistical noise. One has thus to average over events at a given centrality, much as done for $\bbra p_t \kket$:
\begin{equation}
    \bra V_2 \ket = \biggl \bra \frac{1}{N} \sum_i e^{-i2\phi_{p,i}}  \biggr \ket,
\end{equation}
where brackets represent an average in the centrality class. But this does not work. The direction of the impact parameter of the collision is random (uniformly distributed) in a sample of collision events, and thus $\bra V_n \ket$ does vanish upon averaging. The only quantities one can have access to must be necessarily independent of this random phase.

 Forgetting for a moment the dependence on $p_t$, the quantity I am after is the Fourier coefficient, $V_n$, of the single-particle distribution:
\begin{equation}
    \frac{dN}{d\phi} = \sum_{n=-\infty}^{+\infty} V_n e^{-in\phi}.
\end{equation}
Independence of particles implies that the pair distribution can be written as:
\begin{equation}
    \frac{dN_{\rm pairs}}{d\phi_{1}d\phi_{2}} =  \sum_{n_1,n_2=-\infty}^{+\infty} V_{n_1} V_{n_2} e^{-in_1\phi_1} e^{-in_2\phi_2},
\end{equation}
where $\phi_i$ is the direction of the momentum of particle $i$. Introducing $\Delta\phi = \phi_1 - \phi_2$, one obtains:
\begin{equation}
    \frac{dN_{\rm pairs}}{d(\Delta\phi + \phi_2)d\phi_2} = \sum_{n_1,n_2=-\infty}^{+\infty} V_{n_1} V_{n_2} e^{-in_1\Delta\phi} e^{-i(n_1+n_2)\phi_2}.
\end{equation}
Upon integration over $\phi_2$, only the terms $n_1 = - n_2$ remain, but since $V_n=V_{-n}^*$, this implies:
\begin{equation}
    \frac{dN_{\rm pairs}}{d\Delta\phi} = \sum_{n=-\infty}^{+\infty} |V_n|^2 e^{-in\Delta\phi}.
\end{equation}
Hence in the flow paradigm the Fourier coefficients of the distribution of $\Delta\phi$ are equal to $v_n^2 \equiv |V_n|^2$. 

Now, the magnitude $v_n$ does not vanish upon averaging over events. One can thus perform the average over all events in a centrality class to obtain the mean squared $v_n$:
\begin{equation}
    \bra v_n^2 \ket = \bra e^{ n (\phi_1 - \phi_2 ) } \ket,
\end{equation}
where the average is over all pairs of distinct particles (to remove trivial statistical effects) in the centrality bin. Note that, although the correlation does not vanish, the sine components do, because parity-violating effects~\cite{Fukushima:2008xe}, if present, are negligible. Therefore:
\begin{equation}
    \bra e^{ n (\phi_1 - \phi_2 ) } \ket = \bra \cos n ( \phi_1 - \phi_2  ) \ket.
\end{equation}
This observable is naturally affected by nonflow correlations, as discussed above, and thus has to be measured with the implementation of rapidity gaps. An advantage of multi-particle correlations is that they do not require to make distinctions between events. The only event-by-event observable is the multiplicity, which allows one to classify events. Once that is done, the knowledge that a given particle belongs to a given given event becomes redundant in the calculation of $\bra v_n^2 \ket$. 

One can easily show that the previous equations generalize to higher-order quantities. One has to correlate the angles of more and more particles to construct higher-order moment of the distribution of $v_n$. For instance, a four-particle correlation yields the fourth moment:
\begin{equation}
   \bra v_n^4 \ket = \bra e^{ n(\phi_1 +\phi_2 - \phi_3 - \phi_4) } \ket,
\end{equation}
where the average is performed over all distinct quadruplets of particles in the centrality bin. Note that, in a single event, the quadruplets are $N(N-1)(N-2)(N-3)$, which is much larger than the number of pairs. This explains why higher-order moments of the $v_n$ distributions can in fact be measured with great precision. One can further mix different harmonics, e.g.,:
\begin{equation}
    \bra v_2^2 v_3^2 \ket = \bra e^{ 2(\phi_1 - \phi_2) + 3(\phi_3 - \phi_4) } \ket,
\end{equation}
or measure correlations between vectors, so-called \textit{plane correlations}, provided that the phases cancel out:
\begin{equation}
    \bra V_2^2 V_4^* \ket = \bra e^{2(\phi_1 + \phi_2) - 4 \phi_3} \ket. \hspace{30pt}  \bra V_2 V_3 V_5^* \ket = \bra e^{2\phi_1 + 3\phi_2 - 5 \phi_3} \ket,
\end{equation}
and so on. Quantities that are particularly relevant in the phenomenology of heavy-ion collisions are the cumulants of the distribution of $v_n$. They correspond to nontrivial combinations of moments, and they will be used in the next section. I derive their expressions and I say a few words about their meaning in Appendix~\ref{app:B}.

I have thus clarified that, as one can only observe a finite number of hadrons, the phenomenology of the soft sector of heavy-ion collisions is a phenomenology of multi-particle correlations. This represents a new method of analyzing particle collider events, introduced in the context of high-energy nuclear experiments.


\chapter{A matter of shape}
\label{chap:4}

I move on now to the main topic of this manuscript. The discussion of anisotropic flow in the previous chapter should have made clear that the geometric shape of the quark-gluon plasma is of paramount importance in the phenomenology of heavy-ion collisions. The medium is essentially at rest when it is produced, and its transverse expansion is driven by pressure gradients which are determined by the geometry of the system. It is interesting that the study of shape and anisotropy plays such a central role in high-energy nuclear physics. The reason is that the characterization of anisotropy is central as well in low-energy nuclear physics, although applied to the geometric shape of atomic nuclei. Today it is indeed established that the majority of atomic nuclei present in particular a \textit{quadrupole deformation}, which I shall introduce in this chapter.

It is important to appreciate that there are no such things as direct experimental probes of the geometric shape of atomic nuclei. While accurately predicted by models of nuclear structure, the deformation of atomic nuclei from experimental data can only be inferred by means of indirect methods that rely on model-dependent approximations. These models prove successful in the context of low-energy experiments, but what about higher energies? From the previous discussion on the physical origin of $\bra p_t \ket$ and $v_2$ in heavy-ion collisions, it is  natural to expect that these quantities are nontrivially influenced by the shape of the colliding nuclei if they are nonspherical.  If such manifestations of nuclear deformation were observed, then high-energy experiments would provide a novel method to observe direct phenomenological consequences of the deformation of nuclei, and thus to perform spectacular tests of the predictions of nuclear models.

In this chapter, I show that this is indeed the case. I first introduce the concept of nuclear quadrupole deformation, and I explain how this concept is implemented in the modeling of heavy-ion collisions. I perform then a detailed analysis of the current experimental evidence of nuclear deformation in heavy-ion collision experiments. I shall conclude that high-energy nuclear experiments provide in fact a very powerful tool to trigger phenomenological manifestations of the geometric shape of atomic nuclei.

\section{Nuclear quadrupole deformation}

\label{sec:4-1}

A nucleus has a quadrupole deformation if its quadrupole moment does not vanish, i.e.,
    \begin{equation}
\label{eq:quadmoment}
  Q_{2} \propto \bigl \bra Y_2^0 (\Theta, \Phi) r^2 \bigr \ket \neq 0 , 
\end{equation}
where I use spherical coordinates in the intrinsic nuclear frame, ${\bf r}=(\Theta,\Phi,r)$, and the spherical harmonic breaks spherical symmetry: $Y_2^0\propto 3\cos^2 \Theta - 1 $. Angular brackets denote an expectation value with respect to the nuclear wavefunction, expanded in some basis. The concept of deformation is typically associated with the image of a nucleus as a deformed body, like an ellipsoid. Nevertheless, one should be careful when invoking such a picture. For instance, an even-even nucleus presenting a significant quadrupole moment, such as $^{238}$U, has nonetheless a vanishing total angular momentum, $J=0$, and so its wavefunction is invariant under rotations in space. There is a priori no \textit{need} to describe the nucleus as deformed, ellipsoidal object. One could expand the wavefunction in a basis of eigenstates that are spherically symmetric, and then let the quadrupole moment of \equ{quadmoment} emerge solely from correlations between eigenstates. However, the collective nature of nuclear excitations has made clear since long time~\cite{BM} that, especially when dealing with large nuclei, an excellent approximation of the nuclear wavefunction can be obtained with a so-called \textit{rotational} model. Roughly speaking, the idea is that the nucleus is described by an ellipsoidal density of charge, or matter, with a random orientation in space.  Upon averaging over orientations, one finds that the system has the right rotational symmetry. However, in this approach when the wavefunction collapses the positions of the nucleons are determined following the shape of a randomly-oriented ellipsoid. I anticipate that strong evidence of this behavior will be provided by the results discussed later on in Chapter~\ref{chap:5}, in particular by their comparison with experimental data from heavy-ion collisions.

\paragraph{Data-driven approach --} 
Evidence of the quadrupole deformation is provided by rotational spectra, which assume a characteristic form if the nucleus is a rigid rotor. For even-even nuclei, the quantity of interest is the electric quadrupole operator transition probability from the ground state to the first $2^+$ state, a quantity dubbed B(E2)$\uparrow$~\cite{Cline:1986ik,Raman:1201zz}. Other kind of observables can be used for odd nuclei~\cite{Loebner}. In the simple picture of the rotational model, assuming that the electric charge density in the nucleus is a sharp-edged ellipsoid of revolution, one can characterize the quadrupole deformation of the nucleus with the following dimensionless parameter~\cite{ring}:
\begin{equation}
\label{eq:betaexp}
    \beta = \frac{4\pi}{3 Ze R_0^2} \sqrt{ {\rm B(E2)\!\uparrow} },
\end{equation}
where $e$ is the fundamental electric charge, $Z$ is the proton number, and $R_0=1.2A^{1/3}$ is the empirical nuclear radius. The expression of $\beta$ is related to the nuclear quadrupole moment because there is a simple relation between {\rm B(E2)}$\uparrow$ and $Q_2$~\cite{ring}:
\begin{equation}
\label{eq:be2q2}
  {\rm B(E2)}\!\uparrow = \frac{5}{16\pi} |e Q_2|^2.  
\end{equation}
As a matter of fact, as originally pointed out by Kumar~\cite{Kumar:1972zza}, if one considers a uniform ellipsoidal charge density $\rho({\bf r})$ that has the same quadrupole moment as the nucleus, and the same volume (i.e., the same mean squared radius), then the parameter $\beta$ is simply given by the following reduced quadrupole moment:
\begin{equation}
    \beta = \frac{4\pi}{5} \frac{\int_{\bf r} r^2 Y_2^0(\Theta,\Phi) \rho({\bf r})  }{ \int_{\bf r} r^2 \rho({\bf r}) },
\end{equation}
where $r=|{\bf r}|$.  Note the similarity of this expression with that of the two-dimensional eccentricity in \equ{E2}. 
\begin{figure}[t]
    \centering
    \includegraphics[width=.7\linewidth]{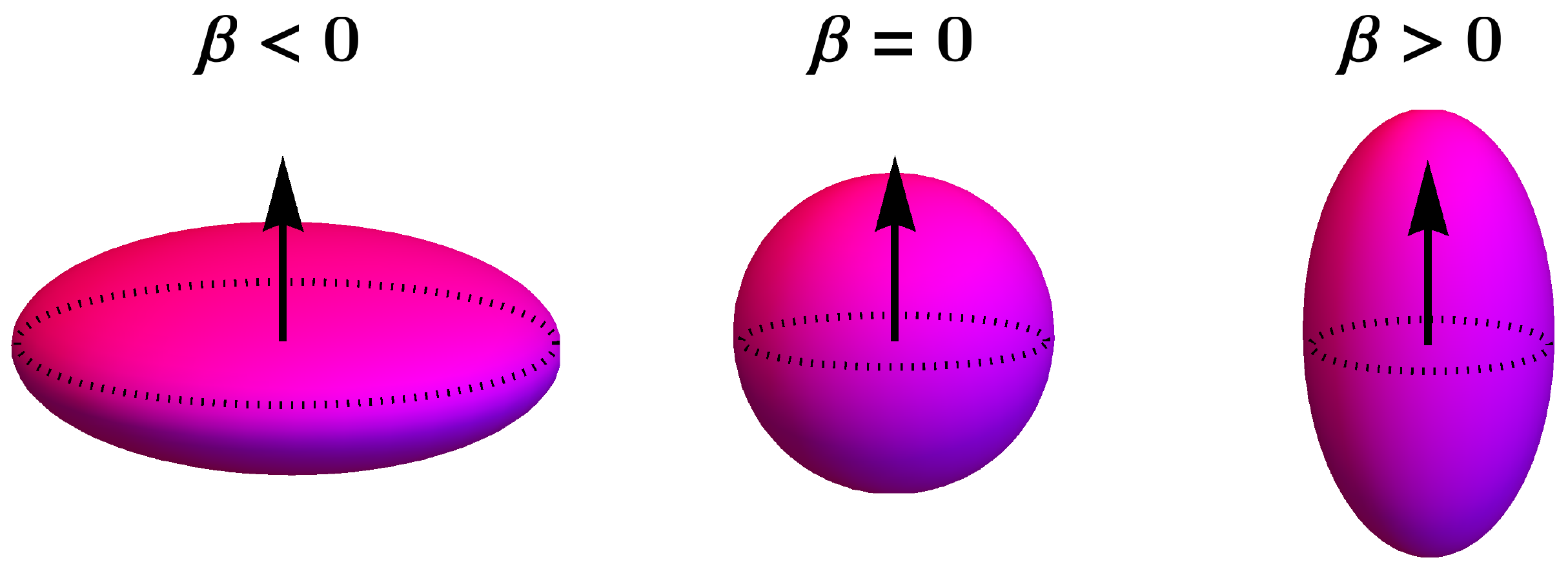}
    \caption{Left: Oblate spheroid ($\beta<0$). Center: Sphere ($\beta=0$). Right: Prolate spheroid ($\beta>0$). The arrow indicates the axis of the nucleus.}
    \label{fig:4-1}
\end{figure}

The value of $\beta$ characterizes the shape of the nucleus under consideration.  A nucleus is spherical if $\beta=0$. It looks instead like an oblate spheroid, squeezed at the poles, when $\beta<0$. When $\beta>0$ on the other hand, the nucleus is a prolate spheroid, and it looks like a rugby ball. These possibilities are illustrated in \fig{4-1}. Note that the experimental determination provided by \equ{betaexp} returns a value of $\beta$ which is positive by construction. One should keep in mind that the simple idea of considering the nucleus as an ellipsoidal object with a well-defined quadrupole deformation is not always justified. It is typically a good approximation for large nuclei that are well-deformed ($\beta>0.2$) but in general one should be careful when assigning shapes to nuclei, as effects related to the \textit{fluctuations} of $\beta$~\cite{Poves:2019byh}, that I shall also briefly mention in the following sections, can also be important.

\paragraph{Mean field estimate --} In a theoretical calculation, the value of $\beta$ requires the knowledge of the quadrupole moment of the nuclear wavefunction. The dynamics of a nucleus is, in an \textit{ab-initio} approach, given by the solution of the Schr\"odinger equation for the full nuclear wavefunction:
\begin{equation}
    H | \psi \ket = E | \psi \ket,
\end{equation}
where $H$ is the nuclear Hamiltonian, $| \psi \ket$ is the nuclear wavefunction, and $E$ is the energy of the system. For large nuclei this problem is intractable. A powerful method that allows one to simplify this matter is the Hartree-Fock, or \textit{mean field}, method. The mean field approach is based on a picture of independent particles, where the nuclear wavefunction is written as a Slater determinant of the system of $A$ fermions, $|\psi_i\ket$, where $i$ labels a nucleon. This Ansatz implies that the nuclear Hamiltonian can be decomposed as $\sum_i h_i$, where $h_i$ is a single-particle Hamiltonian. The dynamics is then given by the solution of $A$ identical Schr\"odinger equations, which are easier to handle:
\begin{equation}
    h_i |\psi_i\ket = E_i |\psi_i\ket.
\end{equation}

The independent particle picture may sound like a rough simplification, especially considering that nuclei are self-bound objects, whose existence depends precisely on nucleon-nucleon interactions. But as a matter of fact this approximation is perfectly justified~\cite{Soma:2018qay}. The main reason is that in low-temperature nuclear matter nucleons have a rather large mean free path, of order of few fm, meaning that two nucleons barely see each other within the nuclear volume, a feature supported by experimental measurements. Second, Fermi statistics and the Pauli principle also tend to keep nucleons far apart.

\begin{figure}[t]
    \centering
    \includegraphics[width=\linewidth]{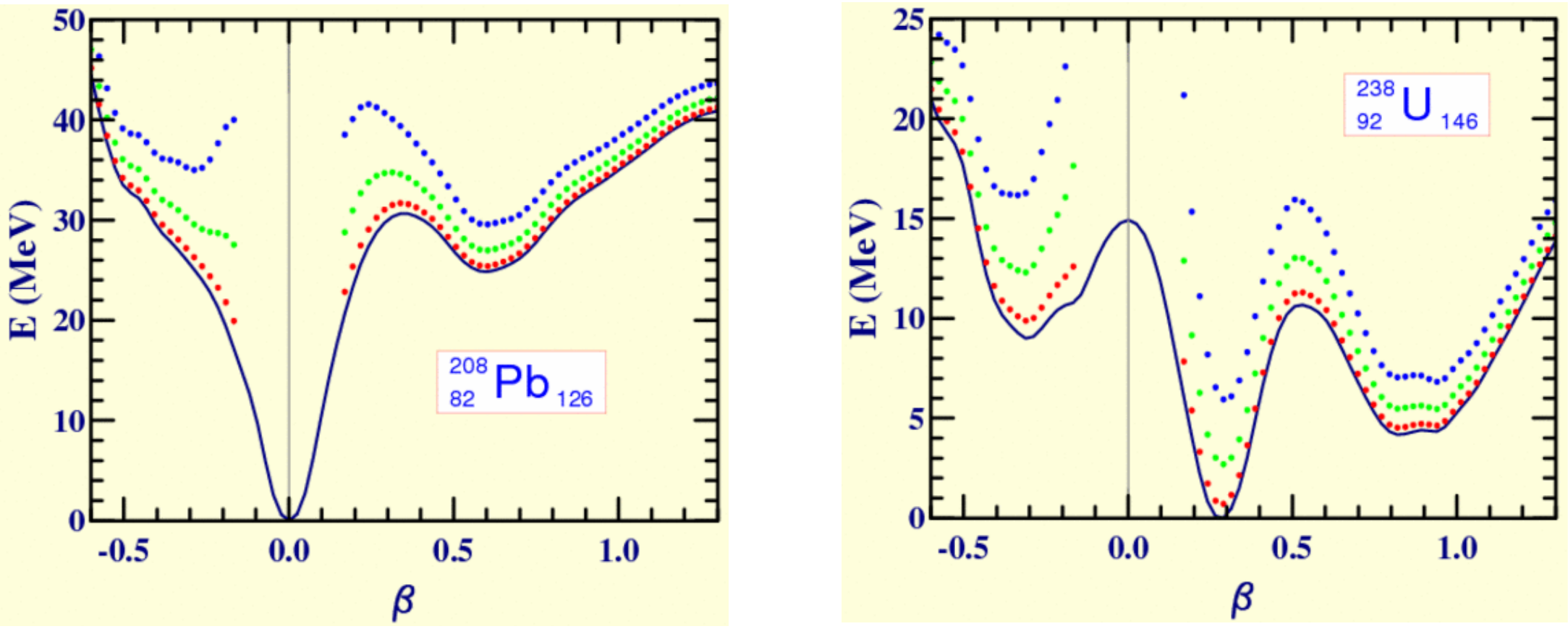}
    \caption{The solid lines represent the potential energy surface of $^{208}$Pb (left) and $^{238}$U (right) obtained within the Hartree-Fock-Bogoliubov calculations of Ref.~\cite{Hilaire:2007tmk}, where the minimization under constraints of \equ{ritzQ2} allows the mean-field wavefunction to break both rotational and particle number symmetry (separately for protons and neutrons). The energy is plotted against the quadrupole deformation parameter, $\beta$. See text for more details. Figures from Ref.~\cite{website}.}
    \label{fig:4-2}
\end{figure}

Variational methods are used to calculate the ground state of the system, say $|\Phi\ket$, at the mean field level. A great advantage of this approach is that it allows one to do so while breaking the symmetries of the exact nuclear wavefunction. To describe a deformed nucleus, for instance, one can perform a minimization under constraints which allows the resulting ground state to explicitly break rotational symmetry. Formally, one performs a variation of the kind:
\begin{equation}
\label{eq:ritzQ2}
    \delta \left (  \bra \Phi | H - \mu Q_2 | \Phi \ket  \right ) = 0,
\end{equation}
 where $\mu$ is a Lagrange multiplier which enforces the ground state returned by the minimization procedure to have a quadrupole moment $Q_2$, i.e., a nonzero value of $\beta$. The minimization can thus be performed by imposing any value of $\beta$. One looks then at the mean-field ground-state energy, $E$, to see where the minimum value of $E$ as a function of $\beta$ lies. This kind of calculation has been performed for hundreds of nuclides in Ref.~\cite{Hilaire:2007tmk}, where the mean-field wavefunction is expanded in a harmonic oscillator basis that allows to break spherical symmetry while preserving axial symmetry. The results of this calculation are collected in the AMEDEE database, that can be found in Ref.~\cite{website}. I report in \fig{4-2} the results for two nuclei in which I am particularly interested.

In the left panel, the procedure is performed for the $^{208}$Pb nucleus. The solid line gives the ground-state energy, $E$, as a function of $\beta$, shifted to $0$ at the minimum. This calculation shows that the potential energy curve has a minimum for $\beta=0$. This minimum is sharp, as the potential curve grows steeply as soon as $\beta \neq 0$, the other minima lying at much higher values of $E$. The fact that the curve grows steeply around the minimum is an indication of a well-define shape, corresponding in this case to a vanishing quadrupole moment. In the right panel of \fig{4-2} I show instead the result of the minimization procedure for $^{238}$U. Here the situation is more interesting. As one varies the magnitude of the quadrupole deformation, one observes a \textit{spontaneous} breaking of rotational symmetry. The minimum of the potential energy curve is around a nonzero value of $\beta$, close to $0.3$, and its sharpness implies again a well-deformed shape. Hence it is right to treat $^{238}$U within the approximations of the rotational model.

I presented two cases of two nuclei possessing a well-defined shape, i.e., a single sharp minimum of $E$ as a function of $\beta$. However, if the minimization procedure yields a minimum that is less prominent, so that the energy difference between different minima is only of order 1~MeV, then the simple mean-field estimate should not be considered a reliable estimate of the value of $\beta$, and one may even need to go beyond the simple approximation of the rotational model. I shall return to this point in the upcoming discussion of $^{197}$Au nuclei.

\section{Relativistic collisions of deformed nuclei}

\label{sec:4-2}

The approach to the structure of nuclei that is used in the modeling of heavy-ion collisions, as described in Sec.~\ref{sec:2-21}, is thus reminiscent of the mean field method. One assumes that the nucleons are independent, and that each nucleon follows a single-particle density of Woods-Saxon form, which I recall here for clarity:
\begin{equation}
\label{eq:ws}
    \rho(r) = \frac{\rho_0}{1+\exp \biggl\{ \frac{1}{a} \biggl[  r - R \biggr] \biggr\} }.
\end{equation}

The modeling of deformed nuclei in the context of heavy-ion collisions goes along the same lines. The only difference is that the single-particle density is no longer given by \equ{ws}, but it breaks spherical symmetry. This is done by adding an angular dependence to the radius of the system. For a surface oscillating around the spherical shape, this can conveniently achieved by means of an expansion in spherical harmonics~\cite{ring}:
\begin{equation}
\label{eq:Rexp}
    R(\Theta,\Phi) \rightarrow R_0 \left ( 1 + c_{00} + \sum_{l=1}^{\infty} \sum_{m=-l}^l c_{lm} Y_l^m(\Theta,\Phi) \right ).
\end{equation}
The constant $c_{00}$ represents a change in the nuclear volume. As nuclear matter is almost incompressible, deformation does not change the volume, and one can find an expression for $c_{00}$ as a function of the other coefficients in order to keep the volume fixed. However, I shall gloss over this detail in the following, and consider densities that yield to a good extent to $A$ upon integration over space. The terms with $l=1$ shift the center of mass of the system, while terms with $l>1$ introduce deviations from spherical symmetry. As the deformation of nuclei originates from long-range collective correlations of nucleons, the most important terms in the expansion are those corresponding to long-wavelength modes, i.e., small values of $l$. Among these, the quadrupole deformation is the most prominent. Hence for all phenomenological purposes, the most relevant term of the expansion is $l=2$, $m=0$, which corresponds to the quadrupole deformation of the long wavelength structures. The other quadrupole terms $l=2$, $m=\pm2$ are less important, although I shall discuss them in some detail in Chapter~\ref{chap:6}. One defines:
\begin{equation}
    c_{20} = \beta,
\end{equation}
where $\beta$ is precisely the quadrupole deformation parameter used in nuclear physics. The density of matter in a deformed nucleus is thus given by an actually minor modification of \equ{ws}:
\begin{equation}
\label{eq:defws}
    \rho(r,\Theta,\Phi) = \frac{\rho_0}{1+\exp \biggl\{ \frac{1}{a} \biggl[  r - R_0 \bigl(1 + \beta Y_{2}^0(\Theta,\Phi) \bigr)  \biggr] \biggr\} },
\end{equation}
The rest of the model is unchanged: \equ{defws} represents the single-particle density, and, in each realization of the nucleus, one samples the positions of the nucleons according to this distribution. From the previous considerations, it is clear that this model is good whenever the nuclei have well-defined shapes, as observed for $^{208}$Pb and $^{238}$U nuclei. 

\begin{figure}[t]
    \centering
    \includegraphics[width=.7\linewidth]{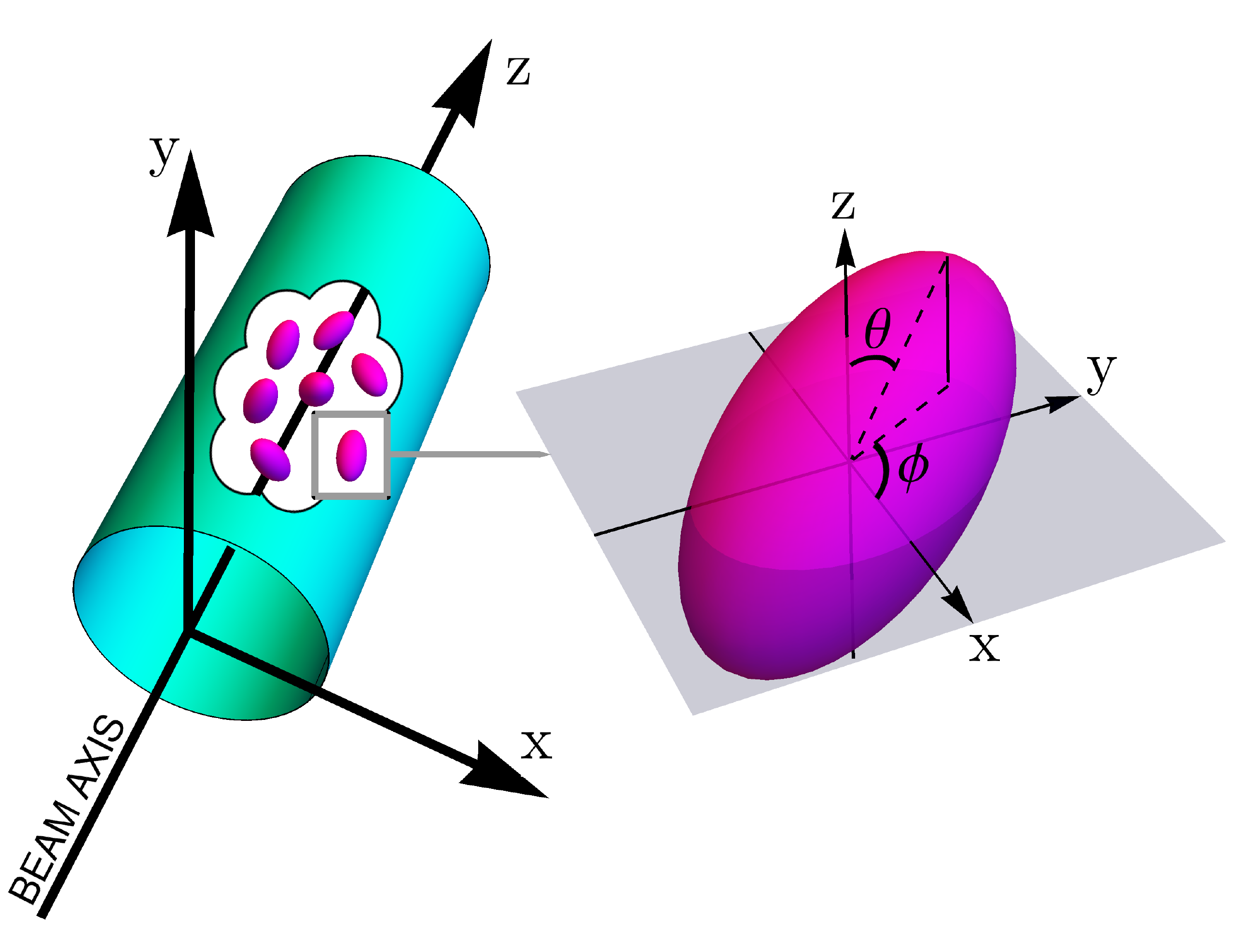}
    \caption{A bunch of deformed nuclei is running in the beam pipe (left). The beam axis corresponds to the $z$ axis. Each nucleus (right) has a random orientation in space, and with respect to the laboratory frame is thus tilted by a polar angle, $\theta$, and by an azimuthal angle, $\phi$. Figure from Ref.~\cite{Giacalone:2019pca}.}
    \label{fig:4-3}
\end{figure}

\paragraph{Deformed nuclei in the beampipe --} The model used in heavy-ion collisions is thus a simple rotational model. The nucleus is an ellipsoidal distribution of matter which is randomly oriented in space. The nuclei running in the beampipe can thus be depicted as in \fig{4-3}. In the laboratory frame, which I recall is defined by a beam axis, $z$, and a plane orthogonal to it, $(x,y)$, each nucleus is randomly oriented, so that the laboratory frame and the intrinsic nuclear frame differ by a polar tilt, $\theta$, and by an azimuthal spin, $\phi$, which I shall refer to as Euler angles. When two nuclei collide, the collision geometry in the laboratory frame is therefore determined by two sets of Euler angles, $\theta_A$, $\theta_B$, and $\phi_A$, $\phi_B$, where $A$ and $B$ label the colliding nuclear bodies.

\paragraph{Body-body and tip-tip collisions --} Collisions of nuclei that are deformed produce nontrivial geometries of overlap. I discuss here two \textit{extreme} kinds of such overlap geometries, showing that a nontrivial phenomenology of the quadrupole parameter, $\beta$, is naturally expected in relativistic collisions of deformed nuclei. 

\begin{figure}[t]
    \centering
    \includegraphics[width=\linewidth]{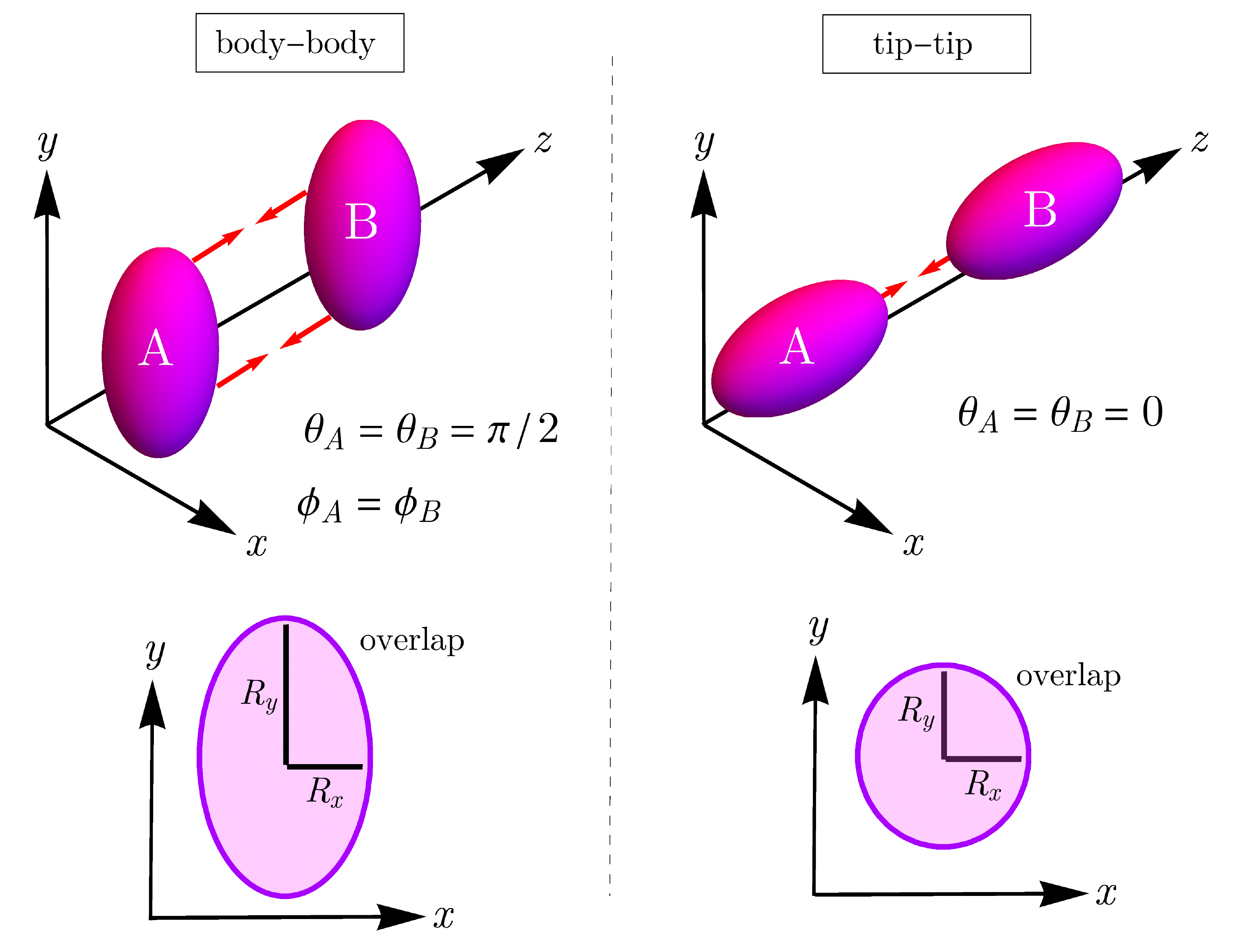}
    \caption{Left: fully-overlapping body-body collision (top). The nuclei collide with their axes aligned orthogonally to the beam axis, $z$ ($\theta_A=\theta_B=\pi/2$), and with the same azimuthal orientation, $\phi$. The transverse area of nuclear overlap in such collisions (bottom) possesses an enhanced quadrupolar asymmetry ($R_y > R_x$). Right: fully-overlapping tip-tip collision (top). The nuclei collide with their axes parallel to the beampipe, $\theta_A=\theta_B=0$.  The area of overlap (bottom) is isotropic in the transverse plane, $R_x = R_y$. Figure from Ref.~\cite{Giacalone:2020awm}.}
    \label{fig:4-4}
\end{figure}

I consider collisions between prolate nuclei, $\beta>0$, and I look at the limit where these nuclei collide at very small impact parameter, i.e., when they are fully overlapping. In this limit, there are two extreme situations, which are illustrated in \fig{4-4}.
\begin{itemize}
    \item There are \textit{body-body} configurations, shown in the left panel of Fig.~\ref{fig:4-4}, in which the axes of the two nuclei are both orthogonal to the beam direction, i.e., $\theta_A=\theta_B=\pi/2$, and both nuclei are rotated by the same azimuthal angle, $\phi_A=\phi_B$. In such configurations the area of overlap in the transverse plane, following the strong Lorentz contraction, has an enhanced elliptical deformation, which originates from the shape of the colliding nuclei.
    \item  There are also \textit{tip-tip} configurations, shown in the right panel of \fig{4-4}, in which the axes of both nuclei are aligned with the beam axis, $z$, or $\theta_A=\theta_B=0$. The resulting area of overlap is circular.
\end{itemize}
The relevant comment is that, while tip-tip collisions produce a medium that has an isotropic shape in the transverse plane ($R_x=R_y$), so that the anisotropy of the created medium is entirely generated by fluctuations, as discussed in Sec.~\ref{sec:3-3}, body-body collisions produce, on the other hand, a medium which has an intrinsic elliptical deformation. Depending on the value of $\beta$, the medium created in body-body collisions possesses an initial eccentricity, $\mathcal{E}_2$, that receives a contribution from the quadrupole deformation of the colliding bodies. For large nuclei, e.g., $A>150$, this contribution is in fact dominant whenever $\beta>0.2$, which unsurprisingly corresponds as well to the limit of well-deformed nuclei. The nontrivial conclusion is that collisions between nonspherical nuclei can thus yield elliptic flow from a genuine elliptic deformation of the created medium even in head-on collisions occurring at zero impact parameter. 

\paragraph{Conclusion --} I have outlined the relevant elements of nuclear structure, as well as their implementation in the modeling of heavy-ion collisions.  I can now look at existing experimental data on elliptic flow in collisions of nonspherical nuclei, to assess whether or not effects of nuclear deformation are visible.

\section{Evidence of deformation at RHIC: $^{238}$U+$^{238}$U collisions}

\label{sec:4-3}

\begin{figure}[t]
    \centering
    \includegraphics[width=.85\linewidth]{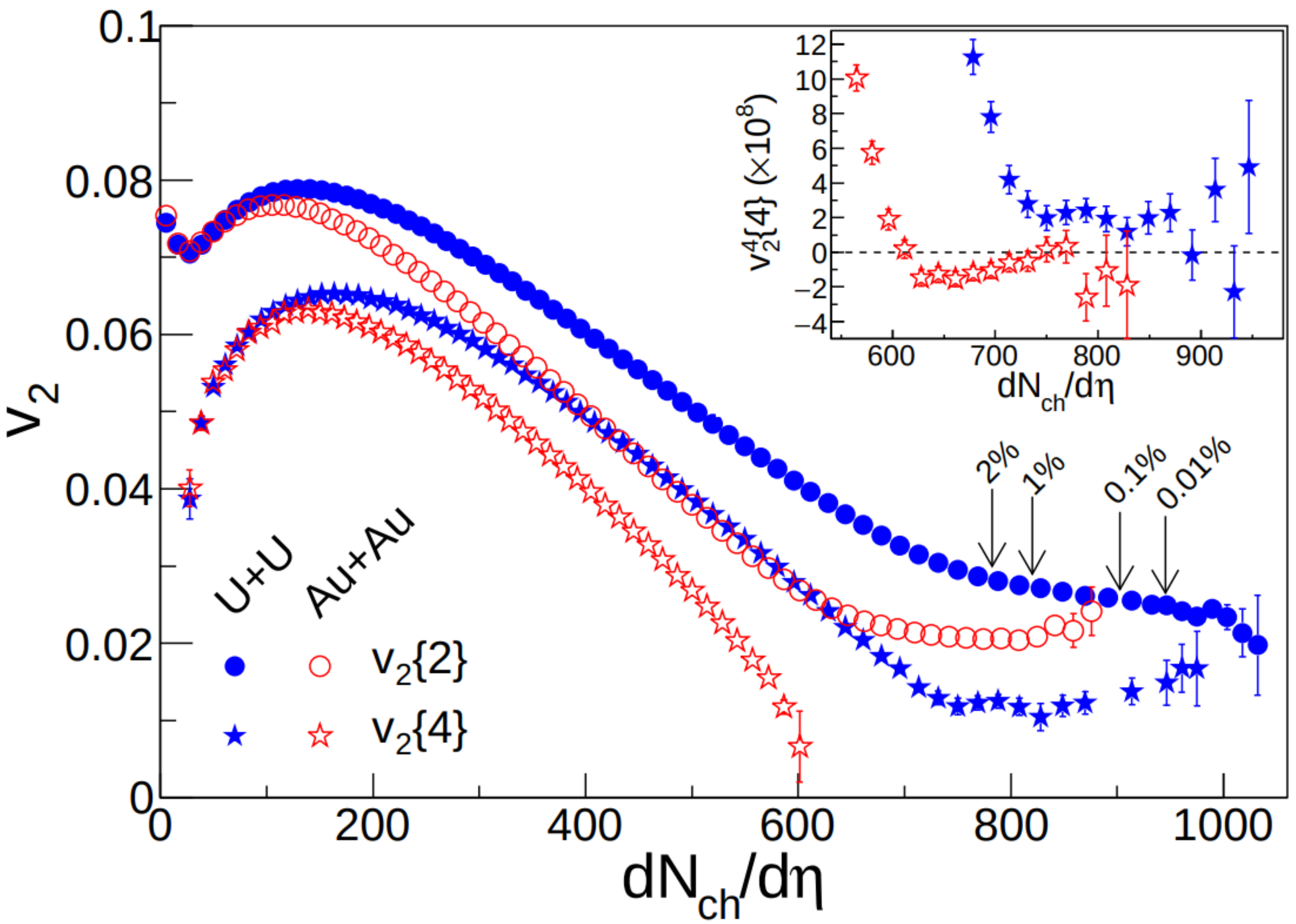}
    \caption{STAR data on the fluctuations of elliptic flow in \auau{} (empty symbols) and \uuuu{} (full symbols) collisions at, respectively, $\sqrt{s_{\rm NN}}=200$~GeV and $\sqrt{s_{\rm NN}}=193$~GeV. Circles represent measurements of the second-order cumulant of elliptic flow, $v_2\{2\}$, while stars represent the cumulant of order four, as defined by \equ{v22v24}. $v_2$ fluctuations are plotted against the charged-particle multiplicity per unit pseudorapidity, which is obtained in the range $-0.5 < \eta < 0.5$. Elliptic flow in ths $\eta$ window is further obtained by imposing an additional cut in the transverse momentum: $0.2 < p_t < 2$~GeV. The inset highlights the differences in the fourth-order cumulant of $v_2$ between ultracentral \auau{} and \uuuu{} collisions. Figure from Ref.~\cite{Adamczyk:2015obl}.}
    \label{fig:4-5}
\end{figure}

In May 2015 the STAR collaboration at RHIC published groundbreaking results, corresponding to the release of data on elliptic flow fluctuations in collisions of deformed $^{238}$U nuclei~\cite{Adamczyk:2015obl}. The results of the STAR collaboration are reported here in \fig{4-5}. Two quantities are displayed:
\begin{align}
\label{eq:v22v24}
    v_2\{2\}^2 &\equiv \bra v_2^2 \ket, \\
    v_2\{4\}^4 &\equiv 2\bra v_2^2 \ket^2 - \bra v_2^4 \ket.
\end{align}
These represent respectively the second- and fourth-order cumulants of the distribution of the magnitude $v_2=|V_2|$, whose expressions are derived in Appendix~\ref{app:B}. For reasons that I shall make clear in the following, to reveal the effect of the deformed nuclear shapes on these observables, it is crucial to perform measurements in central collisions, and by means of fine multiplicity classes. As one can appreciate from \fig{4-5}, this is precisely the path followed by the STAR collaboration, and this is why their results are groundbreaking. My goal in this section is to look at all the experimental results shown by the STAR collaboration in Ref.~\cite{Adamczyk:2015obl}, and discuss their meaning and their implications, with an emphasis on the role played by the deformed shape of the colliding nuclei.

\subsection{Spectacular failure of the two-component Ansatz}

\label{sec:4-31}

\begin{figure}[t]
    \centering
    \includegraphics[width=.85\linewidth]{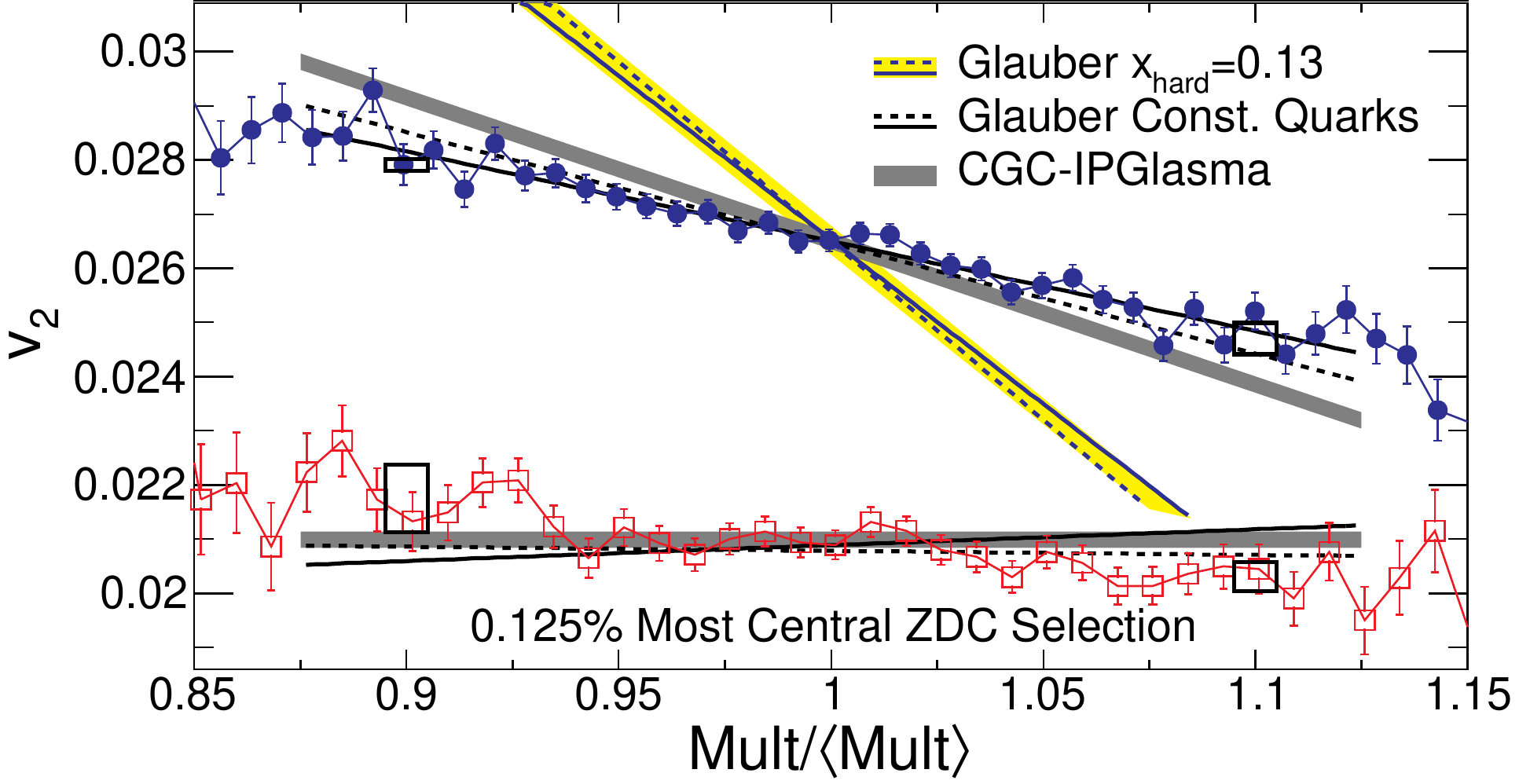}
    \caption{Experimental data on the correlation between $v_2$ and the charged particle multiplicity ${\rm Mult} \equiv dN/d\eta$, with $|\eta|<0.5$, in \uuuu{} collisions (filled circles) and \auau{} collisions (empty squares). The dashed bars represent the prediction of the \ipglasma{} framework~\cite{Schenke:2014tga}. The dashed line is the prediction of the constituent-quark Glauber model, while the dashed line highlighted in yellow is the prediction of the two-component Glauber model. Figure from Ref.~\cite{Adamczyk:2015obl}.}
    \label{fig:4-6}
\end{figure}

I start by discussing an observable from the STAR paper which is not shown in \fig{4-5}. I reproduce it here in \fig{4-6}. The quantity which is plotted is the rms elliptic flow, $v_2$, as a function of the charged-particle multiplicity, $dN/d\eta$, in a given class of ultracentral events, selected by means of the energy deposited in the forward regions of the detector, in the so-called Zero Degree Calorimeters (ZDC). If we assume that the selection based on the ZDC energy allows one to pick events where the two nuclei are fully overlapping, i.e., events at $b=0$, then studying $v_2$ as a function of multiplicity is essentially tantamount to looking at how $v_2$ varies as a function of the total entropy of the system. 

This observable is interesting because a widespread model of particle production utilized in experimental analyses of heavy-ion collisions, the so-called two-component Glauber model, makes a nontrivial prediction for the data points shown in \fig{4-6}. In this model, the number of sources of particles, the so-called \textit{ancestors}, produced in a heavy-ion collision is written as the sum of two terms~\cite{Miller:2007ri}:
\begin{equation}
\label{eq:bs}
     a\times N_{\rm part} + (1-a)\times N_{\rm coll},
\end{equation}
where the parameter $N_{\rm coll}$ corresponds to the number of nucleon-nucleon collisions occurring in the event, while $a$ can be fitted to data on the probability distribution of $dN/d\eta$. Each ancestors produces then a random number of particles following a negative binomial distribution, whose parameter are fitted to data. 

The term with $N_{\rm coll}$ in \equ{bs} has a rather dramatic consequence in the context of collisions between deformed nuclei. I recall \fig{4-4}. Fully-overlapping tip-tip and body-body collisions share the same $N_{\rm part}$, however, the value of $N_{\rm coll}$ is naturally larger in tip-tip collisions, because the density of nucleons per unit area is larger. According to \equ{bs}, a tip-tip collision has a larger density and produces more particle than a body-body collision. By selecting high-multiplicity events, one does effectively select mostly tip-tip events. This implies that one could use high-multiplicity \uuuu{} collisions to produce systems whose temperature and density are larger than that of \auau{} collisions at the same $\sqrt{s_{\rm NN}}$ by as much as 30\% or 40\%.  This very feature was in fact the main motivation driving the run of \uuuu{} collisions at RHIC, and most of the existing pre-2015 literature discussing effects of nuclear deformation at high energy is based on the manifestation of tip-tip geometries in high-multiplicity collisions, see, e.g., Refs.~\cite{Kuhlman:2005ts,Voloshin:2010ut,Hirano:2010jg,Rybczynski:2012av,Goldschmidt:2015kpa}.

But something went wrong. If fully-overlapping events at high-multiplicity correspond to tip-tip events, then, since tip-tip events correspond to the limit of a round geometry of overlap, there should be a negative correlation between $v_2$ and the multiplicity in collisions at $b=0$. This correlation can be calculated quantitatively by assuming that $v_2$ is proportional to the eccentricity of the system, $\varepsilon_2$, and by relating the ZDC energy to the number of \textit{spectator} neutrons, i.e., those neutrons that do not participate in the collision. This calculation was carried out by the STAR collaboration for \uuuu{} collisions implementing $\beta$ from Ref.~\cite{Raman:1201zz}. The resulting curve is showed as a dashed line highlighted in yellow in \fig{4-6}. It must be compared to the experimental measurement, shown as filled blue circles. While both data and the two-component model return a negative slope, the slope of the theoretical estimate is completely off. The STAR collaboration could only but claim the failure of the two-component model.

By means of further model-to-data comparisons, however, the STAR collaboration clarifies where the issue is. Experimental data are compared to two other models. One is the so-called constituent-quark Glauber model~\cite{Loizides:2016djv}. In this model, one does not collide nucleons, but sub-nucleonic constituents (quarks), which are sampled within each nucleon. The multiplicity scales then like the total number of participant quarks. In the notation of \equ{p=0}, this implies a scaling of the form $t_A + t_B$ for the initial entropy density, where I recall that $t_A$ is the linear sum of all the sources of density, i.e., the participant quarks, coming from nucleus $A$. The prediction of this model is shown as a dashed line in \fig{4-6} and is nicely consistent with experimental data. The second model shown by the STAR collaboration is the \ipglasma{} model. Within the \ipglasma{} framework, the observable analyzed by the STAR collaboration was studied in Ref.~\cite{Schenke:2014tga}. The prediction of \ipglasma{} is shown as a dashed line in \fig{4-6}. As discussed in Sec.~\ref{sec:2-21}, the scaling of the multiplicity predicted by the color glass condensate framework is roughly some power of $t_At_B$, and the model turns out to be in good agreement with experimental data.  Additionally, let me stress that this observable is also studied in the original \trento{} publication~\cite{Moreland:2014oya}. It is pointed out in particular that the phenomenological Ansatz for the multiplicity, $\sqrt{t_At_B}$, provides naturally an excellent description of STAR data on the correlation between $v_2$ and the particle number shown in \fig{4-6}.

The common denominator between quark Glauber model, \ipglasma{}, and \trento{}, is that in neither of these models the number of produced particles depends on $N_{\rm coll}$. The negative slope reported by all these calculations, implies that high-multiplicity collisions correspond to some extent to tip-tip geometries, but the effect observed in data is clearly incompatible with the overwhelming effect obtained when the particle production model depends on $N_{\rm coll}$. This is why the two-component Glauber model fails, and why, in my opinion, it should not be considered as a viable model of particle production for heavy-ion collisions. By contrast, this analysis of the STAR collaboration provides a highly nontrivial confirmation of the prediction of the color glass condensate framework, and the idea that particle production in high energy nuclear collisions is a simple \textit{coherent} process, i.e., resulting roughly from the sum of the contributions coming from individual nucleon-nucleon interactions.

That being said, the results shown in \fig{4-7} were considered as a big failure. The problem is that both the \ipglasma{} and the \trento{} frameworks do not provide any simple prescription for discerning body-body and tip-tip geometries in central heavy-ion collisions. The possibility of triggering a phenomenology based on the orientation of the colliding nuclei somehow vanished. The problem was more complicated than expected. Due to this, the \uuuu{} run has not had any follow-up, and no one has even tried to understand in detail the observations made by the STAR collaboration. In Chapter~\ref{chap:5}, I will show that one can in fact overcome the previous difficulty, and that there exists a well-defined method to discern collisions geometries. However, for the moment let me stick to the plan, i.e., explaining in details all measurements shown by the STAR collaboration. 

\subsection{Impact of deformation on the fluctuations of elliptic flow}

\label{sec:4-32}

The second result shown by the STAR collaboration concerns the fluctuations of elliptic flow, shown in \fig{4-6}.  I became aware of these measurements around November 2016. At that time I was performing high-quality comparisons between models of $\varepsilon_n$ and experimental data on $v_n$ by looking at observables that were independent of the linear hydrodynamic response. One such observable was the relative fluctuation of anisotropic flow, corresponding to the ratio $v_n\{4\}/v_n\{2\}$, which in central collisions can be predicted by means of the ratio $\varepsilon_n\{4\}/\varepsilon_n\{2\}$ to assess the goodness of initial-state models. These studies eventually lead to the results published in Ref.~\cite{Giacalone:2017uqx}.

An important result of that paper is the realization that a scaling of the initial density consistent with the color glass condensate framework, as realized e.g. by the \ipglasma{} or \trento{} frameworks, does a very good job in reproducing the centrality dependence of the fluctuations of $v_2$ and $v_3$, quantified by the ratio $v_n\{4\}/v_n\{2\}$. The scaling of the quark Glauber model, $t_A+t_B$, on the other hand, does not work. Hence \trento{} and \ipglasma{} do many things right: They yield histograms of final multiplicity which are in good agreement with data, they predict the right centrality dependence of $v_2\{4\}/v_2\{2\}$ at LHC, and they yield the right correlation between $v_2$ and the charged multiplicity in central collisions of deformed $^{238}$U nuclei. These models should also be able, then, to reproduce the observations made by the STAR collaboration on the fluctuations of elliptic flow in central collisions of deformed nuclei.  
\begin{figure}[t]
    \centering
    \includegraphics[width=.65\linewidth]{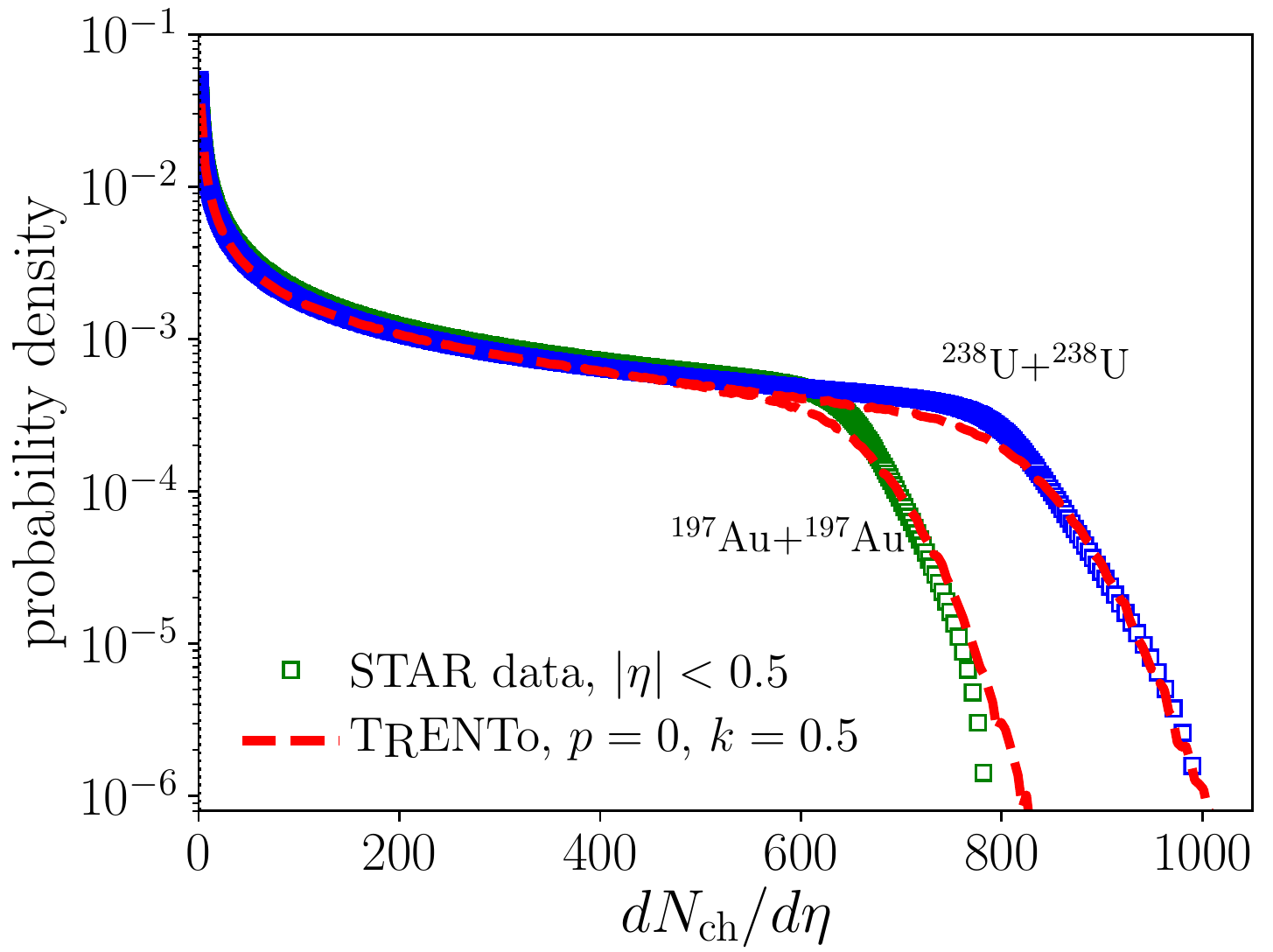}
    \caption{Histograms of $dN_{\rm ch}/d\eta$ in \auau{} collisions (green squares) and \uuuu{} collisions (blue squares), obtained from the STAR collaboration paper~\cite{Adamczyk:2015obl}. The dashed lines represent the rescaled histograms of total entropy provided by the \trento{} model used here. Figure adapted from Ref.~\cite{Giacalone:2018apa}.}
    \label{fig:4-7}
\end{figure}

In Ref.~\cite{Giacalone:2018apa}, I perform such a study within the \trento{} model, which allows for fast large-scale computations. I set up a \trento{} parametrization that allows me to describe STAR data with the best possible accuracy. This turns out to be a nontrivial task, because while \trento{} has been largely tested against LHC data, little has been done concerning RHIC data. I keep the scaling of the density unchanged, i.e., $\sqrt{t_At_B}$, and also the size of the nucleons, $w=0.5$~fm in \equ{rhonboost}. What instead needs to be modified is the fluctuation parameter, $k$, governing the fluctuation of entropy produced at the level of the participant nucleons. In \fig{3-2} I showed that LHC data can be described by implementing $k=2$.  To describe the histograms of multiplicity observed by the STAR collaboration, one needs instead a smaller value, around $k=0.5$. In \fig{4-7}, I show the multiplicity histograms measured by the STAR collaboration in \auau{} and \uuuu{} collisions~\cite{Adamczyk:2015obl}. My \trento{} parametrization, which I compare to experimental data by means of the Bayesian inversion method introduced in Ref.~\cite{Das:2017ned}, is shown as dashed lines. Agreement is good, although it is not as amazing as in the case of LHC data. It should be stressed, though, that the STAR curves do not represent actual experimental data, but parametrizations of $dN/d\eta$ as a function of the collision centrality which are obtained after fitting the measured histograms with the two-component Glauber model. Therefore, it is not clear to me to which extent my results are supposed to reproduce them. Note that the fact that $k=0.5$ works well means in practice that initial-state fluctuations are larger at RHIC energy than at LHC energy, a feature which can be inferred as well from other observables~\cite{Giacalone:2019vwh}.

\paragraph{Variance of $v_2$ fluctuations --}

\begin{figure}[t]
    \centering
    \includegraphics[width=\linewidth]{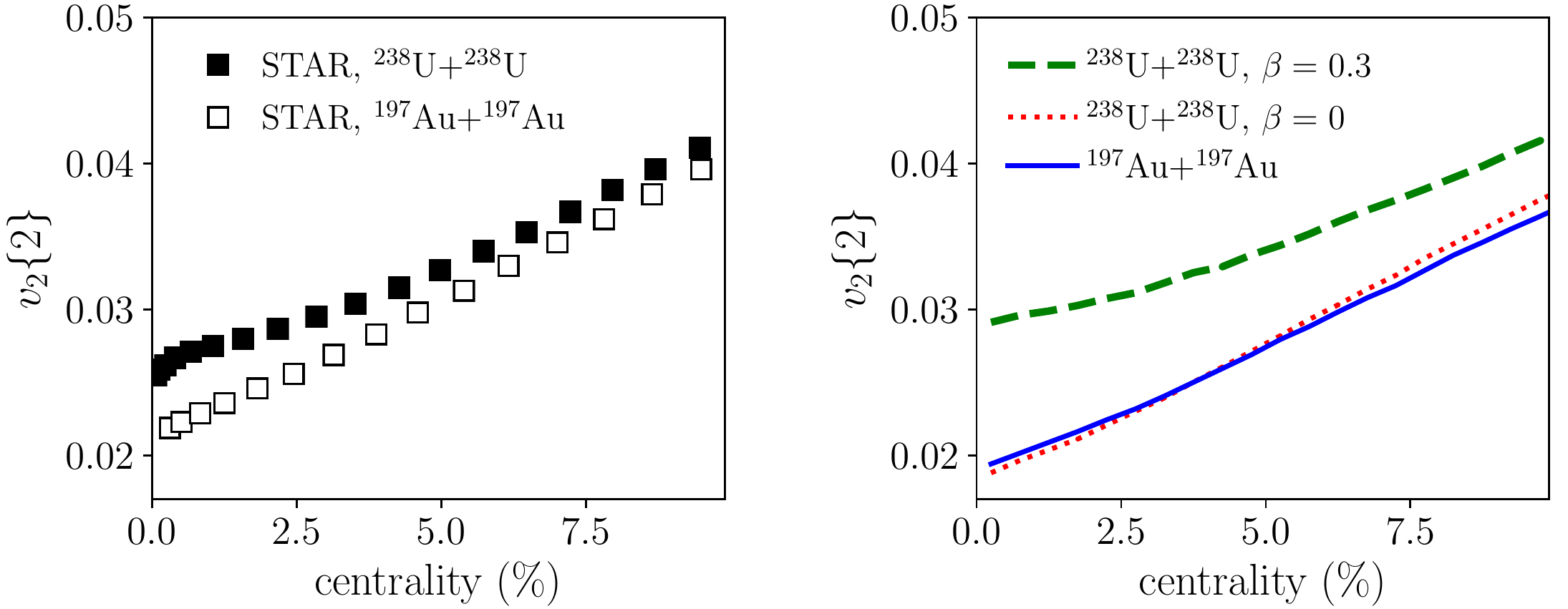}
    \caption{Left: STAR data~\cite{Adamczyk:2015obl} on the rms $v_2$ in 200~GeV \auau{} collisions  (empty symbols) and 193~GeV \uuuu{} collisions (full symbols). Right: Results of the \trento{} model for $\kappa_2\varepsilon_2\{2\}$, where $\kappa_2=0.155$ in \auau{} collisions and $\kappa_2=0.165$ in \uuuu{} collisions. Solid line: \auau{} collisions ($\beta=0$). The dotted line represents \uuuu{} collisions with $\beta=0$, while the dashed line implements $\beta=0.3$.}
    \label{fig:4-8}
\end{figure}

The phenomenological manifestations of nuclear deformation become apparent in the comparison between  \auau{} and \uuuu{} data. As these systems differ in the number of emitted particles, the best way to compare them is to plot observables as a function of collision centrality. I start by analyzing the second-order cumulant of $v_2$. STAR data on this observables is shown in the left panel of \fig{4-8}. One observes that the rms elliptic flow becomes considerably larger in \uuuu{} collisions as one approaches the limit of central collisions. If one forgets any potential effect related to the deformation of nuclei, this result is highly nontrivial. In central collisions, the rms elliptic flow is driven by fluctuations, which are in turn are driven by the number of nucleons. Hence they are smaller in \uuuu{} collisions than in \auau{} collisions. There is no room in hydrodynamics for an rms $v_2$ which is larger by as much as 20\% in \uuuu{} systems than in \auau{} systems. The data thus points clearly to a strong modification of the \uuuu{} results due to some additional features. If the existence of the deformation in $^{238}$U were not known beforehand, we would be in great trouble understanding this data.

The right panel of \fig{4-8} shows the results of the \trento{} model. I evaluate $v_2\{2\}$ as $\varepsilon_2\{2\}$ rescaled by a coefficient $\kappa_2$. This coefficient is $0.165$ for \uuuu{} collisions, and $0.155$ for \auau{} collisions. These numbers are simply taken from Ref.~\cite{Giacalone:2018apa}, and give only a rough indication of their actual values. However, as here I am looking at qualitative difference between systems, the exact normalization of $v_2$ is not relevant for the present discussion. I also make use of a single value for the response coefficient, $\kappa_2$, without addressing the issue that this quantity depends on centrality~\cite{Sievert:2019zjr,Rao:2019vgy,Snyder:2020rdy}. In the 0-10\% centrality range, the coefficient varies by about 10\%, but such kind of corrections to my results would leave my conclusions unchanged. I will however come back to the magnitude of the response coefficients in Chapter~\ref{chap:5}.

The blue solid line in the right panel of \fig{4-8} represents the result for \auau{} collisions. Here I am colliding spherical $^{197}$Au nuclei, where the Woods-Saxon density parameters in \equ{ws} are taken from Ref.~\cite{DeJager:1987qc}. I shall comment on the deformation of this nucleus in the following section. The dotted line represents instead the result for collisions of $^{238}$U nuclei in absence of specific modeling of the deformation, i.e., by simply assuming that $^{238}$U is a spherical charge density whose Woods-Saxon parameters are again given by the fits of Ref.~\cite{DeJager:1987qc}. This simple model leads to wrong results, as the curve for \uuuu{} collisions overlaps with that of \auau{} collisions, showing no sign of the splitting observed in experimental data. 

I add, then, a quadrupole deformation to these nuclei, following \equ{defws}. I implement $\beta=0.3$, which corresponds to the mean-field estimate of Ref.~\cite{Hilaire:2007tmk}, and which is also close to the experimental value of Ref.~\cite{Raman:1201zz}, $\beta\simeq 0.29$. The resulting rms $v_2$ is given by the green dashed line in the right panel of \fig{4-8}. We see that the inclusion of nuclear deformation does precisely create a splitting between \auau{}  and \uuuu{}, in agreement with data. Note that the splitting in the right panel is larger than in the left panel, suggesting that the details of my model might need a little more tweaking.

I explain now why the inclusion of nuclear deformation in the model increases the rms $v_2$. As shown in Appendix~\ref{app:B}, the square of $\varepsilon_2\{2\}$ can in full generality be decomposed as:
\begin{equation}
\label{eq:v22}
    \varepsilon_2\{2\}^2 = \sigma^2 + \mu^2.
\end{equation}
$\sigma^2$ is the variance of the distribution of $\mathcal{E}_2$, and originates from initial-state fluctuations. $\mu$ is instead the average value of the distribution of $\mathcal{E}_2$ along the direction of the impact parameter, and represents the genuine geometric contribution to the eccentricity coming from the elliptical shape of nuclear overlap. For collisions with centralities larger than $5\%$, the impact parameter is sizable enough that the eccentricity becomes dominated by this geometric contribution, so that $\varepsilon_2\{2\} \approx \mu$. This explains why, beyond $5\%$ centrality, the rms $v_2$ observed in data becomes the same for both \auau{} and \uuuu{} collisions, as the value of $\mu$ does not vary between these two systems, and is not affected by the deformed shapes.  In central collisions, on the other hand, $\mu$ vanishes, and $\varepsilon_2\{2\}\approx \sigma$. While the fluctuations due to genuine initial-state effects, such as the number of participant nucleons are larger in \auau{} systems, \uuuu{} systems receive a contribution from the fluctuations of the orientation of the colliding nuclear bodies, which is random. In central collisions, these fluctuations yield in particular body-body geometries, which produce abnormally large values of $v_2$. This naturally enhances the  variance of the distribution of $\varepsilon_2$, lifting up the rms elliptic anisotropy. In Ref.~\cite{Giacalone:2018apa} one can indeed find a plot of $\varepsilon_2\{2\}^2$ as a function of $\beta$ in \uuuu{} collisions at $b=0$, where $\mu=0$ in \equ{v22} by construction. One finds that the mean squared $\varepsilon_2$ grows indeed like $\beta^2$.

\begin{figure}[t]
    \centering
    \includegraphics[width=\linewidth]{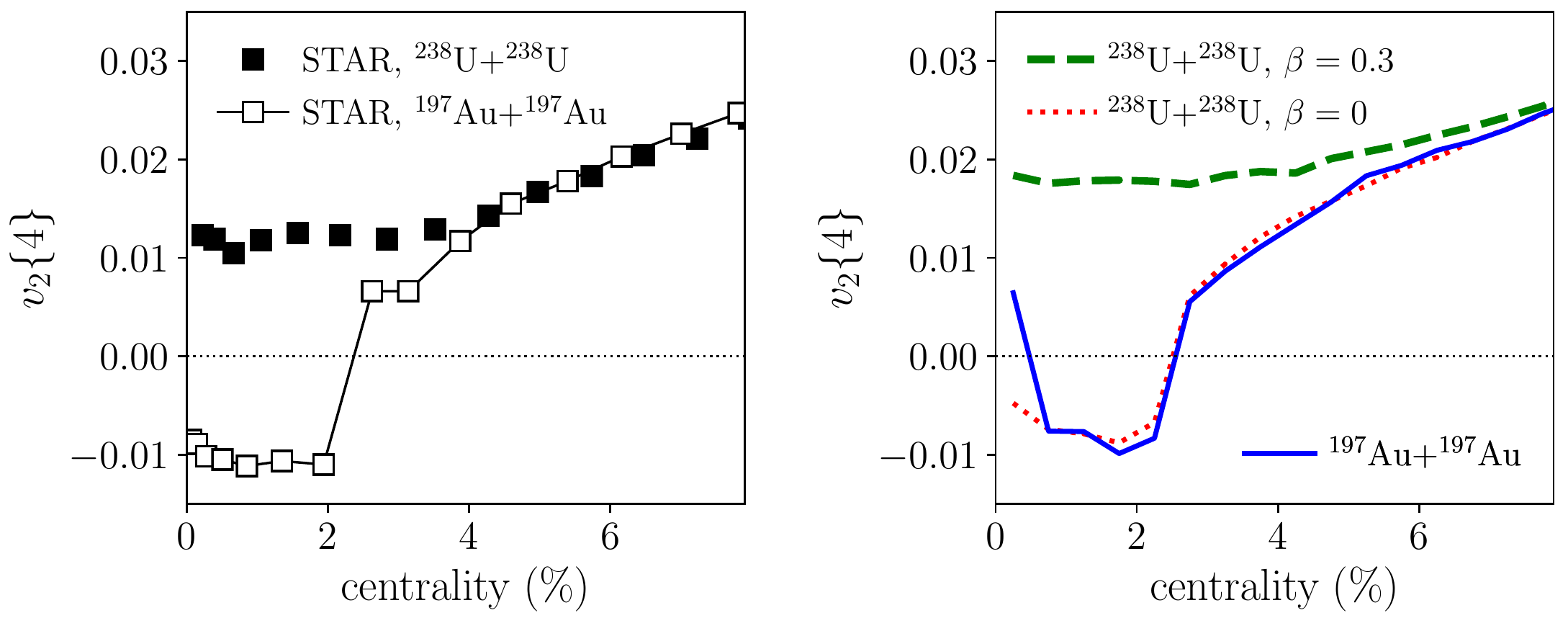}
    \caption{Same as in \fig{4-8}, but for the fourth-order cumulant of elliptic flow, $v_2\{4\}$. The negative values are obtained by calculating $-v_2\{4\}^4$ when the cumulant goes negative, and then taking the fourth root.}
    \label{fig:4-9}
\end{figure}

\paragraph{Non-Gaussianity of $v_2$ fluctuations --} 
The second observable analyzed by the STAR collaboration is even more interesting. I show STAR data on $v_2\{4\}$ as a function of collision centrality in \fig{4-9}. In the most central events, the difference between \auau{} and \uuuu{} collisions is startling. In \auau{} collisions, the cumulant displays a change of sign, occurring around $2.5\%$ centrality. Conversely, the curve for \uuuu{} collisions does never go negative, but flattens around $0.01$ in central collisions. At larger centrality, the two systems overlap.

We can once again understand this behavior from the calculations of Appendix~\ref{app:B}, and the behavior of the fourth-order cumulant of anisotropic flow (or of the eccentricity) in two distinct regimes. When the centrality is larger than typically $5\%$, the cumulant is dominated by the geometric contribution caused by the finite impact parameter, and one has simply 
\begin{equation}
    \varepsilon_2\{4\} = \mu.
\end{equation}
This explains why \auau{} collisions and \uuuu{} collisions have the same $v_2\{4\}$ away from central collisions, as $\beta$ does not affect the average flow along the direction of impact parameter. In the regime of ultracentral collisions, one has instead $\mu\approx0$, which leads to: 
\begin{equation}
\label{eq:v24kurt}
    \varepsilon_2\{4\}^4 = -K,
\end{equation}
where $K$ is the kurtosis of the distribution of $\mathcal{E}_2$, as discussed in Appendix~\ref{app:B}. In central collisions the cumulant is thus dominated by non-Gaussian corrections to the distribution of $\mathcal{E}_2$, which can in fact be either positive or negative. 

In ultracentral collisions of spherical nuclei, the value of $v_2\{4\}$ is negative. This behavior has been observed as well in precision measurements performed in \pbpb{} collisions at LHC energy~\cite{Aaboud:2019sma}. The reason for the negative sign of this cumulant is at present unknown. Theoretical calculations suggest that it is mostly associated to the fluctuations of the number of participant nucleons at a given impact parameter, as it is never observed if one keeps $N_{\rm part}$ fixed~\cite{Zhou:2018fxx}. That being said, calculations of $\varepsilon_2\{4\}$ within initial-state models tuned to data naturally reproduce the experimental observation. The results of the \trento{} model for \auau{} collisions are shown as a solid line in the right panel of \fig{4-9}. I remark a nice similarity between experimental data and the theoretical curve. The model captures the centrality percentile at which experimental data changes sign, and the magnitude of $v_2\{4\}$ in the negative region is fairly compatible with data, except for the most central point, due to a statistical fluctuation. I perform now this calculation for \uuuu{} collisions in the case where there is no quadrupole deformation. The result is shown as a dotted line in \fig{4-9}. It shows once again that, in absence of nuclear deformation, there is no difference between the results of \auau{} and \uuuu{} systems, at variance with the experimental observation.

I repeat, then,  this calculation by correctly implementing $\beta=0.3$ in the Woods-Saxon parametrization of $^{238}$U. The result is shown as a dashed line in \fig{4-9}. One understands now the origin of the puzzling experimental observation, as the cumulant does not change sign but flattens in the same precise way as in the data. From \equ{v24kurt}, this result can be understood as follows. The contribution to the fluctuations of $\varepsilon_2$ which comes from the random orientation of the colliding nuclei increases the width of the distribution of the eccentricity, thus enhancing $v_2\{2\}$, but while doing so, it also modifies the tails, making the distribution of $\varepsilon_2$ less Gaussian. This is shown explicitly in Ref.~\cite{Giacalone:2018apa}, where I plot $\varepsilon_2\{4\}^4$ as a function of $\beta$ in collisions at $b=0$, where $\mu=0$ and $\varepsilon_2\{4\}^4= -K$. One finds indeed that the cumulant grows like $\beta^4$, showing that $\beta$ yields a distribution of $\mathcal{E}_2$ which has tails narrower than a Gaussian.  As the non-Gaussianity of distributions are in general very strongly sensitive to tiny details that modify their tails, it is not surprising that the resulting effect is so sizable in the comparison between \auau{} and \uuuu{}. STAR data on $v_2\{4\}$ represents arguably one of the most striking manifestations of nuclear deformation ever observed in an experiment.

\section{Is gold deformed?}

\label{sec:4-33}

It turns out that we have just uncovered the tip of the iceberg. 

\paragraph{Issue at high energy --} I focus now on \auau{} collisions. Up to this point I have been able to nicely reproduce the experimental results on elliptic flow fluctuations for this collision system by means of a \trento{} parametrization in which the colliding $^{197}$Au nuclei are spherical. However, phenomenological nuclear models, such as the Hartree-Fock-Bogoliubov calculations of Ref.~\cite{Hilaire:2007tmk}, or the comprehensive empirical deductions of the liquid-drop model of Ref.~\cite{Moller:2015fba}, suggest that $^{197}$Au is in fact oblate, with a deformation of order $\beta\approx-0.15$. One can thus ask whether the inclusion of such a value of $\beta$ in \auau{} collisions would change the nice \trento{} model results shown in the previous figures. 
\begin{figure}[t]
    \centering
    \includegraphics[width=.65\linewidth]{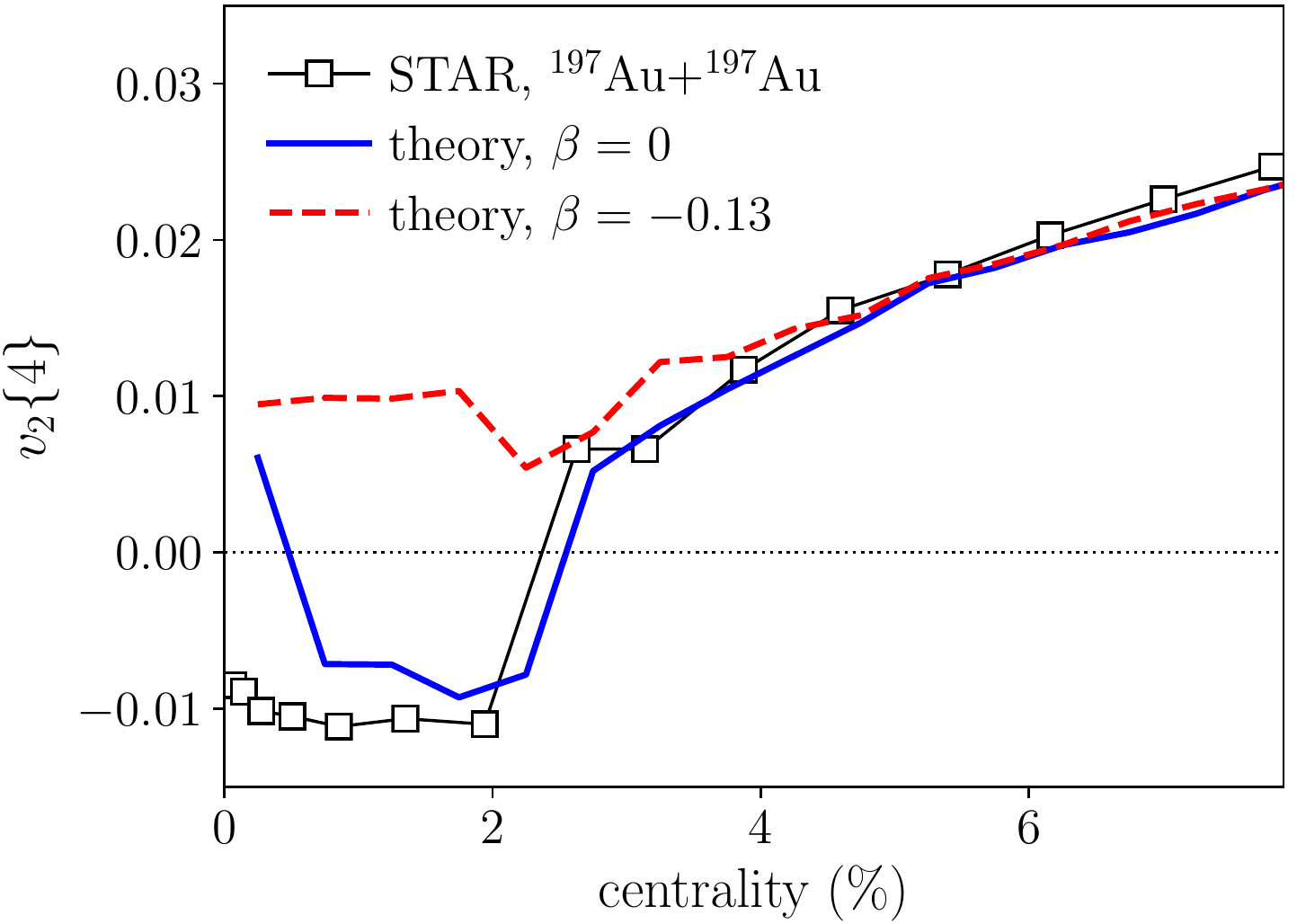}
    \caption{Symbols: STAR data~\cite{Adamczyk:2015obl} on $v_2\{4\}$ in \auau{} collisions at $\sqrt{s_{\rm NN}}=200~{\rm GeV}$. Lines are the results of the \trento{} model, implementing both spherical $^{197}$Au nuclei (solid line) and deformed nuclei with $\beta=-0.13$ (dashed line).}
    \label{fig:4-10}
\end{figure}

I repeat thus the previous calculations by implementing $\beta=-0.13$~\cite{Moller:2015fba} in the Woods-Saxon parametrization of the gold nuclei. The resulting effect on $v_2\{2\}$ is small, and of little interest. However, the effect of the oblate deformation on $v_2\{4\}$ is quite dramatic. The results are shown in \fig{4-10}. The STAR data points and the blue solid line are the same as in \fig{4-9}, showing more explicitly that there is good agreement between data and model for this observable. The new result implementing $\beta=-0.13$ is shown instead as a red dashed line.  One observes the emergence of the the same kind of behavior observed in \uuuu{} collisions. The cumulant flattens towards the limit of central collisions, and does not display a change of sign. In Ref.~\cite{Giacalone:2018apa}, this feature was tested as well within a different initial-state model, namely, the wounded nucleon model, with a very different scaling for the initial density, $t_A+t_B$.  The same result was produced, i.e., a positive sign for $v_2\{4\}$. This wrong sign in presence of $\beta=-0.13$ seems robust, and the discrepancy between data and the \trento{} evaluations implementing such a parameter is so large that it can not be fixed by a simple tweaking of model parameters.

\begin{figure}
    \centering
    \includegraphics[width=.5\linewidth]{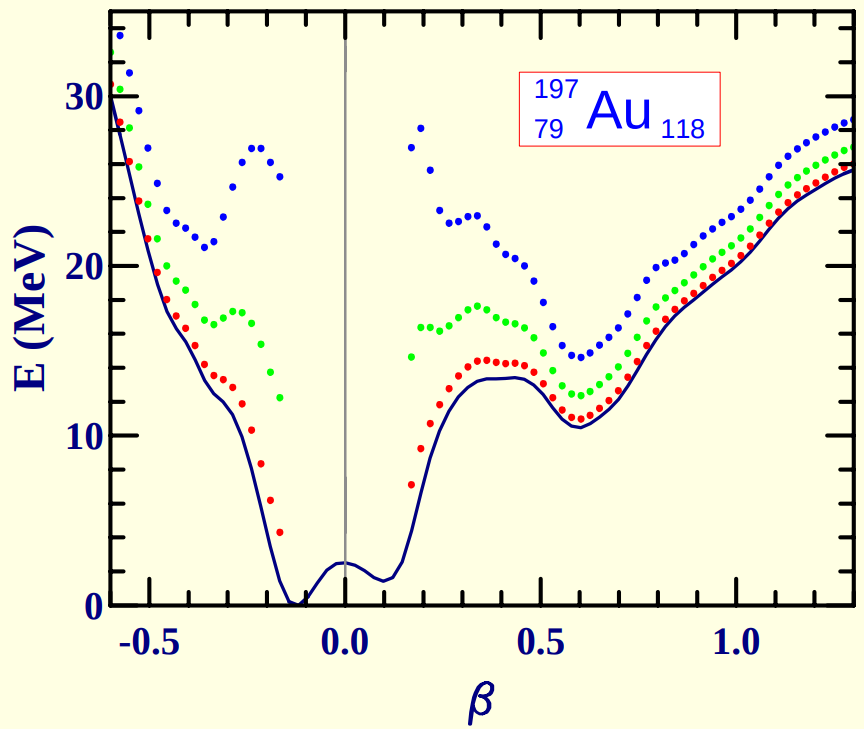}
    \caption{Mean-field potential energy surface for $^{197}$Au. Plot from Ref.~\cite{website}.}
    \label{fig:4-11}
\end{figure}

\paragraph{Nuclear structure solution --} Let me dig, then, into our knowledge of the structure of $^{197}$Au nuclei. First of all, there are no experimental measurements that give an indication of what the deformation of this nucleus should be. From the theory side, the Hartree-Fock-Bogoliubov results of Ref.~\cite{Hilaire:2007tmk} for the potential energy surface of this nucleus are reported in \fig{4-11}. One immediately remarks the difference between this curve and those shown in \fig{4-2} for $^{208}$Pb and $^{238}$U. The minimum at $\beta\approx-0.15$ is not sharp, and the curve is somewhat symmetric around $\beta=0$, displaying a second minimum at $\beta\approx0.15$ lying just a couple of ${\rm MeV}$ above. As I mentioned in the previous sections, this indicates that this nucleus does not possess a well-defined shape, and that the simple mean-field estimate, $\beta\approx-0.15$, is not good enough.

In a situation of this type, a meaningful quantification of the deformation of the nucleus requires going \textit{beyond} the mean field approach. In rough terms, the idea is the following. The minimization under constraints described by \equ{ritzQ2} returns a mean-field wavefunction, $|\Phi\ket$, which explicitly breaks rotational symmetry. This wavefunction describes thus a system which does not possess the same symmetries as the exact nuclear wavefunction, i.e., as the Hamiltonian. To repair this issue, one has to restore symmetry. This is done by projecting the mean-field wavefunction obtained at all values of $\beta$ onto the right quantum numbers, e.g., $J=0$ for an even-even nucleus. One then writes down a \textit{beyond-mean-field} wavefunction that is a mixture of all these states. The contribution of each state to the total wavefunction is weighted depending on the corresponding mean-field energy, $E$, in such a way that to a value of $\beta$ associated with a large $E$ corresponds a small contribution. This approach allows one, hence, to describe the deformation of the nucleus under study by including information coming from the entire potential energy surface, thus going beyond the mean-field estimate where the nucleus has a single value of $\beta$ corresponding to the energy minimum.
\begin{table}[t]
\centering
\begin{tabular}{|c|c|c|c|}
\hline
nuclide & experimental~\cite{Raman:1201zz} & mean field~\cite{Bender:2005ri} & beyond mean field~\cite{Bender:2005ri} \cr
\hline
$^{188}$Pt & 0.186 & -0.18 & 0.04 \\
$^{190}$Pt & 0.149 & -0.16 & -0.02 \\
$^{192}$Pt & 0.153 & -0.15 & -0.04 \\
$^{194}$Pt & 0.151 & -0.15 & -0.05 \\
$^{196}$Pt & 0.129 & -0.14 & -0.05 \\
$^{198}$Pt & 0.114 & -0.12 & -0.06 \\
\hline
\end{tabular}
\caption{Value of $\beta$ for stable platinum isotopes. The isotopes are listed in the first column. The second column from the left reports the values of $\beta$ obtained from the experimental deductions of Ref.~\cite{Raman:1201zz}. The third and the fourth column come both from Ref.~\cite{Bender:2005ri}. The third column reports mean-field estimates, while the last column corresponds to the average value of $\beta$ associated with the full beyond-mean-field wavefunction.}
\label{tab:1}
\end{table}

Data tables with the results of such a calculation for a large number of even-even nuclei have been published by Bender \textit{et al.} in Ref.~\cite{Bender:2005ri}. As expected, for well-deformed nuclei like $^{238}$U the beyond-mean-field estimates are in perfect agreement with the mean-field (and the experimental) results, because the potential energy has a sharp minimum around just one value of $\beta$, which essentially carries all the contribution to the final wavefunction. However, this is not the case for $^{197}$Au, where the evaluation of $\beta$ does include a non-negligible contribution from the minimum of $E$ lying at the opposite value of the deformation. One expects, thus, that by taking into account the full shape of the potential energy surface, the resulting average deformation parameter will be significantly lower in magnitude than $-0.15$, and much closer to zero.

Careful inspection of the literature shows that a very neat example of this phenomenon at play is in fact provided by the values of $\beta$ for the chain of stable platinum isotopes, $^{188,190,192,194,196,198}$Pt. The experimental determination of the value of $\beta$ for these nuclei, as given by Ref.~\cite{Raman:1201zz}, is reported in the second column of \tab{1}. Both the third column and the fourth column of \tab{1} report instead results from the calculations of Ref.~\cite{Bender:2005ri}. The third column reports what is referred to as a mean-field estimate of $\beta$, which for all isotopes turns out to be consistent with the experimental determination, as they both correspond essentially to the same level of approximation. The fourth column of \tab{1} shows instead the average value of $\beta$ in the full beyond-mean-field calculation, following the restoration of symmetry and the mixing of states. As the potential energy surface of these Pt isotopes looks precisely like that in \fig{4-11}, with two minima symmetric about the origin, and separated by a small energy, these estimates turn out to be much closer to zero than those obtained at the mean field level. The same thing should thus occur for $^{197}$Au nuclei. At the beyond-mean-field level, then, these nuclei are on average more spherical than predicted by the simple mean-field estimates. This is in principle good news, as more spherical nuclei will provide an improved agreement between the estimates of the \trento{} model and heavy-ion data on $v_2\{4\}$. On the other hand, within a beyond-mean-field picture all the information from the potential energy surface must be taken into account, and one can not simply employ a single value of $\beta$ in all the realizations of the nuclear wavefunction. Shape-coexistence effects, i.e., fluctuations in the value of $\beta$ in the ground state, are likely to influence the final results even further. A quantitative calculation aimed at assessing the relevance of such phenomena in \auau{} collisions is under way~\cite{benjamin}. 

On the whole, this accurate analysis of RHIC data demonstrates that a detailed understanding of elliptic flow cumulants in central nucleus-nucleus collisions requires state-of-the-art modeling of the structure of the colliding nuclei. This establishes a new important connection between low-energy and high-energy nuclear physics, with potentially far-reaching consequences that are still to explore.

\section{Evidence of deformation at LHC: $^{129}$Xe+$^{129}$Xe collisions}

\label{sec:4-5}

In October 2017, LHC physicists announced a short run of \xexe{} collisions. They asked for predictions from hydrodynamic calculations, and so we sent them results from high-statistics {\small V-USPHYDRO} simulations. However, immediately after doing that, they informed us of an anomalous behavior in the measured $v_2$ in \xexe{} collisions, which was much larger than expected, most probably due to the fact that $^{129}$Xe nuclei were nonspherical. Indeed, although an experimental determination of the quadrupole deformation of this nucleus is not available in nuclear data tables, a detailed analysis performed by the ALICE collaboration~\cite{cdscern}, based on the liquid-drop~\cite{Moller:2015fba} and experimental~\cite{Raman:1201zz} estimates, suggests $\beta=0.18$ for this nucleus. The Hartree-Fock-Bogoliubov estimates of Ref.~\cite{Hilaire:2007tmk} are also in fair agreement with that result, as they report $\beta=0.15$. Much as for $^{197}$Au nuclei, such a value of $\beta$ should have an impact on experimental data. We thus repeated our hydrodynamic calculations by implementing deformed $^{129}$Xe nuclei with $\beta=0.162$, following Ref.~\cite{Moller:2015fba}. We were eventually the first to publish a paper with quantitative hydrodynamic predictions for \xexe{} collisions~\cite{Giacalone:2017dud}.

Contrary to the case of \uuuu{} collisions, \xexe{} collisions were run only for a short time (only for 8 hours on October 12th, 2017), so that only about $15\times10^6$ events were recorded by the ATLAS and CMS detectors, and about 10$^6$ events by the ALICE detector. This somehow prevents one from carrying out high-precision measurements, such as the detailed mapping of $v_2\{4\}$ in central collisions, as reported by the STAR collaboration.

\begin{figure}[t]
    \centering
    \includegraphics[width=.6\linewidth]{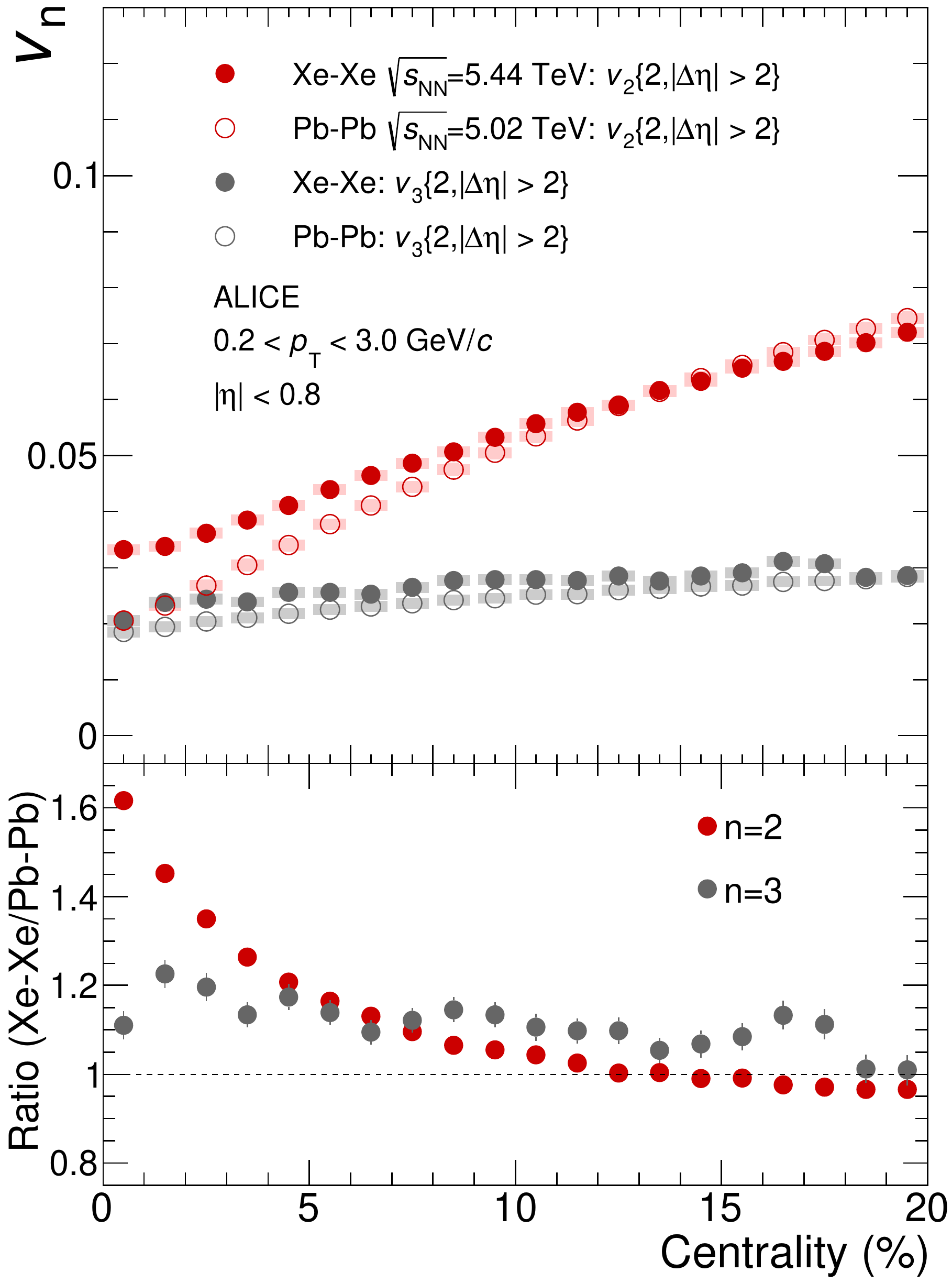}
    \caption{Top: ALICE data on the rms elliptic and triangular flows in central \pbpb{} (empty symbols) and \xexe{} (full symbols) collisions.  The lower panel shows the ratio of the flow coefficients measured in these systems. Figure from Ref.~\cite{Acharya:2018ihu}.}
    \label{fig:4-12}
\end{figure}

\paragraph{Variance --} A clear signature of the deformation of $^{129}$Xe is however visible in the experimental data on $v_2\{2\}$. It emerges in the comparison between \xexe{} data and \pbpb{} data. The corresponding plot made by the ALICE collaboration~\cite{Acharya:2018ihu} is reproduced here in \fig{4-12}. The upper panel of the plot shows the rms $v_2$ and $v_3$ in both \pbpb{} and \xexe{}  collisions, while the lower panel shows the ratio of the flow coefficients in these systems. From the lower panel one sees that in central collisions both $v_2$ and $v_3$ are larger in \xexe{} than in \pbpb{}. This has a simple explanation. In central collisions, flow coefficients, or better the initial anisotropies $\varepsilon_2$ and $\varepsilon_3$, are driven by fluctuations, which are in turn determined by the nucleons involved in the collision. These fluctuations thus scale like $1/\sqrt{A}$, where $A$ is the atomic mass number.  The fluctuation in \xexe{} collisions is thus larger by a factor:
\begin{equation}
    \sqrt{\frac{208}{129}} \approx 1.3.
\end{equation}
However, one can not simply conclude that $v_2$ in \xexe{} is larger by a factor $1.3$. The reason is that \xexe{} collisions present as well a smaller system size. From the dimensional analysis of the Navier-Stokes equation, the hydrodynamic flow of these systems is more damped by the viscous corrections. In formulas, one has:
\begin{equation}
    \frac{v_n [{\rm Xe}]}{v_n [{\rm Pb}]} = \frac{\kappa_n \varepsilon_n [{\rm Xe}]}{\kappa_n \varepsilon_n [{\rm Pb}]} = 1.3 \times \frac{\kappa_2[{\rm Xe}]}{\kappa_2[{\rm Pb}]},
\end{equation}
where dimensional analysis implies:
\begin{equation}
    \frac{\kappa_2[{\rm Xe}]}{\kappa_2[{\rm Pb}]} < 1.
\end{equation}
Our hydrodynamic results with a small $\eta/s$ suggest the the ratio of the $\kappa_2$ coefficient is around 0.95, so that the ratio of the final $v_n$ coefficients is about 1.2. 

The ratio of the triangular flow coefficients in \fig{4-6} is consistent with this analysis, as it is close to 1.2 in central collisions. However, for elliptic flow the situation is completely different. The ratio of the $v_2$ coefficients has a dramatic enhancement in central collisions, reaching a value as large as 1.6. There is only one possible explanation for this observation: $^{129}$Xe nuclei have a significant quadrupole deformation and one is observing its phenomenological manifestation. Note that data has been published as well by the CMS~\cite{Sirunyan:2019wqp} and ATLAS~\cite{Aad:2019xmh} collaborations, reporting equivalent observations. Unfortunately, these collaborations average their events over large intervals of centrality, thus making the manifestation of the deformation much less visible.

\paragraph{Non-Gaussianity --} 

A value of $\beta$ of order 0.2 for $^{129}$Xe nuclei should leave distinct signatures as well in the fourth-order cumulant, $v_2\{4\}$. The situation is however a little different compared to the analysis made for RHIC systems. In \uuuu{} collisions, the positive sign of $v_2\{4\}$ in central collisions is entirely driven by the quadrupole deformation, and, as discussed in \fig{4-9}, the cumulant is negative if $\beta=0$. On the other hand, \xexe{} collisions are affected by larger initial-state fluctuations, associated with the smaller system size, and as a consequence in these systems $v_2\{4\}$ is positive even if $\beta=0$. An effective way to assess the role of $\beta$ is thus to study the behavior of $v_2\{4\}$ as a function of this parameter in simulations, and then check agreement with data. A very good observable for this study is the relative fluctuation of $v_2$, which is quantified by the ratio $v_2\{4\}/v_2\{2\}$~\cite{Giacalone:2017uqx}. If $v_2 = \kappa_2 \varepsilon_2$, then this ratio is equal to the ratio $\varepsilon_2\{4\}/\varepsilon\{2\}$. This is typically a very good approximation up to 20\% centrality. 
\begin{figure}[t]
    \centering
    \includegraphics[width=.65\linewidth]{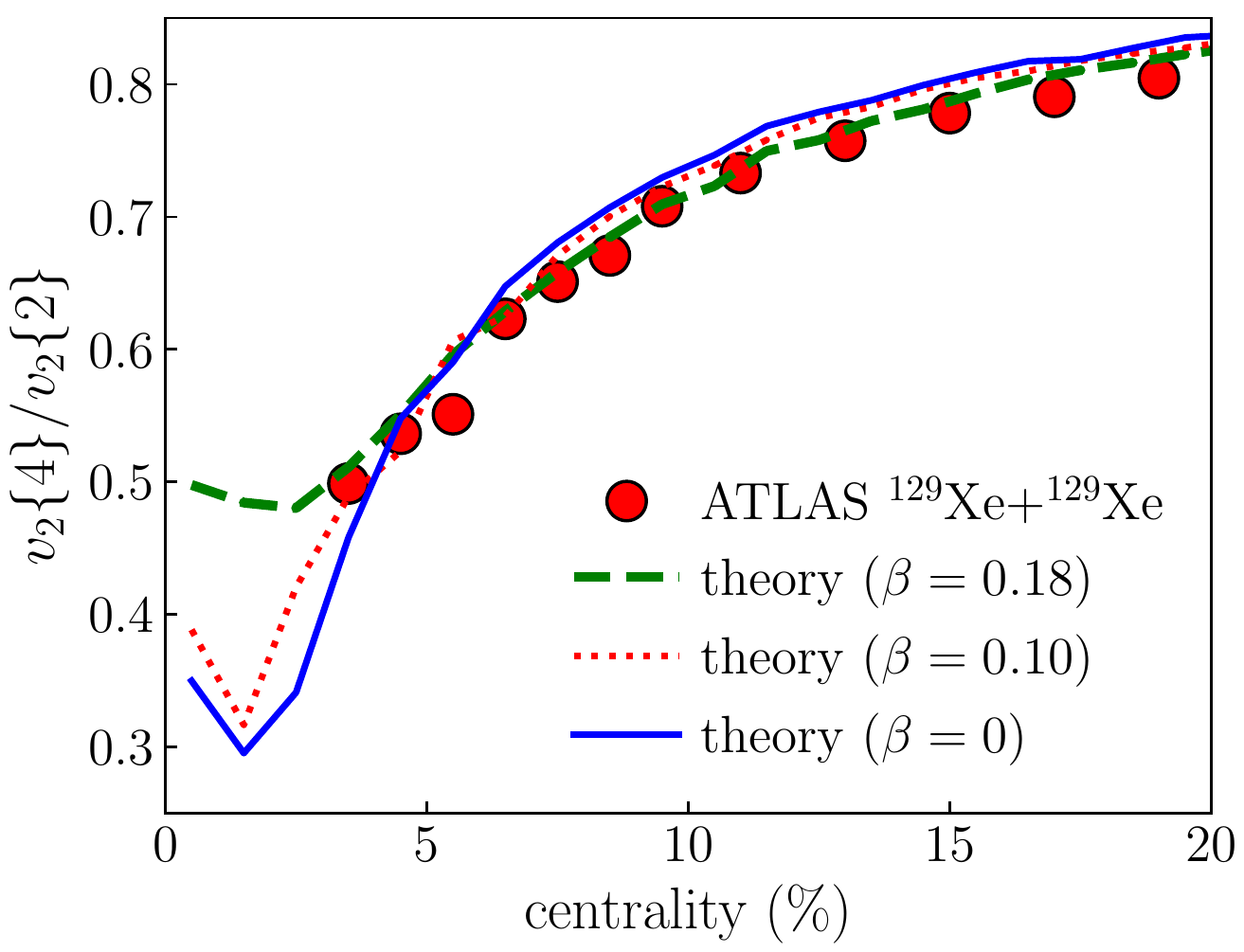}
    \caption{Relative fluctuation of elliptic flow, $v_2\{4\}/v_2\{2\}$ in central \xexe{} collisions at top LHC energy. Symbols: preliminary ATLAS data~\cite{ATLAS:2018iom}. Lines: \trento{} model evaluations, for different values of the quadrupole deformation, namely, $\beta=0$ (solid line), $\beta=0.10$ (dotted line), $\beta=0.18$ (dashed line). Preliminary data is not available below 3.5\% centrality.}
    \label{fig:4-13}
\end{figure}

In \fig{4-13}, I show results for the centrality dependence of the ratio $v_2\{4\}/v_2\{2\}$, estimated from $\varepsilon_2$, in central \xexe{} collisions simulated with the \trento{} model. The Woods-Saxon parametrization for $^{129}$Xe is the same as in the ALICE paper~\cite{Acharya:2018ihu}, while the parameters of the \trento{} models are the same as those used here for LHC \pbpb{} collisions in \fig{3-2}. The calculation is performed for $\beta=0,0.10,0.18$, displayed with different line styles in the figure. 

There are two regimes for $v_2\{4\}/v_2\{2\}$, as discussed in the previous section. For centralities above 5\%, the fourth-order cumulant is equal to the average flow along the impact parameter, i.e., $v_2\{4\} \simeq \mu$. The second-order cumulant constains instead a contribution from both $\mu$ and the fluctuation, $\sigma^2$. In non-central collisions, then, their ratio can be written as:
\begin{equation}
    \frac{v_2\{4\}}{v_2\{2\}} = \frac{\mu}{\sqrt{\mu^2 + \sigma^2}}.
\end{equation}
The growth of $\mu$ due the collision impact parameter drives the growth of the ratio observed in \fig{4-13} towards unity. Since $\beta$ does not modify $\mu$, this explains why the three theoretical estimates in \fig{4-13} do essentially overlap above 5\% centrality. In the opposite limit of central collisions where $\mu=0$, the fourth-order cumulant is a measure of the kurtosis of the $V_2$ distribution, $v_2\{4\}=-K$, while $v_2\{2\}$ originates solely from fluctuations. The ratio becomes:
\begin{equation}
    \frac{v_2\{4\}}{v_2\{2\}} = \frac{-K}{\sigma},
\end{equation}
which is a measure of the standardized kurtosis of the fluctuations of $V_2$~\cite{Abbasi:2017ajp,Bhalerao:2019fzp}. As observed in \auau{} and \uuuu{} collisions, the kurtosis is strongly sensitive to $\beta$. This explains the splitting between the curve for $\beta=0.10$ and the curve for $\beta=0.18$ in \fig{4-13}, for the most central events. 

I compare now these results to experimental data. The most accurate measurement of the relative fluctuation of $v_2$ in \xexe{} collisions has been performed by the ATLAS collaboration. This result has not been published yet, but it can be found in a conference note~\cite{ATLAS:2018iom}. Preliminary ATLAS data are displayed as circles in \fig{4-13}. Unfortunately, preliminary data is available only for centralities larger than 4\%, precisely where the theoretical curves start to overlap, yielding an excellent description of experimental data up to 20\% centrality. One additional data point is thus needed to close the case. If this point falls around 0.5, it will provide a striking indication of the deformed shape of $^{129}$Xe nuclei. This is currently under investigation by the experimental collaboration.
\begin{figure}[t]
    \centering
    \includegraphics[width=.5\linewidth]{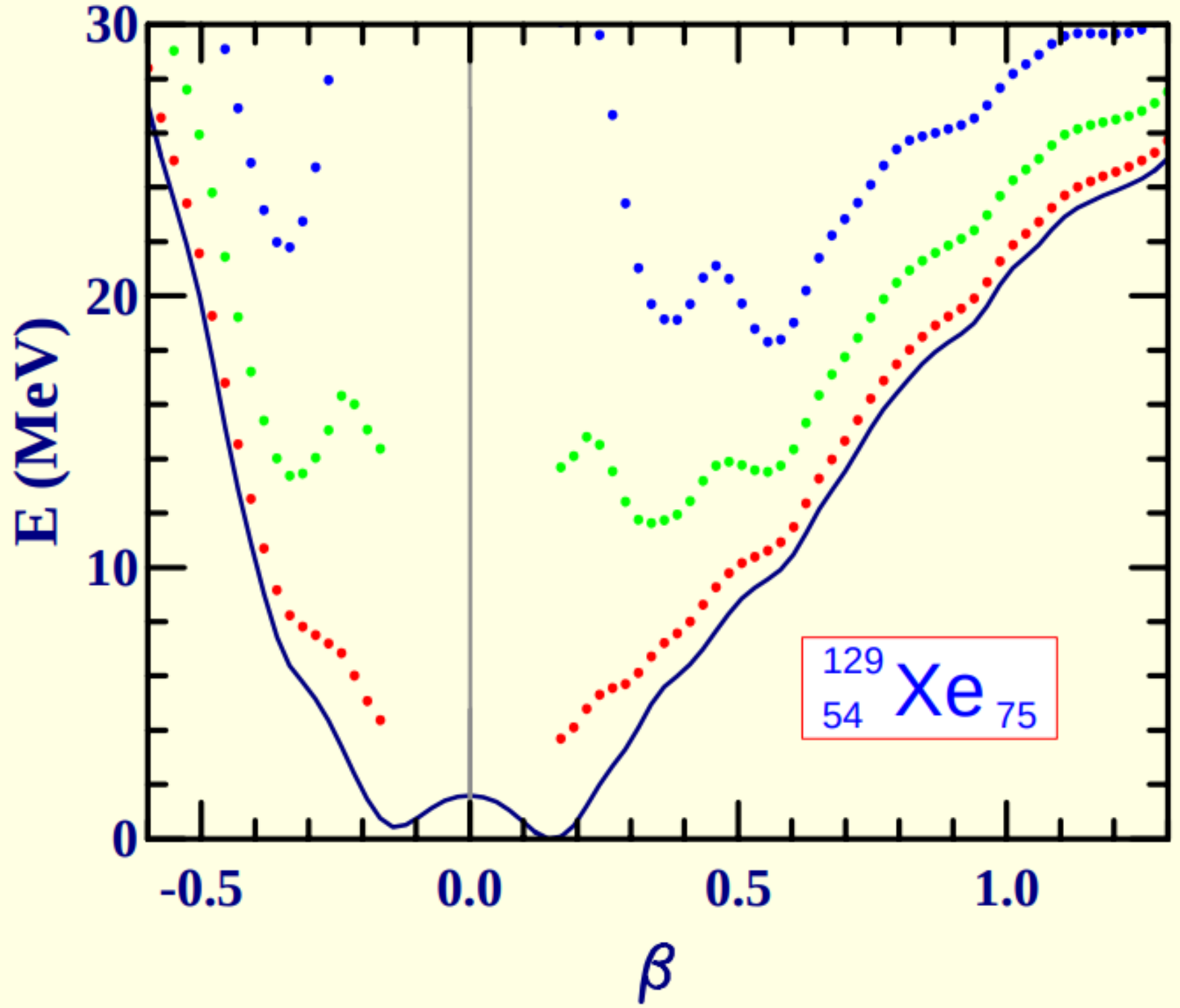}
    \caption{Potential energy surface for $^{129}$Xe returned by the Hartree-Fock-Bogoliubov calculations of Ref.~\cite{Hilaire:2007tmk}.}
    \label{fig:4-14}
\end{figure}

\paragraph{Interpretation beyond the mean field --} Data on $v_2\{2\}$ in \xexe{} collisions provide, thus, neat evidence of the ellipsoidal shape of $^{129}$Xe nuclei, and of a quadrupole deformation parameter close to the prediction of the simple mean field~\cite{Hilaire:2007tmk}, liquid-drop~\cite{Moller:2015fba}, and experimental~\cite{Raman:1201zz} evaluations. But what about the more sophisticated beyond-mean-field estimates that I discussed for $^{197}$Au nuclei?  The potential energy surface of this nucleus resulting from the calculations of Ref.~\cite{Hilaire:2007tmk} is shown in \fig{4-14}. The result is nontrivial. The curve has a minimum around $\beta\approx0.15$, but this minimum is not sharp, nor it is unique, as a symmetric minimum at the opposite value of $\beta$ is also present, at nearly the same value of energy. Much as discussed for $^{197}$Au nuclei, one expects, then, that a more rigorous beyond-mean-field treatment would yield an average deformation parameter close to zero. In the data tables of Ref.~\cite{Bender:2005ri}, one does indeed find that, for $^{128}$Xe and $^{130}$Xe, which also present two symmetric minima close in energy, the mean-field estimates of $\beta$ are respectively 0.17 and 0.14, while for the full beyond-mean-field results the average deformations are 0.05 and 0.04. These latter results are not consistent with heavy-ion collision data, thus opening an interesting question. The energy surface in \fig{4-14} is significantly broader than that of \fig{4-11}, meaning that $^{197}$Au has in general a more well-defined shape than $^{129}$Xe. Hence for this nucleus shape-coexistence effects, related to the fluctuation of $\beta$, are more important. This suggests that the observations made in \xexe{} collisions are not ascribable to the simple fact that these nuclei are ellipsoidal, but rather to the fact that their shape fluctuates around a prolate minimum, a possibility which is currently under investigation~\cite{benjamin}.

Wrapping up, a clear indication of nuclear deformation in heavy-ion collision data does not have an immediate interpretation in nuclear theory, and requires state-of-the-art modeling of the structure of the colliding objects, including nontrivial effects related to the fluctuations of $\beta$ in the ground state. Situations of this kind, motivated by experimental data, do usually lead to nice advances.


\chapter{Discerning collision configurations}

\label{chap:5}

High-energy nuclear experiments thus lead to remarkable phenomenological manifestations of the deformation of atomic nuclei. A nontrivial phenomenology of nuclear deformation has been triggered by the simple measurements of $v_2\{2\}$ and $v_2\{4\}$ in central heavy-ion collisions, leading to results and possibilities which are unprecedented in the context of nuclear experiments. 

The situation remains however a little disappointing. The observables discussed so far are averaged over all events in a given class of centrality.  In the picture of the rotational model, where the colliding nuclei are ellipsoidal objects, this implies that the observables are obtained by averaging over all orientations of the colliding bodies. The net effect is an increase in the fluctuations of $v_2$, which can be significant, however, it would be desirable to have observables that make explicit use of the information on the orientation of the nuclei on an event-by-event basis, i.e., that are sensitive to whether collisions are, e.g., tip-tip-like or body-body-like. This has indeed been the idea driving the \uuuu{} collision run at RHIC, although it has been classified as a hopeless task following the failure~\cite{Adamczyk:2015obl} of the predictions based on the two-component Glauber model.
 
In this chapter, I show that this apparently great difficulty can be overcome. I present a method to make a distinction between body-body and tip-tip events at a given collision centrality. The idea to make a selection of events based on $\bra p_t \ket$, which is a measure of the temperature of these systems. As tip-tip collisions have more compact profiles, they are also hotter at fixed centrality, while the opposite is true for body-body events. This method allows me in particular to construct observables that possess an unparalleled sensitivity to the value of $\beta$.

\section{Discerning body-body and tip-tip geometries}

\label{sec:5-1}

\subsection{The idea}

\label{sec:5-11}

The method is essentially an application of Ref.~\cite{Gardim:2019xjs}, which clarifies the physical origin of the average transverse momentum, $\bra p_t \ket$, and of its fluctuations, in hydrodynamics, as explained in detail in Sec.~\ref{sec:3-2}. The quantity which allows one to discern body-body and tip-tip geometries at a given collision centrality is precisely $\bra p_t \ket$. Let me explain why.

I recall first \fig{4-4}, showing the transverse area of overlap for body-body and tip-tip collisions. The important feature to remark, and which had never occurred to me before September 2019, is that the system size of a body-body collision is considerably larger than the system size of a tip-tip collision as soon as $\beta$ is large enough. Recall then the picture of \fig{3-6}. At the same entropy,  larger-than-average system size implies smaller temperature, and consequently smaller average transverse momentum, $\bra p_t \ket$. Now, combining these two arguments one concludes that, at fixed multiplicity, tip-tip collision yields a larger $\bra p_t \ket$. Summarizing: \vspace{2mm}
\begin{displayquote}
\begin{mdframed}
\textit{At fixed multiplicity, a tip-tip collision produces the same amount of entropy as a body-body collision, but in a smaller volume, resulting in a hotter system that yields a larger average transverse momentum.}
\end{mdframed}
\end{displayquote}
This idea was introduced in Ref.~\cite{Giacalone:2019pca}.

To get an intuitive understanding of the physical effect I am talking about, it is instructive to look at the actual density profile of a body-body collision and of a tip-tip collision that share the same total entropy. I take the profiles shown in Ref.~\cite{Giacalone:2020awm}, which correspond to the initial conditions of the hydrodynamic simulations that I shall discuss in Appendix~\ref{app:C}.  The left panel of \fig{5-1} shows the energy density profile, $e({\bf x},\tau_0)$ of a body-body collision. The right panel shows instead the profile of a tip-tip collision. The most prominent difference between these system is obviously the global geometry, strongly elliptical in the body-body event. However, a second feature that can be inferred essentially by eye is that the tip-tip profile, containing the same total entropy as the body-body profile but within a smaller transverse area, presents on average larger values of energy density (or temperature). For this reason, the tip-tip event yields a larger value of $\bra p_t \ket$ at the end of the hydrodynamic phase. I refer to the figure caption for the actual values of these quantities.
\begin{figure}[t]
    \centering
    \includegraphics[width=\linewidth]{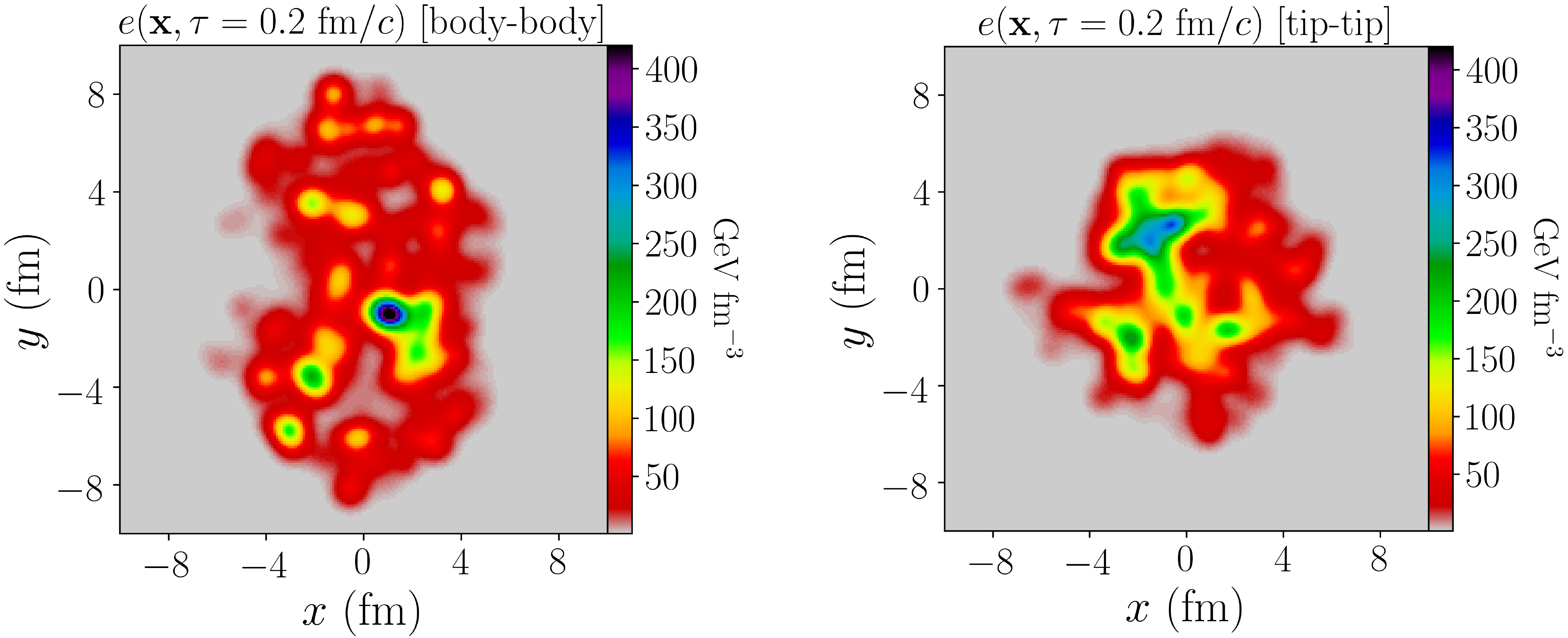}
    \caption{Left: energy density profile at the onset of hydrodynamics ($\tau=0.2~\fmc$) for a body-body collision, with $S=4040$, $E=1294$~GeV, $\varepsilon_2=0.478$, $\varepsilon_3=0.191$, and average temperature $\bra T \ket = 0.433$~GeV. The hydrodynamic evolution of this system yields: $dN_{ch}/d\eta_{|\eta|<1}=1296$, $\bra p_t \ket=0.587$~GeV, $v_2=0.083$, $v_3=0.016$ to the final state. The flow coefficients are calculated by implementing the same kinematic cuts as the STAR collaboration: $|\eta|<1$, and $0.2<p_t<2$~GeV. Right: density profile for a tip-tip collision, with $S=4072$, $E=1429$~GeV, $\varepsilon_2=0.096$, $\varepsilon_3=0.089$, and  $\bra T \ket = 0.475$~GeV. After hydrodynamics this event yields $dN_{ch}/d\eta_{|\eta|<1}=1280$, $\bra p_t \ket=0.651$~GeV, $v_2=0.027$, $v_3=0.009$. Figure from Ref.~\cite{Giacalone:2020awm}.}
    \label{fig:5-1}
\end{figure}

\subsection{Freezing nuclear orientations}

\label{sec:5-12}

I first check that the idea works within the \trento{} model, although any other model of initial conditions would be simply as good for this purpose. My claim is that, at fixed total entropy, collisions at low $\bra p_t \ket$ correspond to body-body events, while collisions at large $\bra p_t \ket$ correspond to tip-tip events. Since in the simulations I have knowledge of the Euler angles of the colliding nuclei, I can check explicitly their average orientation as a function of $\bra p_t \ket$ in a given centrality class. 

\paragraph{Initial-state predictor --} To do so from the initial-state calculation, I need an estimator of $\bra p_t \ket$ and of its fluctuations. In Ref.~\cite{Giacalone:2019pca} I make the simplest choice. I use the system size, $R$, as defined by \equ{R}, as an event-by-event predictor of $\bra p_t \ket$. In the limit of small fluctuations, which is a good approximations for the fluctuations of $R$ at fixed multiplicity, one can use the  thermodynamic identity encountered Sec.~\ref{sec:3-21}:
\begin{equation}
    c_s^2 = \frac{dP}{d\epsilon} = \frac{d\ln T}{d\ln s},
\end{equation}
where $P$, $\epsilon$, $s$, and $T$ are respectively the pressure, the energy density, the entropy density, and the temperature of the system. Dimensional analysis implies that $s \propto R^{-3}$, while I consider that $T$ is proportional to $\bra p_t \ket$ in view of the discussion of Sec.~\ref{sec:3-21}. The relative variation of $R$ is thus related to that of $\bra p_t \ket$ by:
\begin{equation}
\label{eq:Rpred}
   \frac{\bra p_t \ket - \bbra p_t \kket }{\bbra p_t \kket} = -3c_s^2     \frac{R-\bra R \ket}{\bra R \ket},
\end{equation}
where $\bbra p_t \kket$ is the average value of $\bra p_t \ket$ in the centrality class. This equation is used indeed in Ref.~\cite{Giacalone:2019pca} to relate the relative fluctuations of $R$ returned by the \trento{} model to the relative fluctuations of $\bra p_t \ket$. This relation is very accurate, however, later studies showed that the simple choice of $R$ as a predictor of $\bra p_t \ket$ is not good enough. In particular, it does not allow to describe the correlation between $\bra p_t \ket$ and the flow coefficients $v_n$ which is observed in experiments, as I shall show explicitly in Sec.~\ref{sec:5-22}. A predictor which works much better is the quantity used in the discussion of Sec.~\ref{sec:3-2}, i.e., the initial energy of the system, $E$, given by \equ{totE}, divided by the initial entropy, $S$, given by \equ{totS}, whose value is almost fixed in the centrality bin. The predictor becomes:
\begin{equation}
\label{eq:pred2}
   \frac{\bra p_t \ket - \bbra p_t \kket }{\bbra p_t \kket} = \kappa_0     \frac{E/S-\bra E/S \ket}{\bra E/S \ket},
\end{equation}
where $\kappa_0$ is a parameter that can be fixed by imposing that distribution of the relative $E/S$ has the same width as the distribution of the relative $\bra p_t \ket$. A recent paper by the STAR collaboration~\cite{Adam:2019rsf} reports in particular that the relative dynamical fluctuation of $\bra p_t \ket$ in central \auau{} collisions is equal to:
\begin{equation}
\label{eq:sigmadyn}
    \frac { \sigma_{\rm dynamical}(\bra p_t \ket)}{\bbra p_t \kket} = 0.012,
\end{equation}
with a average transverse momentum, $\bbra p_t \kket$, of about $0.57~{\rm GeV}$. The relative fluctuation of $E/S$  in my \trento{} calculation is instead of order $0.03$, so that $\kappa_0 \approx 0.4$ to match my simulations to RHIC data on $\bra p_t \ket$ fluctuations.

\paragraph{Orientation of colliding nuclei --} 
I study the orientation of the colliding nuclei as a function of $\bra p_t \ket/\bbra p_t \kket -1$, estimated from \equ{pred2}. I simulate \uuuu{} collisions at top RHIC energy, implementing $\beta=0.3$, and I focus on a narrow class of ultracentral collisions, corresponding to the $0.4-0.8\%$ range, where the average impact parameter is about $1.5~\fm$. I recall that body-body collisions correspond to $\theta_A=\theta_B=\pi/2$, and correlated azimuthal spins, $\phi_A=\phi_B$, whereas tip-tip collisions correspond to $\theta_A=\theta_B=0$, and uncorrelated azimuthal angles $\phi_A\neq\phi_B$.

\begin{figure}
    \centering
    \includegraphics[width=\linewidth]{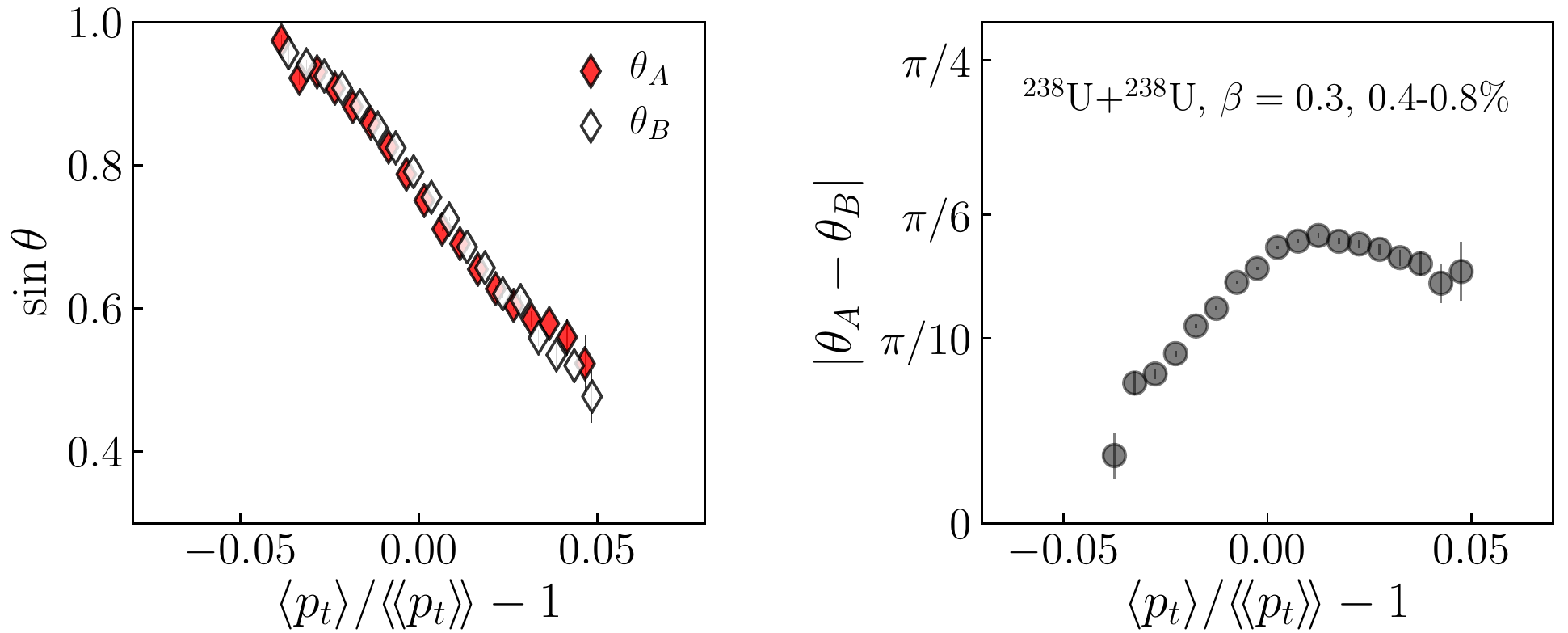}
    \caption{Left:  Sine of $\theta_A$ (full symbols) and $\theta_B$ (empty symbols) as function of the relative variation of $\bra p_t \ket$ in ultracentral \uuuu{} collisions $(\beta=0.3)$ at top RHIC energy. Right: Relative polar angle between the two colliding nuclei.}
    \label{fig:5-2}
\end{figure}

The left panel of \fig{5-2} shows the average value of $\sin \theta$ for both colliding nuclei as a function of the relative variation of $\bra p_t \ket$. The results look excellent. The curve has a rather steep trend. In the limit of small $\bra p_t \ket$, it goes very close to unity, which implies $\theta\simeq\pi/2$ for both nuclei, corresponding precisely to the expectation from body-body collisions. Moving to large values of $\bra p_t \ket$, I find on the other hand that the value of $\sin \theta$ does only get as low as $0.5$, which suggests that the average angle is about $\pi/6$, in contrast with the expectation of tip-tip collisions. In the right panel of \fig{5-2} I show the magnitude of the relative polar angle between the colliding nuclei, where the polar angles are properly defined between 0 and $\pi/2$, so that the difference $|\theta_A-\theta_B|$ is uniquely determined. This quantity should vanish for both body-body and tip-tip collisions. The trend observed in \fig{5-2} is very interesting. At low $\bra p_t \ket$, the relative polar angle is very small, going as low as $\pi/20$, and confirming that the two nuclei have equal polar orientations. The relative angle then grows reaching $\pi/6$ at the average value of $\bra p_t \ket$, corresponding somehow to the most probable value for this relative angle in the centrality bin. This quantity shows then a slow decreasing trend for larger values of $\bra p_t \ket$, meaning that the selection based on $\bra p_t \ket$ is in fact trying to isolate tip-tip configurations, with some difficulty.

To understand these results in greater detail, I study how the two nuclei are aligned in both polar and azimuthal orientation. I quantify the alignment of the axes of the two nuclei in polar angle by means of the following correlator:
\begin{equation}
\label{eq:polar}
1 - \left( \cos 2\theta_A - \cos2\theta_B  \right )^2 / 4.
\end{equation}
This quantity satisfies all the symmetries of the problem: $\theta\leftrightarrow\pi-\theta$,  $\theta_A\leftrightarrow\theta_B$, $\theta\leftrightarrow-\theta$. It is equal to unity for both body-body and tip-tip collisions.  For the alignment of the nuclear axes in azimuthal angle, I use instead:
\begin{equation}
\label{eq:azimuthal}
\cos2(\phi_A-\phi_B),
\end{equation}
which vanishes if the azimuthal directions are uncorrelated. This quantity should be equal to unity in body-body collisions, and equal to zero in tip-tip collisions.

\begin{figure}
    \centering
    \includegraphics[width=.6\linewidth]{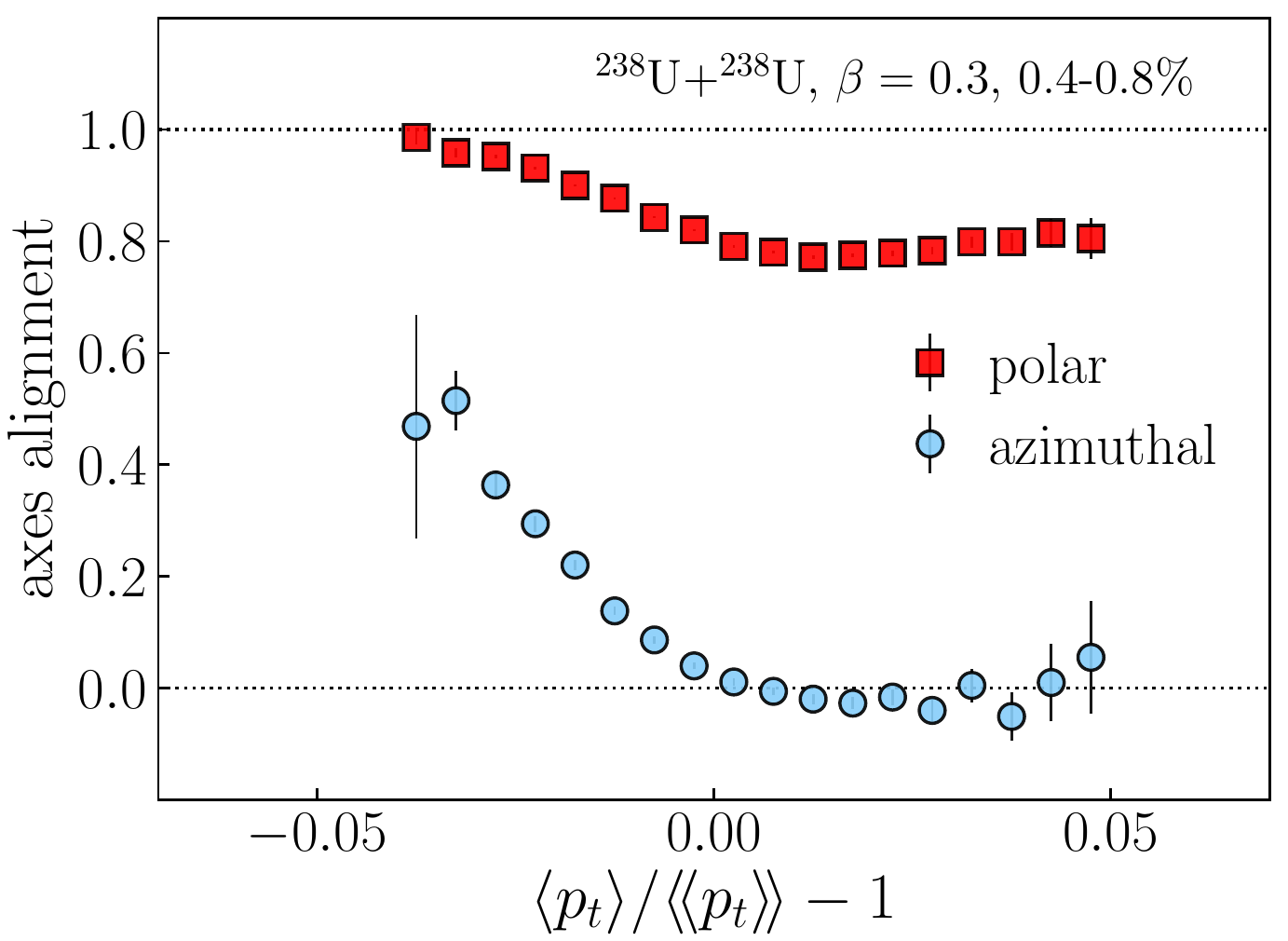}
    \caption{Alignment of the nuclear axes in ultracentral $^{238}$U+$^{238}$U collisions at top RHIC energy. An alignment close to unity implies correlated orientations (nuclear axes are parallel), while the alignment is zero for uncorrelated orientations. Squares: polar orientation, Eq.~(\ref{eq:polar}). Circles: azimuthal orientation, Eq.~(\ref{eq:azimuthal})}
    \label{fig:5-3}
\end{figure}

The results are shown in \fig{5-3}. 
The alignment of the polar orientations is displayed as red squares. At low $\bra p_t \ket$, it is essentially equal to unity, confirming the selection of body-body collisions. The correlator then decreases mildly, reaching a minimum of about 0.8 around the average of $\bra p_t \ket$, and then grows up again, confirming the previous intuition that large-$\bra p_t \ket$ collisions the two nuclei are slowly re-aligning in polar angle towards the limit of tip-tip collisions. As a side remark, it is not surprising that selecting tip-tip events is more difficult than selecting body-body events. The random orientations of the nuclei are sampled such that the distribution of $\cos \theta$ is a uniform distribution. This yields a distribution of $\theta$ which has a peak at $\theta=\pi/2$, and a minimum at $\theta=0$. Tip-tip collisions are thus strongly disfavored from a probabilistic point of view. The fact that the polar correlation remains as high as 0.9 at large $\bra p_t \ket$ with a relative angle lower than $\pi/6$ demonstrates that the selection of events based on $\bra p_t \ket$ in discerning collision geometries is in fact very powerful.

The case is then fully closed by the results on the azimuthal correlation of the nuclear axes, which is shown as circles in \fig{5-3}. Quite remarkably, this quantity grows steeply towards low values of $\bra p_t \ket$, reaching a magnitude as large as 0.6, corresponding to correlated azimuthal orientations. The final conclusion is that events at low $\bra p_t \ket$ correspond indeed to body-body collisions. Conversely, as we move towards high values of $\bra p_t \ket$, the correlator becomes remarkably compatible with zero. As tip-tip collisions have uncorrelated azimuthal angles, this results also confirms that the nearly tip-tip configurations are selected at large $\bra p_t \ket$.

\begin{figure}[t]
    \centering
    \includegraphics[width=.6\linewidth]{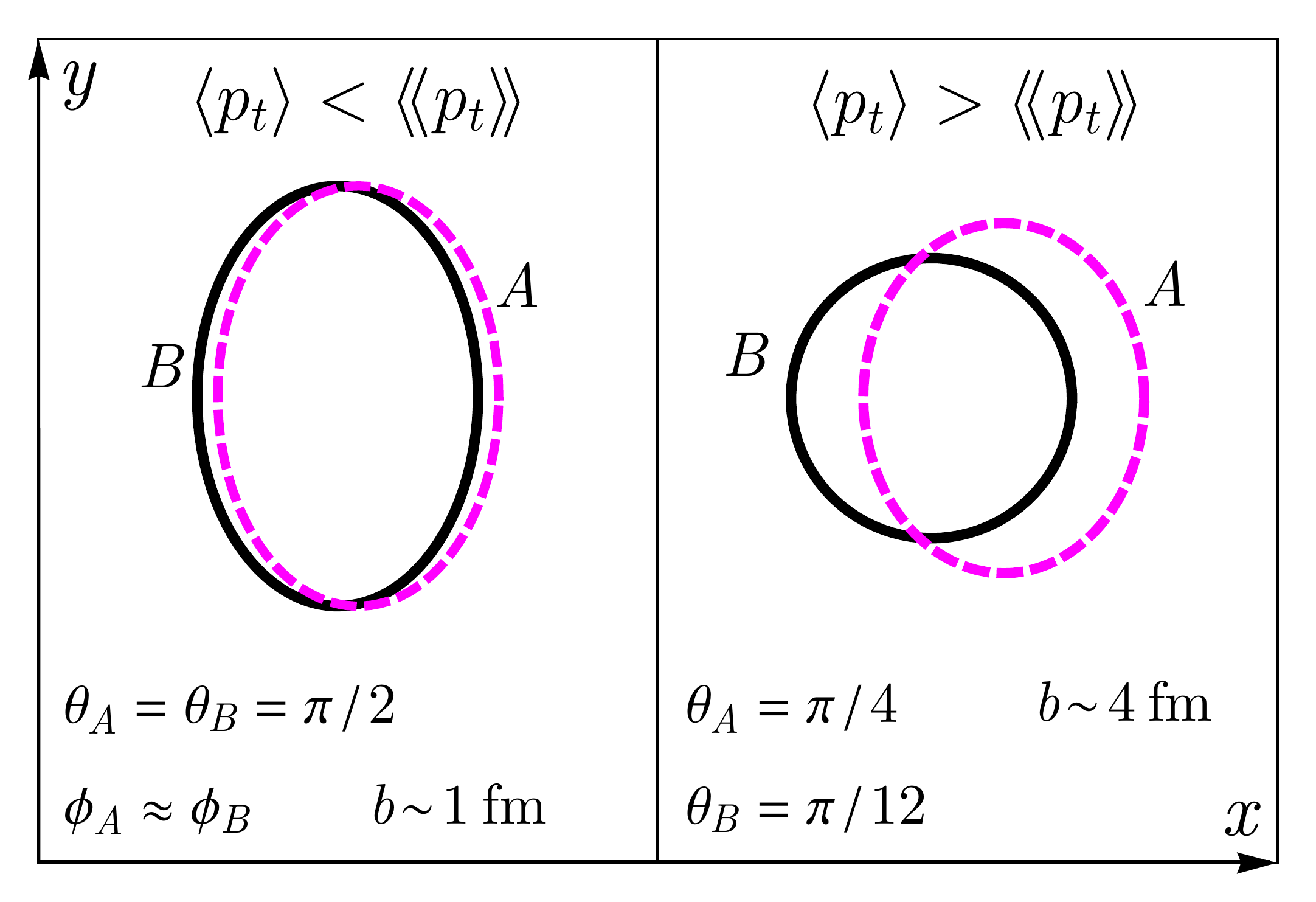}
    \caption{Transverse plane projection of the average geometry of $^{238}$U+$^{238}$U collisions as a function of $\bra p_t \ket$. Left: low $\bra p_t \ket$, corresponding to body-body collisions with correlated orientations and $b\sim1$~fm. Right: high $\bra p_t \ket$, corresponding to nearly-tip-tip collisions with a relative polar tilt of about $\pi/6$, and $b\sim4$~fm. In this case the azimuthal orientations are arbitrary.}
    \label{fig:5-4}
\end{figure}

The conclusions drawn from these results are summarized in \fig{5-4}. The picture is rather clear. As shown in the left panel of the figure, in ultracentral collisions between deformed nuclei events at low $\bra p_t \ket$ correspond to body-body configurations, with an average impact parameter of about 1~fm, and with correlated azimuthal and polar orientations. Events at large $\bra p_t \ket$, shown in the right panel of the figure, correspond instead to nearly tip-tip configurations. The average relative polar angle is around $\pi/6$. As the the actual limit $\theta=0$ is hard to achieve due to probabilistic considerations, a plausible average configuration for these events is thus $\theta_A=\pi/4$ and $\theta_B=\pi/12$, or vice versa. This is consistent with all the results of the previous figures, and corresponds to the right panel of \fig{5-4}. The average impact parameter of these large-$\bra p_t \ket$ events is about 4~fm.

\subsection{Revealing nuclear deformation}

\label{sec:5-13}

At fixed multiplicity, then, a selection of events based on $\bra p_t \ket$ allows to discern body-body and tip-tip configurations. However, experimentally one does not know anything about the Euler angles, hence, if one selects events according to $\bra p_t \ket$, how can they know that this idea works in practice? The answer is obviously that one should look at the elliptic flow. Whenever the colliding nuclei are deformed enough to yield an eccentricity that dominates over the quantum fluctuations in the regime of body-body collisions, the previous selection implies that elliptic flow of events at low $\bra p_t \ket$ should be significantly larger than average. In summary~\cite{Giacalone:2019pca}:

\begin{displayquote}
\begin{mdframed}
\textit{In central collisions of large well-deformed nuclei, the elliptic flow and the average transverse momentum are anticorrelated.}
\end{mdframed}
\end{displayquote}

\begin{figure}[t]
    \centering
    \includegraphics[width=.6\linewidth]{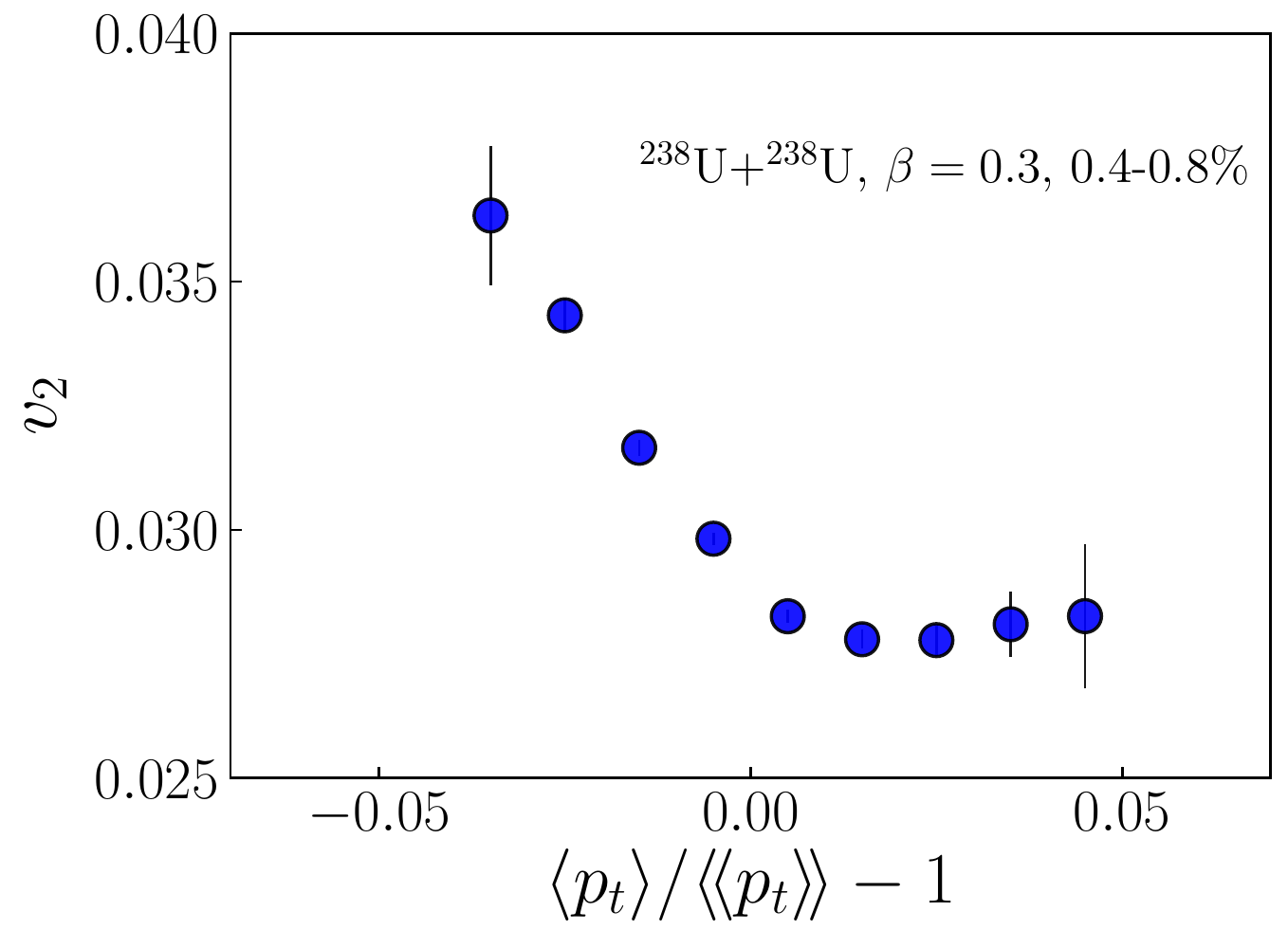}
    \caption{Rms ellptic flow, $v_2=0.165\varepsilon_2$, as a function of the relative variation of $\bra p_t \ket$ in ultracentral \uuuu{} collisions ($\beta=0.3$) at top RHIC energy.}
    \label{fig:5-5}
\end{figure}

I check that this is indeed the case in my \trento{} calculations, by evaluating the eccentricity of the system, $\varepsilon_2$, as a function of relative variation of $\bra p_t \ket$ in the same bunch of ultracentral \uuuu{} collisions used for the previous figures. I rescale the eccentricity by the constant $\kappa_2=0.165$, the same used in Chapter~\ref{chap:4}, to display values of the rms $v_2$ that can be compared to experimental data. The results are shown in \fig{5-5}.

The effect is bright and clear. In collisions of large deformed nuclei, such as \uuuu{} collisions, the effects of the geometry induced by the deformation of these nuclei dominates over the underlying quantum fluctuations, due to the positions of the nucleons, in the limit of body-body events. As a consequence elliptic flow grows steeply as one moves towards low values of $\bra p_t \ket$, the increase depending on the value of $\beta$. In this calculation it increases by almost 40\%. 

This is a new feature of high-energy nuclear physics, and represents a neat prediction to be tested in experiments. However, the result shown in \fig{5-5} can not be considered yet as a quantitative prediction, as I shall discuss in the next section, which is devoted to the derivation of quantitative predictions to be compared with future experimental data. Note also that the conclusions drawn from this analysis have been obtained by means of \trento{} simulations. One can naturally wonder if all the arguments involved in these calculations, such as the fact that body-body events have larger $\bra p_t \ket$ than tip-tip events, or  the value $\kappa_2=0.165$, are in fact consistent with the results of full hydrodynamic evaluations. I perform a number of such checks by means of comprehensive hydrodynamic simulations of body-body and tip-tip collisions in Appendix~\ref{app:C}. 

\section{Quantitative analysis}

\label{sec:5-2}

In a funny turn of events, I obtained these results right before leaving for a visit to the Brookhaven National Laboratory. Once there, I could knock on the door of the STAR physicists, and point out that I had found the way to discern collision geometries, corresponding to a new striking prediction to be tested in \uuuu{} data. The measurement was performed overnight. Although I was not allowed to look at the data, I was confirmed that the effect was there. Official preliminary STAR data were later shown in this year's Winter Workshop on Nuclear Dynamics~\cite{shengli}. The result is that the slope of $v_2$ plotted against $\bra p_t \ket$ is indeed negative in central \uuuu{} collisions, confirming my arguments, and thus the selection of body-body collisions at low $\bra p_t \ket$. 

However, preliminary STAR data looks fairly different from the curve shown in \fig{5-5}, meaning that something is missing in my calculation. The problem is that evaluating the rms $v_2$ as function of $\bra p_t \ket$ requires the knowledge of the mean transverse momentum in each event. As pointed out in Sec.~\ref{sec:3-4}, this is not an issue in a theoretical calculations, however, as the number of particles detected in central \uuuu{} collisions is of order $1000$, the experimental determination of $\bra p_t \ket$ is affected by a significant statistical error. The relative dynamical fluctuation of $\bra p_t \ket$ is only about $1.2\%$ following \equ{sigmadyn}, and this is essentially as large as the relative statistical fluctuation. The effect of these trivial fluctuations should thus be included in my \trento{} evaluations.

In this section, I compute quantitative predictions for the dependence of $v_2$ and $v_3$ on $\bra p_t \ket$ in ultracentral collisions, by properly addressing this problem. Furthermore, I introduce an observable that quantifies the dynamical correlation between $v_n$ and $\bra p_t \ket$. This observable is by construction independent of trivial statistical features, and turns out to possess an amazing sensitivity to the quadrupole deformation of the colliding nuclei.

\subsection{$v_n$ as a function of $\langle p_t \rangle$}

\label{sec:5-21}

The distribution of the relative fluctuations of $\bra p_t \ket$ estimated from the \trento{} model has to be corrected for statistical fluctuations. Performing this correction is straightforward. A simple method to do this is described in Ref.~\cite{Giacalone:2020awm}, and is explained here in Appendix~\ref{app:D}. The correction results in a distribution of $\bra p_t \ket$ that is broader than the original one, as it now includes two independent contributions, and that leads to a depletion of the correlation between $v_n$ and $\bra p_t \ket$.

\begin{figure}[t]
    \centering
    \includegraphics[width=\linewidth]{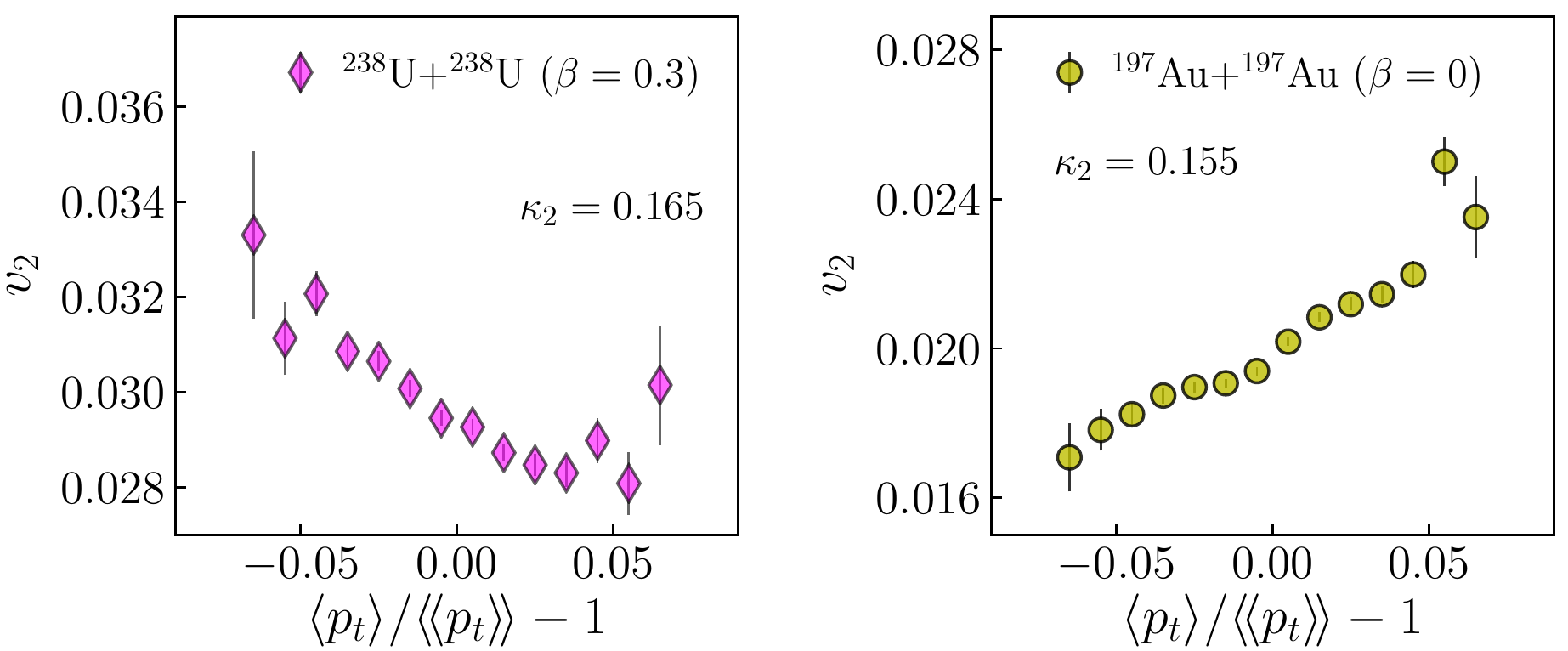}
    \caption{Rms elliptic flow, $v_2$, as a function of $\bra p_t \ket$ in $200~{\rm GeV}$ \uuuu{} collisions (left panel) and \auau{} collisions (right panel). The coefficient of linear hydrodynamic response, $\kappa_2=v_2/\varepsilon_2$, is specified in each panel. Figure adapted from Ref.~\cite{Giacalone:2020awm}.}
    \label{fig:5-6}
\end{figure}

In \fig{5-6} I show quantitative predictions for the rms $v_2$ in ultracentral \uuuu{} collisions (left panel) and \auau{} collisions (right panel) as a function of the corrected relative variation of $\bra p_t \ket$. If one compares the result in the left panel of this figure with that shown in \fig{5-5}, one sees that there is indeed a depletion of the negative slope of $v_2$, which now decreases by only $10\%$ from low to large $\bra p_t \ket$. However, the correlation remains negative, and does not change the overall picture. The effect of the statistical fluctuations on the quantities shown in \fig{5-2} and \fig{5-3} can be found in Ref.~\cite{Giacalone:2020awm}. The right panel of \fig{5-6} shows on the other hand the results for \auau{} collisions. Here I am colliding spherical gold nuclei, following the analysis of Sec.~\ref{sec:4-5}. One sees that in this panel the correlation between $v_2$ and $\bra p_t \ket$ is positive. This is not surprising. As I shall discuss in the next section, the correlation between the mean transverse momentum and $v_2$ has indeed been recently measured in \pbpb{} collisions at top LHC energy by the ATLAS collaboration~\cite{Aad:2019fgl}, and experimental data shows that these quantities are positively correlated in central collisions. The same is thus expected to occur in central \auau{} collisions.

The results of \fig{5-6} can thus be compared to preliminary STAR data~\cite{shengli}. I am not allowed to reproduce this data here, but my predictions turn out to be in good agreement with it. With an accuracy of order $10\%$, the magnitude of elliptic flow is captured, and the slopes of the curves are nicely reproduced by my results, in particular, the negative slope in \uuuu{} collisions, which is a measure of the quadrupole deformation of these nuclei. I emphasize that there are no free parameters in my predictions. The features of the model are all constrained by other sets of data.

\begin{figure}[t]
    \centering
    \includegraphics[width=.65\linewidth]{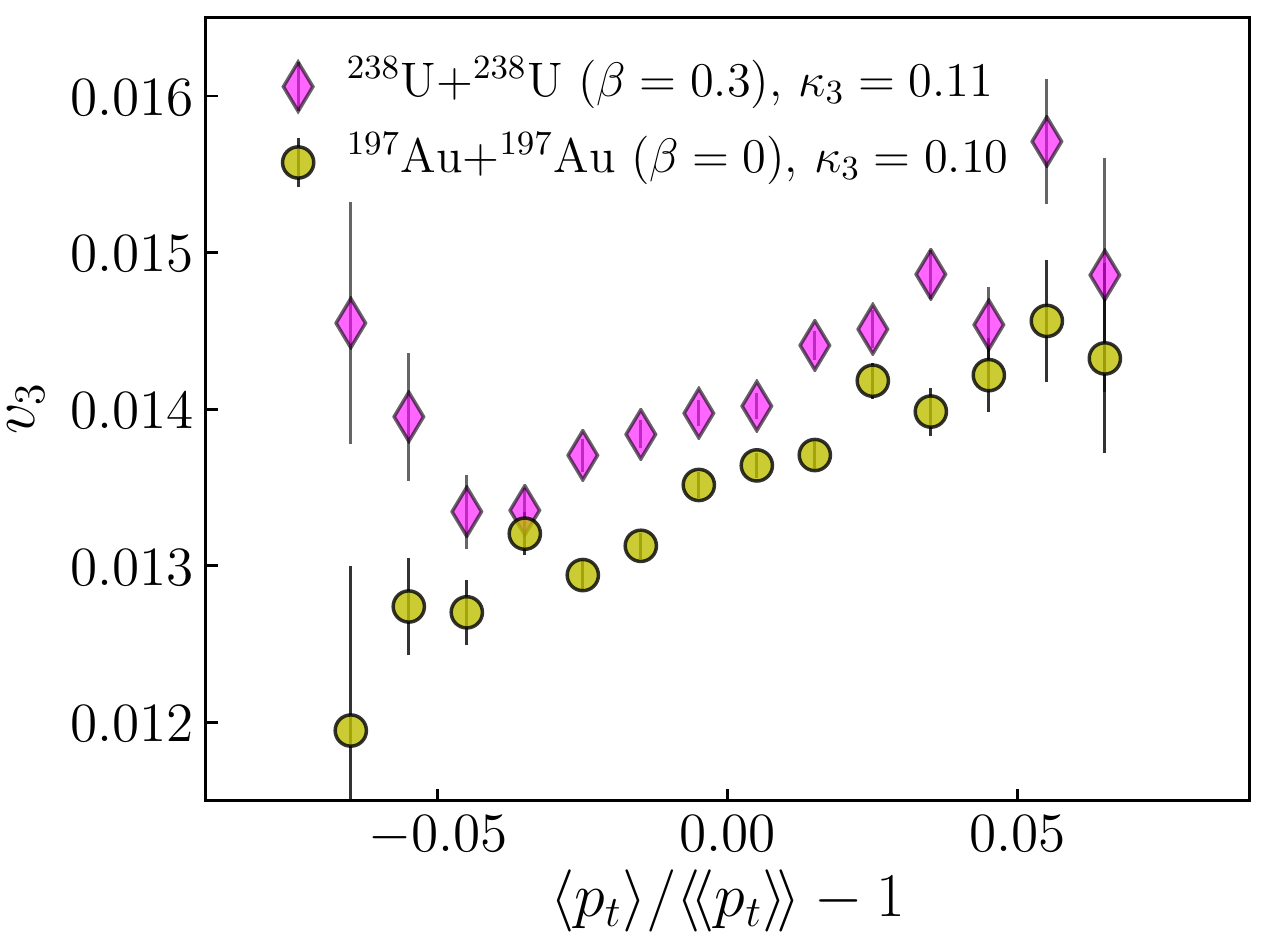}
    \caption{Rms $v_3$ as a function of $\bra p_t \ket$ in $200~{\rm GeV}$ ultracentral \uuuu{} collisions (diamonds) and \auau{} collisions (circles). The response coefficients are specified, and they are consistent with the hydrodynamic simulations reported in Appendix~\ref{app:C}. Figure adapted from Ref.~\cite{Giacalone:2020awm}.}
    \label{fig:5-7}
\end{figure}

Expanding on this latter point, it is particularly insightful to study as well the dependence of $v_3$ on $\bra p_t \ket$ in ultracentral collisions. This is shown in \fig{5-7}. The results from \pbpb{} collisions at top LHC energy from the ATLAS collaboration~\cite{Aad:2019fgl} suggest that the correlation between $v_3$ and $\bra p_t \ket$ should be positive in central collisions. This is also observed in \fig{5-7}, and in preliminary STAR data~\cite{shengli}. The comparison between the results in \fig{5-7} and preliminary STAR data turns out to be very good. This is important. $v_3$ is not affected by the presence of a quadrupole deformation parameter, hence, this result is a nontrivial confirmation of the goodness of the model implementation. The response coefficient for \uuuu{} collisions, $\kappa_3=0.11$, is justified by the hydrodynamic results shown in Appendix~\ref{app:C}, while $\kappa_3=0.10$ in \auau{} collisions is  guessed from the fact that this quantity is slightly damped by viscous corrections.

\subsection{Statistical correlation between $v_2$ and $\bra p_t \ket$}

\label{sec:5-22}

Due to the issue of statistical fluctuations, the measurement of $v_2$ as a function of $\bra p_t \ket$ goes a little against the spirit of the analyses of heavy-ion collisions, described in Sec.~\ref{sec:3-4}, where one typically tries to construct observables that are insensitive to trivial effects due to the finite number of particles. The main point in plotting $v_2$ as a function of $\bra p_t \ket$ is the extraction of the slope of the curve, which is negative in \uuuu{} collisions while positive in \auau{} collisions, and represents essentially a measure of the value of $\beta$. 

The slope can be quantified as the statistical correlation between $v_2$ and $\bra p_t \ket$. This correlation was first analyzed by Teaney and Mazeliauskas in the context of a principal component analysis in Ref.~\cite{Mazeliauskas:2015efa}. Later on, it was reformulated by Bo\.zek~\cite{Bozek:2016yoj} as a simple Pearson correlation coefficient, constructed as a multi-particle observable. The Pearson correlation coefficient between $v_n^2$ and $\bra p_t \ket$ reads:
\begin{equation}
  \label{eq:rhon}
  \rho_n \left ( v_n^2, \bra p_t \ket \right ) = \frac{\left\langle\langle p_t\rangle
    v_n^2\right\rangle
    -\left\langle\langle p_t\rangle\right\rangle\left\langle v_n^2\right\rangle}{\sigma_{p_t}\sigma_{v_n^2} }.
\end{equation}
where as usual outer angular brackets denote an average over events in a given centrality class. The coefficient is normalized by the standard deviations $\sigma_{p_t}$ and $\sigma_{v_n^2}$, given by:
\begin{align}
\label{sigmapt}
\nonumber \sigma_{p_t}&= \sqrt{\left\langle \langle p_t\rangle^2\right\rangle
-\left\langle \langle p_t\rangle\right\rangle^2}, \\
\sigma_{v_n^2}&= \sqrt{\left\langle v_n^4\right\rangle
-\left\langle v_n^2\right\rangle^2}.
\end{align}
In my \trento{} calculations, this quantity is estimated by the correlation of the associated initial-state quantities:
\begin{equation}
  \label{eq:rhoe}
\rho_n \left ( v_n^2, \bra p_t \ket \right ) =    \frac{\left\langle E/S
    \varepsilon_n^2\right\rangle
    -\left\langle E/S\right\rangle\left\langle \varepsilon_n^2\right\rangle}{
    \sigma_{E/S}    \sigma_{\varepsilon_n^2}
}  ,
\end{equation}
with equivalent definitions for the standard deviations in the denominator. 

\begin{figure}
    \centering
    \includegraphics[width=\linewidth]{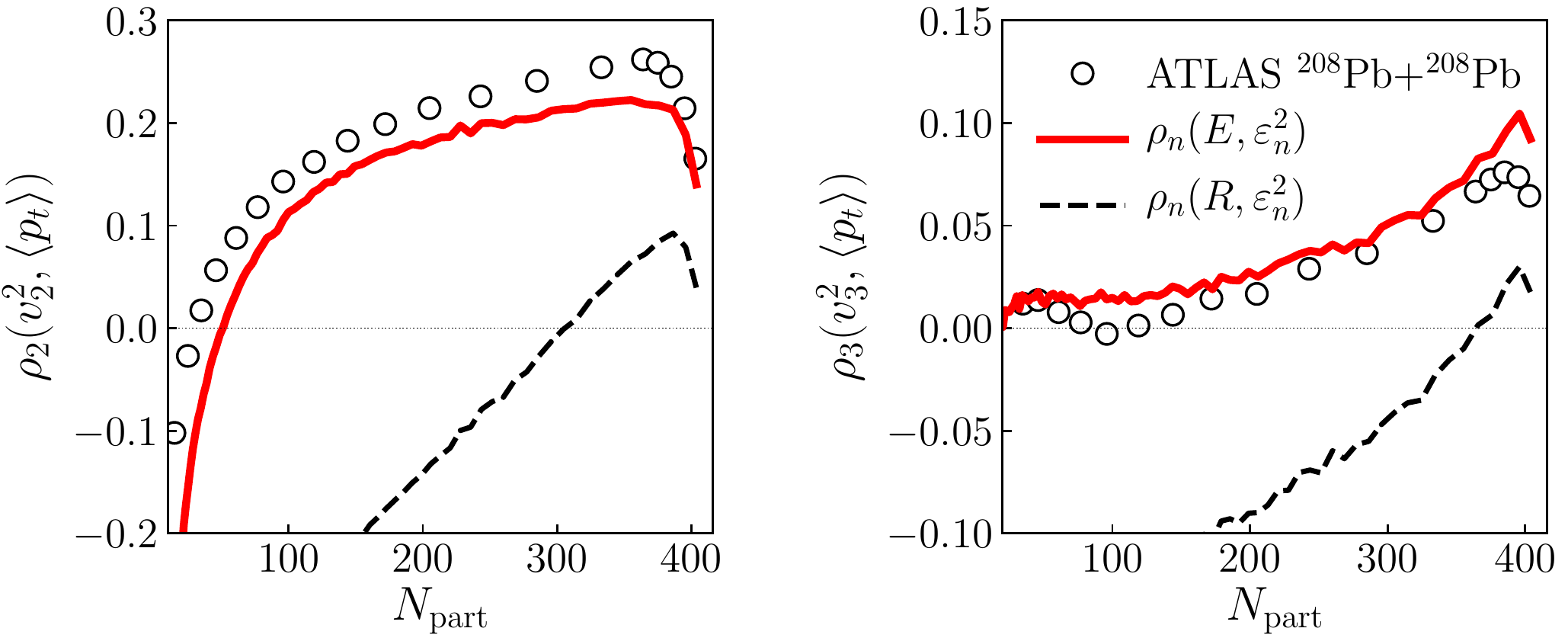}
    \caption{Correlation of $v_n^2$ and $\bra p_t \ket$ for $n=2$ (left panel) and $n=3$ (right panel) in \pbpb{} collisions at top LHC energy. The correlator $\rho_n$ is plotted against the number of participant nucleons, $N_{\rm part}$. Symbols: ATLAS data~\cite{Aad:2019fgl}. Solid lines: initial-state predictor calculated with the \trento{} model and \equ{rhoe}. Dashed lines: initial-state predictor with the initial energy, $E$, replaced by the system size, $R$. Figure adapted from Ref.~\cite{Giacalone:2020dln}.}
    \label{fig:5-8}
\end{figure}

\paragraph{LHC data --} I make here a short digression to show explicitly that the initial-state estimator given in \equ{rhoe} is good. As anticipated in the discussion of \fig{5-6} and \fig{5-7}, the correlation between $v_n$ and $\bra p_t \ket$ has been indeed measured by the ATLAS collaboration in \pbpb{} collisions at top LHC energy~\cite{Aad:2019fgl}. For $n=2$ and $n=3$, the experimental results as a function of the number of participant nucleons are shown as circles in \fig{5-8}. As anticipated, $\rho_n$ is positive for both $n=2$ and $n=3$ in central collisions, and it possesses a nontrivial centrality dependence, which for $n=2$ has been clarified in Ref.~\cite{Schenke:2020uqq}. 

The results of the \trento{} model for the correlator given in \equ{rhoe} come from Ref.~\cite{Giacalone:2020dln}, and are shown here as red solid lines. I remark a nice agreement between the initial-state estimator and data, as both the sign and the nontrivial dependence on centrality of the experimental measurements are captured at the quantitative level. A similar result has also been shown in Ref.~\cite{Schenke:2020uqq}, using \ipglasma{} initial conditions, where a similar initial-state predictor allows to capture accurately the full hydrodynamic evaluation. For completeness, \fig{5-8} reports as well the coefficient $\rho_2$ which is obtained by using the system size, $R$, as initial predictor for $\bra p_t \ket$. One sees that, despite the fact that $\bra p_t \ket$ and $R$ are strongly correlated at fixed multiplicity, this predictor fails in reproducing the experimental observations. This also explains why my original results shown in Ref.~\cite{Giacalone:2019pca}, derived from \equ{Rpred}, can not be used in comparisons with experimental data.

\begin{figure}[t]
    \centering
    \includegraphics[width=.60\linewidth]{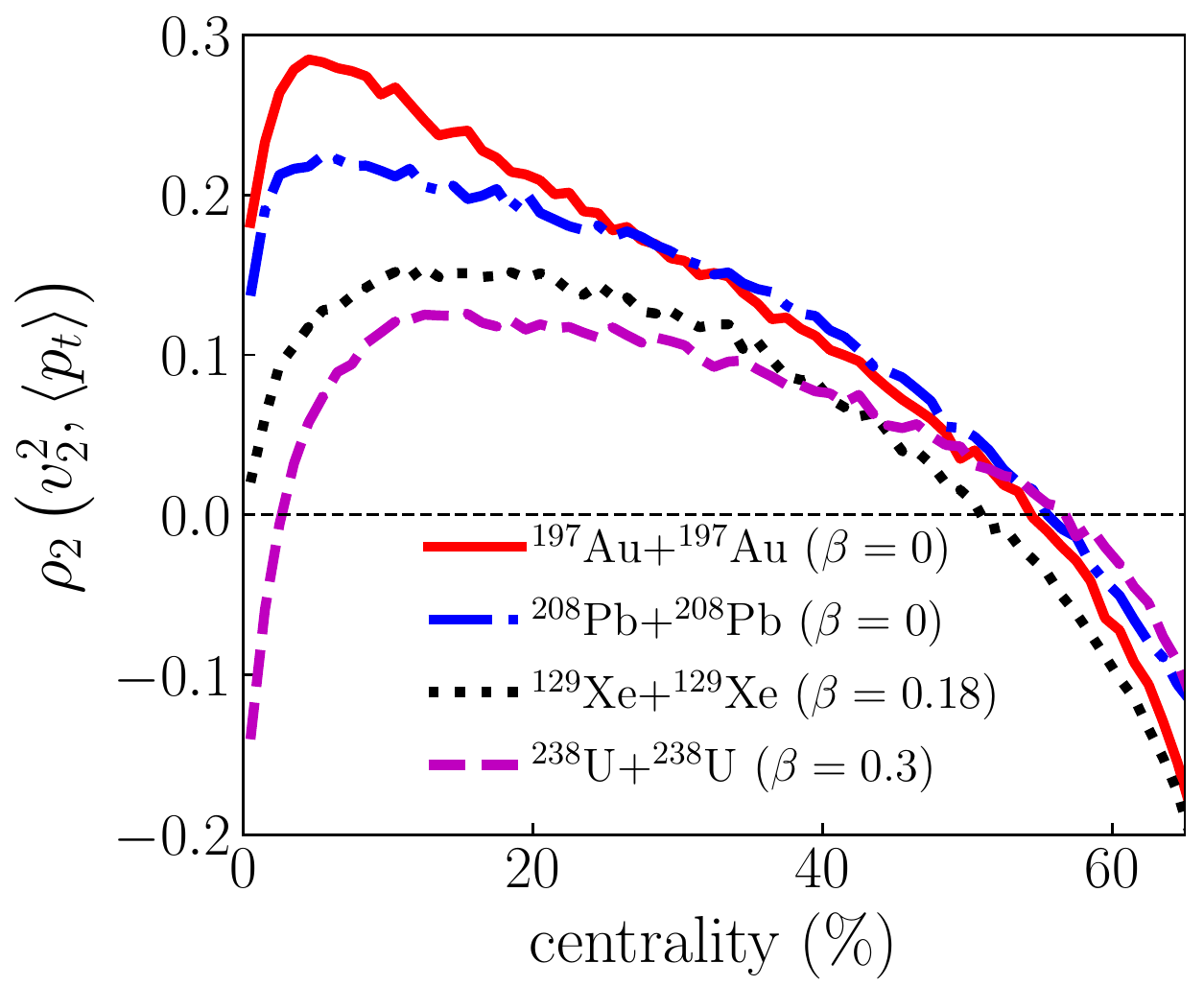}
    \caption{Predictions for $\rho_2\left ( v_2^2,\bra p_t \ket \right)$, calculated from \equ{rhoe}, as a function of collision centrality for RHIC systems, \auau{} collisions (solid line) and \uuuu{} collisions (dashed line), as well as for LHC systems, \pbpb{} collisions (dot-dashed line) and \xexe{} collisions (dotted line). Figure from Ref.~\cite{Giacalone:2020awm}.}
    \label{fig:5-9}
\end{figure}

\paragraph{New predictions --} The results on the correlator $\rho_n$ in \pbpb{} collisions imply that one can use the \trento{} model to perform quantitative predictions for other systems, such as \xexe{} collisions, or RHIC systems. Results for $\rho_2$ as a function of collisions centrality are shown for all these systems in \fig{5-9}. I discuss first \pbpb{} and \auau{} collisions. The two curves overlap to a large extent, as  $\beta=0$ in both these systems. The sole notable difference is that the result for \auau{} collisions reaches higher values in central collision. This is due to the lower value of the fluctuation parameter, $k$, implemented in the \trento{} simulations, and is thus ascribed to the different beam energy, at least in my setup. Note that the curve for \auau{} collisions shown in \fig{5-9} may however provide a wrong description of future data at larger centralities. In Ref.~\cite{Giacalone:2020byk}, we evaluate $\rho_2$ in full hydrodynamic simulations of \auau{} collisions by means of \ipglasma{} calculations. The resulting $\rho_2$ is quite different from that shown here in \fig{5-9}. It is in particular much flatter with centrality, and it does not exhibit a change of sign in peripheral collisions. This comes from the fact that, in \ipglasma{}, the anisotropy $v_2$ of peripheral \auau{} collisions receives an important contribution from a primordial source of anisotropy other than $\varepsilon_2$. This corresponds to the anisotropy induced by the off-diagonal components of the energy-momentum tensor of the system, which are generated by the classical Yang-Mills evolution within the glasma phase. The inclusion of these components increases $\rho_2$ in peripheral collisions, and makes it nearly flat with centrality. The change of sign observed \fig{5-9} is a generic feature of nucleus-nucleus collisions~\cite{Schenke:2020uqq}, so that its absence in future data will provide strong experimental evidence of primordial off-diagonal anisotropic terms predicted by the color glass condensate theory. This issue is under investigation by the STAR collaboration.

Moving on to collisions of deformed nuclei in \fig{5-9}, one can see very clearly the effect of the quadrupole deformation of $^{238}$U nuclei on $\rho_2$. The correlator is negative in central collisions (recall the negative slope of $v_2$ vs. $\bra p_t \ket$ in \fig{5-6}), and then it grows up quickly to positive values, behaving similarly to the other systems. It is remarkable that the splitting between \auau{} collisions and \uuuu{} collisions persists at centralities as large as 30\%. For the fourth-order cumulant of elliptic flow, $v_2\{4\}$, discussed in Chapter~\ref{chap:4}, the splitting between these systems essentially disappears above $5\%$ centrality, when $v_2\{4\}$ becomes dominated by elliptic flow along the direction of impact parameter, $\mu$. This is not however the case for the correlator $\rho_2$, which as a consequence shows a sensitivity to the value of $\beta$ which persists at larger centralities. I will come back to this point in the next section. In \fig{5-9} I show as well results for \xexe{} collisions implementing $\beta=0.18$. The correlator does not get negative, as in \uuuu{} collisions, due to the fact that the quadrupole deformation is not able to compensate for the large quantum fluctuations that affect this system. However, the correlator is close to zero, and thus distinctly lower than in \pbpb{} collisions. This phenomenon is currently under investigation by LHC collaborations.

\section{Constraining the value of $\beta$}

\label{sec:5-23}

The results in \fig{5-9} suggest that $\rho_2$ coefficient is sensitive to the effect of the quadrupole deformation across most of the centrality range. I perform now a study of the sensitivity of this observable to variations in the value of $\beta$. 

I focus on \uuuu{} collisions. To start with, I simply reduce the quadrupole parameter by a factor two, so that I implement $\beta=0.15$. Note that, strictly speaking, this is not allowed. The quadrupole deformation of $^{238}$U is mostly constrained by data on the transition probability for the electric quadrupole operator, following \equ{betaexp}. In particular, the quadrupole moment of the nucleus should satisfy \equ{be2q2}. When varying the value of $\beta$, one should at the same time vary the other Woods-Saxon parameters to ensure that \equ{be2q2} remains true, as done in the careful analysis of Ref.~\cite{Shou:2014eya}. I shall gloss over this feature in the present section, where I am simply concerned with giving an idea of how sensitive observables are to the amount of deformation in the colliding bodies.
\begin{figure}[t]
    \centering
    \includegraphics[width=\linewidth]{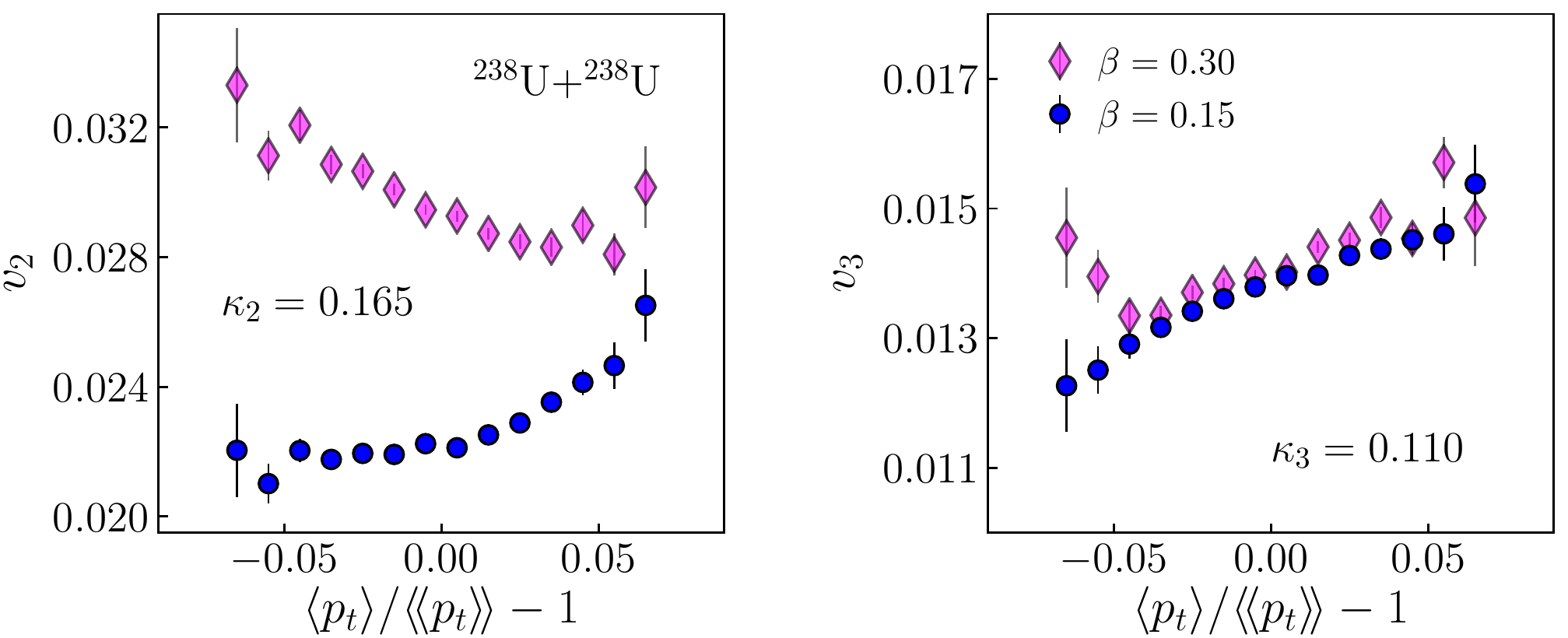}
    \caption{Rms elliptic flow (left panel) and rms triangular flow (right panel) as a function of the relative variation of $\bra p_t \ket$ in ultracentral \uuuu{} collisions, for different choices of the quadrupole deformation parameter. Diamonds: $\beta=0.30$. Circles: $\beta=0.15$. Figure from Ref.~\cite{Giacalone:2020awm}.}
    \label{fig:5-10}
\end{figure}

I show results for the dependence of the rms $v_n$ on $\bra p_t \ket$ in \fig{5-10}. The results for elliptic flow, shown in the left panel, are quite striking. The correlation between $v_2$ and $\bra p_t \ket$ is indeed positive when $\beta=0.15$, in stark disagreement with preliminary STAR data\cite{shengli}. This shows that, as soon as we look at obsrvables that depend so strongly on $\beta$, then heavy-ion collision data can essentially be used to place independent constraints on the quadrupole parameters. The results for $v_3$ as a function of $\bra p_t \ket$ are instead shown in the right panel of the figure. As expected from the previous discussions, the role of $\beta$ in this observable is negligible.

\begin{figure}[t]
    \centering
    \includegraphics[width=.5\linewidth]{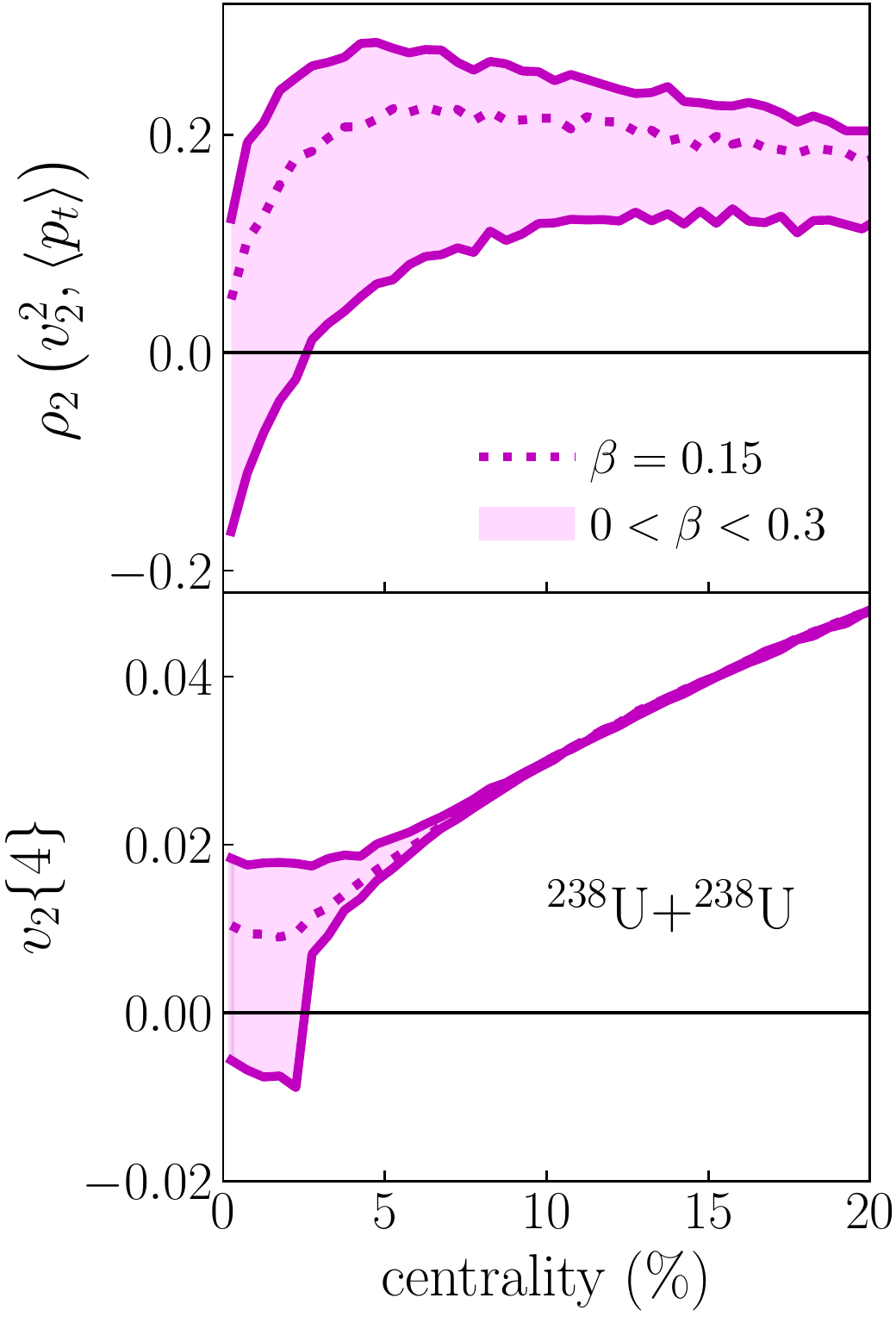}
    \caption{Top: Bo\.zek coefficient $\rho_2$ as a function of collision centrality in \uuuu{} collisions. Bottom: $v_2\{4\}$ as a function of centrality. The dotted curve is obtained by implementing nuclei with $\beta=0.15$. The shaded band corresponds to a variation of $\beta$ in the interval $0<\beta<0.3$. Note that $\beta=0.3$ corresponds to the lower end of the shaded band in the upper panel, and to the upper end of the shaded band in the lower panel.}
    \label{fig:5-11}
\end{figure}

As a final result, which in a sense does summarize all the results related to the deformation of nuclei discussed in this work, I show in \fig{5-11} the centrality dependence of the the Bo\.zek coefficient, $\rho_2$, and of the fourth-order cumulant of elliptic flow, $v_2\{4\}$ from my \trento{} evaluations. The calculation is performed for $\beta=0,0.15,0.30$, and the results are shown as shaded bands to highlight the dependence on $\beta$. The upper panel of \fig{5-11} shows $\rho_2$, while the lower panel shows $v_2\{4\}$. The point I want to make is these observables have a very different sensitivity to the value of $\beta$. Above 5\% centrality, the fourth-order cumulant is dominated by the almond shape of the system induced by the impact parameter of the collision, so that the phenomenological manifestation of $\beta$ is visible only in the most central collisions. The impact of $\beta$ on $\rho_2$ is instead clearly visible up to much larger centrality. At 10\% centrality, for instance, there is still a factor $2$ of difference between the result with $\beta=0$ and that with $\beta=0.3$. This is due to the fact that this observables makes explicit use of the information on the orientation of the colliding objects.

In the business of high-energy nuclear physics, the coefficient $\rho_2$ represents thus the first observable ever found that, for collisions of large and well-deformed nuclei, is almost entirely dominated by the magnitude of $\beta$. If one disregarded effects of nuclear deformation and tried to evaluate $v_2\{4\}$ in hydrodynamics, then, with the exception of the 0-5\% centrality bin, one would get this quantity right. However, if the same were done for the quantity $\rho_2$, hydrodynamics would give the wrong answer across essentially the full range of centrality. This observable makes nuclear deformation effects so sizable to become affordable even for full hydrodynamic simulations, and not only for \trento{}-like evaluations focused on the most central collisions.

\paragraph{Conclusion --} Quantitative studies of the deformation of atomic nuclei can be performed by means of high-energy nuclear experiments. These studies are within the reach of existing high-energy colliders.


\chapter{Conclusion and proposal}

\label{chap:6}

High-energy nuclear physics has undergone great progress over the past 20 years, and is nowadays a mature field of research, possibly, the most active subfield of nuclear science. The flow paradigm, based on an effective hydrodynamic description, has allowed for an accurate understanding of a large wealth of high-quality particle collider data related to the soft sector of relativistic nuclear collision. 

The quark-gluon plasma evolves according to hydrodynamic laws which are governed by a strict causality. If the causes are known, then the effects are also known. This explains why the hydrodynamic framework of heavy-ion collisions is robust and fully predictive, and also why one is able to define initial-state predictors, as I have argued throughout this manuscript, that give a transparent physical understanding of the experimental observations. 

In this work, I have established that among the \textit{causes} that lead to the emergence of the phenomenon of anisotropic flow one can nowadays include the deformation of atomic nuclei. This has been made possible thanks to the great quality of RHIC and LHC data. I summarize the main results:
\begin{itemize}
    \item RHIC data does not show any clear evidence of quadrupole deformation in $^{197}$Au nuclei. This result, at variance with mean-field and empirical estimates, demonstrates the importance of a state-of-the-art modeling of the colliding nuclei for the understanding of high-energy data.
    \item LHC data provides instead evidence of a significant quadrupole deformation in $^{129}$Xe nuclei, suggesting for the first time that high-energy data is largely impacted by shape-coexistence effects in the nuclear ground states.
    \item The observation of a negative correlation between $v_2$ and $\bra p_t \ket$ reported by the STAR collaboration in \uuuu{} collisions shows that experimentally it is possible to \textit{freeze} the orientation of the colliding nuclei, in particular, to isolate body-body collisions. The observable $\rho_2 \left ( v_2^2 , \bra p_t \ket \right )$ displays an unparalleled sensitivity to the prolate deformation of the colliding species.
\end{itemize}
 These nontrivial conclusions have been achieved with just 4 collision systems. It is not possible to foresee the number of new observations and new discoveries that could be made if a larger number of systems were available. I discuss now this possibility in more detail.
 
\section{System-size scan at RHIC}

\label{sec:6-1}

My proposal is thus to collide more nuclear species to observe the emergence of phenomena related to their structure. Such a program should be defined depending on whether certain species are more interesting than others. However, in all cases I think it should be guided by two principles:

\paragraph{1. Collide large enough nuclei --} When I say that nuclei should be large enough, I mean two things. $i)$ If one aims at understanding quantitatively the phenomenological manifestations of nuclear deformation, then the geometric properties of nuclei should somewhat dominate over the quantum fluctuations that affect the geometry of the quark gluon plasma. The analysis of $v_2\{4\}$ in \auau{} collisions makes it clear that one can be sensitive to values of the deformation as low as $\beta=0.13$. However, this is possible because gold nuclei are large, which suppresses the fluctuations associated e.g. with the number of nucleons.  $ii)$ The effects one is after may be small effects, hence, it is important that theory-to-data comparisons are meaningful and the details under control. This implies that these systems should produce enough particles to make non-flow phenomena as irrelevant as possible, which typically requires $\mathcal{O}(10^3)$ particles for $|\eta|<1$. 
\begin{figure}[t]
    \centering
    \includegraphics[width=\linewidth]{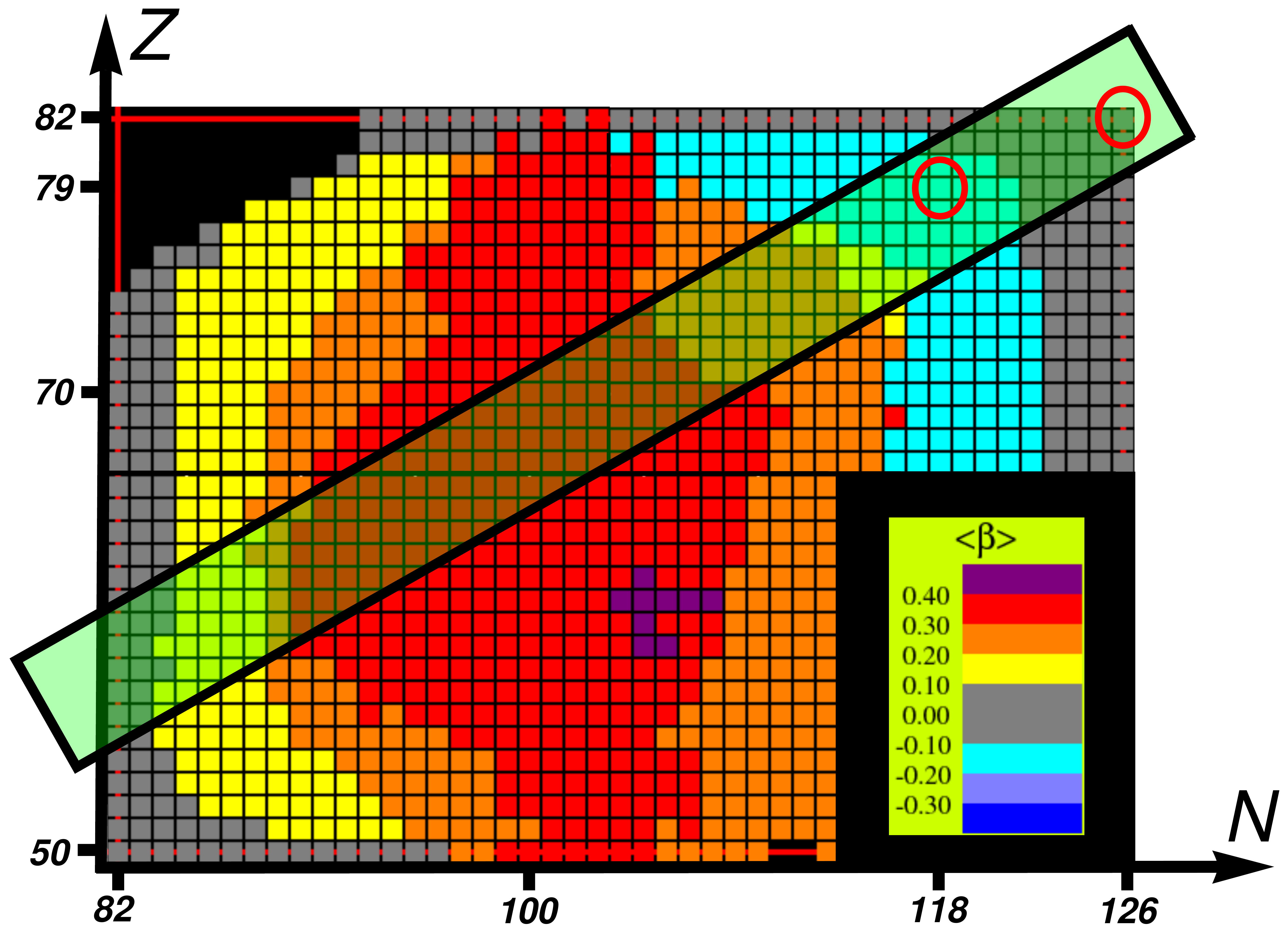}
    \caption{The nuclear chart for $N=82-126$ and $Z=50-82$. Each nucleus is labeled by a color depending on its value of $\beta$. The values of $\beta$ result from the Hartree-Fock-Bogoliubov calculations of Ref.~\cite{Hilaire:2007tmk}. This plot is extracted from the nuclear chart shown in Ref.~\cite{website}. The shaded stripe crossing the graph contains roughly all the stable nuclides that belong to this interval of $Z$ and $N$. The nuclei highlighted with circles are, respectively, $^{197}$Au and $^{208}$Pb.}
    \label{fig:6-1}
\end{figure}

A window of viable nuclei can thus be found in the region of the nuclear chart shown in \fig{6-1}, which is extracted from Ref.~\cite{website}. This region corresponds to $50<Z<82$, and $82<N<126$. It includes in particular the so-called \textit{rare-earth} region, filled with a large number of stable nuclides that are well-deformed (highlighted in red in the figure). The green shaded stripe that I place on top of the figure guides the eye, as it indicates where the stable nuclides lie. 

What seems particularly interesting is the possibility of mapping \textit{transitional} phenomena, i.e., the emergence and disappearance of nuclear deformation, as one moves along the nuclear chart. One could collide e.g. a chain of samarium isotopes, like $^{144,150,152,154}{\rm Sm}$, corresponding, to $\beta \approx 0.05,0.2,0.3,0.35$, some of which have also been recently analyzed in Ref.~\cite{Mustonen:2018ody}, to observe the transition from spherical to deformed nuclei. One could then crosscheck the predictions of nuclear models for the well-deformed shapes of rare-earth nuclei, and then look at collision systems that brings one from prolate to spherical nuclei, in the neighborhood of $^{197}$Au. Nuclear theory results, including those shown in \fig{6-1}, suggest in particular that nuclei are first prolate, then oblate, and then tend become spherical. It is clear from my analysis of $^{197}$Au that this transition from prolate to spherical nuclei requires a fine modeling from a nuclear structure point of view. Relativistic collisions thus offer a chance to test such theoretical models against data.  

\paragraph{2. Make a systematic study possible --} With systematic study, I mean two things. First, that a significant amount of new species should be collided, like 10 or 20, the more the better. Second, these species should represent a reasonably wide range of atomic masses, to look for systematic trends. Besides serving as probes of the shape of nuclei, these experiments would have a great impact on high-energy studies: 
\begin{itemize}
    \item Viscous corrections to the hydrodynamic evolution scale like $1/R$. By scanning over systems with several values of $R$, one can constrain the viscosity of the quark-gluon plasma.  By means of the Bayesian analysis framework developed by the Duke group, for instance, improved constraints on the viscous properties of the quark-gluon plasma, $\eta/s$ and $\zeta/s$, would be obtained by the simultaneous analysis of a large number of different systems. 
    \item The Bayesian analysis could be further modified to include parameters related to the shape of the colliding species, in order to extract their deformation properties with well-defined error bars.
    \item  A system-size scan would also constrain all effects related to the magnetic field created in heavy-ion collisions, which depend essentially on $A$ at fixed collision energy, as I shall further discuss in the next section.
    \item The physics of the hard sector, that will be investigated at RHIC thanks to the upcoming sPHENIX detector, will also be impacted by a systematic system-size scan, as phenomena of energy loss have naturally a dependence on the size of the medium.
\end{itemize}

There are thus many possibilities offered by a systematic system-size scan in high-energy experiments, which are still largely to explore. It would be a pity to miss this opportunity at RHIC over the next decade. One needs about $10^8$ collisions to perform precision measurements, which should amount to a few days of RHIC operation per species.

\section{Further developments}

\label{sec:6-2}

I conclude by pointing out a couple of other potential applications of nuclear deformation to heavy-ion collisions which may become of great relevance in the analysis and interpretation of current and future experimental data.

\subsection{Triaxiality}

\begin{figure}[t]
    \centering
    \includegraphics[width=.3\linewidth]{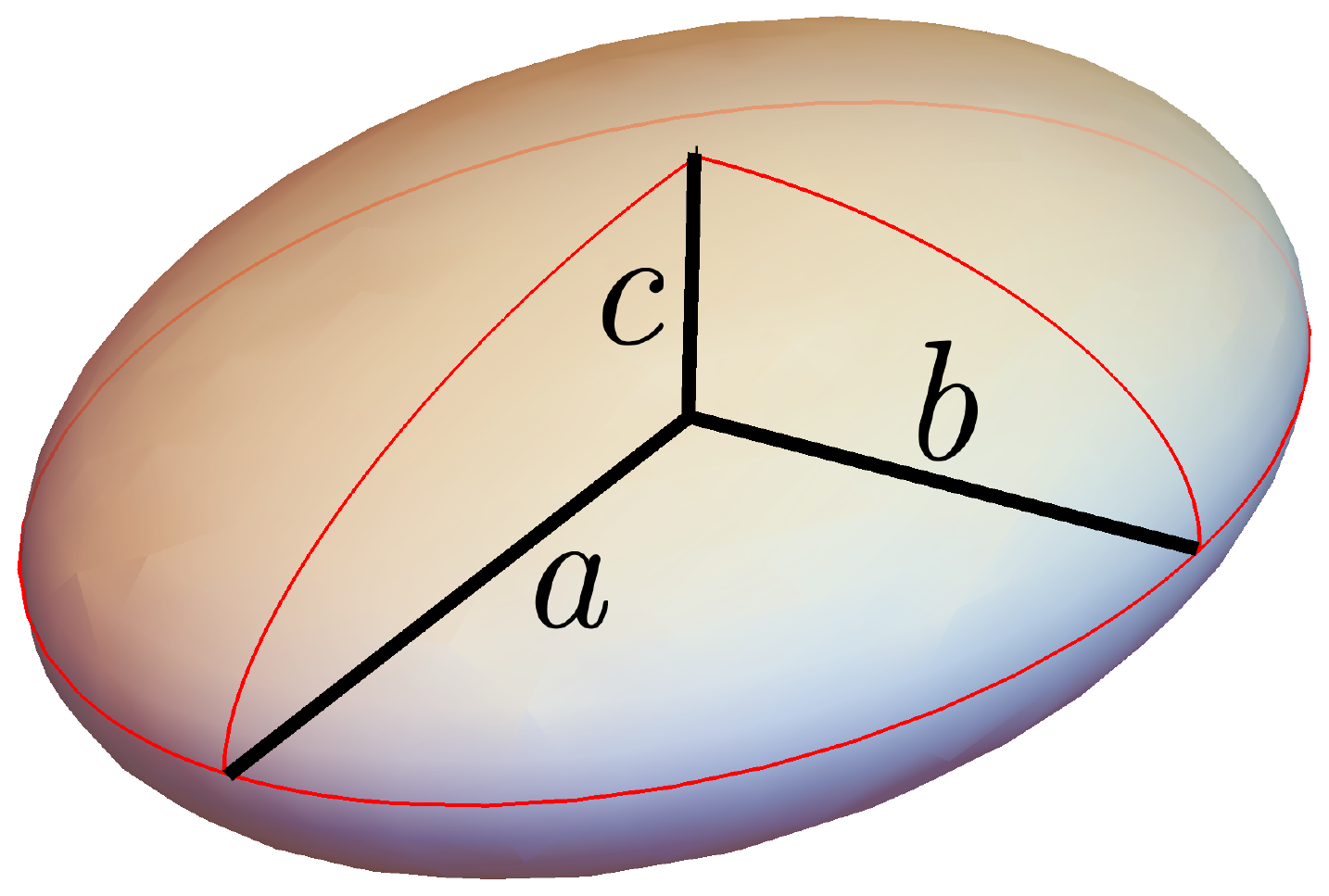}
    \caption{Triaxial spheroid: $a \neq b\neq c$.}
    \label{fig:6-2}
\end{figure}

 So far I have considered only axially-symmetric nuclei that are elongated ($\beta>0$) or squeezed ($\beta<0$) along the direction of the nuclear axis, $z'$, but where the radius of the nucleus is the same along $x'$ and along $y'$, where $(x'y'z')$ is the intrinsic frame of the nucleus. However, in nuclear structure theory the shape of a nucleus is usually not characterized by the sole quadrupole deformation parameter, $\beta$. Nuclei can in fact break axial symmetry, and present an imbalance in their axes, $x'\neq y'$, i.e., they can be triaxial. In theoretical descriptions based on the mean field approach, this feature can be implemented in a rather straightforward manner by simply letting the mean-field wavefunction break axial symmetry in the minimization procedure of \equ{ritzQ2}. The relevant spherical harmonics related to the triaxial shape are $Y_2^{\pm 2}$. The resulting geometry is that of a triaxial spheroid, illustrated in \fig{6-2}. 

In the modeling of heavy-ion collisions, one can include a triaxial deformation via a simple modification of the deformed Woods-Saxon parametrization presented in \equ{defws}. Triaxiality implies that the spherical harmonic $Y_{2,0}$ is no longer sufficient to describe the system, and that one has to keep the $l=2, m=\pm 2$ modes in the expansion of \equ{Rexp}. By doing so the expressions of the coefficients of the spherical harmonic expansion become the following~\cite{ring}:
\begin{align}
    c_{2,0} = \beta\cos\gamma, \\
    c_{2,\pm 2} = \frac{1}{\sqrt{2}}\beta\sin\gamma,
\end{align}
where $\gamma$ is the so-called triaxial deformation parameter, which by symmetry varies between $0$ and $\pi/3$.  The nuclear density used in heavy-ion collisions becomes:
\begin{equation}
    \rho(r,\Theta,\Phi) =   \frac{\rho_0}{1+\exp \left ( \frac{r-R(\Theta,\Phi)}{a}  \right )},
\end{equation}
where this time $R(\Theta,\Phi)$ reads:
\begin{equation}
    R(\Theta,\Phi) = R_0 \left [ 1+\beta \cos \gamma Y_{2,0}(\Theta,\Phi)  + \frac{1}{\sqrt{2}}\beta\sin\gamma \biggl( Y_{2,2}(\Theta,\Phi) + Y_{2,-2}(\Theta,\Phi) \biggr) \right ].
\end{equation}
In the intrinsic frame of the nucleus, the axes lengths of the triaxial spheroid are:
\begin{align}
    \nonumber  R_{x'} &= R_0 \biggl [ 1+ \sqrt{\frac{5}{4\pi}} \beta \cos\biggl(\gamma-\frac{2\pi}{3}\biggr) \biggr ] , \\
    \nonumber  R_{y'} &= R_0 \biggl [ 1+ \sqrt{\frac{5}{4\pi}} \beta \cos\biggl(\gamma+\frac{2\pi}{3}\biggr) \biggr ] , \\
      R_{z'} &= R_0 \biggl [ 1+ \sqrt{\frac{5}{4\pi}} \beta \cos(\gamma) \biggr ] ,
\end{align}
Hence $\gamma$ quantifies the length difference between the $x'$ axis and the $y'$ axis of the deformed ellipsoid.

As pointed out, e.g., in Ref.~\cite{Poves:2019byh}, $\gamma$ is a quantity which is typically characterized by a large degree of softness, meaning that it does not assume a single value, but rather a distribution of values in the ground state. Fluctuations in the triaxiality are typically unimportant for large nuclei, and thus are unlikely to play any role in the phenomenology of e.g. \auau{}, \pbpb{}, or \uuuu{} collisions. However, they may play a role for \xexe{} collisions, as well as in upcoming RHIC data on \zrzr{} and \ruru{} collisions.

\subsection{Phenomenology of the magnetic field}

While playing with the selection of events based on $\bra p_t \ket$ at fixed multiplicity, a funny fact occurred to me.

\begin{figure}[t]
    \centering
    \includegraphics[width=.9\linewidth]{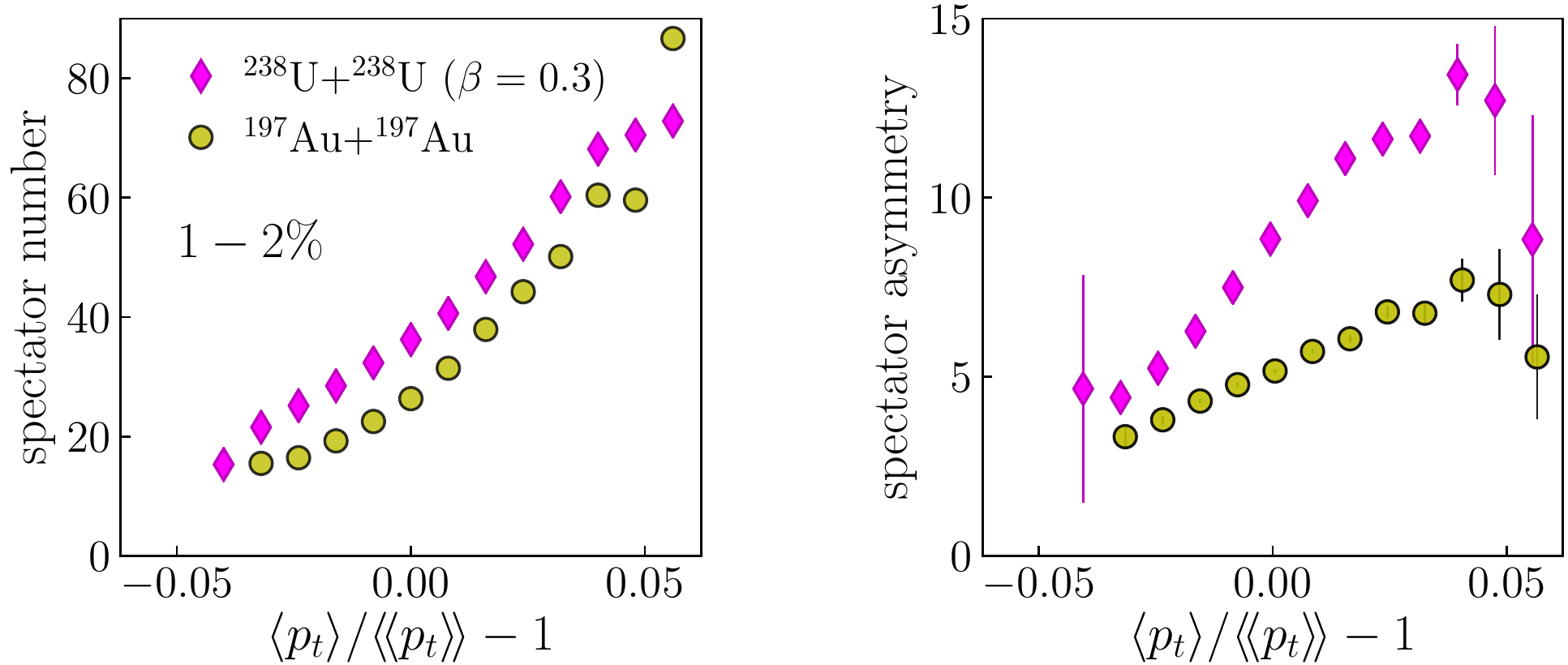}
    \caption{Left: number of spectator nucleons as a function of the relative variation of $\bra p_t \ket$, in central \auau{} collisions (circles) and \uuuu{} collisions (diamonds) at top RHIC energy. Right: spectator nucleon asymmetry, as defined by \equ{asymm}.}
    \label{fig:6-3}
\end{figure}

Among all final-state observables one can think of, $\bra p_t \ket$ turns out to be the one that possesses perhaps the strongest correlation with the value of $N_{\rm part}$ at a given collision centrality. Let me show an explicit example. In \fig{6-3}, I show the average value of the number of \textit{spectator} nucleons in central \auau{} collisions and \uuuu{} collisions. This calculation is done with the same \trento{} setup used in the previous sections. The number of spectator nucleons is simply equal to
\begin{equation}
\label{eq:nspec}
N_{s,A(B)} = A(B) - N_{{\rm part},A(B)},
\end{equation}
where as usual $A$ and $B$ label the colliding nuclei, $N_{{\rm part},A(B)}$ is the number of participant nucleons from nucleus $A(B)$, while $A(B)$ in the right-hand side is the mass number. The left panel of the figure shows indeed a strong positive correlation between this quantity and $\bra p_t \ket$. In both systems, the spectator number does increase by at least a factor $3$ as one moves toward large values of $\bra p_t \ket$. 

Why is this interesting? The point is that there exists a nontrivial phenomenology of spectator and participant nucleons related to the magnetic fields that are produced in high-energy nuclear collisions~\cite{Kharzeev:2015znc,Fukushima:2018grm,Li:2020dwr}. Magnetic fields emerge naturally in nuclear collisions because of the electric current carried by the moving protons. These magnetic fields are the strongest ever created in a laboratory, however, as protons fly away from the interaction region at the speed of light, they exist only for a very short time, and their phenomenological manifestations are somewhat scarce and elusive.  The phenomenology of magnetic fields depend entirely on the number of participant and spectator nucleons that are involved in the collision. The strong correlation observed in the left panel of \fig{6-3} suggests, thus, that $\bra p_t \ket$ could be used as a sort of handle to turn these magnetic fields up and down. I published this simple idea in Ref.~\cite{Giacalone:2020oao}, where I propose to look in particular at the correlation between $\bra p_t \ket$ and the signal of the chiral magnetic effect~\cite{Fukushima:2008xe}. Quantitative calculations are ongoing~\cite{chun}.

Going back to the main topic of this manuscript, one sees in the left panel of \fig{6-3} that there is little difference between \auau{} collisions and \uuuu{} collisions. The larger number of spectators observed in \uuuu{} events is simply a consequence of the larger number of nucleons in $^{238}$U nuclei. The number of spectators does not seem to know anything about the fact that the nuclei are deformed and that one is selecting tip-tip configurations at large $\bra p_t \ket$, or body-body configurations at low $\bra p_t \ket$.

However, a quantity that is instead aware of the deformation of $^{238}$U is the spectator \textit{asymmetry}, which corresponds to the imbalance of forward- and backward-going spectator nucleons. In the context of the phenomenology of magnetic fields in heavy-ion collisions, the study of this quantity was originally proposed by Chatterjee and Tribedy~\cite{Chatterjee:2014sea}. The spectator asymmetry is defined by:
\begin{equation}
\label{eq:asymm}
    {\rm spectator~asymmetry} = | N_{s,A} - N_{s,B} |,
\end{equation}
where I do not make any distinction between spectator protons or spectator neutrons.  The right panel of \fig{6-3} shows the average spectator asymmetry in these collision systems. The asymmetry grows as a function of $\bra p_t \ket$ in both systems. However, this growth is visibly steeper in \uuuu{} collisions, where the asymmetry is also much higher in magnitude. One can easily understand this behavior from the fact that, for $\bra p_t \ket > \bbra p_t \kket$, the nuclei approach the tip-tip configuration, but keeping a sizable relative polar angle, as discussed around \fig{5-4}. The configuration proposed for events at large $\bra p_t \ket$, i.e, $\theta_A=\pi/12$ and $\theta_B=\pi/4$, is essentially a \textit{body-tip} collisions, with the nucleus in the body position tilted by 45$^\circ$. This gives rise to a collision geometry that exhibits an intrinsic spectator asymmetry, hence the steeper growth of this quantity for \uuuu{} systems.  

The fact that the magnetic field produced by the spectator nucleons is stronger along one direction may have nontrivial influence on its phenomenological manifestations, which was in fact the original point made in Ref.~\cite{Chatterjee:2014sea}. It would be therefore interesting to assess in a theoretical calculation whether observables sensitive to the magnetic field are more or less correlated with $\bra p_t \ket$ when the colliding species are well-deformed, especially in central collisions.

\begin{appendices}

\chapter{Cumulant expansion for anisotropy}

\label{app:A}

I derive the expressions of the spatial anisotropies $\mathcal{E}_2$, $\mathcal{E}_1$, and $\mathcal{E}_3$, shown respectively in \equ{E2} and \equ{E1E3}. These quantities were originally derived in a famous paper by Teaney and Yan~\cite{Teaney:2010vd}. The derivation proposed here differs from theirs, and is in fact simpler. It has been shown to me by Prof. Matt Luzum. Parts of the derivation are scattered across the literature (e.g. \cite{Sousa:2020cwo}), and it is reproduced in its entirety here for the first time.

Consider the density of energy deposited in the transverse plane following a heavy-ion collision, $T^{00}({\bf x}) \equiv \epsilon({\bf x})$. The idea of Teaney and Yan is that, since the hydrodynamic expansion of the system  (much as the collective dynamics of nucleons that lead to the deformation of nuclei) is mostly driven by long-wavelength properties, what really matters for the phenomenology of anisotropic flow, like elliptic flow, is the anisotropy of the large-scale structures of the fluid. The natural way to discern long- and short-wavelength structures is to take a Fourier transform:
\begin{equation}
\label{eq:B0}
    \rho({\bf k}) = \int_{\bf x} \rho({\bf x}) e^{i{\bf k}\cdot{\bf x}}.
\end{equation}
Small values of $k\equiv|{\bf k}|$ correspond to large-scale structure, while large $k$ represents small scales. Assuming then that the Fourier transform is sufficiently well-behaved to be expanded in a Maclaurin series around $k=0$, one can write:
\begin{equation}
\label{eq:B1}
    \rho({\bf k}) = \sum_{m=0}^{\infty} \rho_m (\phi_k) k^m,
\end{equation}
where the coefficient $\rho_m$ does not depend on $k$, and I have introduced polar coordinates where $\phi_k$ represents the azimuthal angle in momentum space. Now, if $\rho({\bf k})$ is the Fourier transform of the initial density, one can also write:
\begin{equation}
\label{eq:B2}
    \rho({\bf k}) = \int_{\bf x} \rho({\bf x}) \sum_{m=0}^\infty \frac{1}{m!} (i{\bf k}\cdot{\bf x})^m = \sum_{m=0}^\infty \frac{i^m}{m!} \int_{\bf x} \rho({\bf x}) r^m \cos^m(\phi_k-\phi),
\end{equation}
where $r\equiv|{\bf r}|$, and I have again made use of polar coordinates. Matching powers of $m$ in \equ{B1} and \equ{B2}, one obtains:
\begin{equation}
\label{eq:B3}
    \rho_m(\phi_k) = \frac{i^m}{m!} \int_{\bf x} r^m \rho({\bf x}) \cos^m (\phi_k - \phi).
\end{equation}
Now, since one is eventually interested in the anisotropy of the density profile, one has to further expand the density in Fourier series. The idea of Teaney and Yan is that of performing the Fourier expansion mode-by-mode, for each value of $m$. From the Taylor expansion in \equ{B1}, this amounts to expanding the coefficients of the power series with respect to the angle $\phi_k$. The expression of the $\rho({\bf k})$ thus becomes:
\begin{equation}
\label{eq:B4}
    \rho({\bf k}) = \sum_{m=0}^\infty \sum_{n=-\infty}^{+\infty} \rho_{n,m} k^m e^{-in\phi_k}.
\end{equation}
The label $n$ is the order of the Fourier harmonic, and represents the rotational property of the mode labeled by $m$, which represents instead the wavelength. 

For a given $m$, the $n$-th order Fourier coefficient finally reads:
\begin{equation}
\label{eq:B5}
    \rho_{n,m} =  \frac{1}{2\pi}  \int d\phi_k \rho_m(\phi_k)  e^{in\phi_k},
\end{equation}
which according to \equ{B3} can be written as:
\begin{equation}
\label{eq:B6}
    \rho_{n,m} = \frac{i^m}{m!} \int_{\bf x} \int d\phi_k r^m \rho({\bf x}) \cos(\phi_k-\phi) e^{-in\phi_k}.
\end{equation}
The integral over $\phi_k$ vanish unless $m<|n|$ and $m -|n|$ is an even number. If these conditions are met, the final expression reads:
\begin{equation}
\label{eq:B7}
    \rho_{n,m} = \frac{i^m}{m!} \frac{1}{2^m} \frac{m!}{ \bigl(\frac{m+n}{2} \bigr)! \bigl(\frac{m-n}{2} \bigr)! } \int_{\bf x} r^m \rho({\bf x}) e^{i n \phi},
 \end{equation}
where I recall that $\phi$ in the last term is the azimuthal angle in spatial coordinates. The coefficient $\rho_{n,m}$ is in principle the quantity one is after, i.e., the $n$-th order Fourier harmonic of the modes $m$, which for the lowest values of $m$ correspond to the large-scale structures of the system.

These coefficients can not however be the end of the story. The problem is that they depend on the center of coordinates, while the Fourier harmonics, $V_n$, to which one wants to match them, are translation-invariant. This is an important property that one should require. Small $k$ does not in fact represent large-scale structures, but rather structures that are far from the chosen center of coordinates. Teaney and Yan realized that this issue can be overcome if, instead of $\rho({\bf k})$, the expansion in Fourier modes is performed on $\ln \rho({\bf k})$, which corresponds, with a slight abuse of language, to the cumulant generating function of the distribution $\rho({\bf x})$. One writes:
\begin{equation}
\label{eq:B8}
    \rho({\bf k}) = e^{W({\bf k})} ,
\end{equation}
so that formally $W({\bf k})$ represents the cumulant generating function of $\rho({\bf x})$. The idea is then to take the Fourier series of the Maclaurin coefficients of this quantity:
\begin{equation}
\label{eq:B12}
    W({\bf k}) = \sum_{m=0}^\infty \sum_{n=-\infty}^{+\infty} W_{n,k} e^{-in\phi_k},
\end{equation}
and to characterize anisotropy through the coefficients $W_{n,m}$. As anticipated the advantage is that $W({\bf k})$ has good properties under translations of $\rho({\bf x})$. Suppose to perform a shift of the system:
\begin{equation}
\label{eq:B9}
\rho({\bf x}) \longrightarrow \rho({\bf x} + {\bf b}).
\end{equation}
The Fourier transform becomes:
\begin{equation}
\label{eq:B10}
\rho({\bf k}) \longrightarrow    \int_{\bf x} \rho({\bf x} + {\bf b}) e^{i{\bf k}\cdot {x}} = \int_{\bf x} \rho({\bf x}) e^{i{\bf k}\cdot({\bf x} - {\bf b}) } = \rho({\bf k})e^{-i{\bf k}\cdot{\bf b}}.
\end{equation}
After taking the logarithm, the transformation of $W({\bf k})$ thus reads:
\begin{equation}
\label{eq:B11}
    W({\bf k}) \longrightarrow W({\bf k}) - i k b \cos(\phi_k-\phi).
\end{equation}
This means in particular that the quantities $W_{n,m}$ with $m>1$ are invariant under translations of $\rho({\bf x})$, thus representing genuine short- or large-scale features of the density profile, depending on the value of $m$.

The coefficients $W_{n,m}$ can be obtained iteratively from the expression of $\rho_{n,m}$ derived in \equ{B7}. $\rho({\bf k})$ can be viewed as the moment generating function of $\rho({\bf x})$, and so it does satisfy the following identity:
\begin{equation}
\label{eq:B13}
    \rho({\bf k}) = \sum_{s=0}^\infty \frac{k^s}{s!} c_s, 
\end{equation}
where $c_s$, the moments of $\rho({\bf x})$, are defined by $ c_s= \partial^{(m)}_{\bf k} \rho({\bf k})|_{{\bf k}=0}$. Combining with \equ{B8}, one obtains:
\begin{align}
\label{eq:B14}
\nonumber   \rho({\bf k})  = e^{W(0)} \biggl [ &1 + k W'(0) + \frac{k^2}{2} \left ( W''(0) + W'(0)^2 \right ) \\ 
   &+\frac{k^3}{6} \left ( W'''(0) + W'(0)^3 + 3 W'(0)W''(0) \right ) + \ldots \biggr].
\end{align}
This can now be matched, order by order in $k$, to the formal expansion of $\rho({\bf k})$ in \equ{B4}, which gives:
\begin{align}
\label{eq:B15}
\nonumber    \rho_{0,0} = e^{W(0)},& \hspace{15pt}  \sum_n \frac{\rho_{n,1}}{\rho_{0,0}} e^{-in\phi_k} = W'(0), \hspace{15pt} \sum_n \frac{\rho_{n,2}}{\rho_{0,0}} e^{-in\phi_k} = \frac{1}{2} \left ( W''(0) +  W'(0)^2  \right ), \\
 &\sum_n \frac{\rho_{n,3}}{\rho_{0,0}} e^{-in\phi_k} = \frac{1}{6} \left ( W'''(0) +  W'(0)^3 + 3 W'(0)W''(0)  \right ),
\end{align}
and so on. From \equ{B12}, one further has:
\begin{equation}
\label{eq:B16}
    \nonumber W^{(l)}(0) = \sum_n W_{n,l} e^{-in\phi_k},
\end{equation}
which inserted into \equ{B15} yields:
\begin{align}
\label{eq:B17}
    \nonumber &\rho_{0,0} = e^{W(0)} , \hspace{30pt}   \sum_n \frac{\rho_{n,1}}{\rho_{0,0}} e^{-in\phi_k} =  \sum_n W_{n,1} e^{-in\phi_k}, \\
\nonumber &\sum_n \frac{\rho_{n,2}}{\rho_{0,0}} e^{-in\phi_k} = W_{2,0} + W_{1,-1}W_{1,1} \\
\nonumber &\hspace{40pt}+ e^{i2\phi_k} \left (  W_{2,2} + \frac{1}{2} W_{1,1}^2 \right ) + e^{-i2\phi_k} \left (  W_{-2,2} + \frac{1}{2} W_{-1,1}^2 \right )  , \\
 \nonumber &\sum_n  \frac{\rho_{n,3}}{\rho_{0,0}} e^{-in\phi_k} = e^{i\phi_k} \biggl[ W_{1,3} + W_{-1,1}W_{2,2} + W_{1,1}W_{0,2} + \frac{1}{2} W_{1,1}^2 W_{-1,1} \biggr] \\
 \nonumber &\hspace{20pt}+ e^{i3\phi_k} \biggl [ W_{3,3}+ W_{1,1}W_{2,2} + \frac{1}{6} W_{1,1}^3 \biggr] + e^{-i3\phi_k} \biggl [ W_{-3,3}+ W_{-1,1}W_{-2,2} + \frac{1}{6} W_{-1,1}^3 \biggr] \\
  &\hspace{20pt}+ e^{-i\phi_k} \biggl[ W_{-1,3} + W_{1,1}W_{-2,2} + W_{-1,1}W_{0,2} + \frac{1}{2} W_{-1,1}^2 W_{1,1} \biggr],
\end{align}
and so on. Matching each azimuthal harmonics at each order, substituting lower order solutions into higher order equations, substituting the expressions of $\rho_{n,m}$ from \equ{B8}, and with the notation:
\begin{equation}
    \label{eq:B18}
\bra \ldots \ket = \frac{\int_{{\bf x}} \rho({\bf x}) \ldots}{\rho_{0,0}}  = \frac{\int_{{\bf x}} \rho({\bf x}) \ldots }{ \int_{{\bf x}} \rho({\bf x}) },
\end{equation}
one arrives at the final expressions of the Fourier coefficients. The term with $m=0$, $\rho_{0,0}$, is the integral of the density, which physically corresponds to the total energy of the system.  For $m=1$:
\begin{equation}
    W_{1,1} = \frac{i}{2} \bra r e^{i\phi} \ket, \hspace{30pt}
    W_{-1,1} = \frac{i}{2} \bra r e^{i\phi} \ket.
\end{equation}
For $m=2$:
\begin{align}
\nonumber    W_{2,2} &= \frac{i^2}{8} \biggl[ \bra r^2 e^{i2\phi} \ket - \bra r e^{i\phi} \ket^2 \biggr ]  , \\
\nonumber    W_{0,2} &= \frac{i^2}{4} \biggl [ \bra r^2 \ket -  \bra r e^{-i\phi} \ket\bra r e^{i\phi} \ket \biggr] , \\
    W_{-2,2} &= \frac{i^2}{8} \biggl[ \bra r^2 e^{-i2\phi} \ket - \bra r e^{-i\phi} \ket^2 \biggr ].  
\end{align}
For $m=3$:
\begin{align}
\nonumber    W_{3,3} &= \frac{i^3}{48} \biggl[ \bra r^3 e^{i3\phi} \ket + \bra r e^{i\phi} \ket \left( 3\bra r^2 e^{i2\phi} \ket - 2 \bra r e^{i\phi} \ket^2  \right) \biggr]  , \\
\nonumber    W_{1,3} &= \frac{i^3}{16} \biggl [  \bra r^3 e^{i\phi} \ket  - \bra r^2 e^{i2\phi} \ket  \bra r e^{-i\phi}  \ket -2\bra r^2 \ket \bra r^{i\phi} \ket + 2\bra r e^{i\phi} \ket \bra re^{-i\phi} \ket  \biggr], \\
\nonumber    W_{-1,3} &= \frac{i^3}{16} \biggl [  \bra r^3 e^{-i\phi} \ket  - \bra r^2 e^{-i2\phi} \ket  \bra r e^{i\phi}  \ket -2\bra r^2 \ket \bra r^{-i\phi} \ket + 2\bra r e^{-i\phi} \ket \bra re^{i\phi} \ket  \biggr], \\
\nonumber    W_{3,3} &= \frac{i^3}{48} \biggl[ \bra r^3 e^{-i3\phi} \ket + \bra r e^{-i\phi} \ket \left( 3\bra r^2 e^{-i2\phi} \ket - 2 \bra r e^{-i\phi} \ket^2  \right) \biggr]  .
\end{align}
Two comments are in order. First, coefficients with negative $n$ are trivially related to those with positive $n$, and so they are redundant. Second, and more important, the coefficients with $m=1$ correspond essentially to the center of mass of the system. If one shifts the system in such a way that the center of mass vanishes, i.e.,
\begin{equation}
    \int_{\bf x} {\bf x} \rho({\bf x}) = 0,
\end{equation}
then the quantities $W_{n,1}$ vanish, and the previous expressions simplify a lot. Since the choice of the center of coordinates is arbitrary, one can always re-center the system before performing the cumulant expansion, and thus consider simplified expressions.

Eventually, one is left with:
\begin{equation}
    W_{0,2} = \bra r^2 \ket,
\end{equation}
which corresponds to the mean squared radius of the system, while the anisotropy is carried by the other quantities:
\begin{equation}
 W_{2,2} = \bra r^2 e^{i2\phi} \ket, \hspace{30pt}
  W_{1,3} = \bra r^3 e^{i\phi} \ket, \hspace{30pt}
    W_{3,3} = \bra r^3 e^{i3\phi} \ket.
\end{equation}
These correspond, respectively, to the quadrupole, the octupole, and the dipole asymmetry of the density profile. However, while these quantities possess the same translational and rotational properties as the Fourier harmonics $V_n$, the latter coefficients are dimensionless, as well as bounded from above. Both these missing conditions can be fulfilled by normalizing $W_{n,m}$ with $\bra r^m \ket$. This eventually leads to the final expressions:
\begin{equation}
\label{eq:B25}
    \mathcal{E}_{2,2} = \frac{\bra r^2 e^{i2\phi}\ket}{\bra r^2 \ket},
 \hspace{30pt}
    \mathcal{E}_{1,3} = \frac{\bra r^3 e^{i\phi} \ket}{\bra r^3 \ket}, \hspace{30pt}
    \mathcal{E}_{3,3} = \frac{\bra r^3 e^{i3\phi} \ket }{\bra r^3 \ket},
\end{equation}
which correspond to \equ{E2} and \equ{E1E3}. Note that these complex quantities are in magnitude lower than unity. 

In summary, the idea of the Teaney-Yan derivation is that the final anisotropies $V_n$ can be decomposed as:
\begin{equation}
\label{eq:A25}
    V_n = \sum_{m=n}^{\infty} \kappa_{n,m} \mathcal{E}_{n,m} ~~+ ~~{\rm higher~orders}, 
\end{equation}
where $n$ represents the rotational property of the harmonic, $m$ labels the wavelength of the considered modes in the density profile, $\kappa_{n,m}$ is a real coefficient that depends now on both $n$ and $m$, and ``higher-orders'' means all terms of the form $\mathcal{E}_{n-l,m'}\mathcal{E}_{n+l,m''}$, or even more complicated, that are allowed by rotational symmetry, but that are associated with higher-order modes of the system. As the gradients that drive the hydrodynamic expansion are mostly due to large-scale structures, one can truncate \equ{A25} at the lowest orders. This leads to the eccentricities of \equ{B25}, which are excellent predictors of the final $V_n$ coefficients in hydrodynamic simulations.

\chapter{Cumulants of flow fluctuations}

\label{app:B}

I derive the expressions of the cumulants of anisotropic flow, $V_n$, that are measured in experimental analyses. I derive expressions for $n=2$, although equivalent expressions can be derived for any value of $n$. 

One has first to choose a sample of events, typically, a class of events at a given multiplicity. In this sample there is a probability distribution for the flow vector:
\begin{equation}
    P(V_2) = P(v_x,v_y),
\end{equation}
where $(x,y)$ corresponds to an appropriate choice of the frame. A useful choice for $n=2$, and also the most standard, is that of considering $x$ as the direction of the impact parameter of the collision. The reason is that, as soon as the impact parameter, $b$, is sizable, elliptic flow goes preferably along the direction of $b$, and thus the probability distribution $P(v_x,v_y)$ has a nonzero average value along $x$, the so-called elliptic flow in the reaction plane.

The cumulant generating function of the distribution of $V_2$ is defined by:
\begin{equation}
    \ln \bigl  \bra e^{{\bf k} \cdot V_2} \bigr \ket .
\end{equation}
One can now write ${\bf k}\cdot V_2 = k v_2 \sin\theta $, where $k=|{\bf k}|$, $v_2=|V_2|$, and realize that the angle $\theta$, like the reaction plane angle, is random with a uniform distribution. One can thus average over this angle, and write:
\begin{equation}
\label{eq:average}
\ln \bigl \bra e^{{\bf k} \cdot V_2} \bigr \ket  = \ln \biggl \bra \int_0^{2\pi} \frac{d\theta}{2\pi} e^{kv_2\sin\theta} \biggr \ket = \ln \bigl \bra I_0(kv_2) \bigr \ket, 
\end{equation}
where $I_0$ is the modified Bessel function, which, as a consequence of the random averaging over orientations, is an even function. The cumulants of the distribution of $v_2$, whose standard notation is $v_2\{m\}^m$, are then defined by the formal equivalence:
\begin{equation}
\label{eq:cumull}
    \ln \bigl \bra I_0(kv_2) \bigr \ket = \sum_{m=0}^{\infty} c_m k^m v_2\{m\}^m,
\end{equation}
where $m\geq2$ is an even number, and the coefficients $c_m$ are the coefficients of the power series of the function $I_0$, i.e., $c_2=1/4$, $c_4=-1/64$, $c_6=1/576$, etc. . The left-hand side of \equ{cumull} can in fact be written as:
\begin{equation}
    \ln \bigl \bra I_0(kv_2) \bigr \ket = \ln \biggl ( 1 + \frac{\bra v_2^2 \ket k^2}{4} - \frac{\bra v_2^4 \ket k^4}{64}  + \frac{\bra v_2^6 \ket k^6}{576} + \ldots \biggr).
\end{equation}
Expanding $\ln(1+\ldots)$ in powers of $k$, and matching to the right-hand side of \equ{cumull}, one obtains:
\begin{align}
\label{eq:cumulants}
\nonumber    v_2\{2\}^2 &= \bra v_2 \ket, \\
\nonumber    v_2\{4\}^4 &= 2\bra v_2^2 \ket - \bra v_2^4 \ket, \\
    v_2\{6\}^6 &= \frac{1}{4} \bigl ( \bra v_2^6 \ket - 9 \bra v_2^4 \ket \bra v_2^2 \ket + 12 \bra v_2^2 \ket^3  \bigr),
\end{align}
and so on. I recall that angular brackets denote an average over events in a given centrality class. Historically, higher-order cumulants have been introduced because they are to a large extent insensitive to nonflow contribution (see e.g. Fig.~6 of Ref.~\cite{Adare:2018zkb}), and isolate the genuine multi-particle correlations observed in the final state~\cite{Borghini:2001vi}.  Expressions for the cumulants up to order 16 can be found in Ref.~\cite{Moravcova:2020wnf}. In a recent paper, Taghavi~\cite{Taghavi:2020toe} generalizes these expressions to include correlations between harmonics of different order. The idea is to calculate the cumulants of the joint probability distribution $P(V_2,V_3,\ldots)$. 

As realized in Ref.~\cite{Giacalone:2016eyu} for the first time, more insightful expressions can be obtained if one makes use of the full two-dimensionality of the distribution of $V_2$, i.e., without performing the average over $\theta$ in \equ{average}. I follow the derivations of Ref.~\cite{Bhalerao:2018anl}. The cumulants of the distribution of $V_2$ can in fact be defined by:
\begin{equation}
\label{eq:expa}
    \ln \bra e^{k_xv_x+k_yv_y} \ket = \sum_{n_x,n_y} \frac{k_x^{n_x}k_y^{n_y}}{n_x!n_y!} \kappa_{n_x,n_y},
\end{equation}
where now there is a double sequence of cumulants $\kappa_{n_x,n_y}$, which correspond to the cumulants of the joint probability distribution of $v_x$ and $v_y$.  Neglecting effects of parity violation, and considering that $x$ is the direction of impact parameter, the distribution of $V_2$ must be an even function of $v_y$, meaning that all terms where $n_y$ is an odd number vanish. Up to fourth order, the expansion of the right-hand side of \equ{expa} gives:
\begin{equation}
  \label{eq:generating}
  \ln \left\bra  e^{k_xv_x+k_yv_y}\right\ket =   k_x\kappa_{10}+  \frac{k_x^2}{2}\kappa_{20}+\frac{k_y^2}{2}\kappa_{02} + \frac{k_x^3}{6}\kappa_{30}+\frac{k_xk_y^2}{2}\kappa_{12}+ \frac{k_x^4}{24}\kappa_{40}+ \frac{k_y^4}{24}\kappa_{04}+
  \frac{k_x^2k_y^2}{4}\kappa_{22},
\end{equation}
where:
\begin{align}
\nonumber\kappa_{10}&=\langle v_x\rangle \equiv \mu, \hspace{46pt}
\kappa_{20}=\left\langle( v_x-\mu)^2\right\rangle,\\
\nonumber\kappa_{30}&=\left\langle( v_x-\mu)^3\right\rangle, \hspace{30pt}
\kappa_{40}=\left\langle( v_x-\mu)^4\right\rangle-3\kappa_{20}^2, 
\end{align}
and similarly for $\kappa_{0,n_y}$, with $n_y=2,4$, while for the mixed terms:
\begin{equation}
    \kappa_{12} = \bra v_x v_y^2 \ket, \hspace{30pt}
 \kappa_{22} = \bra (v_x-\mu)^2 v_y^2 \ket -  \kappa_{20}\kappa_{02}.  
\end{equation}

What one would like to do, then, is to find a way to match this expansion to the measured cumulants of $v_2$ in \equ{cumulants}, and perhaps use experimental data to gain insight about the detailed features of the two-dimensional distribution of $V_2$. The idea is to move to polar coordinates, and then average over $\theta$ inside the two-dimensional expansion. The procedure is the following. 
\begin{itemize}
    \item Substitute $k_x=k\cos\theta$ and $k_y=k\sin\theta$ in \equ{generating}.
    \item Exponentiate the resulting equation, and expand it in powers of $k$.
    \item Average over the value of $\theta$, as done in \equ{average}.
    \item Expand the logarithm of the resulting expression in powers of $k$.
    \item Match to the right-hand side of \equ{cumull}.
\end{itemize}
This yields for the first three cumulants:
\begin{align}
  \label{eq:exact}
\nonumber v_2\{2\}^2&=\mu^2+\kappa_{20}+\kappa_{02}, \\
\nonumber v_2\{4\}^4&=\mu^4 +2 \mu^2(\kappa_{02}-\kappa_{20}) -4 \mu(\kappa_{30}+\kappa_{12})- (\kappa_{20}-\kappa_{02})^2 - (\kappa_{04}+\kappa_{40}+2\kappa_{22}), \\
\nonumber v_2\{6\}^6 &= \mu^6 + 3\mu^4(\kappa_{02}-\kappa_{20}) -2\mu^3(2\kappa_{30}+3\kappa_{12}) +\frac{3}{2}\mu^2(\kappa_{40}-\kappa_{04}) \\
\nonumber&- 6\mu\kappa_{30}(\kappa_{02}-\kappa_{20}) +\frac{3}{2}(\kappa_{04}-\kappa_{40})(\kappa_{02}-\kappa_{20})
\nonumber+\frac{5}{2} \kappa_{30}^2+ 3\kappa_{30}\kappa_{12} +\frac{9}{2} \kappa_{12}^2 \\
&+\frac{3}{2}\mu(\kappa_{50}+\kappa_{14}+2\kappa_{32})+\frac{3}{4}(\kappa_{24}+\kappa_{42})+\frac{\kappa_{60}}{4}+\frac{\kappa_{06}}{4}.
\end{align}
These relations can not be inverted, however, they can help one understand in deeper detail the results shown in this manuscript. 

If the distribution of elliptic flow is a two-dimensional Gaussian~\cite{Voloshin:2007pc}:
\begin{equation}
    P(v_x,v_y) = \frac{1}{2 \pi \kappa_{20} \kappa_{02} } e^{- \frac{ (v_x - \mu)^2}{2\kappa_{20}} - \frac{vy}{2\kappa_{02}}},
\end{equation}
then \equ{exact} yields:
\begin{equation}
  v_2\{2\} = \sqrt{ \mu^2 + \kappa_{20} + \kappa_{02} }, \hspace{30pt}
 v_2\{4\} = v_2\{6\} = \ldots = \mu.
\end{equation}
Remarkably enough, this corresponds to the experimental observations. As soon as the centrality percentile is of order 5\%, the value of $\mu$ starts to dominate over the other terms, and experimental data shows precisely~\cite{Sirunyan:2017fts,Acharya:2018lmh,Aaboud:2019sma} that $v_2\{2\}>v_2\{4\} \simeq v_2\{6\} \simeq v_2\{8\}=\ldots$. The experimental fact that in non-central collisions the distribution of $V_2$ is close to a Gaussian explains in particular why, above 5\% centrality, the value of $v_2\{4\}$ in \auau{} collisions is the same as in \uuuu{} collisions, and also why the observable $v_2\{4\}$ loses any sensitivity to the value of the deformation parameter, as explicitly shown in \fig{5-11}.

For studies of nuclear deformation, hence, the limit of central collisions, where $\mu=0$, is more interesting. In this limit, one finds in particular:
\begin{equation}
    v_2\{2\} = \kappa_{20} + \kappa_{02} \equiv \sigma^2,
\end{equation}
which can be found in \equ{v22}, and is equal to the sum of the variances of the projections of the distribution of $V_2$. Elliptic flow is in this case generated solely by fluctuations. The presence of nuclear deformations increase both $\kappa_{20}$ and $\kappa_{02}$, thus showing why the rms elliptic flow is larger in central \uuuu{} collisions than in central \auau{} collisions. For $\mu=0$, the fourth-order cumulant becomes on the other hand:
\begin{equation}
    v_2\{4\} = - (\kappa_{04} + \kappa_{40} + 2\kappa_{22}) \equiv -K, 
\end{equation}
shown in \equ{v24kurt}. This is the sum of the coefficients of kurtosis, and implies in particular that, in this limit, the cumulant in central collisions originates as a non-Gaussian correction to the distribution of $V_2$. Nuclear deformation modifies the tails, and thus the kurtosis of the distribution of $V_2$, thus leading to sizable effects on the value of $v_2\{4\}$ for central collisions. As shown in \fig{4-9} and \fig{4-10}, it can in fact change the sign of this cumulant. 

As a final remark, if the initial anisotropy $\mathcal{E}_2$ satisfies $V_2 = \kappa_2 \mathcal{E}_2$, then all the previous derivations are equivalent for the cumulants of $\mathcal{E}_2$. The final quantities are simply rescaled by appropriate powers of the coefficient $\kappa_2$.

\chapter{Hydrodynamic study}

\label{app:C}

The results obtained in Chapter~\ref{chap:4} and Chapter~\ref{chap:5} are obtained by means of \trento{} simulations, which are then compared to experimental data on the basis of approximations, such as the scaling $v_n\propto\varepsilon_2$, or the fact that the collisions at large $E$ correspond to collisions at large $\bra p_t \ket$. These approximations are solid, however, it is relatively cheap to perform some explicit checks of their goodness by means of full hydrodynamic evaluations. I do so in this appendix, which follows closely Appendix~B of Ref.~\cite{Giacalone:2020awm}. 

I perform hydrodynamic simulations of central \uuuu{} collisions with the aim of checking the following points:
\begin{itemize}
    \item  The correlation between $\bra p_t \ket$ and $E$ is strong  in viscous hydrodynamics.
    \item Elliptic flow is indeed larger in body-body collisions than in tip-tip collisions.
    \item The response coefficients $\kappa_n = v_n / \varepsilon_n$ used in the phenomenological applications of this paper are consistent with full hydrodynamic results.
    \item Tip-tip collisions do in fact yield larger $\bra p_t \ket$ than body-body collisions.
    \item Whether or not the response coefficient $\kappa_n$ has a dependence on $\bra p_t \ket$.
\end{itemize}
To do this, I evolve hydrodynamically profiles of entropy density generated with the \trento{} model:
\begin{equation}
\label{eq:strent}
s({\bf x}, \tau_0) = \frac{N_0}{\tau_0} \sqrt[]{t_A({\bf x})t_B({\bf x})},
\end{equation}
where $t_{A,B}$ is defined by \equ{tA}. I fix the orientation of the colliding nuclei to impose body-body and tip-tip configurations. I let the impact parameter of these body-body and tip-tip configurations fluctuate, and I select events according to their total entropy, consistent with a realistic selection of ultracentral events. For both choices, I select and then evolve 60 initial conditions. I consider that the selected profiles correspond to the initial condition of hydrodynamics at proper time $\tau_0=0.2~\fmc$, while the overall multiplicative factor in \equ{strent} is $N_0 = 21.6$. The profiles thus obtained belong to the $0.78-0.96\%$ centrality class, which corresponds to the narrow interval of total initial entropy per unit rapidity $4023<S<4078$.

 I neglect the effects of the pre-equilibrium~\cite{Kurkela:2018wud,Schlichting:2019abc} phase of the system. I carry out a boost-invariant evolution of these initial conditions by means the \music{} hydrodynamic code~\cite{Schenke:2010nt,Schenke:2010rr,Paquet:2015lta}. I use the equation of state of lattice QCD~\cite{Borsanyi:2013bia}, and I implement a freeze-out temperature $T=0.15~{\rm GeV}$. The hydrodynamic expansion is viscous, and the viscous corrections are chosen such that they are pretty sizable, for instance, they increase the total entropy of the system during the hydrodynamic evolution by nearly 40\%. For the shear viscosity over entropy ratio, I implement a temperature-independent $\eta/s=0.16$. The implementation of the bulk viscosity over entropy ratio, $\zeta/s$, requires instead more thinking, because I am essentially the first to implement this quantity in simulations at RHIC energy with \trento{} initial conditions. The situation in the literature is at present a little paradoxical~\cite{Byres:2019xld}. Calculations that implement  \ipglasma{} initial conditions tend to yield too large values of $\bra p_t \ket$ at the end of the expansion, and solve this issues by implementing a bulk viscosity that has a peak around $\zeta/s\approx0.3$, at both RHIC and LHC energy~\cite{Ryu:2015vwa,McDonald:2016vlt,Ryu:2017qzn,Schenke:2019ruo}. Calculations that start with \trento{} initial conditions, available only for simulations of LHC collisions, produce on the other hand values of $\bra p_t \ket$ that agree with experimental data even in absence of a bulk viscosity~\cite{Bernhard:2016tnd}, hence, they implement a $\zeta/s$ that is smaller by almost one order of magnitude at the peak~\cite{Bernhard:2019bmu}. As I want to describe RHIC collisions with \trento{} initial conditions, I make up a sort of hybrid scenario. The bulk viscosity has the same temperature profile as in the \ipglasma{} papers, but the value of $\zeta/s$ at the peak is reduced by a factor 10. Finally, the corrections $\delta f_{\eta}$ and $\delta f_{\zeta}$ to the equilibrium distribution at freeze-out are chosen following Ref.~\cite{Ryu:2017qzn}. All hadronic resonances can be formed at freeze-out~\cite{Alba:2017hhe}, and I take into account their decay to stable hadrons. The outcome is a boost-invariant spectrum of charged hadrons, $\frac{dN}{d^2{\bf p}_t}$, which is used to calculate the charged-particle multiplicity, the average transverse momentum, and the flow coefficients, following \equ{multiplicity}, \equ{mpt}, and \equ{V2}.

\begin{figure}
    \centering
    \includegraphics[width=.6\linewidth]{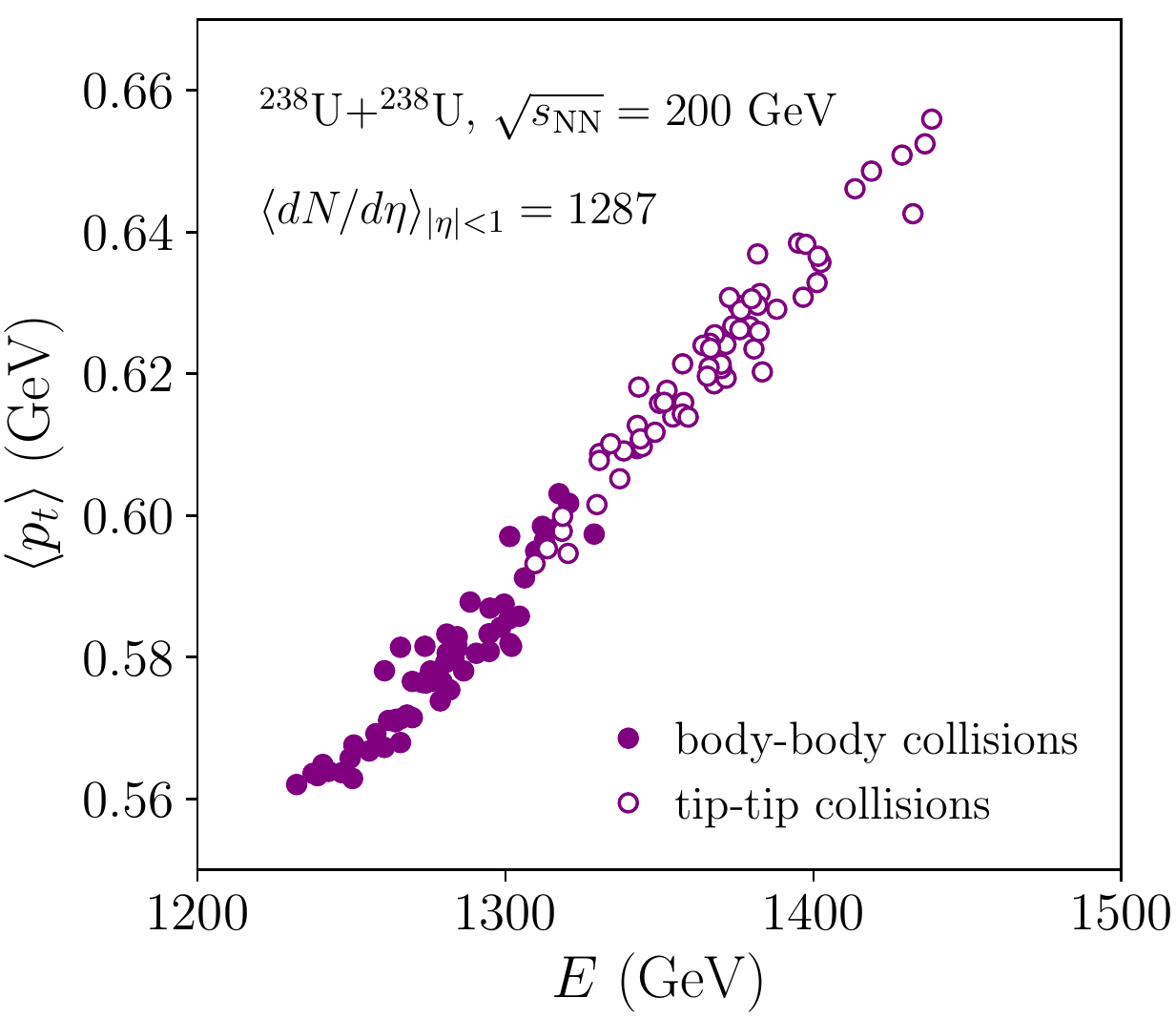}
    \caption{Correlation between the average transverse momentum, $\bra p_t \ket$ and the initial energy, $E$, in viscous hydrodynamic simulations of ultracentral \uuuu{} collisions at top RHIC energy. Full symbols: body-body collisions. Empty symbols: tip-tip collisions. These events present $\bra dN/d\eta \ket = 1287$ for $|\eta| < 1$, which corresponds approximately to the $1\%$ centrality cut in the STAR analysis. Figure from Ref.~\cite{Giacalone:2020awm}.}
    \label{fig:C-1}
\end{figure}

In \fig{C-1}, I show my results for $\bra p_t \ket$ as a function of $E$. One  notes that there is a strong correlation between these two quantities. I recall that the same plot is shown in \fig{3-5}, although for the ideal hydrodynamic expansion of events at fixed entropy and fixed impact parameter. The strong correlation observed in \fig{C-1} implies then that the inclusion of viscous corrections and some fluctuations in impact parameter do not disrupt the physical picture about the origin of $\bra p_t \ket$. The second remarkable result observed in \fig{C-1} is the fact that body-body and tip-tip collisions, while falling on the same line, cover distinct regions in $E$ and $\bra p_t \ket$. This shows that the tip-tip produce indeed larger $\bra p_t \ket$ than body-body events at the same entropy, thus confirming the idea that $\bra p_t \ket$ can be used to discern collision geometries.

\begin{figure}
    \centering
    \includegraphics[width=\linewidth]{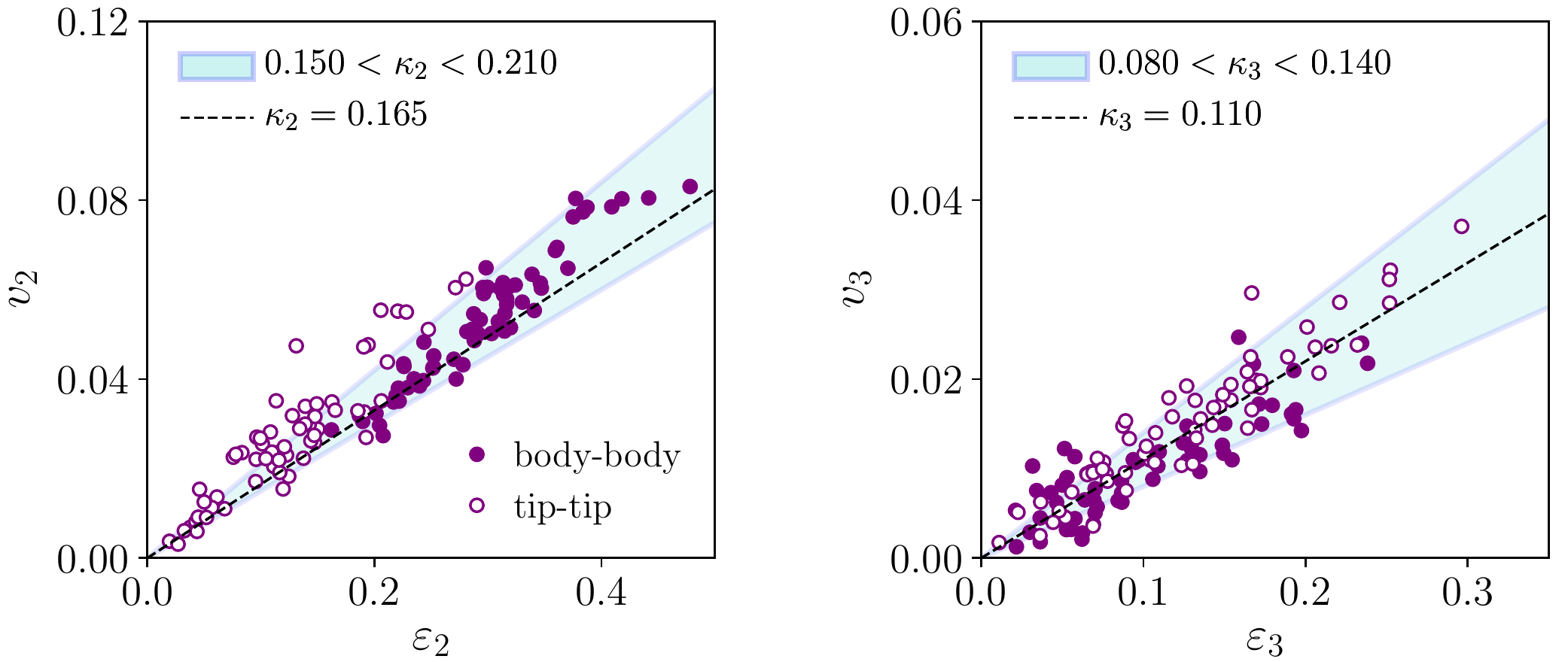}
    \caption{Flow harmonics $v_2$ (left panel) and $v_3$ (right panel) as a function of the corresponding spatial anisotropies, $\varepsilon_2$ and $\varepsilon_3$, in body-body and tip-tip \uuuu{} collisions. The $v_n$  coefficients are obtained with STAR kinematic cuts, $|\eta|<1$ and $0.2 < p_t < 2$~GeV. The shaded bands provide the viable range of values for $\kappa_n=v_n/\varepsilon_n$. The values of $\kappa_n$ chosen in this manuscript are shown with dashed line. Figure from Ref.~\cite{Giacalone:2020awm}.}
    \label{fig:C-2}
\end{figure}

I calculate then the flow coefficients, $v_n$. They are plotted as a function of the corresponding initial anisotropy, $\varepsilon_n$, in \fig{C-2}. 

The left panel contains results for $n=2$. First of all, I confirm the strong linear correlation between $v_2$ and $\varepsilon_2$. Second, as expected, one sees that body-body and tip-tip collisions are separated, and cover distinct regions in both $\varepsilon_2$ and $v_2$. The value of the response coefficient is essentially given by the slope of the scatter plot. The shaded band shows the range of viable values for $\kappa_2 = v_2 / \varepsilon_2$. The dashed line corresponds to the value used throughout this manuscript, $\kappa_2=0.165$, suggested by the \trento{} results of Ref.~\cite{Giacalone:2018apa}. One sees that this value is fairly reasonable, although a bit small, suggesting that the predictions that depend on $\kappa_2$ shown in this manuscript are underestimated by about $5\%$. Body-body and tip-tip collisions appear then to fall almost on the same curve, indicating that $\kappa_2$ has a very mild dependence on $\bra p_t \ket$. The right panel of the figure shows instead results for $n=3$. One notes both $\varepsilon_3$ and $v_3$ are slightly larger in the case of tip-tip collisions, as expected. Remarkably, the value $\kappa_3=0.110$ chosen in \fig{5-7}, and which yield an rms $v_3$ in good agreement with STAR preliminary data, is fully consistent with the correlation between $\varepsilon_3$ and $v_3$ shown in \fig{C-2}

The fact that the coefficients $\kappa_2$ and $\kappa_3$ chosen for \uuuu{} collisions are reasonable implies that the choices made for \auau{} collisions, $\kappa_2=0.155$ and $\kappa_3=0.100$, are also reasonable. The $\kappa_n$ coefficients are damped by viscous corrections, and should be indeed a little smaller in \auau{} systems, due to the smaller system size. However, a difference of order 5\% or 10\% between \auau{} and \uuuu{} sounds a little large, hence, this estimate could probably be improved if actual hydrodynamic simulations of ultracentral \auau{} events were available.

\chapter{Realistic estimate of $\langle p_t \rangle$ fluctuations}

\label{app:D}

Evaluating an observable as a function of $\bra p_t \ket$, as done for \fig{5-5}, requires the knowledge of the mean transverse momentum in each event. As discussed in Sec.~\ref{sec:3-4}, while this is not an issue in a hydrodynamic simulation, where the output of an event is a continuous spectrum in momentum space, the situation is problematic in an experiment, where the integration in \equ{mpt} corresponds to a discrete average:
\begin{equation}
    \bra p_t \ket = \frac{1}{N}\sum_{i=1}^{N} p_{t,i},
\end{equation}
where $N$ is the multiplicity of the event, and $p_{t,i}$ is the transverse momentum of particle $i$. The multiplicity is of order $1000$ in central \uuuu{} collisions, implying that the determination of $\bra p_t \ket$ is affected by a relative statistical fluctuation, proportional to $1/\sqrt{N}$, which is as large as the relative dynamical fluctuation of $\bra p_t \ket$, which is only about $1.2\%$ following the STAR measurement~\cite{Adam:2019rsf}. To include the presence of statistical fluctuations in the \trento{} calculation, one has to add an artificial decorrelation between the value of $\bra p_t \ket$ and that of $E/S$, which is the initial-state predictor of the average transverse momentum considered in this work. \fig{5-5} is obtained by assuming that $E/S$ and $\bra p_t \ket$ are in a one-to-one correspondence. The idea is to disrupt this correspondence in a realistic way to mimic the trivial finite-$N$ effect.

Within the flow paradigm explained in Sec.~\ref{sec:3-4}, particles are emitted independently from the decoupling surface at the end of the hydrodynamic phase. Each particle has a random value of $p_t$, chosen, independently for each particle, from an underlying probability distribution, i.e., the $p_t$ spectrum, which is a measurable observable at a given collision centrality. The magnitude of the trivial statistical fluctuations can thus be evaluated from law of large numbers:
\begin{equation}
\label{eq:sigmastats}
    \sigma_{\rm stat}=\frac{1}{\sqrt{N}}\sqrt{\bra p_t^2 \ket - \bra p_t \ket^2},
\end{equation}
where $\bra \ldots \ket = \frac{1}{N} \left [ \int \ldots \frac{dN}{d^2{\bf p}_t} \right ]$. At top RHIC energy in the full acceptance of the STAR detector, $|\eta|<1$, one detects $N \approx 1000$ particles. Evaluation of the moments of the $p_t$ distribution from the spectrum measured in central \auau{} collisions~\cite{Back:2003qr} then yields:
\begin{equation}
    \sigma_{\rm stat} = 0.01~{\rm GeV}.
\end{equation}
Normalizing by the average value of $\bra p_t\ket$, one obtains the relative statistical fluctuation:
\begin{equation}
\label{eq:statrel}
    \sigma_{\rm stat}/\bbra p_t \kket = 0.18,
 \end{equation}
where $\bbra p_t \kket=0.57$~GeV~\cite{Adams:2005ka}. This corresponds to the the relative fluctuation of the average transverse momentum originating from the simple fact that $N$ is finite.

Now, comparing \equ{statrel} with \equ{sigmadyn}, one finds:
\begin{equation}
\label{eq:dynstat}
    \frac{\sigma_{\rm stat} (\bra p_t \ket)}{\bbra p_t \kket} = 1.5 \times \frac{\sigma_{\rm dynamical} (\bra p_t \ket)}{\bbra p_t \kket},
\end{equation}
showing that the relative statistical fluctuation is larger than the relative dynamical fluctuation, and that as a consequence any theoretical estimate of observables the require the evaluation of $\bra p_t \ket$ on an event-by-event basis must include this statistical smearing before they can be compared to experimental data.

It is however simple to include the effect of these trivial fluctuations in the theoretical calculation. The flow paradigm helps out. As particles are emitted independently from the freezeout hypersurface, the number of emitted particles in a sample of events follows a Poisson distribution, or more simply a Gaussian distribution, since $N \gg 1$. I can thus readily correct the results of the \trento{} calculations for statistical fluctuations. First, from \equ{pred2} I compute the distribution of $\bra p_t \ket/\bbra p_t \kket$, by choosing the coefficient $\kappa_0$ that allows me to reproduce the magnitude of the dynamical fluctuations given in \equ{sigmadyn}, which I dub $\sigma$. Each entry of the distribution is then multiplied by a number sampled from a Gaussian distribution of unit mean and standard deviation equal to $1.5 \times \sigma$, in agreement with \equ{dynstat}. The resulting fictitious distribution of $\bra p_t \ket/\bbra p_t \kket$ thus properly includes a decorrelation between $E/S$ and $\bra p_t \ket$ due the finite number of particles. Observables can now be computed as functions of this corrected relative variation of $\bra p_t \ket$, and can be genuinely compared to experimental data. This is how the quantitative predictions shown in Sec.~\ref{sec:5-21} and in \fig{5-10} are obtained.

\end{appendices}

\newpage

\thispagestyle{plain}
\thispagestyle{empty}
\ifthispageodd{\newpage\thispagestyle{empty}\null\newpage}{}
\thispagestyle{empty}
\newgeometry{top=.4cm, bottom=1.1cm, left=1.7cm, right=1.7cm}
\fontsize{12}{14}\selectfont
\fontfamily{rm}\selectfont

\lhead{}
\rhead{}
\rfoot{}
\cfoot{}
\lfoot{}

\noindent 
\includegraphics[height=2.0cm]{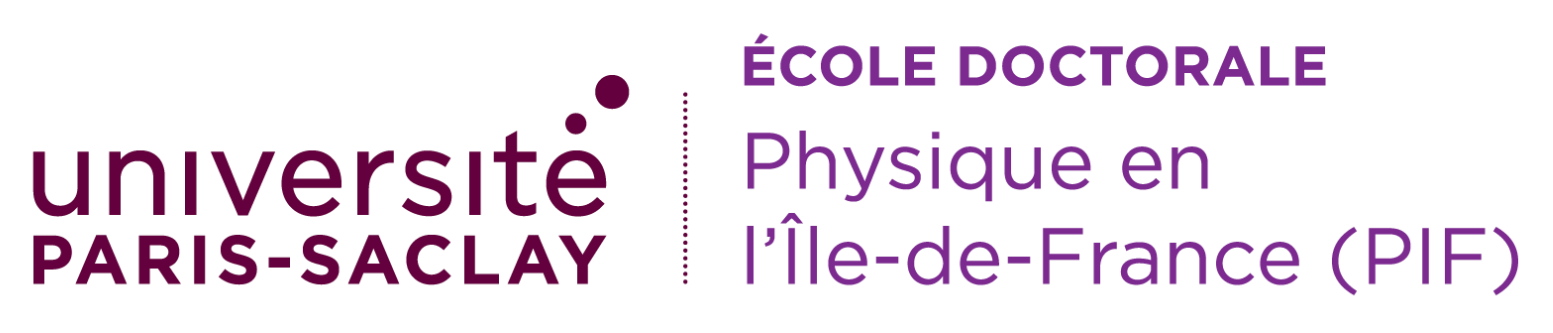}
\vspace{.25cm}

\begin{mdframed}[linecolor=Prune,linewidth=1]
\vspace{-.25cm}
\paragraph*{Titre:} {\normalsize Question de forme: observer la d\'eformation des noyaux atomiques aux collisionneurs des hautes \'energies}

\begin{small}
\vspace{-.25cm}
\paragraph*{Mots clés:} ions lourds, quark-gluon plasma, flot elliptique, déformation nucl\'eaire

\vspace{-.5cm}
\begin{multicols}{2}
\paragraph*{Résumé:} 
Les exp\'eriences conduites au collisionneurs de particules BNL RHIC et CERN LHC montrent que l'\'emission azimutale de hadrons vers l'\'etat final des collisions relativistes noyau-noyau est fortemente anisotrope. Cette observation est compatible avec un paradigme hydrodynamique, selon lequel les hadrons observ\'es dans l'\'etat final sont \'emis \`a l'issue de l'expansion d'un milieu fluidiforme cr\'e\'e dans la r\'egion d'interaction. Ce paradigme pr\'edit notamment que l'anisotropie de l'\'emission des particules est sensible \`a la d\'eformation de l'\'etat fondamental des noyaux interagissants. \`A travers des comparaisons de haute pr\'ecision entre les donn\'ees des exp\'eriences et le mod\`ele hydrodynamique, j'\'etudie les manifestations ph\'enomenologiques de la d\'eformation des noyaux atomiques dans les collisions \auau{}, \uuuu{}, et \xexe{}. Cette analyse d\'emontre qu'une compr\'ehension approfondie de la structure des ions interagissants est n\'ecessaire pour l'interpr\'etation des donn\'ees aux hautes \'energies. Les donn\'ees du RHIC confirment que la g\'eometrie du noyau $^{238}$U est bien celle d'un ellipso\"ide, tandis que le noyau $^{197}$Au appara\^it \^etre presque sph\'erique, ce qui est en d\'esaccord avec les pr\'edictions des mod\`eles nucl\' eaires empiriques et de champ moyen. Le donn\'ees du LHC indiquent ensuite la pr\'esence de d\'eformation quadrupolaire dans l'\'etat fondamental du $^{129}$Xe, ce qui pourrait indiquer la premi\`ere observation d'effets de coexistence de forme en physique nucl\'eaire des hautes \'energies. J'introduis une m\'ethode pour isoler les configurations de collision o\`u l'orientation des noyaux d\'eformes brise la symmetrie azimutale du syst\`eme d'une fa\c con maximale. Cela me permet de d\'efinir une nouvelle cat\'egorie d'observables sensibles \`a la d\'eformation des noyaux qu'on utilise, en ouvrent ainsi le chemin vers des \'etudes quantitatives de la structure des noyaux atomiques en physique des hautes \'energies.

\end{multicols}
\end{small}
\end{mdframed}

\begin{mdframed}[linecolor=Prune,linewidth=1]
\vspace{-.25cm}
\paragraph*{Title:} A matter of shape: seeing the deformation of atomic nuclei at high-energy colliders

\begin{small}
\vspace{-.25cm}
\paragraph*{Keywords:} heavy-ion physics, quark-gluon plasma, elliptic flow, nuclear deformation

\vspace{-.5cm}
\begin{multicols}{2}
\paragraph*{Abstract:} Collider experiments conducted at the BNL RHIC and at the CERN LHC show that the the emission of particles following the interaction of two nuclei at relativistic energy is highly anisotropic in azimuthal angle. This observation is compatible with a hydrodynamic paradigm, according to which the final-state hadrons are emitted following the expansion of a fluidlike system created in the interaction region. Within this paradigm, anisotropy in the emission of particles is enhanced whenever the colliding nuclei have deformed ground states. By means of high-quality comparisons between the predictions of hydrodynamic models and particle collider data, I study the phenomenological manifestations of the quadrupole deformation of atomic nuclei in relativistic \auau{}, \uuuu{}, and \xexe{} collisions. This analysis demonstrates that a deep understanding of the structure of the colliding ions is required for the interpretation of data in high-energy experiments. RHIC data confirms in particular the well-known fact that the geometry of $^{238}$U nuclei is that of a well-deformed ellipsoid, while indicating that $^{197}$Au nuclei are nearly spherical, a result which is at odds with the estimates of mean-field and empirical nuclear models. LHC data brings instead evidence of quadrupole deformation in the ground state of $^{129}$Xe nuclei, ascribable to the first visible manifestation of shape coexistence phenomena in high-energy nuclear experiments. I introduce a simple method to isolate collision configurations that maximally break azimuthal symmetry due to the orientation of the deformed nuclei. This allows me to define observables with an unprecedented sensitivity to the deformation of the colliding species, thus paving the way for quantitative studies of nuclear structure at high energy.
\end{multicols}
\end{small}
\end{mdframed}

\vspace{.4cm} 

\fontfamily{fvs}\fontseries{m}\selectfont
\noindent \begin{tabular}{p{14cm}r}
\multirow{3}{16cm}[+0mm]{{\color{Prune} {\small \textbf{Université Paris-Saclay}\\
Espace Technologique / Immeuble Discovery\\
Route de l’Orme aux Merisiers RD 128 / 91190 Saint-Aubin, France}}} & \multirow{3}{2.1cm}[+2mm]{\includegraphics[height=1.5cm]{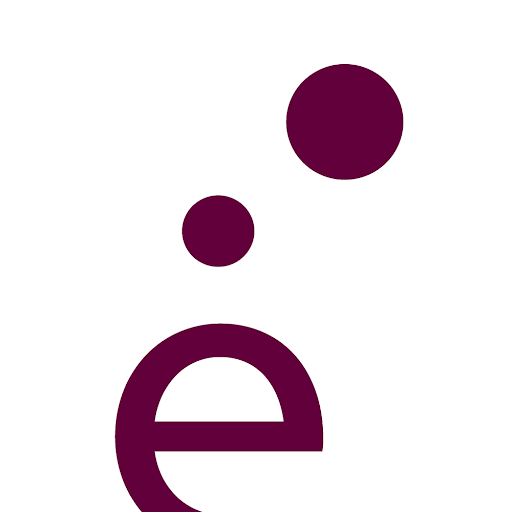}}\\
\end{tabular}

\end{document}